\documentstyle[12pt]{book}
\newcommand{\sh}{\sinh}
\def\der{{\buildrel\,\leftarrow\over D}\!\!}
\newcommand{\bea}{\begin{eqnarray}}
\newcommand{\eea}{\end{eqnarray}}
\renewcommand{\a}{\alpha}
\newcommand{\ad}{\dot{\alpha}}
\renewcommand{\b}{\beta}
\newcommand{\bd}{\dot{\beta}}
\newcommand{\q}{\theta}
\newcommand{\pa}{\partial}
\renewcommand{\l}{\lambda}
\newcommand{\g}{\gamma}
\newcommand{\f}{\phi}
\newcommand{\F}{\Phi}
\newcommand{\s}{\sigma}
\newcommand{\bD}{\bar{D}}
\newcommand{\tr}{{\rm tr}}
\newcommand{\Str}{{\rm Str}}
\newcommand{\D}{{\cal D}}
\renewcommand{\L}{\Lambda}
\newcommand{\hf}{\frac{1}{2}}
\newcommand{\e}{\epsilon}
\renewcommand{\O}{\Omega}

\begin{document}
\immediate\write16{<<WARNING: LINEDRAW macros work with emTeX-dvivers
                    and other drivers supporting emTeX \special's
                    (dviscr, dvihplj, dvidot, dvips, dviwin, etc.) >>}

%% Macros for drawing Feynman graphs and other complex diagrams
%% Designed by A.V.Voronin (1993); modified in 1995
%% Steklov Math. Inst., e-mail: av@voronin.mian.su
%%
\newdimen\Lengthunit       \Lengthunit  = 1.5cm
\newcount\Nhalfperiods     \Nhalfperiods= 9
\newcount\magnitude        \magnitude = 1000

\catcode`\*=11
\newdimen\L*   \newdimen\d*   \newdimen\d**
\newdimen\dm*  \newdimen\dd*  \newdimen\dt*
\newdimen\a*   \newdimen\b*   \newdimen\c*
\newdimen\a**  \newdimen\b**
\newdimen\xL*  \newdimen\yL*
\newdimen\rx*  \newdimen\ry*
\newdimen\tmp* \newdimen\linwid*

\newcount\k*   \newcount\l*   \newcount\m*
\newcount\k**  \newcount\l**  \newcount\m**
\newcount\n*   \newcount\dn*  \newcount\r*
\newcount\N*   \newcount\*one \newcount\*two  \*one=1 \*two=2
\newcount\*ths \*ths=1000
\newcount\angle*  \newcount\q*  \newcount\q**
\newcount\angle** \angle**=0
\newcount\sc*     \sc*=0

\newtoks\cos*  \cos*={1}
\newtoks\sin*  \sin*={0}

\catcode`\[=13

\def\rotate(#1){\advance\angle**#1\angle*=\angle**
\q**=\angle*\ifnum\q**<0\q**=-\q**\fi
\ifnum\q**>360\q*=\angle*\divide\q*360\multiply\q*360\advance\angle*-\q*\fi
\ifnum\angle*<0\advance\angle*360\fi\q**=\angle*\divide\q**90\q**=\q**
\def\sgcos*{+}\def\sgsin*{+}\relax
\ifcase\q**\or
 \def\sgcos*{-}\def\sgsin*{+}\or
 \def\sgcos*{-}\def\sgsin*{-}\or
 \def\sgcos*{+}\def\sgsin*{-}\else\fi
\q*=\q**
\multiply\q*90\advance\angle*-\q*
\ifnum\angle*>45\sc*=1\angle*=-\angle*\advance\angle*90\else\sc*=0\fi
\def[##1,##2]{\ifnum\sc*=0\relax
\edef\cs*{\sgcos*.##1}\edef\sn*{\sgsin*.##2}\ifcase\q**\or
 \edef\cs*{\sgcos*.##2}\edef\sn*{\sgsin*.##1}\or
 \edef\cs*{\sgcos*.##1}\edef\sn*{\sgsin*.##2}\or
 \edef\cs*{\sgcos*.##2}\edef\sn*{\sgsin*.##1}\else\fi\else
\edef\cs*{\sgcos*.##2}\edef\sn*{\sgsin*.##1}\ifcase\q**\or
 \edef\cs*{\sgcos*.##1}\edef\sn*{\sgsin*.##2}\or
 \edef\cs*{\sgcos*.##2}\edef\sn*{\sgsin*.##1}\or
 \edef\cs*{\sgcos*.##1}\edef\sn*{\sgsin*.##2}\else\fi\fi
\cos*={\cs*}\sin*={\sn*}\global\edef\gcos*{\cs*}\global\edef\gsin*{\sn*}}\relax
\ifcase\angle*[9999,0]\or
[999,017]\or[999,034]\or[998,052]\or[997,069]\or[996,087]\or
[994,104]\or[992,121]\or[990,139]\or[987,156]\or[984,173]\or
[981,190]\or[978,207]\or[974,224]\or[970,241]\or[965,258]\or
[961,275]\or[956,292]\or[951,309]\or[945,325]\or[939,342]\or
[933,358]\or[927,374]\or[920,390]\or[913,406]\or[906,422]\or
[898,438]\or[891,453]\or[882,469]\or[874,484]\or[866,499]\or
[857,515]\or[848,529]\or[838,544]\or[829,559]\or[819,573]\or
[809,587]\or[798,601]\or[788,615]\or[777,629]\or[766,642]\or
[754,656]\or[743,669]\or[731,681]\or[719,694]\or[707,707]\or
\else[9999,0]\fi}

\catcode`\[=12

\def\GRAPH(hsize=#1)#2{\hbox to #1\Lengthunit{#2\hss}}

\def\Linewidth#1{\global\linwid*=#1\relax
\global\divide\linwid*10\global\multiply\linwid*\mag
\global\divide\linwid*100\special{em:linewidth \the\linwid*}}

\Linewidth{.4pt}
\def\sm*{\special{em:moveto}}
\def\sl*{\special{em:lineto}}
\let\moveto=\sm*
\let\lineto=\sl*
\newbox\spm*   \newbox\spl*
\setbox\spm*\hbox{\sm*}
\setbox\spl*\hbox{\sl*}

\def\mov#1(#2,#3)#4{\rlap{\L*=#1\Lengthunit
\xL*=#2\L* \yL*=#3\L*
\xL*=\xscale\xL* \yL*=\yscale\yL*
\rx* \the\cos*\xL* \tmp* \the\sin*\yL* \advance\rx*-\tmp*
\ry* \the\cos*\yL* \tmp* \the\sin*\xL* \advance\ry*\tmp*
\kern\rx*\raise\ry*\hbox{#4}}}

\def\rmov*(#1,#2)#3{\rlap{\xL*=#1\yL*=#2\relax
\rx* \the\cos*\xL* \tmp* \the\sin*\yL* \advance\rx*-\tmp*
\ry* \the\cos*\yL* \tmp* \the\sin*\xL* \advance\ry*\tmp*
\kern\rx*\raise\ry*\hbox{#3}}}

\def\lin#1(#2,#3){\rlap{\sm*\mov#1(#2,#3){\sl*}}}

\def\arr*(#1,#2,#3){\rmov*(#1\dd*,#1\dt*){\sm*
\rmov*(#2\dd*,#2\dt*){\rmov*(#3\dt*,-#3\dd*){\sl*}}\sm*
\rmov*(#2\dd*,#2\dt*){\rmov*(-#3\dt*,#3\dd*){\sl*}}}}

\def\arrow#1(#2,#3){\rlap{\lin#1(#2,#3)\mov#1(#2,#3){\relax
\d**=-.012\Lengthunit\dd*=#2\d**\dt*=#3\d**
\arr*(1,10,4)\arr*(3,8,4)\arr*(4.8,4.2,3)}}}

\def\arrlin#1(#2,#3){\rlap{\L*=#1\Lengthunit\L*=.5\L*
\lin#1(#2,#3)\rmov*(#2\L*,#3\L*){\arrow.1(#2,#3)}}}

\def\dasharrow#1(#2,#3){\rlap{{\Lengthunit=0.9\Lengthunit
\dashlin#1(#2,#3)\mov#1(#2,#3){\sm*}}\mov#1(#2,#3){\sl*
\d**=-.012\Lengthunit\dd*=#2\d**\dt*=#3\d**
\arr*(1,10,4)\arr*(3,8,4)\arr*(4.8,4.2,3)}}}

\def\clap#1{\hbox to 0pt{\hss #1\hss}}

\def\ind(#1,#2)#3{\rlap{\L*=.1\Lengthunit
\xL*=#1\L* \yL*=#2\L*
\rx* \the\cos*\xL* \tmp* \the\sin*\yL* \advance\rx*-\tmp*
\ry* \the\cos*\yL* \tmp* \the\sin*\xL* \advance\ry*\tmp*
\kern\rx*\raise\ry*\hbox{\lower2pt\clap{$#3$}}}}

\def\sh*(#1,#2)#3{\rlap{\dm*=\the\n*\d**
\xL*=\xscale\dm* \yL*=\yscale\dm* \xL*=#1\xL* \yL*=#2\yL*
\rx* \the\cos*\xL* \tmp* \the\sin*\yL* \advance\rx*-\tmp*
\ry* \the\cos*\yL* \tmp* \the\sin*\xL* \advance\ry*\tmp*
\kern\rx*\raise\ry*\hbox{#3}}}

\def\calcnum*#1(#2,#3){\a*=1000sp\b*=1000sp\a*=#2\a*\b*=#3\b*
\ifdim\a*<0pt\a*-\a*\fi\ifdim\b*<0pt\b*-\b*\fi
\ifdim\a*>\b*\c*=.96\a*\advance\c*.4\b*
\else\c*=.96\b*\advance\c*.4\a*\fi
\k*\a*\multiply\k*\k*\l*\b*\multiply\l*\l*
\m*\k*\advance\m*\l*\n*\c*\r*\n*\multiply\n*\n*
\dn*\m*\advance\dn*-\n*\divide\dn*2\divide\dn*\r*
\advance\r*\dn*
\c*=\the\Nhalfperiods5sp\c*=#1\c*\ifdim\c*<0pt\c*-\c*\fi
\multiply\c*\r*\N*\c*\divide\N*10000}

\def\dashlin#1(#2,#3){\rlap{\calcnum*#1(#2,#3)\relax
\d**=#1\Lengthunit\ifdim\d**<0pt\d**-\d**\fi
\divide\N*2\multiply\N*2\advance\N*\*one
\divide\d**\N*\sm*\n*\*one\sh*(#2,#3){\sl*}\loop
\advance\n*\*one\sh*(#2,#3){\sm*}\advance\n*\*one
\sh*(#2,#3){\sl*}\ifnum\n*<\N*\repeat}}

\def\dashdotlin#1(#2,#3){\rlap{\calcnum*#1(#2,#3)\relax
\d**=#1\Lengthunit\ifdim\d**<0pt\d**-\d**\fi
\divide\N*2\multiply\N*2\advance\N*1\multiply\N*2\relax
\divide\d**\N*\sm*\n*\*two\sh*(#2,#3){\sl*}\loop
\advance\n*\*one\sh*(#2,#3){\kern-1.48pt\lower.5pt\hbox{\rm.}}\relax
\advance\n*\*one\sh*(#2,#3){\sm*}\advance\n*\*two
\sh*(#2,#3){\sl*}\ifnum\n*<\N*\repeat}}

\def\shl*(#1,#2)#3{\kern#1#3\lower#2#3\hbox{\unhcopy\spl*}}

\def\trianglin#1(#2,#3){\rlap{\toks0={#2}\toks1={#3}\calcnum*#1(#2,#3)\relax
\dd*=.57\Lengthunit\dd*=#1\dd*\divide\dd*\N*
\divide\dd*\*ths \multiply\dd*\magnitude
\d**=#1\Lengthunit\ifdim\d**<0pt\d**-\d**\fi
\multiply\N*2\divide\d**\N*\sm*\n*\*one\loop
\shl**{\dd*}\dd*-\dd*\advance\n*2\relax
\ifnum\n*<\N*\repeat\n*\N*\shl**{0pt}}}

\def\wavelin#1(#2,#3){\rlap{\toks0={#2}\toks1={#3}\calcnum*#1(#2,#3)\relax
\dd*=.23\Lengthunit\dd*=#1\dd*\divide\dd*\N*
\divide\dd*\*ths \multiply\dd*\magnitude
\d**=#1\Lengthunit\ifdim\d**<0pt\d**-\d**\fi
\multiply\N*4\divide\d**\N*\sm*\n*\*one\loop
\shl**{\dd*}\dt*=1.3\dd*\advance\n*\*one
\shl**{\dt*}\advance\n*\*one
\shl**{\dd*}\advance\n*\*two
\dd*-\dd*\ifnum\n*<\N*\repeat\n*\N*\shl**{0pt}}}

\def\w*lin(#1,#2){\rlap{\toks0={#1}\toks1={#2}\d**=\Lengthunit\dd*=-.12\d**
\divide\dd*\*ths \multiply\dd*\magnitude
\N*8\divide\d**\N*\sm*\n*\*one\loop
\shl**{\dd*}\dt*=1.3\dd*\advance\n*\*one
\shl**{\dt*}\advance\n*\*one
\shl**{\dd*}\advance\n*\*one
\shl**{0pt}\dd*-\dd*\advance\n*1\ifnum\n*<\N*\repeat}}

\def\l*arc(#1,#2)[#3][#4]{\rlap{\toks0={#1}\toks1={#2}\d**=\Lengthunit
\dd*=#3.037\d**\dd*=#4\dd*\dt*=#3.049\d**\dt*=#4\dt*\ifdim\d**>10mm\relax
\d**=.25\d**\n*\*one\shl**{-\dd*}\n*\*two\shl**{-\dt*}\n*3\relax
\shl**{-\dd*}\n*4\relax\shl**{0pt}\else
\ifdim\d**>5mm\d**=.5\d**\n*\*one\shl**{-\dt*}\n*\*two
\shl**{0pt}\else\n*\*one\shl**{0pt}\fi\fi}}

\def\d*arc(#1,#2)[#3][#4]{\rlap{\toks0={#1}\toks1={#2}\d**=\Lengthunit
\dd*=#3.037\d**\dd*=#4\dd*\d**=.25\d**\sm*\n*\*one\shl**{-\dd*}\relax
\n*3\relax\sh*(#1,#2){\xL*=\xscale\dd*\yL*=\yscale\dd*
\kern#2\xL*\lower#1\yL*\hbox{\sm*}}\n*4\relax\shl**{0pt}}}

\def\shl**#1{\c*=\the\n*\d**\d*=#1\relax
\a*=\the\toks0\c*\b*=\the\toks1\d*\advance\a*-\b*
\b*=\the\toks1\c*\d*=\the\toks0\d*\advance\b*\d*
\a*=\xscale\a*\b*=\yscale\b*
\rx* \the\cos*\a* \tmp* \the\sin*\b* \advance\rx*-\tmp*
\ry* \the\cos*\b* \tmp* \the\sin*\a* \advance\ry*\tmp*
\raise\ry*\rlap{\kern\rx*\unhcopy\spl*}}

\def\wlin*#1(#2,#3)[#4]{\rlap{\toks0={#2}\toks1={#3}\relax
\c*=#1\l*\c*\c*=.01\Lengthunit\m*\c*\divide\l*\m*
\c*=\the\Nhalfperiods5sp\multiply\c*\l*\N*\c*\divide\N*\*ths
\divide\N*2\multiply\N*2\advance\N*\*one
\dd*=.002\Lengthunit\dd*=#4\dd*\multiply\dd*\l*\divide\dd*\N*
\divide\dd*\*ths \multiply\dd*\magnitude
\d**=#1\multiply\N*4\divide\d**\N*\sm*\n*\*one\loop
\shl**{\dd*}\dt*=1.3\dd*\advance\n*\*one
\shl**{\dt*}\advance\n*\*one
\shl**{\dd*}\advance\n*\*two
\dd*-\dd*\ifnum\n*<\N*\repeat\n*\N*\shl**{0pt}}}

\def\wavebox#1{\setbox0\hbox{#1}\relax
\a*=\wd0\advance\a*14pt\b*=\ht0\advance\b*\dp0\advance\b*14pt\relax
\hbox{\kern9pt\relax
\rmov*(0pt,\ht0){\rmov*(-7pt,7pt){\wlin*\a*(1,0)[+]\wlin*\b*(0,-1)[-]}}\relax
\rmov*(\wd0,-\dp0){\rmov*(7pt,-7pt){\wlin*\a*(-1,0)[+]\wlin*\b*(0,1)[-]}}\relax
\box0\kern9pt}}

\def\rectangle#1(#2,#3){\relax
\lin#1(#2,0)\lin#1(0,#3)\mov#1(0,#3){\lin#1(#2,0)}\mov#1(#2,0){\lin#1(0,#3)}}

\def\dashrectangle#1(#2,#3){\dashlin#1(#2,0)\dashlin#1(0,#3)\relax
\mov#1(0,#3){\dashlin#1(#2,0)}\mov#1(#2,0){\dashlin#1(0,#3)}}

\def\waverectangle#1(#2,#3){\L*=#1\Lengthunit\a*=#2\L*\b*=#3\L*
\ifdim\a*<0pt\a*-\a*\def\x*{-1}\else\def\x*{1}\fi
\ifdim\b*<0pt\b*-\b*\def\y*{-1}\else\def\y*{1}\fi
\wlin*\a*(\x*,0)[-]\wlin*\b*(0,\y*)[+]\relax
\mov#1(0,#3){\wlin*\a*(\x*,0)[+]}\mov#1(#2,0){\wlin*\b*(0,\y*)[-]}}

\def\calcparab*{\ifnum\n*>\m*\k*\N*\advance\k*-\n*\else\k*\n*\fi
\a*=\the\k* sp\a*=10\a*\b*\dm*\advance\b*-\a*\k*\b*
\a*=\the\*ths\b*\divide\a*\l*\multiply\a*\k*
\divide\a*\l*\k*\*ths\r*\a*\advance\k*-\r*\dt*=\the\k*\L*}

\def\arcto#1(#2,#3)[#4]{\rlap{\toks0={#2}\toks1={#3}\calcnum*#1(#2,#3)\relax
\dm*=135sp\dm*=#1\dm*\d**=#1\Lengthunit\ifdim\dm*<0pt\dm*-\dm*\fi
\multiply\dm*\r*\a*=.3\dm*\a*=#4\a*\ifdim\a*<0pt\a*-\a*\fi
\advance\dm*\a*\N*\dm*\divide\N*10000\relax
\divide\N*2\multiply\N*2\advance\N*\*one
\L*=-.25\d**\L*=#4\L*\divide\d**\N*\divide\L*\*ths
\m*\N*\divide\m*2\dm*=\the\m*5sp\l*\dm*\sm*\n*\*one\loop
\calcparab*\shl**{-\dt*}\advance\n*1\ifnum\n*<\N*\repeat}}

\def\arrarcto#1(#2,#3)[#4]{\L*=#1\Lengthunit\L*=.54\L*
\arcto#1(#2,#3)[#4]\rmov*(#2\L*,#3\L*){\d*=.457\L*\d*=#4\d*\d**-\d*
\rmov*(#3\d**,#2\d*){\arrow.02(#2,#3)}}}

\def\dasharcto#1(#2,#3)[#4]{\rlap{\toks0={#2}\toks1={#3}\relax
\calcnum*#1(#2,#3)\dm*=\the\N*5sp\a*=.3\dm*\a*=#4\a*\ifdim\a*<0pt\a*-\a*\fi
\advance\dm*\a*\N*\dm*
\divide\N*20\multiply\N*2\advance\N*1\d**=#1\Lengthunit
\L*=-.25\d**\L*=#4\L*\divide\d**\N*\divide\L*\*ths
\m*\N*\divide\m*2\dm*=\the\m*5sp\l*\dm*
\sm*\n*\*one\loop\calcparab*
\shl**{-\dt*}\advance\n*1\ifnum\n*>\N*\else\calcparab*
\sh*(#2,#3){\xL*=#3\dt* \yL*=#2\dt*
\rx* \the\cos*\xL* \tmp* \the\sin*\yL* \advance\rx*\tmp*
\ry* \the\cos*\yL* \tmp* \the\sin*\xL* \advance\ry*-\tmp*
\kern\rx*\lower\ry*\hbox{\sm*}}\fi
\advance\n*1\ifnum\n*<\N*\repeat}}

\def\*shl*#1{\c*=\the\n*\d**\advance\c*#1\a**\d*\dt*\advance\d*#1\b**
\a*=\the\toks0\c*\b*=\the\toks1\d*\advance\a*-\b*
\b*=\the\toks1\c*\d*=\the\toks0\d*\advance\b*\d*
\rx* \the\cos*\a* \tmp* \the\sin*\b* \advance\rx*-\tmp*
\ry* \the\cos*\b* \tmp* \the\sin*\a* \advance\ry*\tmp*
\raise\ry*\rlap{\kern\rx*\unhcopy\spl*}}

\def\calcnormal*#1{\b**=10000sp\a**\b**\k*\n*\advance\k*-\m*
\multiply\a**\k*\divide\a**\m*\a**=#1\a**\ifdim\a**<0pt\a**-\a**\fi
\ifdim\a**>\b**\d*=.96\a**\advance\d*.4\b**
\else\d*=.96\b**\advance\d*.4\a**\fi
\d*=.01\d*\r*\d*\divide\a**\r*\divide\b**\r*
\ifnum\k*<0\a**-\a**\fi\d*=#1\d*\ifdim\d*<0pt\b**-\b**\fi
\k*\a**\a**=\the\k*\dd*\k*\b**\b**=\the\k*\dd*}

\def\wavearcto#1(#2,#3)[#4]{\rlap{\toks0={#2}\toks1={#3}\relax
\calcnum*#1(#2,#3)\c*=\the\N*5sp\a*=.4\c*\a*=#4\a*\ifdim\a*<0pt\a*-\a*\fi
\advance\c*\a*\N*\c*\divide\N*20\multiply\N*2\advance\N*-1\multiply\N*4\relax
\d**=#1\Lengthunit\dd*=.012\d**
\divide\dd*\*ths \multiply\dd*\magnitude
\ifdim\d**<0pt\d**-\d**\fi\L*=.25\d**
\divide\d**\N*\divide\dd*\N*\L*=#4\L*\divide\L*\*ths
\m*\N*\divide\m*2\dm*=\the\m*0sp\l*\dm*
\sm*\n*\*one\loop\calcnormal*{#4}\calcparab*
\*shl*{1}\advance\n*\*one\calcparab*
\*shl*{1.3}\advance\n*\*one\calcparab*
\*shl*{1}\advance\n*2\dd*-\dd*\ifnum\n*<\N*\repeat\n*\N*\shl**{0pt}}}

\def\triangarcto#1(#2,#3)[#4]{\rlap{\toks0={#2}\toks1={#3}\relax
\calcnum*#1(#2,#3)\c*=\the\N*5sp\a*=.4\c*\a*=#4\a*\ifdim\a*<0pt\a*-\a*\fi
\advance\c*\a*\N*\c*\divide\N*20\multiply\N*2\advance\N*-1\multiply\N*2\relax
\d**=#1\Lengthunit\dd*=.012\d**
\divide\dd*\*ths \multiply\dd*\magnitude
\ifdim\d**<0pt\d**-\d**\fi\L*=.25\d**
\divide\d**\N*\divide\dd*\N*\L*=#4\L*\divide\L*\*ths
\m*\N*\divide\m*2\dm*=\the\m*0sp\l*\dm*
\sm*\n*\*one\loop\calcnormal*{#4}\calcparab*
\*shl*{1}\advance\n*2\dd*-\dd*\ifnum\n*<\N*\repeat\n*\N*\shl**{0pt}}}

\def\hr*#1{\L*=\xscale\Lengthunit\ifnum
\angle**=0\clap{\vrule width#1\L* height.1pt}\else
\L*=#1\L*\L*=.5\L*\rmov*(-\L*,0pt){\sm*}\rmov*(\L*,0pt){\sl*}\fi}

\def\shade#1[#2]{\rlap{\Lengthunit=#1\Lengthunit
\special{em:linewidth .001pt}\relax
\mov(0,#2.05){\hr*{.994}}\mov(0,#2.1){\hr*{.980}}\relax
\mov(0,#2.15){\hr*{.953}}\mov(0,#2.2){\hr*{.916}}\relax
\mov(0,#2.25){\hr*{.867}}\mov(0,#2.3){\hr*{.798}}\relax
\mov(0,#2.35){\hr*{.715}}\mov(0,#2.4){\hr*{.603}}\relax
\mov(0,#2.45){\hr*{.435}}\special{em:linewidth \the\linwid*}}}

\def\dshade#1[#2]{\rlap{\special{em:linewidth .001pt}\relax
\Lengthunit=#1\Lengthunit\if#2-\def\t*{+}\else\def\t*{-}\fi
\mov(0,\t*.025){\relax
\mov(0,#2.05){\hr*{.995}}\mov(0,#2.1){\hr*{.988}}\relax
\mov(0,#2.15){\hr*{.969}}\mov(0,#2.2){\hr*{.937}}\relax
\mov(0,#2.25){\hr*{.893}}\mov(0,#2.3){\hr*{.836}}\relax
\mov(0,#2.35){\hr*{.760}}\mov(0,#2.4){\hr*{.662}}\relax
\mov(0,#2.45){\hr*{.531}}\mov(0,#2.5){\hr*{.320}}\relax
\special{em:linewidth \the\linwid*}}}}

\def\vdot{\rlap{\kern-1.9pt\lower1.8pt\hbox{$\scriptstyle\bullet$}}}
\def\vtimes{\rlap{\kern-3pt\lower1.8pt\hbox{$\scriptstyle\times$}}}
\def\vDot{\rlap{\kern-2.3pt\lower2.7pt\hbox{$\bullet$}}}
\def\vTimes{\rlap{\kern-3.6pt\lower2.4pt\hbox{$\times$}}}

\def\arc(#1)[#2,#3]{{\k*=#2\l*=#3\m*=\l*
\advance\m*-6\ifnum\k*>\l*\relax\else
{\rotate(#2)\mov(#1,0){\sm*}}\loop
\ifnum\k*<\m*\advance\k*5{\rotate(\k*)\mov(#1,0){\sl*}}\repeat
{\rotate(#3)\mov(#1,0){\sl*}}\fi}}

\def\dasharc(#1)[#2,#3]{{\k**=#2\n*=#3\advance\n*-1\advance\n*-\k**
\L*=1000sp\L*#1\L* \multiply\L*\n* \multiply\L*\Nhalfperiods
\divide\L*57\N*\L* \divide\N*2000\ifnum\N*=0\N*1\fi
\r*\n*  \divide\r*\N* \ifnum\r*<2\r*2\fi
\m**\r* \divide\m**2 \l**\r* \advance\l**-\m** \N*\n* \divide\N*\r*
\k**\r* \multiply\k**\N* \dn*\n* 
\advance\dn*-\k** \divide\dn*2\advance\dn*\*one
\r*\l** \divide\r*2\advance\dn*\r* \advance\N*-2\k**#2\relax
\ifnum\l**<6{\rotate(#2)\mov(#1,0){\sm*}}\advance\k**\dn*
{\rotate(\k**)\mov(#1,0){\sl*}}\advance\k**\m**
{\rotate(\k**)\mov(#1,0){\sm*}}\loop
\advance\k**\l**{\rotate(\k**)\mov(#1,0){\sl*}}\advance\k**\m**
{\rotate(\k**)\mov(#1,0){\sm*}}\advance\N*-1\ifnum\N*>0\repeat
{\rotate(#3)\mov(#1,0){\sl*}}\else\advance\k**\dn*
\arc(#1)[#2,\k**]\loop\advance\k**\m** \r*\k**
\advance\k**\l** {\arc(#1)[\r*,\k**]}\relax
\advance\N*-1\ifnum\N*>0\repeat
\advance\k**\m**\arc(#1)[\k**,#3]\fi}}

\def\triangarc#1(#2)[#3,#4]{{\k**=#3\n*=#4\advance\n*-\k**
\L*=1000sp\L*#2\L* \multiply\L*\n* \multiply\L*\Nhalfperiods
\divide\L*57\N*\L* \divide\N*1000\ifnum\N*=0\N*1\fi
\d**=#2\Lengthunit \d*\d** \divide\d*57\multiply\d*\n*
\r*\n*  \divide\r*\N* \ifnum\r*<2\r*2\fi
\m**\r* \divide\m**2 \l**\r* \advance\l**-\m** \N*\n* \divide\N*\r*
\dt*\d* \divide\dt*\N* \dt*.5\dt* \dt*#1\dt*
\divide\dt*1000\multiply\dt*\magnitude
\k**\r* \multiply\k**\N* \dn*\n* \advance\dn*-\k** \divide\dn*2\relax
\r*\l** \divide\r*2\advance\dn*\r* \advance\N*-1\k**#3\relax
{\rotate(#3)\mov(#2,0){\sm*}}\advance\k**\dn*
{\rotate(\k**)\mov(#2,0){\sl*}}\advance\k**-\m**\advance\l**\m**\loop\dt*-\dt*
\d*\d** \advance\d*\dt*
\advance\k**\l**{\rotate(\k**)\rmov*(\d*,0pt){\sl*}}%
\advance\N*-1\ifnum\N*>0\repeat\advance\k**\m**
{\rotate(\k**)\mov(#2,0){\sl*}}{\rotate(#4)\mov(#2,0){\sl*}}}}

\def\wavearc#1(#2)[#3,#4]{{\k**=#3\n*=#4\advance\n*-\k**
\L*=4000sp\L*#2\L* \multiply\L*\n* \multiply\L*\Nhalfperiods
\divide\L*57\N*\L* \divide\N*1000\ifnum\N*=0\N*1\fi
\d**=#2\Lengthunit \d*\d** \divide\d*57\multiply\d*\n*
\r*\n*  \divide\r*\N* \ifnum\r*=0\r*1\fi
\m**\r* \divide\m**2 \l**\r* \advance\l**-\m** \N*\n* \divide\N*\r*
\dt*\d* \divide\dt*\N* \dt*.7\dt* \dt*#1\dt*
\divide\dt*1000\multiply\dt*\magnitude
\k**\r* \multiply\k**\N* \dn*\n* \advance\dn*-\k** \divide\dn*2\relax
\divide\N*4\advance\N*-1\k**#3\relax
{\rotate(#3)\mov(#2,0){\sm*}}\advance\k**\dn*
{\rotate(\k**)\mov(#2,0){\sl*}}\advance\k**-\m**\advance\l**\m**\loop\dt*-\dt*
\d*\d** \advance\d*\dt* \dd*\d** \advance\dd*1.3\dt*
\advance\k**\r*{\rotate(\k**)\rmov*(\d*,0pt){\sl*}}\relax
\advance\k**\r*{\rotate(\k**)\rmov*(\dd*,0pt){\sl*}}\relax
\advance\k**\r*{\rotate(\k**)\rmov*(\d*,0pt){\sl*}}\relax
\advance\k**\r*
\advance\N*-1\ifnum\N*>0\repeat\advance\k**\m**
{\rotate(\k**)\mov(#2,0){\sl*}}{\rotate(#4)\mov(#2,0){\sl*}}}}

\def\gmov*#1(#2,#3)#4{\rlap{\L*=#1\Lengthunit
\xL*=#2\L* \yL*=#3\L*
\rx* \gcos*\xL* \tmp* \gsin*\yL* \advance\rx*-\tmp*
\ry* \gcos*\yL* \tmp* \gsin*\xL* \advance\ry*\tmp*
\rx*=\xscale\rx* \ry*=\yscale\ry*
\xL* \the\cos*\rx* \tmp* \the\sin*\ry* \advance\xL*-\tmp*
\yL* \the\cos*\ry* \tmp* \the\sin*\rx* \advance\yL*\tmp*
\kern\xL*\raise\yL*\hbox{#4}}}

\def\rgmov*(#1,#2)#3{\rlap{\xL*#1\yL*#2\relax
\rx* \gcos*\xL* \tmp* \gsin*\yL* \advance\rx*-\tmp*
\ry* \gcos*\yL* \tmp* \gsin*\xL* \advance\ry*\tmp*
\rx*=\xscale\rx* \ry*=\yscale\ry*
\xL* \the\cos*\rx* \tmp* \the\sin*\ry* \advance\xL*-\tmp*
\yL* \the\cos*\ry* \tmp* \the\sin*\rx* \advance\yL*\tmp*
\kern\xL*\raise\yL*\hbox{#3}}}

\def\Earc(#1)[#2,#3][#4,#5]{{\k*=#2\l*=#3\m*=\l*
\advance\m*-6\ifnum\k*>\l*\relax\else\def\xscale{#4}\def\yscale{#5}\relax
{\angle**0\rotate(#2)}\gmov*(#1,0){\sm*}\loop
\ifnum\k*<\m*\advance\k*5\relax
{\angle**0\rotate(\k*)}\gmov*(#1,0){\sl*}\repeat
{\angle**0\rotate(#3)}\gmov*(#1,0){\sl*}\relax
\def\xscale{1}\def\yscale{1}\fi}}

\def\dashEarc(#1)[#2,#3][#4,#5]{{\k**=#2\n*=#3\advance\n*-1\advance\n*-\k**
\L*=1000sp\L*#1\L* \multiply\L*\n* \multiply\L*\Nhalfperiods
\divide\L*57\N*\L* \divide\N*2000\ifnum\N*=0\N*1\fi
\r*\n*  \divide\r*\N* \ifnum\r*<2\r*2\fi
\m**\r* \divide\m**2 \l**\r* \advance\l**-\m** \N*\n* \divide\N*\r*
\k**\r*\multiply\k**\N* \dn*\n* \advance\dn*-\k** \divide\dn*2\advance\dn*\*one
\r*\l** \divide\r*2\advance\dn*\r* \advance\N*-2\k**#2\relax
\ifnum\l**<6\def\xscale{#4}\def\yscale{#5}\relax
{\angle**0\rotate(#2)}\gmov*(#1,0){\sm*}\advance\k**\dn*
{\angle**0\rotate(\k**)}\gmov*(#1,0){\sl*}\advance\k**\m**
{\angle**0\rotate(\k**)}\gmov*(#1,0){\sm*}\loop
\advance\k**\l**{\angle**0\rotate(\k**)}\gmov*(#1,0){\sl*}\advance\k**\m**
{\angle**0\rotate(\k**)}\gmov*(#1,0){\sm*}\advance\N*-1\ifnum\N*>0\repeat
{\angle**0\rotate(#3)}\gmov*(#1,0){\sl*}\def\xscale{1}\def\yscale{1}\else
\advance\k**\dn* \Earc(#1)[#2,\k**][#4,#5]\loop\advance\k**\m** \r*\k**
\advance\k**\l** {\Earc(#1)[\r*,\k**][#4,#5]}\relax
\advance\N*-1\ifnum\N*>0\repeat
\advance\k**\m**\Earc(#1)[\k**,#3][#4,#5]\fi}}

\def\triangEarc#1(#2)[#3,#4][#5,#6]{{\k**=#3\n*=#4\advance\n*-\k**
\L*=1000sp\L*#2\L* \multiply\L*\n* \multiply\L*\Nhalfperiods
\divide\L*57\N*\L* \divide\N*1000\ifnum\N*=0\N*1\fi
\d**=#2\Lengthunit \d*\d** \divide\d*57\multiply\d*\n*
\r*\n*  \divide\r*\N* \ifnum\r*<2\r*2\fi
\m**\r* \divide\m**2 \l**\r* \advance\l**-\m** \N*\n* \divide\N*\r*
\dt*\d* \divide\dt*\N* \dt*.5\dt* \dt*#1\dt*
\divide\dt*1000\multiply\dt*\magnitude
\k**\r* \multiply\k**\N* \dn*\n* \advance\dn*-\k** \divide\dn*2\relax
\r*\l** \divide\r*2\advance\dn*\r* \advance\N*-1\k**#3\relax
\def\xscale{#5}\def\yscale{#6}\relax
{\angle**0\rotate(#3)}\gmov*(#2,0){\sm*}\advance\k**\dn*
{\angle**0\rotate(\k**)}\gmov*(#2,0){\sl*}\advance\k**-\m**
\advance\l**\m**\loop\dt*-\dt* \d*\d** \advance\d*\dt*
\advance\k**\l**{\angle**0\rotate(\k**)}\rgmov*(\d*,0pt){\sl*}\relax
\advance\N*-1\ifnum\N*>0\repeat\advance\k**\m**
{\angle**0\rotate(\k**)}\gmov*(#2,0){\sl*}\relax
{\angle**0\rotate(#4)}\gmov*(#2,0){\sl*}\def\xscale{1}\def\yscale{1}}}

\def\waveEarc#1(#2)[#3,#4][#5,#6]{{\k**=#3\n*=#4\advance\n*-\k**
\L*=4000sp\L*#2\L* \multiply\L*\n* \multiply\L*\Nhalfperiods
\divide\L*57\N*\L* \divide\N*1000\ifnum\N*=0\N*1\fi
\d**=#2\Lengthunit \d*\d** \divide\d*57\multiply\d*\n*
\r*\n*  \divide\r*\N* \ifnum\r*=0\r*1\fi
\m**\r* \divide\m**2 \l**\r* \advance\l**-\m** \N*\n* \divide\N*\r*
\dt*\d* \divide\dt*\N* \dt*.7\dt* \dt*#1\dt*
\divide\dt*1000\multiply\dt*\magnitude
\k**\r* \multiply\k**\N* \dn*\n* \advance\dn*-\k** \divide\dn*2\relax
\divide\N*4\advance\N*-1\k**#3\def\xscale{#5}\def\yscale{#6}\relax
{\angle**0\rotate(#3)}\gmov*(#2,0){\sm*}\advance\k**\dn*
{\angle**0\rotate(\k**)}\gmov*(#2,0){\sl*}\advance\k**-\m**
\advance\l**\m**\loop\dt*-\dt*
\d*\d** \advance\d*\dt* \dd*\d** \advance\dd*1.3\dt*
\advance\k**\r*{\angle**0\rotate(\k**)}\rgmov*(\d*,0pt){\sl*}\relax
\advance\k**\r*{\angle**0\rotate(\k**)}\rgmov*(\dd*,0pt){\sl*}\relax
\advance\k**\r*{\angle**0\rotate(\k**)}\rgmov*(\d*,0pt){\sl*}\relax
\advance\k**\r*
\advance\N*-1\ifnum\N*>0\repeat\advance\k**\m**
{\angle**0\rotate(\k**)}\gmov*(#2,0){\sl*}\relax
{\angle**0\rotate(#4)}\gmov*(#2,0){\sl*}\def\xscale{1}\def\yscale{1}}}

\newcount\CatcodeOfAtSign
\CatcodeOfAtSign=\the\catcode`\@
\catcode`\@=11
\def\@arc#1[#2][#3]{\rlap{\Lengthunit=#1\Lengthunit
\sm*\l*arc(#2.1914,#3.0381)[#2][#3]\relax
\mov(#2.1914,#3.0381){\l*arc(#2.1622,#3.1084)[#2][#3]}\relax
\mov(#2.3536,#3.1465){\l*arc(#2.1084,#3.1622)[#2][#3]}\relax
\mov(#2.4619,#3.3086){\l*arc(#2.0381,#3.1914)[#2][#3]}}}

\def\dash@arc#1[#2][#3]{\rlap{\Lengthunit=#1\Lengthunit
\d*arc(#2.1914,#3.0381)[#2][#3]\relax
\mov(#2.1914,#3.0381){\d*arc(#2.1622,#3.1084)[#2][#3]}\relax
\mov(#2.3536,#3.1465){\d*arc(#2.1084,#3.1622)[#2][#3]}\relax
\mov(#2.4619,#3.3086){\d*arc(#2.0381,#3.1914)[#2][#3]}}}

\def\wave@arc#1[#2][#3]{\rlap{\Lengthunit=#1\Lengthunit
\w*lin(#2.1914,#3.0381)\relax
\mov(#2.1914,#3.0381){\w*lin(#2.1622,#3.1084)}\relax
\mov(#2.3536,#3.1465){\w*lin(#2.1084,#3.1622)}\relax
\mov(#2.4619,#3.3086){\w*lin(#2.0381,#3.1914)}}}

\def\bezier#1(#2,#3)(#4,#5)(#6,#7){\N*#1\l*\N* \advance\l*\*one
\d* #4\Lengthunit \advance\d* -#2\Lengthunit \multiply\d* \*two
\b* #6\Lengthunit \advance\b* -#2\Lengthunit
\advance\b*-\d* \divide\b*\N*
\d** #5\Lengthunit \advance\d** -#3\Lengthunit \multiply\d** \*two
\b** #7\Lengthunit \advance\b** -#3\Lengthunit
\advance\b** -\d** \divide\b**\N*
\mov(#2,#3){\sm*{\loop\ifnum\m*<\l*
\a*\m*\b* \advance\a*\d* \divide\a*\N* \multiply\a*\m*
\a**\m*\b** \advance\a**\d** \divide\a**\N* \multiply\a**\m*
\rmov*(\a*,\a**){\unhcopy\spl*}\advance\m*\*one\repeat}}}

\catcode`\*=12

\newcount\n@ast

\def\n@ast@#1{\n@ast0\relax\get@ast@#1\end}
\def\get@ast@#1{\ifx#1\end\let\next\relax\else
\ifx#1*\advance\n@ast1\fi\let\next\get@ast@\fi\next}

\newif\if@up \newif\if@dwn
\def\up@down@#1{\@upfalse\@dwnfalse
\if#1u\@uptrue\fi\if#1U\@uptrue\fi\if#1+\@uptrue\fi
\if#1d\@dwntrue\fi\if#1D\@dwntrue\fi\if#1-\@dwntrue\fi}

\def\halfcirc#1(#2)[#3]{{\Lengthunit=#2\Lengthunit\up@down@{#3}\relax
\if@up\mov(0,.5){\@arc[-][-]\@arc[+][-]}\fi
\if@dwn\mov(0,-.5){\@arc[-][+]\@arc[+][+]}\fi
\def\lft{\mov(0,.5){\@arc[-][-]}\mov(0,-.5){\@arc[-][+]}}\relax
\def\rght{\mov(0,.5){\@arc[+][-]}\mov(0,-.5){\@arc[+][+]}}\relax
\if#3l\lft\fi\if#3L\lft\fi\if#3r\rght\fi\if#3R\rght\fi
\n@ast@{#1}\relax
\ifnum\n@ast>0\if@up\shade[+]\fi\if@dwn\shade[-]\fi\fi
\ifnum\n@ast>1\if@up\dshade[+]\fi\if@dwn\dshade[-]\fi\fi}}

\def\halfdashcirc(#1)[#2]{{\Lengthunit=#1\Lengthunit\up@down@{#2}\relax
\if@up\mov(0,.5){\dash@arc[-][-]\dash@arc[+][-]}\fi
\if@dwn\mov(0,-.5){\dash@arc[-][+]\dash@arc[+][+]}\fi
\def\lft{\mov(0,.5){\dash@arc[-][-]}\mov(0,-.5){\dash@arc[-][+]}}\relax
\def\rght{\mov(0,.5){\dash@arc[+][-]}\mov(0,-.5){\dash@arc[+][+]}}\relax
\if#2l\lft\fi\if#2L\lft\fi\if#2r\rght\fi\if#2R\rght\fi}}

\def\halfwavecirc(#1)[#2]{{\Lengthunit=#1\Lengthunit\up@down@{#2}\relax
\if@up\mov(0,.5){\wave@arc[-][-]\wave@arc[+][-]}\fi
\if@dwn\mov(0,-.5){\wave@arc[-][+]\wave@arc[+][+]}\fi
\def\lft{\mov(0,.5){\wave@arc[-][-]}\mov(0,-.5){\wave@arc[-][+]}}\relax
\def\rght{\mov(0,.5){\wave@arc[+][-]}\mov(0,-.5){\wave@arc[+][+]}}\relax
\if#2l\lft\fi\if#2L\lft\fi\if#2r\rght\fi\if#2R\rght\fi}}

\catcode`\*=11

\def\Circle#1(#2){\halfcirc#1(#2)[u]\halfcirc#1(#2)[d]\n@ast@{#1}\relax
\ifnum\n@ast>0\L*=\xscale\Lengthunit
\ifnum\angle**=0\clap{\vrule width#2\L* height.1pt}\else
\L*=#2\L*\L*=.5\L*\special{em:linewidth .001pt}\relax
\rmov*(-\L*,0pt){\sm*}\rmov*(\L*,0pt){\sl*}\relax
\special{em:linewidth \the\linwid*}\fi\fi}

\catcode`\*=12

\def\wavecirc(#1){\halfwavecirc(#1)[u]\halfwavecirc(#1)[d]}
\def\dashcirc(#1){\halfdashcirc(#1)[u]\halfdashcirc(#1)[d]}

\def\xscale{1}

\def\yscale{1}

\def\Ellipse#1(#2)[#3,#4]{\def\xscale{#3}\def\yscale{#4}\relax
\Circle#1(#2)\def\xscale{1}\def\yscale{1}}

\def\dashEllipse(#1)[#2,#3]{\def\xscale{#2}\def\yscale{#3}\relax
\dashcirc(#1)\def\xscale{1}\def\yscale{1}}

\def\waveEllipse(#1)[#2,#3]{\def\xscale{#2}\def\yscale{#3}\relax
\wavecirc(#1)\def\xscale{1}\def\yscale{1}}

\def\halfEllipse#1(#2)[#3][#4,#5]{\def\xscale{#4}\def\yscale{#5}\relax
\halfcirc#1(#2)[#3]\def\xscale{1}\def\yscale{1}}

\def\halfdashEllipse(#1)[#2][#3,#4]{\def\xscale{#3}\def\yscale{#4}\relax
\halfdashcirc(#1)[#2]\def\xscale{1}\def\yscale{1}}

\def\halfwaveEllipse(#1)[#2][#3,#4]{\def\xscale{#3}\def\yscale{#4}\relax
\halfwavecirc(#1)[#2]\def\xscale{1}\def\yscale{1}}

\catcode`\@=\the\CatcodeOfAtSign

%\begin{titlepage}
\thispagestyle{empty}

%\vspace*{3cm}

\begin{center}
{\LARGE\bf Quantum superfield supersymmetry}

\vspace*{3cm}

{A.Yu. Petrov}

\footnotesize
{\it
 Departamento de F\'{i}sica,\\
Universidade Federal de Paraiba,\\
Jo\~{a}o Pessoa, PB, Brazil}\\

\end{center}

\vfill

This is a preprint of the following work: Albert Petrov, Quantum Superfield Supersymmetry, 2021, Springer, reproduced with permission of Springer
  Nature Switzerland AG 2021. The final authenticated version is available
  online at:\\ https://doi.org/10.1007/978-3-030-68136-4.

\newpage

\thispagestyle{empty}

\tableofcontents

\newpage

\chapter{Introduction}

\begin{flushright}
{\small We prove, once and for all, that people who don't use superspace are really out of it.}\\
{\it ``Stuperspace''}
\end{flushright}

This review presents itself as a collection of the lecture notes on superfield
supersymmetry based on lectures given at Instituto de F\'{i}sica, Universidade de S\~{a}o Paulo,
Instituto de F\'{i}sica, Universidade Federal do Rio Grande do Sul
(Porto Alegre) and Departamento de F\'{i}sica, Universidade Federal da Paraiba
(Jo\~{a}o Pessoa).

The idea of supersymmetry is now considered as one of the basic
concepts of theoretical high energy physics (see f.e. \cite{GSH}). 
The supersymmetry, being a
fundamental symmetry of bosons and fermions, provides possibilities to
construct theories with much better renormalization properties
since some bosonic and fermionic divergent contributions cancel each other.
Moreover, there are essentially finite supersymmetry theories {\bf without}
higher derivatives, f.e. $N=4$ super-Yang-Mills theory (for the detailed discussion of the finiteness of this theory see, f.e., \cite{Kovacz} and references therein), and, probably, $N=8$  supergravity \cite{Kosower} and $N=6$, $N=8$ Chern-Simons theories \cite{Aharony}.
Now most specialists in quantum field theory suggest that unified
theory of all interactions must be supersymmetric.

Concept of supersymmetry was introduced in known papers by Volkov and
Akulov \cite{VA} and Golfand and Lichtman \cite{GL} in early 70s and
received further development in \cite{WZ,WZYM} (the history of arising of
the concept of the supersymmetry is well described in the book
\cite{sworld}). The essential 
breakthrough in supersymmetric field theory was achieved with introducing
the idea of a superfield \cite{WZ1} (see also \cite{SS,FWZ}). The reason for it consists in the fact that the superfield formulation, first, allows to maintain manifest supersymmetric covariance on all steps of calculations, with all calculations are realized in a very compact manner (indeed, any supergraph corresponds to a set of the Feynman diagrams in components), second, automatically takes into account the famous ``miraculous cancellations" of ultraviolet divergences responsible for an essential improvement of the renormalization behaviour of the supersymmetric field theories \cite{miracle}. Moreover, in the context of the noncommutative field theory it turns out to be that the supersymmetry is responsible for cancellation not only of the ultraviolet divergences, as usual, but also of the dangerous infrared divergences arising due to the UV/IR mixing mechanism (see f.e. \cite{ours}). Further, the concept of the extended supersymmetry has been introduced, and the corresponding extended superspace formalism has been elaborated. Within this lecture course, nevertheless, we concentrate on the $N=1$ superfield formalism which is known as one of the most universal tools for studying the supersymmetric field theories.

Now, let us briefly describe the main steps in development of the superfield methodology. The first example of the successful application of the superfield concept was the model proposed by Wess and Zumino in their seminal paper \cite{WZ1} where the simplest superfield model, further denominated as the Wess-Zumino model, has been formulated. Further, in \cite{WZYM} they introduced the superfield gauge model, that is, the super-Yang-Mills theory. Intensive studies of different issues related to the quantum aspects of these theories began to be discussed, in particular, the finiteness of $N=4$ super-Yang-Mills theory was proved \cite{Brink} (for discussions of finiteness of the supersymmetric gauge theories see also \cite{Piguet}), which implied in a strongest interest to the supergauge theories, and the superfield supergravity was formulated \cite{SUGRAWZ}. In 1984, the first consistent methodology possessing an explicit extended ($N=2$) supersymmetry has been developed, that is, the harmonic superspace \cite{GIKOS1,GIKOS2}. At the same year, the superfield approach has been successfully applied to the superstring theory \cite{GSstring}, which emphasized the importance of the this approach within the string context. A bit earlier, the superfield formulation has been developed for the supersymmetric theories in a three-dimensional space-time \cite{Siegel,SGRS}, which further manifested itself as a very convenient laboratory for study of different issues related to the supersymmetry due to its simplicity.

A new epoch for studies of the superfield theories began in 1991 when the chiral contributions to the effective action were discussed first time in the papers \cite{West3,West4}. These papers called an attention to the superfield methodology for evaluating the effective potential whose development has been carried out in a series of papers initiated by the paper \cite{Buch1}, where this methodology has been successfully applied to the Wess-Zumino model. Further, the success of the paper \cite{SWgauge} strongly increased the interest to the supergauge theories. 

Among the most successful application of the superfield methodology, also its application to the noncommutative supersymmetric field theories deserves to be mentioned. The famous paper \cite{Popp} established the fact that, since the space-time noncommutativity does not affect the Grassmannian coordinates of superfields, the superfield methodology can be used as a main instrument to deal with the famous problem of the UV/IR mixing \cite{Minw} known for generating the new infrared divergences which break the perturbative expansion. It was shown in \cite{Popp} that the superfield methodology, known for improving he ultraviolet behaviour, also allows to remove the dangerous UV/IR infrared divergences. Further, the non(anti)commutativity was introduced also to the fermionic sector \cite{Sei}.

The $N=1$ superfield methodology in supersymmetric quantum field theory, being an universal tool for a great number of supersymmetric models, is a main topic of these lectures. We consider the superfield description both of three- and four-dimensional supersymmetric field theories, including the noncommutative generalization for some cases. In the chapter devoted to three-dimensional theories we use the notations and conventions introduced in \cite{SGRS}, while in the chapter devoted to four-dimensional theories -- those ones introduced in \cite{WB,BK0}.

Within this review, we are going not only to describe the superfield formalism in three- and four-dimensional space-times and give the superfield formulation for the most important examples of the supersymmetric field theories, but also to consider in details calculation of the loop corrections within the superfield description. The review is focused, principally, in evaluating the effective action in different supersymmetric field theories (because of the restricted value of the lecture course and of this review, we do not discuss here the superstring and supergravity issues, for which we recommend \cite{GSH} and \cite{BK0} respectively). 

The structure of this review looks like follows. In the Chapter 2, we give a basic description of an effective action and the loop expansion. In the Chapter 3 we discuss the superfield formalism, supergraph technique and methods of calculation of the effective action in three-dimensional space-time. In the Chapter 4, the four-dimensional superfield methodology is described on many examples. In the Chapter 5, we present the introduction to the problem of a spontaneous supersymmetry breaking. Finally, in Summary we discuss the perspectives of supersymmetry.

\newpage

\chapter[Effective action]{Effective action and loop expansion: general description}

An effective action is a central object of quantum field theory. Studying
of an effective action allows to investigate problems of vacuum stability,
Green functions, spontaneous symmetry breaking, anomalies and
many other problems. In the description of its general structure we follow the methodology described in \cite{CW,BO}.

An effective action (in particular, in a superfield theory) is defined, as usual, as a
generating functional of one-particle-irreducible Green functions. It
is obtained as a Legendre transform for generating functional of
connected
Green functions:
\bea
\Gamma[\Phi]=W[J]-\int dz J(z)\Phi(z).
\eea
Here $\Gamma[\Phi]$ is an effective action, $dz$ denotes integral over
the corresponding space or its subspace (in some cases, we omit the sign of the integral together with the coordinate dependence of the fields, i.e. $J\Phi\equiv\int dz J(z)\Phi(z)$),  $J(z)$ is the classical source, $\Phi(z)=\frac{\delta W[J]}{\delta J(z)}$ is a so called
mean (super)field or background (super)field, which is an essentially classical object (in principle, one can consider a set of background (super)fields as well, so, in general, $\Phi$ is a column vector); in terms of the path integral one has $$\Phi=\frac{\int D\phi e^{iS[\phi]+i\phi J}\phi}{\int D\phi e^{iS[\phi]+i\phi J}},$$ 
which is exactly the definition of the vacuum expected value in the path integral formalism, and $W[J]=\frac{1}{i}\log Z[J]$ is a generating functional of
the connected Green functions. It is easy to see that the $\Gamma[\Phi]$ satisfies the
equation
$$
\frac{\delta\Gamma[\Phi]}{\delta\Phi(x)}=-J(x).
$$
The effective action can be expressed in the form of the path integral
\cite{BO}:
\bea
\label{efact}
e^{\frac{i}{\hbar}\Gamma[\Phi]}=\int D\phi e^{\frac{i}{\hbar}
(S[\phi]+\phi J-\Phi J)}.
\eea
Here $S[\phi]$ is a classical action of the corresponding theory. Note
that $\phi$ is an integration variable, and $\Phi$ is a function of a
classical source $J$ which does not depend on $\phi$. We introduced the Planck constant
$\hbar$
by dimensional reasons (indeed, the Planck constant has a dimension of the action) and in order to obtain the loop  expansion. We note that, to distinguish quantum effects from the classical ones, we treat $\hbar$ as a small parameter, thus, the effective action is a power series in $\hbar$.
To calculate this integral we make change of variables of integration:
$$
\phi\to\Phi+\sqrt{\hbar}\phi.
$$
If we have several fields we can unite them into a column vector, and
all consideration is quite analogous.
The integral (\ref{efact}) after this change takes the form
\bea
\label{intmov}
e^{\frac{i}{\hbar}\Gamma[\Phi]}=\int D\phi e^{\frac{i}{\hbar}
(S[\Phi+\sqrt{\hbar}\phi]+\sqrt{\hbar}\phi J)} .
\eea
Our aim consists here of the expansion of $\Gamma[\Phi]$ in power series
in $\hbar$ following the approach described in \cite{BO}.

First, we expand factor in the exponential of the above expression into power series in $\hbar$:
\bea
& &\frac{i}{\hbar}S[\Phi+\sqrt{\hbar}\phi]+\frac{i}{\sqrt{\hbar}}\phi J=
\frac{i}{\hbar}\Big(S[\Phi]+(S'[\Phi]+J)\sqrt{\hbar}\phi+
\frac{\hbar}{2}S^{''}[\Phi]\phi^2+\ldots+\nonumber\\&+&
\frac{\hbar^{n/2}}{n!}S^{(n)}[\Phi]
\phi^n+\ldots
\Big).
\eea
Here $S^{(n)}[\Phi]$ denotes $n$-th variational derivative of
the classical action with respect to $\Phi$ (integration over
corresponding space is assumed).
This expansion can be substituted into (\ref{intmov}). We can introduce the quantum part of the effective action,
$\bar{\Gamma}[\Phi]=\Gamma[\Phi]-S[\Phi]$, which can be expanded into power
series in $\hbar$, beginning from the first order:
$\bar{\Gamma}=\sum\limits_{n=1}^{\infty}\hbar^n\Gamma^{(n)}$.
As a result we have
\bea
\label{expol}
e^{\frac{i}{\hbar}\bar{\Gamma}[\Phi]}=\int D\phi \exp\Big[
\frac{i}{\hbar}\Big(\sqrt{\hbar}(S'[\Phi]+J)\phi+
\frac{\hbar}{2}S^{''}[\Phi]\phi^2+\ldots+\frac{\hbar^{n/2}}{n!}S^{(n)}[\Phi]
\phi^n+\ldots\Big)\Big].
\eea
Then the first block,  $\frac{i}{\sqrt{\hbar}}(S'[\Phi]+J)\phi$, can lead only to
one-particle-reducible supergraphs since its contribution with one
quantum field $\phi$ can form only one propagator during the contraction of the quantum fields, thus it does not contribute to the effective action. Hence we can omit
this term. Then we can expand the exponent into power series in $\hbar$:
\bea
\label{exph0}
& &e^{\frac{i}{\hbar}\bar{\Gamma}[\Phi]}=\int D\phi e^{
\frac{i}{2}S^{''}[\Phi]\phi^2}
\Big(
1+\frac{i\sqrt{\hbar}}{3!}S^{(3)}[\Phi]\phi^3+\frac{i\hbar}{4!}S^{(4)}
[\Phi]\phi^4+\nonumber\\&+&
\left(\frac{i\sqrt{\hbar}}{3!}\right)^2(S^{(3)}[\Phi]\phi^3)^2+\ldots
\Big).
\eea 
At the same time, after substituting the expansion of
$\bar{\Gamma}$ in the left-hand side of (\ref{expol}) 
into power series in $\hbar$ (which we assume to be small) we get: 
$$\exp(\frac{i}{\hbar}\bar{\Gamma}[\Phi])=e^{i(\Gamma^{(1)}[\Phi]+\hbar\Gamma^{(2)}[\Phi]+\ldots)}=e^{i\Gamma^{(1)}[\Phi]}
(1+i\hbar\Gamma^{(2)}[\Phi]+\ldots).$$ 
Substituting this expansion into
(\ref{exph0}) and comparing equal powers of $\hbar$ we see that any 
correction $\Gamma^{(n)}$ corresponds to some correlator. For example, a
one-loop correction is defined from the equation
\bea
\label{1loop}
\exp(i\Gamma^{(1)}[\Phi])=\int D\phi e^{\frac{i}{2}S^{\prime\prime}[\Phi]\phi^2},
\eea 
and two-loop one -- from the equation
\bea
\label{2loop}
\Gamma^{(2)}=\frac{1}{i}\frac{
\int D\phi e^{(\frac{i}{2}S^{''}[\Phi]\phi^2)}\Big(
\frac{i}{4!}S^{(4)}[\Phi]\phi^4-\frac{1}{(3!)2}(S^{(3)}[\Phi]
\phi^3)^2\Big) }{\int D\phi \exp(\frac{i}{2}S^{''}[\Phi]\phi^2)}.
\eea
Here, as usual, integration over coordinates in expressions of the form
$S^{(n)}[\Phi]\phi^n$ is assumed. The denominator of this expression serves to eliminate the one-particle-reducible contributions.

We can see that:

(i) All non-integer orders in $\hbar$ vanish since they correspond to 
$\int D\phi\phi^{2n+1}\exp(\frac{i}{2}S^{''}[\Phi]\phi^2)$. By the symmetry
reasons this integral is equal to zero.

(ii) All terms beyond first order in $\hbar$ are expressed in the form of
some correlators.

(iii) The one-loop correction (\ref{1loop}) can be expressed in the form of
functional determinant since
\bea
\int D\phi\exp(\frac{i}{2}S^{''}[\Phi]\phi^2)={\rm Det}^{-1/2}S^{''}[\Phi],
\eea
which leads to
\bea
\Gamma^{(1)}=\frac{i}{2} {\rm Tr}\log S^{''}[\Phi].
\eea
The $S^{''}[\Phi]\equiv \Delta$ is a some
operator. In many cases it has the form $\Delta=\Box+(\ldots)$, with dots are for derivative-independent terms, for example, mass (in particular, background dependent). 
We can express one-loop effective action in terms of functional (super)trace
\bea
\Gamma^{(1)}=\frac{i}{2} {\rm Tr} \int_0^{\infty}\frac{ds}{s}e^{is\Delta}.
\eea  
This expression is called Schwinger representation for the one-loop
effective action. The sign ${\rm Tr}$
denotes both matrix trace ${\rm tr}$ (if $\Delta$ possesses matrix
indices) and
functional trace, i.e.
$$
{\rm Tr} e^{is\Delta}={\rm tr} \int d^n z_1 d^n z_2 \delta^n(z_1-z_2)
e^{is\Delta}\delta^n(z_1-z_2).
$$
Here $n$ is a dimension of the corresponding (super)space.
The calculation of $e^{is\Delta}$ in field theories is carried out
with use of a special procedure called Schwinger-De Witt method or the
proper time method \cite{BSD}. This method will be discussed further.

Let us consider higher loop corrections. From (\ref{exph0}) it is easy
to see that all loop corrections beyond one-loop order have the form of
some correlators, i.e. they include
\bea
\int D\phi \exp(\frac{i}{2}S^{''}[\Phi]\phi^2)\prod_n(S^{(n)}[\Phi]\phi^n).
\eea
Such correlators can be calculated in the way analogous to the standard
perturbative methodology. We can use the identity
\bea
\int D\phi \phi^n e^{i\frac{1}{2}\phi\Delta\phi}=(\frac{1}{i}
\frac{\delta}{\delta j})^n\int D\phi e^{i(\frac{1}{2}\phi\Delta\phi+j\phi)}|_{j=0},
\eea
which allows to introduce a diagram technique in which the role of vertices
is played by $\frac{S^{(n)}[\Phi]\phi^n}{n!}$, and role of propagators
-- by $\Delta^{-1}$. However since $\Delta=S^{''}[\Phi]$ is background
dependent (see above)
we arrive at background dependent propagators $<\phi(z_1)\phi(z_2)>=
\Delta^{-1}\delta^8(z_1-z_2)$. These propagators, generally, 
can be found exactly
only in some special cases, the most important of them are: first,
constant in space-time background superfields, second, the
background superfields are only chiral. Further we consider some
examples.

Let us turn again to (\ref{exph0}). We see that each quantum superfield
corresponds to $\hbar^{-1/2}$, and each vertex -- to $\hbar^{-1}$ (which
provides $\hbar^{n/2-1}S^{(n)}[\Phi]\phi^n$). An arbitrary (super)graph
with $P$ propagators and $V$ vertices contains $2P$ quantum superfields
(indeed, each propagator is formed by contraction of two
superfields). Therefore if this (super)graph contain vertices 
$S^{(n_1)}[\Phi]\phi^{n_1},S^{(n_2)}[\Phi]\phi^{n_2},\ldots,
S^{(n_V)}[\Phi]\phi^{n_V}$, its power in $\hbar$ is $\sum\limits_{i=1}^V
(\frac{n_i}{2}-1)=\frac{1}{2}\sum\limits_{i=1}^V n_i -V$. However,
$\sum\limits_{i=1}^V n_i$ is just the number of quantum fields associated with
all vertices which is equal to $2P$. Therefore the correlator described by
this (super)graph
has power of $\hbar$ equal to $P-V=L-1$, with $L$ is number of
loops. But any correlator of the
form (\ref{exph0}) is a contribution to $\frac{\Gamma}{\hbar}$, hence 
contribution from $L$-loop (super)graph to $\Gamma$ is proportional to 
$\hbar^L$. Hence we found that the order in $\hbar$ from an arbitrary
(super)graph is just the number of loops in it, and the expansion in powers of
$\hbar$ is called loop expansion. As a result we see that loop
corrections can be calculated on the base of special (super)field technique.

In this review we will evaluate the effective action in the superfield formalism both in three- and four-dimensional superspace, calculating one- and two-loop contributions to the effective actions of different superfield theory models in an explicit form.

\newpage

\chapter[Three-dimensional theories]{Superfield description of three-dimensional supersymmetric theories}

\section{Definitions and conventions}

The general principle of supersymmetric field theory consists in
existence of some essentially new symmetry transformations with
a fermionic parameter which mix fermionic and bosonic dynamical variables
of the theory. To provide a nontrivial connection of the supersymmetry 
transformations with usual Poincare transformations, we also suggest that the anticommutator of two
supersymmetry transformations is nontrivial, in the most used, and simplest, versions it is a space-time translation. 

In this chapter we follow conventions of \cite{SGRS}: in the three-dimensional space-time we choose the Minkowski metric of the form $\eta_{mn}=
{\rm diag}(-++)$, and the Dirac matrices are the $2\times 2$  matrices whose explicit form is: $(\gamma^0)^{\a}_{\phantom{a}\b}=-i\s^2,
(\gamma^1)^{\a}_{\phantom{a}\b}=\s^1,(\gamma^2)^{\a}_{\phantom{a}\b}=\s^3$, with 
$\{\gamma^m,\gamma^n\}=2\eta^{mn}$. We use bispinor notations based on converting any vector index into two spinor indices by the rule
$A^m\to A^{\a\b}=A^m(\gamma_m)^{\a\b}$. We note that the gamma
matrices with two lower indices are $(\gamma^m)_{\a\b}=(-{\bf
1}_2,-\s^3,\s^1)$ are the symmetric matrices hence all vectors (in
particular, coordinates $x^m$ and corresponding derivatives $\pa_m$) are
represented by symmetric bispinors. To raise and lower the spinor indices
we use the Hermitian Levi-Civita-like $C$ symbol: 
$C_{\a\b}=-i\epsilon_{\a\b}=
\left(
\begin{array}{cc}0 &-i\\
i&0\end{array}\right)=-C^{\a\b}$, with $\psi^{\a}=C^{\a\b}\psi_{\b}$, 
$\psi_{\a}=\psi^{\b}C_{\b\a}$ (``north-western'' convention). Also, one has $C^{\alpha\beta}C_{\beta\gamma}=-\delta^{\alpha}_{\gamma}$, and $C^{\alpha\beta}C_{\alpha\beta}=2$. We define 
$\psi^2=\frac{1}{2}\psi^{\a}\psi_{\a}$ (the reader should note the presence of $\frac{1}{2}$
factor!!! This is the difference of this formulation from the version of the three-dimensional superspace proposed by Ruiz Ruiz and Nieuwenhuizen in \cite{RR}), and use the identity $A_{[\a}B_{\b]}=-C_{\a\b}A^{\g}B_{\g}$ following from the properties of the irreducible representations of the Lorentz group.

To formulate the superspace, we start with introducing spinor coordinates
$\theta^{\a}$ with $\a=1,2$. These coordinates are transformed under the spinor
representation of three-dimensional Lorentz group, i.e. $SO(1,2)$ group (the spinor representation of the Lorentz group in the 
three-dimensional space-time is given by the $SL(2,R)$ group).
These spinor coordinates $\theta^{\a}$, satisfying the Grassmannian anticommutation condition
$\{\theta^{\a},\theta^{\b}\}=0$, together with the usual bosonic coordinates $x^m$ parametrize the superspace \footnote{We do not consider here the alternative anticommutation relations for the spinor coordinates, however, we will briefly discuss the simplest deformation of the anticommutation relation, that is, $\{\theta^{\a},\theta^{\b}\}=C^{\a\b}$, further, in that case the spinor coordinates form the Clifford algebra rather than the Grassmann algebra; this methodology has been originally introduced in \cite{Sei}.}.

To develop field theory on superspace we must introduce integration
and differentiation on superspace, i.e. with respect to Grassmann
coordinates \cite{ber}. We can introduce left $\pa_L$ and right $\pa_R$ 
derivatives with respect to Grassmann coordinates as
\bea
\label{derdef}
\frac{\pa_L}{\pa\q^{\alpha_i}}
(\q^{\a_1}\q^{\a_{i-1}}\q^{\a_i}\q^{\a_{i+1}}\ldots \q^{\a_n})&=&
{(-1)}^{i-1}
(\q^{\a_1}\q^{\a_{i-1}}\q^{\a_{i+1}}\ldots \q^{\a_n});
\nonumber\\
\frac{\pa_R}{\pa\q^{\alpha_i}}
(\q^{\a_1}\q^{\a_{i-1}}\q^{\a_i}\q^{\a_{i+1}}\ldots \q^{\a_n})&=&
{(-1)}^{n-i+1}
(\q^{\a_1}\q^{\a_{i-1}}\q^{\a_{i+1}}\ldots \q^{\a_{n}}).
\eea
Therefore these derivatives differ only by a sign factor. We can
choose one of them, for example, the left one, and use it henceforth.
So, we define the derivatives with respect to the Grassmannian  coordinates as
follows:
\bea
\pa_{\a}&\equiv&\frac{\pa}{\pa\theta^{\a}};\quad\, 
\pa_{\a}\theta^{\b}=\delta_{\a}^{\b};\nonumber\\
\pa_{\a\b}x^{\gamma\delta}&=&\frac{1}{2}
\delta_{(\a}^{\gamma}\delta_{\b)}^{\delta};\quad\, 
\Box=\frac{1}{2}\pa_{\a\b}\pa^{\a\b}.
\eea
Here $A_{(\a}B_{\b)}\equiv A_{\a}B_{\b}+A_{\b}B_{\a}$. 

The superfield is a (general) function of the superspace coordinates which can
be introduced in the form the Taylor series in
$\theta$ which is finite because of the anticommuting nature of
$\theta^{\a}$:
\bea
f(x,\theta)=f_0(x)+f_1^{\a}(x)\theta_{\a}+f_2(x)\theta^2,
\eea
where $\theta^2=\frac{1}{2}\theta^{\a}\theta_{\a}$. In a general
case, the $f_0(x),f_1(x),f_2(x)$ can be also tensors of different ranks. 
The integral over the Grassmann coordinates is defined as
\bea
\int d\theta_{\a}\theta^{\b}=\delta_{\a}^{\b},
\eea
which, with use of the definition
$d^2\theta=\frac{1}{2}d\theta^{\a}d\theta_{\a}$, implies 
\bea
\int d^2\theta \theta^2=-1.
\eea
This integral allows to define the Grassmannian delta function
\bea
\delta^2(\theta)=-\theta^2,
\eea
which satisfies the common condition
\bea
\int d^2\theta_1 f(\theta_1)\delta^2(\theta_1-\theta_2)=f(\theta_2).
\eea

The generators of the supersymmetric transformations (which are the
translations in the superspace), also called the supercharges, are defined as
\bea
Q_{\a}=i\pa_{\a}+\theta^{\b}\pa_{\b\a},
\eea
they evidently commute with the generators of ``common'' bosonic
translations $P_{\a\b}=i\pa_{\a\b}$, while the anticommutator of two supercharges is
\bea
\label{anti}
\{Q_{\a},Q_{\b}\}=2P_{\a\b}.
\eea
The supersymmetry transformation for the given superfield
$\Phi(x,\theta)$ with the infinitesimal parameter $\epsilon^{\a}$ is defined as
\bea
\delta\Phi(x,\theta)=-i\epsilon^{\a}Q_{\a}\Phi(x,\theta).
\eea
Projecting this transformation to components, we can obtain the
transformation laws for components of any superfield.

The (super)covariant derivative $D_{\a}$ of an superfield must be transformed also as a superfield. To satisfy this condition, the $D_{\alpha}$ must anticommute
with the supersymmetry generators $Q_{\b}$, i.e. 
$\{D_{\a},Q_{\b}\}$=0, and
commute with the translation generators $P_{\a\b}$. Also, spinor supercovariant derivatives must be linear in the simple derivatives $\pa_{\a},\pa_{\a\b}$ to satisfy the Leibnitz rule.
All these properties are satisfied if $D_{\a}$ looks like
\bea
D_{\a}=\pa_{\a}+i\theta^{\b}\pa_{\b\a}.
\eea
The spinor supercovariant derivatives $D_{\a}$ defined in such a way possess the following properties:
\bea
\label{com}
\{D_{\a},D_{\b}\}=2i\pa_{\a\b};\quad\, [D_{\a},D_{\b}]=-2C_{\a\b}D^2.
\eea
After summation of two these expressions we arrive at the following
very important relation
\bea
\label{key}
D_{\a}D_{\b}=i\pa_{\a\b}-C_{\a\b}D^2.
\eea
This identity is employed to carry out $D$-algebra transformations used
for simplifying the forms of the contributions of the supergraphs (see
examples further). One can also notice that the first expression in (\ref{com}) can be interpreted 
in the sense that the superspace possesses a fundamental torsion.

Let us derive some other useful identities (cf. \cite{RR}; note, however, that our conventions differ from those ones used in \cite{RR}). 
First of all, let us note
that the totally antisymmetric object with three spinor indices should
vanish (really, such an object vanishes if any two of its indices
coincide, and since spinor indices can take only two values, that is, 1 and 2,
this object should have at least two coinciding indices). If such an object is
constructed by antisymmetrization of the product $D_{\a}D_{\b}D_{\g}$,
we get
\bea
D_{\a}D_{\b}D_{\g}+D_{\b}D_{\g}D_{\a}+D_{\g}D_{\a}D_{\b}-
D_{\a}D_{\g}D_{\b}-D_{\b}D_{\a}D_{\g}-D_{\g}D_{\b}D_{\a}=0.
\eea
Contracting this expression to $C^{\b\g}$ we get
\bea
\label{i0}
2(D_{\a}D_{\b}D^{\b}+D_{\b}D^{\b}D_{\a}+D^{\b}D_{\a}D_{\b})=0.
\eea
Then, in the first term we substitute the identity
$D_{\a}D_{\b}=-D_{\b}D_{\a}+\{D_{\a},D_{\b}\}$, and in the second one we substitute
the identity $D^{\b}D_{\a}=-D_{\a}D^{\b}+\{D^{\b},D_{\a}\}$. Afterwards, the
anticommutator terms cancel each other, and we rest with
\bea
\label{id1}
D^{\b}D_{\a}D_{\b}=0.
\eea
This is a very important identity which role is similar to the
properties of the projecting operators in four-dimensional superfield
supersymmetry (see further). Applying this identity to (\ref{i0}) we find another
important identity
\bea
\label{id2}
\{D_{\a},D^2\}=0.
\eea
Applying the (\ref{key}) and the (\ref{id2}) we can derive one more
important identity
\bea
\label{id3}
(D^2)^2=\Box.
\eea
These properties of the supercovariant derivatives can be used for
constructing of the superfields.

One more very important property of the supercovariant derivatives
and delta function which can be checked by a direct applying the spinor supercovariant deruvatives on the delta function is
\bea
\label{delta}
\delta_{12}D^2\delta_{12}=\delta_{12}.
\eea
It is straightforward to verify as well that
\bea
\label{delta1}
\delta_{12}D_{\a}\delta_{12}=0.
\eea

The most important superfields used in the known field theory models in the three-dimensional 
superspace are the scalar and the spinor ones.
The scalar superfield is defined in the form of the following $\theta$ expansion:
\bea
\phi(x,\theta)=\varphi(x)+\theta^{\a}\psi_{\a}(x)-\theta^2 F(x).
\eea
Its components can be also defined as the projections:
\bea
\label{projscal}
\varphi(x)&=&\phi(x,\theta)|;\nonumber\\
\psi_{\a}(x)&=&D_{\a}\phi(x,\theta)|;\nonumber\\
F(x)&=&D^2\phi(x,\theta)|.
\eea
Here and further the symbol $|$ means that the Grassmannian coordinates $\theta$ in the
corresponding expression are put to zero after the differentiation. 
In the expression above, the $\varphi(x)$ is the usual scalar field, $\psi_{\a}(x)$ is the spinor one, and $F(x)$ is the auxiliary field, whose equations of motion in the usual scalar superfield theory model are just constraints, so, it can be completely eliminated from the theory on the mass shell.
The supersymmetry transformations for the component fields can be obtained from the projections:
\bea
\delta\varphi(x)&=&\delta\phi(x,\theta)|=-i\epsilon^{\alpha}Q_{\alpha}\phi(x,\theta)|=\epsilon^{\alpha}D_{\alpha}\phi(x,\theta)|=\epsilon^{\alpha}\psi_{\alpha}(x);\nonumber\\
\delta\psi_{\alpha}(x)&=&\delta D_{\a}\phi(x,\theta)|=-i\epsilon^{\beta}Q_{\beta}D_{\alpha}\phi(x,\theta)|=-\epsilon^{\beta}D_{\alpha}D_{\beta}\phi(x,\theta)|=
\nonumber\\&=&
-\epsilon^{\beta}(i\partial_{\alpha\beta}-C_{\alpha\beta}D^2)\phi(x,\theta)|=-\epsilon^{\beta}(i\partial_{\alpha\beta}\varphi(x)-C_{\alpha\beta}F(x));\nonumber\\
\delta F(x)&=&\delta D^2\phi(x,\theta)|=-i\epsilon^{\alpha}Q_{\alpha}D^2\phi(x,\theta)|=\epsilon^{\alpha}D_{\alpha}D^2\phi(x,\theta)|=\nonumber\\
&=&-\frac{1}{2}\epsilon^{\alpha}\{D_{\alpha},D_{\beta}\}D^{\beta}\phi(x,\theta)|=-i\epsilon^{\alpha}\partial_{\alpha\beta}\psi^{\beta}.
\eea
We conclude that the spinor field $\psi_{\alpha}(x)$ related with the scalar one $\varphi(x)$ through a supersymmetry transformation, thus, we refer to the $\psi_{\alpha}$ as to the superpartner of $\phi$.

Other important superfield is the spinor one introduced as
\bea
A_{\a}(x,\theta)&=&\chi_{\a}(x)-\theta_{\a}B(x)+i\theta^{\b}V_{\b\a}(x)
-2\theta^2[\lambda_{\a}(x)+\frac{i}{2}\pa_{\a\b}\chi^{\b}(x)],
\eea
therefore its components are defined as
\bea
\label{spi}
\chi_{\a}(x)&=&A_{\a}(x,\theta)|;\nonumber\\
B(x)&=&\frac{1}{2}D^{\a}A_{\a}(x,\theta)|;\nonumber\\
V_{\a\b}(x)&=&-\frac{i}{2}D_{(\a}A_{\b)}(x,\theta)|;\nonumber\\
\lambda_{\a}&=&\frac{1}{2}D^{\b}D_{\a}A_{\b}(x,\theta)|.
\eea
Here $V_{\a\b}(x)$ is a bispinor form of the usual vector (in the most interesting case -- gauge) field, $\lambda_{\a}(x)$ is its superpartner (photino), and $\chi_{\a}(x)$ and $B(x)$ are the auxiliary fields.

These scalar and spinor superfields are the basic ingredients for
constructing the most popular field theory models in the
three-dimensional superspace. In principle, other superfields (for example bispinor ones) can be also introduced.

\section{Field theory models}

Now, let us introduce the models involving the scalar and spinor superfields. In this section, we proceed mostly in a manner similar to \cite{SGRS}.

{\bf 1.} The action for the 
models involving only scalar fields have the common form 
(here and further we denote $d^5z=d^3xd^2\theta$):
\bea
\label{scaf}
S=\int d^5 z [-\frac{1}{2}\Phi D^2\Phi+\frac{m}{2}\Phi^2+f(\Phi)]
\eea
where $f(\Phi)$ is an arbitrary function of the scalar superfield (by the
reasons of renormalizability it must have no more than fourth order in
$\Phi$). The case of the complex scalar superfield does not
essentially differ.

The component form of this action can be obtained in the following way: since the integration and the differentiation are equivalent, $\int d^2\theta f=D^2f|$, after integrating by parts in the kinetic term one has
\bea
S=\int d^3 x [-\frac{1}{2}D^2(\Phi D^2\Phi)+\frac{m}{2}D^2(\Phi^2)+D^2f(\Phi)]|,
\eea
that is,
\bea
S&=&\int d^3 x [-\frac{1}{2}(D^2\Phi D^2\Phi+\Phi\Box\Phi+\frac{1}{2}D^{\alpha}\Phi D_{\alpha}D^2\Phi)+\frac{m}{2}(2\Phi D^2\Phi+D^{\alpha}\Phi D_{\alpha}\Phi)+\nonumber\\
&+&
(\frac{1}{2}f^{\prime\prime}(\Phi)D^{\alpha}\Phi D_{\alpha}\Phi+f^{\prime}(\Phi)D^2\Phi)]|,
\eea
which, with use of (\ref{projscal}), yields
\bea
S&=&\int d^3 x
[-\frac{1}{2}F^2+\frac{1}{2}i\psi_{\a}\pa^{\a\b}\psi_{\b}-
\frac{1}{2}\varphi\Box \varphi+\nonumber\\&+&
m (\psi^2+\varphi F)+ 
f^{\prime\prime}(\varphi)\psi^2+f^{\prime}(\varphi)F].
\eea
This is the general approach for reduction of a superfield action to components.
As we have already mentioned, the auxiliary field $F$ can be eliminated with use of the equation of
motion
\bea
F=m\varphi+f^{\prime}(\varphi).
\eea
which gives
\bea
S&=&\int d^3 x
[-\frac{1}{2}(m\varphi+f^{\prime}(\varphi))^2
+\frac{1}{2}i\psi_{\a}\pa^{\a\b}\psi_{\b}-
\frac{1}{2}\varphi\Box \varphi+\nonumber\\&+&
m (\psi^2+\varphi (m\varphi+f^{\prime}(\varphi)))+ 
f^{\prime\prime}(\varphi)\psi^2+
f^{\prime}(\varphi)(m\varphi+f^{\prime}(\varphi))]
\eea
This is the theory of the self-coupled scalar field $\varphi$ interacting also with the spinor $\psi$. In other words, it is a supersymmetric extension of the scalar field theory.
In particular, for $f(\Phi)=\lambda\Phi^4$ (the higher possible
renormalizable self-coupling of the scalar superfield, remind that the
mass dimension of $\Phi$ is $\frac{1}{2}$ as well as of the derivative
$D_{\alpha}$, and of $d^5 z$ is $-2$) we get the usual renormalizable coupling
$\lambda\varphi^6$ as one of the terms of interaction which is present in the theory.

{\bf 2.} The construction of action for the spinor superfield is more
involved. The reason is that the spinor multiplet turns out to describe supersymmetric three-dimensional gauge models, hence we must formulate the gauge invariant model 
for this superfield, since it contains the vector
$V_{\a\b}(x)$ as one of the components.

We start with introduction of the three-dimensional gauge
transformations for the scalar superfield $\Phi$ (cf. \cite{SGRS}):
\bea
\label{tra}
\Phi\to e^{iK}\Phi,\quad\,\bar{\Phi}\to \bar{\Phi}e^{-iK},
\eea
where $K$ is a parameter of the gauge transformation. For a constant $K$
the kinetic term $\frac{1}{2}\int d^5z D^{\a}\bar{\Phi} D_{\a}\Phi$ is invariant under
these transformations.
Then we introduce a covariant derivative
\bea
\nabla_{\a}\Phi=(D_{\a}+iA_{\a})\Phi,
\eea
which under the transformation (\ref{tra}) carried out together with 
the following transformations for the $A_{\a}$ superfield:
\bea
\label{tra1}
A_{\a}\to A_{\a}-D_{\a}K
\eea 
is transformed as
\bea
\label{tra2}
\nabla_{\a}\Phi\to e^{iK}\nabla_{\a}\Phi.
\eea
The complex conjugate expression $(\nabla_{\a}\Phi)^*$ is, in a similar manner,
transformed by the factor $e^{-iK}$.
Therefore the expression 
\bea
\nabla^{\a}\Phi{(\nabla_{\a}\Phi)}^*
\eea
is invariant under the transformations (\ref{tra},\ref{tra1}). It is natural to consider it as a simplest Lagrangian for the scalar field coupled to the gauge one, introducing thus the following action:
\bea
\label{matterint}
S_m=\int d^5 z [\frac{1}{2}(D^{\a}\Phi+i\Phi A^{\a})
(D_{\a}\bar{\Phi}-i A_{\a}\bar{\Phi})+m\Phi\bar{\Phi}],
\eea
Taking into account (\ref{tra},\ref{tra2}) we can write down the following formal 
transformation law for the $\nabla_{\a}$:
\bea
\nabla_{\a}\to e^{iK}\nabla_{\a}e^{-iK}.
\eea
The covariant derivatives $\nabla_{\a}$ represent themselves as a base
for constructing the superfield strengths.

We impose the following anticommutation relation which is a
straightforward covariant generalization of first relation in (\ref{com}):
\bea
\label{idcov}
\{\nabla_{\a},\nabla_{\b}\}=2i\nabla_{\a\b}.
\eea
If we suggest for any covariant derivative $\nabla_A$ (with both
$A=\a$ and $A=\a\b$) the relation $\nabla_A=D_A+i\Gamma_A$, we get 
$\Gamma_{\a}=A_{\a}$ and $\Gamma_{\a\b}=-\frac{i}{2}D_{(\a}A_{\b)}$.

Then we suggest that the Bianchi identities on the $\nabla_A$ are valid:
\bea
\label{bia}
[\nabla_{[A},[\nabla_B,\nabla_{C\} }\}\}=0,
\eea
where the anticommutator (and symmetrization over indices) is suggested between two fermionic objects
whereas the commutator (and antisymmetrization over indices) -- in all other cases.
We also introduce the general curvature-torsion definition:
\bea
\label{ct}
[\nabla_A,\nabla_B\}=T_{AB}^C\nabla_C-iF_{AB},
\eea
where $T_{AB}^C$ is a torsion (note that unlike of the ``common'' flat
space the superspace has the nontrivial intrinsic torsion even for ``simple''
covariant derivatives $D_{\a},\pa_{\a\b}$), and $F_{AB}$ is a curvature.
Suggesting in (\ref{bia}) the set $A,B,C=\a,\b,\g$ we get
\bea
\label{bia1}
[\nabla_{\a},\{\nabla_{\b},\nabla_{\g}\}]+
[\nabla_{\b},\{\nabla_{\g},\nabla_{\a}\}]+
[\nabla_{\g},\{\nabla_{\a},\nabla_{\b}\}]=0,
\eea
which implies
\bea
[\nabla_{(\a},\nabla_{\b\g)}]\equiv -iF_{(\a,\b\g)}=0.
\eea
Also, it follows from (\ref{bia1}) that $T_{\a,\b\g}^D=0$.
Then, splitting the $F_{\a,\b\g}$ into the irreducible
representations we get
\bea
F_{\a,\b\g}=\frac{1}{6}F_{(\a,\b\g)}-
\frac{1}{3}C_{\a(\b|}F^{\delta}_{,\phantom{\delta}\delta|\g )},
\eea
where the symbol $|$ means that the index $\delta$ is not involved
into the symmetrization. Since $F_{(\a,\b\g)}=0$ we get 
$F_{\a,\b\g}=iC_{\a(\b}W_{\g)}$ with 
$W_{\g}=\frac{i}{3}F^{\delta}_{,\phantom{\delta}\delta\g}$
Applying the definition of $F_{AB}$ through the covariant derivatives
(\ref{ct}) and taking into account that $T_{\a,\b\g}^D=0$ which naturally follows from the Bianchi identities above we arrive at
the following expression for $W_{\a}$ (to which we further refer as to
the superfield strength):
\bea
\label{wdef}
W_{\a}=\frac{1}{2}D^{\b}D_{\a}A_{\b}.
\eea
This object is invariant under the transformations (\ref{tra1}) which
with taking into account the component structure of $A_{\b}$ superfield
(\ref{spi}) correspond to the following variation of the component
fields:
\bea
\label{tra11}
\delta\chi_{\a}=-\s_{\a},\quad\,\delta B=-\tau,\quad\,
\delta V_{\a\b}=-\pa_{\a\b}\omega,\quad\,\delta\lambda_{\a}=0,
\eea
where for the given superfield gauge parameter $K$ its components are
defined as $\omega=K|$, $\s_{\a}=D_{\a}K|$, $\tau=D^2K|$.
The remarkable fact that the gauge transformation of the superfield $A_{\a}$
corresponds to the common gradient transformation for its vector
component $V_{\a\b}$. We also note that by an appropriate choice of the gauge
parameter $K$ (and hence its components $\s_{\a},\tau$) in a proper
way we can completely gauge away the components $\chi_{\a}$ and
$B$. Such a gauge choice providing $\chi_{\a}=B=0$ is called the
Wess-Zumino gauge. Its advantage consists in vanishing of all terms
involving third and higher powers of the $A_{\a}$ superfield itself
(note however that the terms involving derivatives of $A_{\alpha}$ must be considered in a more careful manner! In other words, for example, the non-Abelian term $\{A^{\alpha},A^{\beta}\}\{A_{\alpha},A_{\beta}\}$ would vanish while $\{A^{\alpha},D^{\gamma}A^{\beta}\}\{A_{\alpha},D_{\gamma}A_{\beta}\}$ would not) in the vertices of interaction
but it implies in breaking of supersymmetry, with only some residual
supersymmetry persists in this case. 

The component structure of the $W_{\a}$ strength looks as follows:
\bea
\label{w0}
W_{\a}=\lambda_{\a}+\theta^{\b}f_{\a\b}+i\theta^2\pa_{\a\b}\lambda^{\b},
\eea
i.e. the $W_{\a}$ involves only strength tensor $f_{\a\b}=
\frac{1}{2}(\s^{mn})_{\a\b}F_{mn}$ (here $F_{mn}=\pa_m V_n-\pa_n V_m$
is a common strength tensor, and $\s^{mn}=[\gamma^m,\gamma^n]$, notice
that $\s^{mn}$ with two upper or two lower spinor indices is a symmetric
matrix with respect to the spinor indices! This follows from the fact that for the Dirac matrices used within this section the following property is valid: $\gamma^m\gamma^n=\eta^{mn}-\epsilon^{mnl}\gamma_l$). This strength tensor is evidently gauge invariant, its component expansion
is not modified even after imposing of the Wess-Zumino gauge.
Its components can be defined by
\bea
\lambda_{\a}&=&W_{\a}|;\nonumber\\
f_{\a\b}&=&D_{\a}W_{\b}|.
\eea
We note the transversality of the $W^{\a}$ following from
(\ref{wdef}):
\bea
D^{\a}W_{\a}=0,
\eea
which is a superfield analogue of the known property of the dual vector $F^m=
\frac{1}{2}\epsilon^{mnl}F_{nl}$:
\bea
\pa_m F^m=0.
\eea

The most natural, gauge invariant kinetic term for the action of the spinor
superfield is hence
\bea
\label{qed}
S_g=\frac{1}{2g^2}\int d^5 z W^{\a}W_{\a},
\eea
which, because of (\ref{w0}), can be easily shown to give in components
\bea
S_g=\int d^3 x (\frac{1}{2}f^{\a\b}f_{\a\b}+\lambda^{\a}i\pa_{\a\b}\lambda^{\b}).
\eea
Coupling of the gauge superfield to matter is given by the term
\bea
\label{smat}
S_m&=&\frac{1}{2}\int d^5 z \nabla^{\a}\Phi({\nabla_{\a}\Phi})^*=\frac{1}{2}\int d^5 z (D^{\a}+iA^{\a})\Phi(D_{\a}-iA_{\a})\Phi^*,
\eea
whose component content is
\bea
S_m&=&\int d^3 x
\big[-F\bar{F}+i\bar{\psi}_{\a}\pa^{\a\b}\psi_{\b}-
\bar{\varphi}\Box \varphi-\nonumber\\&-&
i V^{\a\b}\varphi\stackrel{\leftrightarrow}{\pa}_{\a\b}\bar{\varphi}
+\frac{1}{2}\varphi V^{\a\b}V_{\a\b}\bar{\varphi}+
V^{\a\b}(\psi_{\a}\bar{\psi}_{\b}-\bar{\psi}_{\a}\psi_{\b})
-\lambda^{\a}(\varphi\bar{\psi}_{\a}-\bar{\varphi}\psi_{\a})
\nonumber\\&+&
(\psi^{\a}\bar{F}-\bar{\psi}^{\a}F)\chi_{\a}+
(\pa_{\a\b}\psi^{\a}\bar{\varphi}-\pa_{\a\b}\bar{\psi}^{\a}\varphi-
\pa_{\a\b}\bar{\varphi}\psi^{\a}+\pa_{\a\b}\varphi\bar{\psi}^{\a})\chi^{\b}+
\nonumber\\&+&\frac{1}{2}(\varphi B^2\bar{\varphi}-
\varphi\bar{\varphi}\chi^{\a}\lambda_{\a}+\chi^{\a}\chi_{\a}
(F\bar{\varphi}+\varphi\bar{F}))\big].
\eea
We note that only the terms in two first lines do not vanish 
in the Wess-Zumino gauge. Presence of the additional terms (or, as is
the same -- additional vertices) implies in
the known difference between the results obtained within superfield
approach and component approach -- note that most of papers devoted to
the component calculations use the ``simplified'' supersymmetric
formulation of the theories which is effectively 
obtained by imposing of the Wess-Zumino gauge -- the reason is that
such a theory really possesses (residual) supersymmetry and contains
much less terms, however, such a formulation cannot be obtained by
direct projecting of the superfield action into components.
This difference is a quite known phenomenon, f.e. in
four-dimensional super-Yang-Mills theory imposing of the Wess-Zumino
gauge allows to truncate nonpolynomial expansion of the action. However, in this case only the residual supersymmetry survives, which makes the superfield description to be a bit senseless (for a discussion of noncovariant gauges in superfield theories see f.e. \cite{noncov}).

We should mention also one more gauge invariant construction in
superspace which is the superfield analogue of the Chern-Simons term:
\bea
\label{cs}
S_{CS}=\frac{m}{2g^2}\int d^5 z A^{\a}W_{\a},
\eea
which component structure is 
\bea
S_{CS}=\frac{m}{g^2}\int d^3 x (V^{\a\b}f_{\a\b}-\lambda^{\a}\lambda_{\a}).
\eea
We notice that $f_{\a\b}$ is a vector dual to the stress tensor.

{\bf 3.} There is also an alternative free action for the spinor superfield initially introduced in \cite{ourdual}. In this case, we have two spinor fields, $\Psi^{\alpha}$ and $\bar{\Psi}^{\alpha}$ whose component structure is
\bea
\Psi^{\alpha}&=&\psi^{\alpha}+\theta^{\alpha}b+i\theta_{\beta}b^{\beta\alpha}-\theta^2\phi^{\alpha};\nonumber\\
\bar{\Psi}^{\alpha}&=&\bar{\psi}^{\alpha}+\theta^{\alpha}\bar{b}+i\theta_{\beta}\bar{b}^{\beta\alpha}-\theta^2\bar{\phi}^{\alpha}.
\eea
We can introduce the Dirac-like action for these fields:
\bea
\label{dirac}
S=-\int d^5z\bar{\Psi}^{\alpha}(i\partial_{\alpha\beta}-MC_{\alpha\beta})\Psi^{\beta}.
\eea
The corresponding action for the component fields looks like
\begin{equation}
S_M \, = \, S_M ^ {(1/2)} + S_M ^ {(1)},
\end{equation}
where
\begin{equation}
S_M ^ {(1/2)}\,=\,\int d^{3}x \,
\left[ \overline{\varphi} \left(i\,\gamma^a\partial_a-M\right) \psi
+\overline{\psi} \left(i\,\gamma^a\partial_a-M\right) \varphi \right]\,,\label{eq:7}
\end{equation}
\begin{equation}
S_M ^ {(1)}\,=\,-\int d^{3}x  \,\left[
\frac{1}{2} \varepsilon^{abc}\overline{b}_{a}\partial_{b}b_{c}
+\frac{M}{2}\,\overline{b}^{a}b_{a}
+\overline{b} \partial^a b_a + {b} \partial^a \overline{b}_a - 2 M \overline{b} b
\right]\,,\label{eq:8}
\end{equation}
We can eliminate the auxiliary field $b$ using its equation of motion, thus
\begin{equation}
S_M ^ {(1)}\,=\,-\int d^{3}x  \,\left[
\frac{1}{2} \varepsilon^{abc}\overline{b}_{a}\partial_{b}b_{c}
+\frac{M}{2}\,\overline{b}^{a}b_{a}
- \frac{1}{2M} \left( \partial^a \overline{b}_a  \right) \left( \partial^b b_b  \right)
\right]\,,\label{eq:8b}
\end{equation}
This action can be treated as a gauge-fixed Chern-Simons theory, with a Proca mass term. However, the theory (\ref{dirac}), up to now, has been discussed only within the duality context \cite{ourdual}.

\section{Non-Abelian gauge models}

Here we generalize the results for the gauge theories to the case of
the non-Abelian gauge group. To develop it we suggest that the
gauge superfield $A^{\beta}$ takes value in some Lie algebra:
$A^{\beta}=A^{\beta a}T^a$, where $T^a$ are the generators of the corresponding Lie group. We suggest the scalar (matter) superfield also
to be the Lie-algebra valued. Again, we follow the methodology of \cite{SGRS}.
We start with the action of matter coupled to the gauge superfield in the non-Abelian case can be introduced
as
\bea
\label{matcoup}
S_m=\frac{1}{2}\int d^5 z (D^{\a}\Phi+i[\Phi,A^{\a}])
(D_{\a}\bar{\Phi}-i[A_{\a},\bar{\Phi}]),
\eea
with the gauge transformations are
\bea
\label{trana}
\Phi\to e^{-iK}\Phi e^{iK},\quad\,\bar{\Phi}\to
e^{-iK}\bar{\Phi}e^{iK}, \quad A_{\a}\to e^{-iK}A_{\a}e^{iK}-i
e^{-iK}(D_{\a}e^{iK}).
\eea
This is the case of the coupling of the matter to the gauge field in the adjoint representation where the matter is Lie-algebra valued as well as the gauge field.
The transformation above for the $A_{\a}$ field in the infinitesimal form
looks like
\bea
\delta A_{\a}=D_{\a}K+i[A_{\a},K].
\eea
The introduction of the covariant derivatives is carried out as above:
\bea
\nabla^{\a}\Phi=D^{\a}\Phi+i[\Phi,A^{\a}].
\eea
Applying the identities (\ref{idcov}), (\ref{bia1}) as above,
we again arrive at
\bea
W_{\g}=-\frac{i}{3}C^{\a\b}F_{\a,\b\g},
\eea
with the stress tensor $F_{AB}$ is defined from (\ref{ct}).
However, now we should take into account that the $A^{\a}$ superfields
and their derivatives no more (anti)commute now, hence the 
$F_{\a,\b\g}=i[\nabla_{\a},\nabla_{\b\g}]$ is nonlinear in $A_{\a}$:
since it follows from (\ref{idcov}) that
\bea
\Gamma_{\a\b}=-\frac{i}{2}(D_{(\a}A_{\b)}-i\{A_{\a},A_{\b}\})
\eea
we get
\bea
\label{wnab}
W_{\a}=\frac{1}{2}D^{\b}D_{\a}A_{\b}-\frac{i}{2}[A^{\b},D_{\b}A_{\a}]-
\frac{1}{6}[A^{\b},\{A_{\b},A_{\a}\}].
\eea
The $W_{\a}$ defined in such a way is covariant but not invariant under
transformation (\ref{trana}), it is transformed as
\bea
\label{preobr}
W_{\a}\to e^{-iK}W_{\a}e^{iK},
\eea
or, in the infinitesimal form,
\bea
\label{preobr1}
W_{\a}\to W_{\a}+i[W_{\a},K].
\eea
Now, let us consider the Bianchi identity
\bea
\{\nabla_{\a},[\nabla_{\b},\nabla_{\g\delta}]\}+
\{\nabla_{\b},[\nabla_{\g\delta},\nabla_{\a}]\}+
[\nabla_{\g\delta},\{\nabla_{\a},\nabla_{\b}\}]=0.
\eea
After the contractions with the symbols $C^{\alpha\gamma}$ and $C^{\beta\delta}$ it gives
\bea
\label{div}
\{\nabla^{(\a},[\nabla^{\b)},\nabla_{\a\b}]\}=-6\{\nabla^{\a},W_{\a}\}=0.
\eea
This is the non-Abelian transversality condition.

The action invariant under the gauge transformations (\ref{preobr}),
similarly to the Abelian case, looks like
\bea
\label{sym}
S=\frac{1}{2g^2}{\rm tr}\int d^5 z W^{\a}W_{\a}.
\eea
Using the expression (\ref{wnab}) and adding the gauge-fixing action 
\bea
S_{gf}=-\frac{1}{4g^2\xi}{\rm tr}
\int d^5 z (D^{\alpha}A_{\alpha})D^2(D^{\beta}A_{\beta}),
\eea
one can write down the exact form of total action of $A_{\a}$ superfield:
\bea
\label{sqed}
S_{total}&=&{\rm tr}\frac{1}{2g^2}\int d^5 z
\Big[\frac{1}{2}(1+\frac{1}\xi)A^\alpha
\Box A_\alpha
-\frac{1}{2}(1-\frac{1}\xi )A^\alpha i\pa_{\alpha\beta}D^2A^\beta \Big]\,.
+\nonumber\\&+&
\frac{1}{g^2}\int d^5 z\Big[-\frac{i}{4}D^{\gamma}D^\alpha
A_{\gamma}
[A^\beta ,D_\beta A_\alpha ]-\frac{1}{12}D^{\gamma}D^\alpha A_{\gamma}
[A^\beta ,\{A_\beta ,A_\alpha \}]-\nonumber\\&-&
\frac{1}{8}[A^{\gamma},D_{\gamma}A^\alpha ][A^\beta ,D_\beta A_\alpha
]+\frac{i}{12}
[A^{\gamma},D_{\gamma}A^\alpha ][A^\beta ,\{A_\beta ,A_\alpha \}]+
\nonumber\\&+&\frac{1}{72}
[A^{\gamma},\{A_{\gamma},A^\alpha \}][A^\beta ,\{A_\beta ,A_\alpha \}]
\Big].
\eea
The action (\ref{sym}) formally replays the structure of the action for the supersymmetric Abelian gauge theory. For the
Chern-Simons theory, however, the action is not a ``direct'' non-Abelian
generalization of (\ref{cs}). Really, under the infinitesimal transformation
(\ref{preobr1}), the ``naive'' non-Abelian generalization of the Chern-Simons action (\ref{cs}), obtained by a direct promotion of the Abelian strength $W_{\alpha}$ to its non-Abelian analogue (\ref{wnab}), acquires a
variation
\bea
\delta S=-\frac{m}{2g^2} {\rm tr} \int d^5 z K D^{\a}W_{\a},
\eea
but $D^{\a}W_{\a}\neq 0$ in the non-Abelian case, see (\ref{div}). To
cancel this variation we should add to the (\ref{cs}) some new
terms. As a result, the gauge invariant action takes the form
\bea
\label{csna}
S=\frac{m}{2g^2}{\rm tr}\int d^5 z
(A^{\a}W_{\a}+\frac{i}{6}\{A^{\a},A^{\b}\}D_{\b}A_{\a}+
\frac{1}{12}\{A^{\a},A^{\b}\}\{A_{\a},A_{\b}\}).
\eea
The ghost action in both these theories, that is, super-Yang-Mills and non-Abelian supersymmetric Chern-Simons theories is the same since it can be obtained from the same gauge-fixing function $\chi=D^{\alpha}A_{\alpha}$ by use of the Faddeev-Popov prescription. It looks like
\bea
\label{gh}
S_{gh}=\frac{1}{2g^2}{\rm tr}\int d^5 z c'D^{\a}(D_{\a}c+i\{A_{\a},c\}).
\eea
Here $c,c'$ are the Faddeev-Popov ghosts, they are fermionic superfields by the general rules. We note that the last term in the expression above must include an anticommutator to provide its vanishing for the Abelian gauge group.

Now we are in position to develop the quantum formulation for the
theories. 

\section{Quantum description for the superfield models}

Our aim here consists in the development of the perturbative approach
for the theories described above: scalar superfield model, super-Yang-Mills and
super-Chern-Simons field theories.

We start with the introduction of the superfield propagators. The
definition of the propagator is the following one: if the operator
defining the free theory (second functional derivative of the free
action) is $\Delta$, the propagator $G(z_1-z_2)$ satisfies the equation
\bea
\Delta G(z_1-z_2)=i\delta^5(z_1-z_2).
\eea
For the scalar field theory (\ref{scaf}) the propagator is
\bea
G(z_1,z_2)=<\Phi(z_1)\Phi(z_2)>=i\frac{D^2+m}{\Box-m^2}\delta(z_1-z_2).
\eea
The ghost propagator is very similar:
\bea
G^{gh}(z_1,z_2)=<c(z_1)c'(z_2)>=-ig^2\frac{D^2}{\Box}\delta(z_1-z_2).
\eea
However, we note that any ghost loop, despite similarity of the ghost and scalar propagators, will carry an additional minus sign.

For the super-Yang-Mills theory (and, in particular, QED), because of the
gauge invariance, we must fix the gauge by adding to the action (\ref{qed}) the gauge fixing term
\bea
S^{QED}_{GF}=-\frac{1}{4\xi g^2}\int d^5 z (D^{\a}A_{\a})D^2(D^{\b}A_{\b}),
\eea
which gives the propagator
\bea
\label{propqed}
G^{\a\b}_{QED}(z_1,z_2)=<A^{\a}(z_1)A^{\b}(z_2)>=\frac{ig^2}{2\Box^2}
[D^2D^{\b}D^{\a}-\xi D^2D^{\a}D^{\b}]\delta(z_1-z_2),
\eea
or, after applying the identity (\ref{key}), the same propagator takes
the form
\bea
\label{propqed1}
&&G^{\a\b}_{QED}(z_1,z_2)=<A^{\a}(z_1)A^{\b}(z_2)>=\nonumber\\&=&\frac{ig^2}{2}
\Big[C^{\a\b}\frac{1}{\Box}(\xi+1)-\frac{1}{\Box^2}(\xi-1)i\pa^{\a\b}D^2
\Big]\delta(z_1-z_2).
\eea
The most important gauges are: $\xi=1$, the Feynman gauge, where the propagator does not involve spinor derivatives; $\xi=-1$, where only second term of (\ref{propqed1}) survives, and $\xi=0$, the Landau gauge which makes the propagator to be transversal, i.e. $D^{\alpha}G_{\alpha\beta}|_{\xi=0}=0$. Nevertheless, sometimes other gauges are also important, for example, it was shown in \cite{oured} that the supersymmetric three-dimensional QED is finite in all loop orders at $\xi=-8$.

For the Chern-Simons theory we also must add to the free action (\ref{cs})
the gauge fixing term which we choose to be in the form
\bea
\label{gfcs}
S^{CS}_{GF}=-\frac{m}{4\xi}\int d^5 z (D^{\a}A_{\a})(D^{\b}A_{\b}),
\eea
implying in the propagator
\bea
\label{pcs}
G^{\a\b}_{CS}(z_1,z_2)=\frac{i}{2m\Box}[D^{\b}D^{\a}+\xi
D^{\a}D^{\b}]\delta(z_1-z_2). 
\eea
Its equivalent form is
\bea
G^{\a\b}_{CS}(z_1,z_2)=<A^{\a}(z_1)A^{\b}(z_2)>=\frac{i}{2m\Box}
\Big[C^{\a\b}D^2(1-\xi)+(1+\xi)i\pa^{\a\b}\Big]\delta(z_1-z_2).
\eea
Some authors, instead of the gauge-fixing term (\ref{gfcs}), suggest to add another gauge-fixing term
\bea
\label{gfcs1}
S^{CS}_{GF2}=-\frac{m}{4\xi}\int d^5 z (D^{\a}A_{\a})D^2(D^{\b}A_{\b}),
\eea
with $\xi$ is now a parameter with a non-zero mass dimension.
However, this case, implying propagator 
\bea
G^{\a\b}_{CS2}(z_1,z_2)=\frac{i}{2m\Box}[D^{\b}D^{\a}+\xi
\frac{D^2}{\Box} D^{\a}D^{\b}]\delta(z_1-z_2). 
\eea
with the same $\xi$-independent part as that one of (\ref{pcs}), does not essentially differ.

It is instructive here to give the inverse operator for the generic one of the form $\Delta^{\a\b}=A_1 D^{\alpha}D^{\beta}+A_2 D^{\beta}D^{\alpha}$. We suggest here that both $A_1$ and $A_2$ commute with the product $D_{\alpha}D_{\beta}$, being either constants or functions of $D^2$ and the space-time derivatives. Supposing that the inverse operator $Q_{\alpha\beta}$ by definition satisfying the relation $\Delta^{\alpha\beta}Q_{\beta\gamma}=\delta^{\alpha}_{\gamma}$, we express
\bea
Q_{\beta\gamma}=X_1 D_{\beta}D_{\gamma}+X_2 D_{\gamma}D_{\beta}.
\eea
We carry out the straightforward multiplication of $\Delta$ and $Q$. The property $D^{\alpha}D_{\beta}D_{\alpha}=0$ cancels two from four terms in this product. We simplify remaining ones with the use of the key identity (\ref{key}), compare the factors accompanying $\delta^{\alpha}_{\gamma}$ and $\partial^{\alpha}_{\gamma}$ in both sides of the equation and arrive at
\bea
&&2\Box(A_1X_1+A_2X_2)=-1;\nonumber\\
&&A_1X_1-A_2X_2=0.
\eea
From this system, we find the coefficients $X_1$ and $X_2$, writing down the operator $Q_{\beta\gamma}$ as
\bea
Q_{\beta\gamma}=-\frac{1}{4\Box}(\frac{1}{A_1} D_{\beta}D_{\gamma}+\frac{1}{A_2} D_{\gamma}D_{\beta}).
\eea
It is easy to see that the cases of the QED and the Chern-Simons theory given above match this expression.

The quantum contributions as usual are described by the supergraphs
arising from the standard path integral:
\bea
\int D\Phi e^{i(\Phi\frac{\Delta}{2}\Phi+V(\Phi)+\Phi J)}\equiv
e^{iV(\frac{1}{i}\frac{\delta}{\delta J})}e^{-\frac{i}{2}J\Delta^{-1}J}
\eea
To describe the general divergence structure we can apply the
superficial degree of divergences. First, because of the
(\ref{delta}) one spinor supercovariant derivative contributes $\frac{1}{2}$.
Then, the propagator of scalar superfield
contributes $-1$ as well as the propagator of the Chern-Simons
superfield, that one of gauge superfield for QED (and similarly --
Maxwell-Chern-Simons and super-Yang-Mills) contributes $-2$. Each loop
contributes 2 since any integration over $d^3k$ contributes 3, but number
of the $D$-factors which can be converted to momenta by the rule
(\ref{key}) is decreased by 2 in any loop because of the shrinking of any
loop to a point by the identity (\ref{delta}). 
Denoting number of vertices involving $i$ $A^{\a}$ gauge superfields
and no other superfields as $V_A^{(i)}$, number of vertices involving
ghosts as $V_c$, number of vertices involving scalar superfields with
one and none spinor supercovariant derivatives as $V_{\f}^{D}$ (which
contributes $\frac{1}{2}$) and
$V_{\f}^0$ respectively -- here we suggest that the matter is coupled only to
the gauge superfield just in the form given in
(\ref{matcoup}). Summation and use of the topological identity $L+V-P=1$
allows to find that the superficial degree of divergence in the
super-Yang-Mills theory with a scalar matter looks like (cf. \cite{ours}):
\bea
\label{sdd1}
\omega=2-2V_A^{(6)}-\frac{3}{2}V_A^{(5)}-V_A^{(4)}-
\frac{1}{2}(V_A^{(3)}+V_c)-\frac{1}{2}E_{\f}-\frac{1}{2}V^D_{\f}-V^0_{\f},
\eea
whereas in the Chern-Simons theory -- like
\bea
\label{sdd2}
\omega=2-\frac{1}{2}(E_A+E_{\f}).
\eea
Here the $E_A$ and $E_{\f}$ are the numbers of the external gauge and
scalar legs. 
Therefore we find that the SYM theory is super-renormalizable (and
finite beyond two loops) whereas the Chern-Simons theory is renormalizable.
Moreover, in the framework of the dimensional regularization both
theories are one-loop finite (since the integral
$\int\frac{d^3k}{k^2+m^2}$ is finite within this approach, and the one-loop logarithmic divergences in the odd-dimensional theories vanish by symmetry reasons, except of the cases of the very specific effective theories where the propagators proportional to $\frac{1}{\sqrt{k^2}}$ are possible, the typical examples of such theory are the nonlinear sigma model and the $CP^{N-1}$ model \cite{ours}).

The contributions from the supergraphs are evaluated with help of the
D-algebra transformations whose aim consists in the reduction of the
contribution to the supergraph to the single integral over
$d^2\theta$ (the possibility of this reduction can be treated as a some kind of the ``nonrenormalization theorem'' originally introduced in the four-dimensional case \cite{West}). 
These transformations are based on Leibnitz rule and use 
identities (\ref{delta},\ref{delta1}) and the similar ones. We note
that for tadpole graphs we cannot put $D^2\delta(0)=0$, really the
expression of this type is treated as $D^2\delta(0)\equiv
D^2\delta(\theta_1-\theta_2)|_{\theta_1=\theta_2}=1$. We also use the
relation
\bea
D_{\a}(\theta_1,k)\delta(\theta_1-\theta_2)=-D_{\a}(\theta_2,-k)
\delta(\theta_1-\theta_2).
\eea

Let us give the typical example of the D-algebra transformations 
\cite{ours}.
The matter contribution to the two-point function spinor field
$A_{\a}$ arising in the model (\ref{matterint}) is formed by the two diagrams shown in Fig. 1.

\vspace*{2mm}

\hspace{4.5cm}
\Lengthunit=1cm
%\Linewidth{1.2pt}
\GRAPH(hsize=3){\Circle(2)\mov(-1,0){\wavelin(-1,0)}\mov(1,0){\wavelin(1,0)}\ind(0,-15){a}
\mov(4.5,0){\Circle(2)\mov(-1,-1){\wavelin(2,0)} }
\ind(45,-15){b} \ind(20,-20){Fig.1.}}

\vspace*{2mm}

The first graph, depicted in  Fig. 1$a$, gives the following contribution:
\begin{eqnarray}
iS_{1a}(p)&=&-\frac{1}{4}\int d^2\theta_1 d^2\theta_2 \int\frac{d^3k}{(2\pi)^3}A^{\a}(-p,
\theta_1)
A^{\b}(p,\theta_2)
\\&&\times\Big[
D_{\a 1}<\phi(-k,\theta_1)\bar{\phi}(k,\theta_2)>(<\bar{\phi}(k+p,\theta_1)\phi(-k-p,\theta_2)>\der_{\phantom a \b 2})\nonumber\\
&&-
(D_{\a 1}<\phi(-k,\theta_1)\bar{\phi}(k,\theta_2)>\der_{\phantom a \b 2})
<\bar{\phi}(k+p,\theta_1)\phi(-k-p,\theta_2)>
\Big],\nonumber
\end{eqnarray}
where the notation $D_{\gamma i}$ was used to indicate that the supercovariant derivative is applied to the field whose Grassmannian argument is $\theta_i$.
Taking into account the explicit form of the propagators, we have 
\begin{eqnarray}
iS_{1a}(p)&=& \frac{1}{4}\int d^2\theta_1 d^2\theta_2 \int\frac{d^3k}{(2\pi)^3}A^{\a}
(-p,\theta_1)
A^{\b}(p,\theta_2)\nonumber\\&\times&\Big[
\frac{D_{\a 1}(D^2_1+m)}{k^2+m^2}\delta_{12}
\frac{(D^2_1+m)D_{\b 2}}{(k+p)^2+m^2}\delta_{12}
\nonumber\\&-&
\frac{D_{\a 1}(D^2_1+m)D_{\b 2}}{k^2+m^2}\delta_{12}
\frac{D^2_1+m}{(k+p)^2+m^2}\delta_{12}
\Big].
\end{eqnarray}
Integrating by parts some of the spinor derivatives and by using the identity~$D_{\b 2}(k,\theta_2)\delta_{12}=-D_{\b 1}(-k,\theta_1)\delta_{12}$
we arrive at 
\begin{eqnarray}
\label{expr}
iS_{1a}(p)&=&\frac{1}{4}
\int d^2\theta_1 d^2\theta_2 \int\frac{d^3k}{(2\pi)^3}
I(k,p)
\nonumber\\&\times&\Big[
2(D^2_1+m)\delta_{12}
D_{\a 1}(D^2_1+m)D_{\b 1}\delta_{12}
A^{\a}(-p,\theta_1) A^{\b}(p,\theta_2)
\nonumber\\&+&
(D^2_1+m)\delta_{12} (D^2_1+m)D_{\b 1}\delta_{12}
(D^{\a}A_{\a})(-p,\theta_1) A^{\b}(p,\theta_2)
\Big].
\end{eqnarray}
where
\bea
I(k,p)=\frac{1}{(k^2+m^2)[(k+p)^2+m^2]}.
\eea

It is convenient to separate $S_{1a}$ into two parts,
$S_{1a}=S_{1a}^{(1)}+S_{1a}^{(2)}$, where $iS^{(1)}_{1a}$ and
$iS_{1a}^{(2)}$ are associated to the two terms in the large brackets
of (\ref{expr}).  Consider first $iS^{(1)}_{1a}$ which, after
transporting $D^2$ from one of the propagators to the other factors,
becomes
\begin{eqnarray}
iS^{(1)}_{1a}(p)&=&\frac{1}{4} \int d^2\theta_1 d^2\theta_2 \int\frac{d^3k}{(2\pi)^3} I(k,p)
\nonumber\\&\times&\Big[
2m\delta_{12}
D_{\a 1}(D^2_1+m)D_{\b 1}\delta_{12}
A^{\a}(-p,\theta_1) A^{\b}(p,\theta_2)
\nonumber\\&+&2\delta_{12}D^2_1\Big( D_{\a 1}(D^2_1+m)D_{\b 1}\delta_{12}
A^{\a}(-p,\theta_1)\Big) A^{\b}(p,\theta_2)
\Big].
\end{eqnarray}
Now we employ the identity $\{D_{\a 1}, D^2_1\}=0$ which leads to
\begin{eqnarray}
iS^{(1)}_{1a}(p)&=&\frac{1}{4}  \int d^2\theta_1 d^2\theta_2 \int\frac{d^3k}
{(2\pi)^3} I(k,p)
\\&\times&\Big[
2\delta_{12}(k^2+m^2)D_{\a 1}D_{\b 1}
\delta_{12}A^{\a}(-p,\theta_1) A^{\b}(p,\theta_2)\nonumber\\&+&
2\delta_{12}(-D^2_1+m)D_{\a 1}D_{\b 1}\delta_{12}
(D^2A^{\a}(-p,\theta_1)) A^{\b}(p,\theta_2)
\Big].\nonumber
\end{eqnarray}
The use of  the relationship (\ref{key}) now provides
\begin{eqnarray}
iS^{(1)}_{1a}(p)&=&\frac{1}{2} \int d^2\theta_1 d^2\theta_2 \int\frac{d^3k}{(2\pi)^3} 
I(k,p)
\nonumber\\&\times&\Big[
\delta_{12}(k^2+m^2)(k_{\a\b}-C_{\a\b}D^2)
\delta_{12}A^{\a}(-p,\theta_1) A^{\b}(p,\theta_2)
\nonumber\\&+&
\delta_{12}(-D^2+m)(k_{\a\b}-C_{\a\b}D^2)\delta_{12}
(D^2A^{\a}(-p,\theta_1)) A^{\b}(p,\theta_2)
\Big].
\end{eqnarray}

The only terms giving non-zero contributions are those containing just
one $D^2$ since $\delta_{12}D^2\delta_{12}=\delta_{12}$.  Indeed, by
employing this identity and after integrating over $\theta_2$ with the
help of the delta function, we obtain
\begin{eqnarray}\label{1a}
iS^{(1)}_{1a}(p)&=&-\frac{1}{2} \int d^2\q  \int\frac{d^3k}{(2\pi)^3} I(k,p)
\\&\times&\Big[
(k^2+m^2)C_{\a\b} A^{\a}(-p,\q)A^{\b}(p,\q)
+(k_{\a\b}+mC_{\a\b})(D^2A^{\a}(-p,\q)) A^{\b}(p,\q)
\Big].\nonumber
\end{eqnarray}

The second term of (\ref{expr}) is
\begin{eqnarray}
iS^{(2)}_{1a}(p)&=&\frac{1}{4} \int d^2\theta_1 d^2\theta_2 \int\frac{d^3k}{(2\pi)^3}
I(k,p)
\nonumber\\&\times&\Big[
(D^2_1+m)\delta_{12} (D^2_1+m)D_{\b 1}\delta_{12}
(D^{\a}A_{\a})(-p,\theta_1) A^{\b}(p,\theta_2)
\Big].
\end{eqnarray}
In this expression we must keep only the  term proportional to 
$D^2_1\delta_{12}(D^2_1+m)D_{\b1}\delta_{12}$ (the remaining part is a
trace of an  odd number of derivatives and therefore vanishes).
Thus, after manipulations similar to those done for 
$S^{(1)}_{1a}$, we find
\begin{eqnarray}
iS^{(2)}_{1a}(p)=-\frac{1}{4}\int d^2\q  \int\frac{d^3k}{(2\pi)^3} I(k,p)
\Big[D^{\gamma}D^{\a}A_{\a}(-p,\q)
(k_{\gamma\b}+mC_{\gamma\b})A^{\b}(p,\q)
\Big].\label{23d}
\end{eqnarray}
\noindent
By adding (\ref{1a}) and (\ref{23d}) we can write the total contribution from  
Fig. 1$a$  as
\begin{eqnarray}
\label{s10}
iS_{1a}(p)&=&-\frac{1}{2}\int d^2\q \int \frac{d^3k}{(2\pi)^3} 
I(k,p)
\nonumber\\&\times&\Big[
(k^2+m^2)C_{\a\b} A^{\a}(-p,\q)A^{\b}(p,\q)
+(k_{\a\b}+mC_{\a\b})(D^2A^{\a}(-p,\q)) A^{\b}(p,\q)
\nonumber\\&+&\hf D^{\gamma}D^{\a}A_{\a}
(-p,\theta)(k_{\gamma\b}+mC_{\gamma\b})A^{\b}(p,\q)
\Big].
\end{eqnarray}

The algebraic manipulations for the graph  Fig.1b are much more 
simpler and yield
 \begin{eqnarray}
\label{s20}
iS_{1b}(p)&=&\frac{1}{2}\int\frac{d^3k}{(2\pi)^3}\frac{1}{(k+p)^2+m^2}
C_{\a\b} A^{\a}(-p,\q)A^{\b}(p,\q).
\end{eqnarray}

The complete two point vertex  function for the $A_\alpha$ field is the sum of (\ref{s10}) and (\ref{s20}) and
therefore reads
\begin{eqnarray}
\label{stot}
iS_1(p)&=&-\frac{1}{2} \int d^2\q \int \frac{d^3k}{(2\pi)^3} 
\frac{1}{(k^2+m^2)[(k+p)^2+m^2]}(k_{\gamma\beta}+mC_{\gamma\beta})
\nonumber\\&\times&
\Big[(D^2A^{\gamma}(-p,\q)) A^{\b}(p,\q)
+\hf D^{\gamma}D^{\a}A_{\a}(-p,\q) A^{\b}(p,\q)
\Big].
\end{eqnarray}

One can observe that the linear divergences presented in $S_{1a}$ and $S_{1b}$ were cancelled in
the above result, and the logarithmic ones vanish by symmetry reasons, therefore this contribution is finite. However, this finiteness is based on the gauge symmetry rather than on the supersymmetry. Indeed, after integration over the internal momentum we find that this expression is proportional to the gauge-invariant expression
\bea
iS_1(p)=\int d^2\theta f(p)(W^{\alpha}(-p)W_{\alpha}(p)+2mW^{\alpha}(p)A_{\alpha}(p)),
\eea
where 
$$f(p)=\int \frac{d^3k}{(2\pi)^3} 
\frac{1}{(k^2+m^2)[(k+p)^2+m^2]}.$$ 
This expression, at $p\to 0$ (notice that $f(p)|_{p\to 0}=\frac{1}{8\pi|m|}$), reproduces the expression for the quadratic Maxwell-Chern-Simons action. As a result, even if we consider the spinor $A^{\alpha}$ superfield as a purely external one, it acquires a nontrivial dynamics due to the one-loop correction. This is a key effect which also takes place for the supersymmetric $CP^{N-1}$ model explicitly studied in \cite{cpn} in the commutative case, and in \cite{ours} in the noncommutative one. We note that the only difference of the noncommutative case consists in the modification of the factor $f(p)$, see (\ref{stot1}).

\section[Effective action]{Effective action of the three-dimensional superfield theories and the proper-time method}

In this section we develop a prescription for calculating the superfield effective action within the three-dimensional superfield formalism. Our description of this approach is based on the method proposed in \cite{ourep3}. 

We start with the following three-dimensional superfield theory which is described by a general scalar superfield action (see f.e.~\cite{SGRS}):
\begin{eqnarray}
\label{sfa}
S[\Phi]=\int d^5z \left[\frac{1}{2}\Phi D^2\Phi-V(\Phi)\right],
\end{eqnarray}
where $\Phi$ is a scalar superfield. We will use again the loop expansion methodology. To do it, we make a shift in the field $\Phi$,
\begin{eqnarray}
\Phi\to\Phi+\sqrt{\hbar} \, \phi,
\end{eqnarray}
where $\Phi$ is now a background (super)field, and $\phi$ is a quantum one, which is contracted into propagators.
As a result, the classical action (\ref{sfa}) takes the form
\begin{eqnarray}
\label{2}
S[\Phi,\phi]&=&S[\Phi]+\int d^5z\left(\hbar\frac{1}{2}\phi[D^2-V^{\prime\prime}(\Phi)]\phi-\hbar^{3/2}\frac{1}{3!}V^{\prime\prime\prime}(\Phi)\phi^{3}-
\hbar^2\frac{1}{4!}V^{(IV)}(\Phi)\phi^{4}
\right)\nonumber\\
&&+\ldots,
\end{eqnarray}
where dots are for terms which are irrelevant in the two-loop approximation. 
Following the definitions given in the Chapter 2, we can find that the corresponding effective action $\Gamma[ \Phi]$ is defined by the expression 
\begin{eqnarray}
\exp\left(\frac{i}{\hbar}\Gamma[\Phi]\right)\,= \,\int D\phi \, \exp\left(\frac{i}{\hbar}S[\Phi,\phi]\right).
\end{eqnarray}
The general structure of the effective action can be cast in a form of the derivative expansion:
\begin{eqnarray}
\Gamma[\Phi]=\int d^5z \, K(\Phi)+\int d^5z \, F(D_{\alpha}\Phi,D^2\Phi;\Phi)+\ldots,
\end{eqnarray}
where the $K(\Phi)$ is the term which we call the K\"{a}hlerian effective potential by analogy with four-dimensional studies (see the next chapter), depending only on the superfield $\Phi$ but not on its derivatives, and $F$ is called auxiliary fields effective potential whose key property is its vanishing in the case when all derivatives of the superfields are equal to zero (these definitions have been firstly introduced in \cite{Buch1} for the four-dimensional superfield theories), and dots are for terms involving space-time derivatives of superfields. It is easy to see that $F$ is at least of the second order in the auxiliary field of the scalar supermultiplet. 
It can be explicitly written as
\begin{eqnarray}
F(D_{\alpha}\Phi,D^2\Phi;\Phi)\,=\,F_{2.1}(\Phi)D^2\Phi+F_{2.2}(\Phi)D^{\alpha}\Phi D_{\alpha}\Phi+\ldots,
\end{eqnarray}
where the $F_{2.1}(\Phi)$ and $F_{2.2}(\Phi)$ are functions of $\Phi$ only but not of its derivatives, and the dots correspond to terms with four or more supercovariant derivatives. It is clear that this approach does not require to impose the condition $D_{\alpha}\Phi=0$ which is known to imply in strong difficulties in the interpretation of the results (see f.e. \cite{Burgess}). 

We will work with a loop expansion for the effective action $\Gamma$, 
\begin{eqnarray}
\label{4}
\Gamma[ \Phi]=S[ \Phi]+\hbar \Gamma^{(1)}[ \Phi]+\hbar^2\Gamma^{(2)}[ \Phi]+\ldots,
\end{eqnarray}
for the K\"ahlerian potential $K$,
\begin{eqnarray}
K(\Phi)=-V(\Phi)+\sum_{L=1}^{\infty}\hbar^LK_L(\Phi),
\end{eqnarray}
and similarly for $F$. 

We start by considering the one-loop effective action in the form
\begin{eqnarray}
\label{g1}
\Gamma^{(1)}=\frac{i}{2}{\rm Tr}\ln[D^2-V^{\prime\prime}(\Phi)].
\end{eqnarray}
As a first approximation, let us consider the K\"{a}hlerian effective action. From a formal viewpoint this corresponds to disregarding all terms depending on derivatives of $\Phi$ (both common and spinor ones), and allows us to calculate the quantum corrections to $V(\Phi)$. In this case, we add a constant equal to $\frac{i}{2}{\rm Tr}\ln (D^2)$ and write
\begin{eqnarray}
\label{g12}
\Gamma^{(1)}=\frac{i}{2}{\rm Tr}\ln[\Box-V^{\prime\prime}(\Phi)D^2].
\end{eqnarray}
This expression can be represented via the Schwinger proper-time representation \cite{Ojima,ourpt}:
\begin{eqnarray}
\Gamma^{(1)}&=&\frac{i}{2}{\rm Tr}\int_0^{\infty}\frac{ds}{s}e^{is[\Box-V^{\prime\prime}(\Phi)D^2]} \nonumber\\
&=&\frac{i}{2}\int d^5z\int_0^{\infty}\frac{ds}{s}e^{is[\Box-V^{\prime\prime}(\Phi)D^2]}\delta^5(z-z')|_{z=z'}\,.
\end{eqnarray}
Again, since we are calculating only the K\"{a}hlerian part of the effective action, we have
\begin{eqnarray}
\label{gamma1}
\Gamma^{(1)}= \frac{i}{2} {\rm Tr}\int d^5z\int_0^{\infty}\frac{ds}{s}e^{-isV^{\prime\prime}(\Phi)D^2}e^{is\Box}\delta^5(z-z')|_{z=z'}\,,
\end{eqnarray}
or, using that $(D^2)^2=\Box$, 
\begin{eqnarray}
\label{expa}
e^{-isV^{\prime\prime}(\Phi)D^2}=\sum_{n=0}^{\infty}\frac{\left[-isV^{\prime\prime}(\Phi)\right]^{2n+1}}{(2n+1)!}\Box^nD^2+\ldots.
\end{eqnarray}
Here the dots stand for terms which do not contribute to the integral. 
At this point, we can clearly emphasize the difference between the calculation of $\Gamma^{(1)}$ in four- and three-dimensional space-times. In four dimensions~\cite{BK0,WZ}, $\Gamma^{(1)}$ is given by an expression similar to Eq.~(\ref{gamma1}), but there are more independent structures involving supercovariant derivatives and chiral and antichiral background superfields. The calculation of the exponential similar to Eq.~(\ref{expa}) involves the solving of a coupled set of differential equations, whose solutions can be found but are of rather cumbersome form. In three dimensions the number of independent structures is much smaller, actually only terms involving a $D^2$ will be relevant to the calculation of the K\"{a}hlerian effective action. We will shortly show that these terms can be directly summed, thus providing a closed-form expression for $\Gamma^{(1)}$.

Let us now consider a function $U(x,x';s)=e^{is\Box}\delta^3(x-x')$. Its key property is that
\begin{eqnarray}
i\frac{\partial U}{\partial s}=-\Box U,
\end{eqnarray}
which allows us to obtain
\begin{eqnarray}
&&\Box^n U(x,x';s)|_{x=x'}\equiv \Box^ne^{is\Box}\delta^3(x-x')|_{x=x'}=\frac{\sqrt{i}}{8\pi^{3/2}}\left(-i\frac{d}{ds}\right)^n\frac{1}{s^{3/2}}=\nonumber\\&=&\frac{i^{n+1/2}}{8\pi^{3/2}}\frac{(2n+1)!!}{2^ns^{3/2+n}}.
\end{eqnarray}
From Eq.~(\ref{expa}), after calculating the trace using that $D^2\delta^2(\theta-\theta')|_{z=z'}=1$ and $\frac{(2n+1)!!}{(2n+1)!}=\frac{1}{(2n)!!}=\frac{1}{2^nn!}$, we obtain
\begin{eqnarray}
\Gamma^{(1)}=\frac{i}{16\pi^{3/2}}
\int d^5z\int_0^{\infty}\frac{ds}{s}\sum_{n=0}^{\infty}\frac{\left[-i\sqrt{i}\,V^{\prime\prime}(\Phi)\right]^{2n+1}}{4^nn!}s^{n-1/2}.
\end{eqnarray}
By performing the summation, we end up with
\begin{eqnarray}\label{eq13d}
\Gamma^{(1)}=-\frac{i\sqrt{i}}{16\pi^{3/2}}\int d^5z V^{\prime\prime}(\Phi)\int_0^{\infty}\frac{ds}{s^{3/2}}e^{-is[\frac{(V^{\prime\prime}(\Phi))^2}{4}]}.
\end{eqnarray}
After an appropriate analytic continuation, we find that (\ref{eq13d}) is proportional to a gamma function, and finally arrive at
\begin{eqnarray}
\label{finresd3ep}
\Gamma^{(1)}=\frac{1}{16\pi}\int d^5z \left[V^{\prime\prime}(\Phi)\right]^2.
\end{eqnarray}
This is our final expression for the one-loop K\"ahlerian effective action. It is positively defined, as it should be in a supersymmetric theory. We note its finiteness, and we can also observe that this expression holds also in the noncommutative case. Indeed, since the background superfield is constant in the space-time, the Moyal product of these superfields reduces to the usual one (in the higher loops, this is not so -- the noncommutative deformation of the two-loop effective potential is discussed in \cite{ourep3}). It is interesting to note that this result, up to a constant multiplier, can be obtained without calculations. Indeed, it follows already from (\ref{g1}) that the one-loop effective potential is a function of $V^{\prime\prime}(\Phi)$ only. Since our result is naturally finite (in three-dimensional theories the one-loop results diverge only for exotic effective dynamics), it cannot depend on any arbitrary parameter like the cutoff scale $\mu$. Therefore, by dimensional reasons, the only form of the effective potential is $a\left[V^{\prime\prime}(\Phi)\right]^2$, where $a$ is a some number.

We note that in principle one could elaborate the expression (\ref{g1}), for the constant background, by a much more straightforward way. Indeed, if one denotes $-V^{\prime\prime}=\Psi$, the one-loop effective action becomes
$\Gamma^{(1)}=\frac{i}{2}{\rm Tr}\ln[D^2+\Psi]$, which is a function of $\Psi$. Then, one can consider
\bea
\frac{d\Gamma^{(1)}}{d\Psi}=\frac{i}{2}{\rm Tr}\frac{1}{D^2+\Psi},
\eea
which, after finding the inverse operator together with the Fourier transform, becomes
\bea
\frac{d\Gamma^{(1)}}{d\Psi}=-\frac{i}{2}\int d^5z\int\frac{d^3k}{(2\pi)^3}\frac{D^2-\Psi}{k^2+\Psi^2}\delta^5(z-z')|_{z=z'}.
\eea
The D-algebra is trivial, and after Wick rotation we have
\bea
\frac{d\Gamma^{(1)}}{d\Psi}=-\frac{1}{2}\int d^5z\int\frac{d^3k}{(2\pi)^3}\frac{1}{k^2+\Psi^2}=\int d^5z\frac{|\Psi|}{8\pi},
\eea
Integrating this equation with respect to $\Psi$, we arrive at the expression (\ref{finresd3ep}) obtained above. Nevertheless, we remind that the proper-time method we formulated can be applied for a wide class of superfield theory models, and its significance is not exhausted by scalar superfield theory considered here, further in this section we discuss the case of contributions depending on derivatives of background superfields. 

Now, let us go to two loops. Applying the expansion given in Eqs.~(\ref{2}--\ref{4}), we can find that the expression for the two-loop effective action $\Gamma^{(2)}$ looks like,
\begin{eqnarray}
\Gamma^{(2)}=-i\int D\phi \exp\left(\frac{i}{2}\phi[D^2-V^{\prime\prime}(\Phi)]\phi\right)
\left[\frac{1}{2}\left(\frac{1}{3!}V^{\prime\prime\prime}(\Phi)\phi^{3}\right)^2-
\frac{1}{4!}V^{(IV)}(\Phi)\phi^{4}\right].
\end{eqnarray}

The two-loop contributions are given by two supergraphs,

%\Lengthunit=1mm
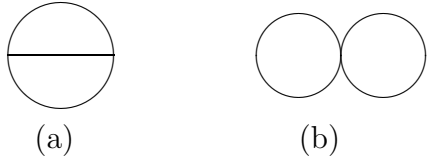
\begin{center}
\begin{picture}(200,60)
\put(30,30){\circle{40}}
\put(10,30){\line(1,0){40}}
\put(20,-5){(a)}
\put(120,30){\circle{30}}
\put(152,30){\circle{30}}
\put(120,-5){(b)}
\end{picture}
\end{center}

Since we are interested in calculating the two-loop contribution to the K\"ahlerian effective action, we can effectively assume that $D^\alpha \Phi = 0$, so that the background field dependent mass $\Psi=-V^{\prime\prime}(\Phi)$ is independent of $\theta$, thus the simple propagator
\begin{eqnarray}
<\phi(z_1)\phi(z_2)>=-i\frac{D^2-\Psi}{\Box-\Psi^2}\,\delta^5 (z_1 - z_2),
\end{eqnarray}
can be used. We also remind that the background superfield is constant. 

Thus, the contributions from diagram (a) and (b) respectively, after trivial D-algebra transformations, look like
\begin{eqnarray}
\Gamma^{(2)}_{a}&=&-\frac{1}{4}\int d^5z\,
\left[V^{\prime\prime\prime}(\Phi)\right]^2\Psi\int\frac{d^3kd^3l}{(2\pi)^6}\frac{1}
{(k^2+\Psi^2)(l^2+\Psi^2)[(k+l)^2+\Psi^2]},
\end{eqnarray}
and
\begin{eqnarray}
\Gamma^{(2)}_{b}&=&-\frac{1}{4}\int d^5z \, V^{(IV)}(\Phi)\int\frac{d^3kd^3l}{(2\pi)^6}\frac{1}
{(k^2+\Psi^2)(l^2+\Psi^2)}.
\end{eqnarray}

After integration, these expressions look like
\begin{eqnarray}
\label{g1ac}
\Gamma_{a}^{(2)}&=&\frac{1}{128\pi^2}\int d^5z\, \left[V^{\prime\prime\prime}(\Phi)\right]^2\Psi
\left[
\frac{1}{\epsilon}+\ln\frac{\Psi^2}{\mu^2}\right].
\end{eqnarray}
and
\begin{eqnarray}
\label{g2c}
\Gamma_{b}^{(2)}&=&-\frac{1}{32\pi^2}\int d^5z \, V^{(IV)}(\Phi)\Psi^2.
\end{eqnarray}
We can add the corresponding counterterm to cancel the divergence.

Up to now, we have considered only the K\"ahlerian effective action. 
Let us describe the general procedure to obtain the one-loop effective potential taking into account the supercovariant derivatives of the background superfield. As we have already noticed, the one-loop effective action (\ref{g12}), up to the irrelevant additive constant, reads
\begin{eqnarray}
\Gamma^{(1)}=\frac{i}{2}{\rm Tr}\ln(D^2+\Psi).
\end{eqnarray}
Using the Schwinger representation, we can write this effective action as
\begin{eqnarray}\label{eq:1}
\Gamma^{(1)}=\frac{i}{2}\int d^5z\int\frac{ds}{s}e^{is(D^2+\Psi)}\delta^5(z-z')|_{z=z'}.
\end{eqnarray}
We then introduce the operator
\begin{eqnarray}
\label{expo}
\Omega(s)=e^{is(D^2+\Psi)}\,,
\end{eqnarray}
which can be expanded in a power series in the supercovariant derivatives as
\begin{eqnarray}\label{Oexp}
\Omega(s)=1+c_0(s)+c_1^{\alpha}(s)D_{\alpha}+c_2(s)D^2~.
\end{eqnarray}
We note that higher degrees of the spinor derivatives can be reduced to the structures  which are already present in Eq.~(\ref{Oexp}) by using the rules $D_{\alpha}D_{\beta}=i\partial_{\alpha\beta}-C_{\alpha\beta}D^2$, $(D^2)^2=\Box$ and $D_{\alpha}D^2=-i\partial_{\alpha\beta}D^{\beta}$. The coefficient functions $c_0,c_1,c_2$ depend analytically on $s$, the superfield $\Psi$ and its supercovariant derivatives, and the space-time derivatives $\partial_{\alpha\beta}$, which act on the delta function appearing in Eq.~(\ref{eq:1}).

The operator $\Omega(s)$ satisfies the differential equation
\begin{eqnarray}
\frac{1}{i}\frac{d\Omega}{ds}=\Omega(D^2+\Psi)\,.
\end{eqnarray}
Substituting here the explicit form for $\Omega(s)$ in Eq.~(\ref{expo}), we obtain a coupled set of differential equations for the coefficient functions $c_0,c_1,c_2$,
\begin{eqnarray}
\label{sys3d}
\frac{1}{i}\frac{dc_0}{ds}&=&c_0\Psi+c_2(\Box+D^2\Psi)+c_1^{\alpha}(D_{\alpha}\Psi)+\Psi,\nonumber\\
\frac{1}{i}\frac{dc_1^{\alpha}}{ds}&=&-ic_{1\gamma}\partial^{\gamma\alpha}+c_1^{\alpha}\Psi+c_2D^{\alpha}\Psi,\nonumber\\
\frac{1}{i}\frac{dc_2}{ds}&=&c_0+c_2\Psi+1.
\end{eqnarray}

As $\Omega(s=0)=1$, the initial conditions are $c_0(0)=c_1^{\alpha}(0)=c_2(0)=0$. Since this is a linear inhomogeneous system of differential equations, the solution is of the form $c_i(s)=b_ie^{i\omega s}+d_i$, where $b_i$ and $d_i$ are some $s$-independent coefficients. Substituting this ansatz into the equations (\ref{sys3d}), one finds for the solution of the homogeneous equation,
\begin{eqnarray}
\label{sys1}
(\omega-\Psi)b_0&=&b_2(\Box+D^2\Psi)+b_1^{\alpha}(D_{\alpha}\Psi),\nonumber\\
(\omega-\Psi)b_1^{\alpha}&=&-ib_{1\beta}\partial^{\beta\alpha}+b_2D^{\alpha}\Psi,\nonumber\\
(\omega-\Psi)b_2&=&b_0,
\end{eqnarray} 
and for the particular solution of the inhomogeneous one,
\begin{eqnarray}
&&d_0\Psi+d_2(\Box+D^2\Psi)+d_1^{\alpha}(D_{\alpha}\Psi)+\Psi=0,\nonumber\\
&&-id_{1\gamma}\partial^{\gamma\alpha}+d_1^{\alpha}\Psi+d_2D^{\alpha}\Psi=0,\nonumber\\
&&d_0+d_2\Psi+1=0.
\end{eqnarray}

The system~(\ref{sys1}), after some simplifications, implies in the following equation:
\begin{equation}
b_{1\gamma}\left[ \left( \omega-\Psi+\frac{1}{2}\frac{D^{\beta}\Psi D_{\beta}\Psi}{(\omega-\Psi)^2-\Box-(D^2\Psi)}\right) C^{\alpha\gamma}-i\partial^{\alpha\gamma}
\right]=0.
\end{equation}
Since $b_{1\gamma}\neq 0$ (otherwise the solution is trivial), the $\omega$'s can be found requiring the $2\times 2$ matrix $\Delta^{\gamma\alpha}$ defined as
\begin{equation}
\Delta^{\gamma\alpha}=\left(\omega-\Psi+\frac{1}{2}\frac{D^{\beta}\Psi D_{\beta}\Psi}{(\omega-\Psi)^2-\Box-(D^2\Psi)}\right)C^{\alpha\gamma}-i\partial^{\alpha\gamma}
\end{equation}
to have zero determinant. This condition is solvable in principle, but its finding is technically difficult, and we do not carry out the solution.  The corresponding results would be extremely complicated. 

Let us also briefly discuss the calculation of the one-loop effective potential via the more traditional method, that is, via summation of the supergraphs. In the simplest case of the purely scalar superfield theory, we can start with the expression (\ref{g1}) and expand it in the power series in the background field $\Psi=-V^{\prime\prime}(\Phi)$:
\begin{eqnarray}
\Gamma^{(1)}&=&\frac{i}{2}{\rm Tr}\ln[D^2+\Psi]=\frac{i}{2}{\rm Tr}\ln(D^2)+\frac{i}{2}{\rm Tr}\ln[1+\Psi D^2\Box^{-1}]=\nonumber\\
&=&
\frac{i}{2}{\rm Tr}\ln(D^2)+\frac{i}{2}{\rm Tr}\sum\limits_{n=1}^{\infty}\frac{(-1)^{n+1}}{n}[\Psi \frac{D^2}{\Box}]^n,
\end{eqnarray}
where we can evaluate the trace and calculate the sum, arriving at the same result (\ref{finresd3ep}). However, in the case of the scalar superfield coupled to the gauge one the situation is much more complicated. Let us consider for example the scalar QED whose action is
\bea
\label{scalQED}
S_m=\int d^5 z\left[ \frac{1}{2}(D^{\a}\Phi+i\Phi A^{\a})
(D_{\a}\bar{\Phi}-iA_{\a}\bar{\Phi})+\frac{1}{2g^2} W^{\a}W_{\a}\right].
\eea
First of all, one can find that the triple gauge-matter vertices in the one-loop order diagrams arise within the following fragments of the corresponding diagrams:

\vspace*{3mm}

\hspace{6.0cm}
\Lengthunit=2cm
\GRAPH(hsize=2){\lin(-.5,0)\lin(0,1)\wavelin(2,0)\ind(-1,0){|}\ind(1,0){|}\ind(-2,3){D_{\alpha}}\ind(8,3){D^{\beta}D^{\alpha}+\xi D^{\alpha}D^{\beta}}\mov(2,0){\lin(0,1)\lin(1,0)}\ind(0,-2){\alpha}\ind(20,-2){\beta}
}

\vspace*{3mm}

Here the factor $D^{\beta}D^{\alpha}+\xi D^{\alpha}D^{\beta}$ originates from the propagator (there is no essential difference between the QED and Chern-Simons theories in this aspect since this factor commutes with $D^2$). The indices $\alpha$ and $\beta$ are the indices of gauge fields contracted into this propagator. We see that if we move the derivative $D_{\alpha}$, originated from the interaction vertex, to the propagator of the gauge fields proportional to $D^{\beta}D^{\alpha}+\xi D^{\alpha}D^{\beta}$, we will annihilate its gauge-independent part, and the gauge-dependent part vanishes at $\xi=0$. Thus, we impose the gauge $\xi=0$ (an analogue of the Landau gauge \cite{SGRS}) to simplify radically the calculations, i.e. to remove from considerations all diagrams with "improper" vertices $(\Phi A^{\alpha}D_{\alpha}\bar{\phi}-\bar{\Phi}A^{\alpha}D_{\alpha}\phi)$, where the external field is denoted via the capital letter, that is, triple ones contributing only to the quadratic part (we will refer to the vertices as to the "proper" one if it contains three and more quantum fields since such a vertex cannot be absorbed into redefinition of the propagator in the framework of the background field approach, otherwise, if the vertex involves only two quantum fields, it is qualified as an "improper" one). As a result, all triple vertices are ruled out (the similar situation occurs in Landau gauge also in other three-dimensional gauge theories with a scalar matter), and hence all vertices in the one-loop diagrams are quartic ones, and the one-loop diagrams composed only of the gauge propagators. However, in the QED, the gauge propagator (\ref{propqed}) involves four derivatives, hence one-loop diagrams with $n$ propagators will contain $4n$ spinor derivatives, and the expression involving $4n$ spinor derivatives turns out to be proportional to $(D^2D^2)^n$ whose trace is evidently equal to zero. Therefore, we have just proved that the one-loop K\"{a}hlerian effective potential in the scalar QED vanishes in the Landau gauge.

In this subsection, we developed a superfield method for calculation of the effective potential in three-dimensional supersymmetric field theories. We succeeded to obtain explicit expressions for the K\"ahlerian effective potential (which depends on superfield $\Phi$ but not on its derivatives) up to two loops. In principle, our approach can be directly generalized for higher loops. We have also shown the approach for the much more difficult calculation of the non-K\"ahlerian contributions to the effective action and the effective potential. The generalization for the noncommutative case is straightforward (it is clear that at the one-loop order, the result for the effective potential in commutative and noncommutative cases is the same, the difference begins with the two-loop order, see \cite{ourep3} for the examples).

\section[Noncommutativity]{Supersymmetry in three-dimensional space-time and noncommutativity}

The concept of the space-time noncommutativity has attracted a great attention recently. One of the principal motivations for it is the hope that introducing the noncommutative coordinates obeying the relation \cite{Snyder}
\bea
\label{sny}
[x^{\mu},x^{\nu}]=i\Theta^{\mu\nu},
\eea
where $\Theta^{\mu\nu}$ is a constant matrix, can essentially improve the renormalization properties of the field theories. The most adequate formulation \cite{Filk} allowing to implement these relations within the framework of the quantum field theory is based on the replacement of the usual product of the fields by their Moyal product defined as
\bea
\label{moyal}
&&\phi_1(x)*\ldots\phi_n(x)=\nonumber\\&=&\prod\limits_{l=1}^n\int\frac{d^dk_i}{(2\pi)^d}(2\pi)^d e^{i(k_1+\ldots+k_n)x}\tilde{\phi}_1(k_1)\ldots\tilde{\phi}_n(k_n)\exp(i\sum\limits_{i<j\leq n}k_i\wedge k_j),
\eea
where $k_i\wedge k_j=\Theta^{\mu\nu}k_{i\mu}k_{j\nu}$.

However, it turns to be that the famous UV/IR mixing effect \cite{Minw} implies in the partial conversion of the ultraviolet divergences to the infrared singularities. Indeed, if one considers the three-dimensional (nonsupersymmetric) noncommutative $\phi^4$ model, the following simplest contribution to the two-point function of the $\phi$ field will arise

\vspace*{2mm}

\hspace{5cm}
\Lengthunit=1cm
%\Linewidth{1.2pt}
\GRAPH(hsize=3){\Circle(2)\mov(-1.5,-1){\lin(3,0)}%\ind(0,-15){a}
}

\vspace*{2mm}

The contribution of this graph in $d$ space-time dimensions is (see f.e. \cite{Ar,Alv})
\bea
S_2(p)&=&\frac{\lambda}{6}\int\frac{d^dk}{(2\pi)^d}\frac{2+\cos(2k\wedge p)}{k^2+m^2}\phi(-p)\phi(p)=-\frac{\lambda}{3}[\frac{\Gamma(1-d/2)}{(4\pi)^{d/2}(m^2)^{1-d/2}}+\nonumber\\&+&
\frac{1}{2(2\pi)^{d/2}}(m^2)^{d/2-1}\frac{K_{d/2-1}(\sqrt{m^2\tilde{p}^2})}{(\sqrt{m^2\tilde{p}^2})^{d/2-1}}],
\eea
where $\tilde{p}^m=\Theta^{mn}p_n$.
We see that the first term of this expression is similar to the common UV divergent term different from the commutative case only by the overall numerical coefficient (this difference, from a formal viewpoint, can be explained by the fact that, when the noncommutativity is introduced, some of the contractions of the fields continue to be planar ones, i.e. do not acquire the phase factor), while the second one is singular, and, since $K_{|n|}(x)|_{x\to 0}\propto \frac{1}{x^n}$, we arrive at the infrared singularity of the order $d-2$ in the external momentum $p_m$, in particular, at $d=3$ we obtain the linear singularity, and at $d=4$ -- the quadratic one. A remarkable fact that this singularity presents in the massive case as well as in the massless one where the typical momentum integral looks like
\bea
\int\frac{d^dk}{(2\pi)^d}\frac{\cos(2k\wedge p)}{k^2}=\frac{\Gamma(d/2-1)}{(4\pi)^{d/2}(\tilde{p}^2)^{d/2-1}},
\eea
which displays the same infrared singularities.

The natural hope to solve this problem was related with the well-known fact that the supersymmetry improves the renormalization behaviour of the field theories eliminating some ultraviolet divergences \cite{miracle}. Therefore, it is natural to expect that the situation with the infrared singularities can also be cured, at least partially. Moreover, since the commutation relation (\ref{sny}) affects only the bosonic coordinates, it is natural to expect that introducing the Moyal product into superfield theories will not affect the supersymmetry algebra, therefore the noncommutative extension of the supersymmetric field theories formulated in the superfield language will be straightforward! This idea was originally proposed in \cite{Popp} for the four-dimensional Wess-Zumino model, however, it is clear that this idea can be straightforwardly applied also to the three-dimensional superspace. The first consistent three-dimensional example of the noncommutative supersymmetric field theory is the nonlinear supersymmetric sigma model studied within the component approach in \cite{sigcomp}. It was shown in that paper that while both the $O(N)$ noncommutative nonlinear sigma model and the noncommutative $O(N)$ Gross-Neveu model within the $\frac{1}{N}$ expansion display the nonintegrable (linear) infrared singularities, the supersymmetric noncommutative nonlinear sigma model formulated within the component approach involving both these models as ingredients displays the explicit cancellation of such singularities involving only harmless logarithmic infrared divergences. 

As a first example, let us study the scalar field coupled to the external gauge field. The action, for the adjoint form of the coupling, is (see f.e. \cite{cpn})
\bea
S=\int d^5z \Big[\frac{1}{2}(D^{\alpha}\bar{\phi}-ig[\bar{\phi},A^{\alpha}])*(D_{\alpha}\phi-ig[A_{\alpha},\phi])+m^2\phi\bar{\phi}\Big],
\eea
which is a straightforward analogue of the theory (\ref{matcoup}), but with the algebraic commutators replaced by the Moyal ones. Due to the Moyal commutators, the vertices will acquire the phase factors looking like
\bea
V_3&=&g\sin(k_2\wedge k_3)A^{\alpha}(k_1)(\phi(k_2)D_{\alpha}\bar{\phi}(k_3)-D_{\alpha}\phi(k_2)\bar{\phi}(k_3));\nonumber\\
V_4&=&2g^2\sin(k_1\wedge k_2)\sin(k_3\wedge k_4)\phi(k_1)A^{\alpha}(k_2)A_{\alpha}(k_3)\bar{\phi}(k_4).
\eea
The corresponding supergraphs are just those ones depicted at Fig. 1. The only modification of our calculations with respect to those ones carried out in section 3.4 will consist in arising the additional factor $4\sin^2(k\wedge p)$ in all contributions, as a result, the final expression for the two-point function of the gauge field will look like
\begin{eqnarray}
\label{stot1}
iS_3(p)&=&-2 \int d^2\q \int \frac{d^3k}{(2\pi)^3} 
\frac{\sin^2(k\wedge p)}{(k^2+m^2)[(k+p)^2+m^2]}
\\&\times&
(k_{\gamma\beta}+mC_{\gamma\beta})\Big[(D^2A^{\gamma}(-p,\q)) A^{\b}(p,\q)
+\hf D^{\gamma}D^{\a}A_{\a}(-p,\q) A^{\b}(p,\q)
\Big].\nonumber
\end{eqnarray}
This expression can be shown to imply in the action similar to the Maxwell-Chern-Simons form \cite{ours}, whose explicit form is 
\bea
iS_3(p)=\int d^2\theta f(p)(W^{\alpha}(-p)W_{\alpha}(p)+2mW^{\alpha}(p)A_{\alpha}(p)),
\eea
where 
$$f(p)=\int \frac{d^3k}{(2\pi)^3} 
\frac{\sin^2(k\wedge p)}{(k^2+m^2)[(k+p)^2+m^2]}.$$ 
This expression, in the case of the noncommutative supersymmetric $CP^{N-1}$ model \cite{ours}, implies in the nontrivial effective dynamics for the originally purely external $A^{\alpha}$ field. Indeed, if we consider, instead of one scalar superfield $\phi$, a set of $N$ scalar superfields $\phi_i$, after adding an appropriate gauge-fixing term, we find the following effective propagator for the $A_{\alpha}$ field:
\bea
<A_{\alpha}(-p,\theta_1)A_{\beta}(p,\theta_2)>=\frac{4\pi i}{Nf(p)}\left[\frac{(D^2-2m)D_{\beta}D_{\alpha}}{p^2(p^2+4m^2)}+\xi\frac{D^2D_{\alpha}D_{\beta}}{p^4}
\right]\delta_{12}.
\eea
This propagator decreases as $\frac{1}{p}$, at large momenta since $f(p)\simeq \frac{\pi}{\sqrt{p^2}}$ in this limit. This implies that the $CP^{N-1}$ theory is not finite but only renormalizable, in the lower order of $\frac{1}{N}$ expansion (and, probably, also in higher orders). The similar asymptotics of the effective propagator of the auxiliary $\Sigma$ field takes place in the supersymmetric nonlinear noncommutative sigma model \cite{sig} which is, however, renormalizable in all orders.

However, we should note that the cancellation of the linear singularity arising from these graphs (as well as from the graphs in the noncommutative supersymmetric QED, Chern-Simons and Maxwell Chern-Simons theories which will be considered below) occurs here due to the gauge symmetry rather that the supersymmetry. Moreover, the logarithmic divergences also vanish (however, we should note that, due to peculiarities of the odd-dimensional space, the one-loop logarithmic divergences can arise only in theories with highly unusual effective dynamics like the noncommutative sigma model \cite{sig} where one of the propagators is proportional to $\frac{1}{\sqrt{k^2}}$).

Further, the renormalizability of the three-dimensional supersymmetric noncommutative gauge theories was studied in great details \cite{ours}. Let us give a brief review on renormalizability of these theories. First, let us write down the following contributions to the two-point function of the gauge superfield.

%\vspace*{2mm}

\hspace{3cm}
\Lengthunit=1cm
%\Linewidth{1.2pt}
\GRAPH(hsize=3){\wavecirc(2)\mov(-1,0){\wavelin(-1,0)}\mov(1,0){\wavelin(1,0)}\ind(0,-15){a}
\mov(4.5,0){\wavecirc(2)\mov(-1,-1){\wavelin(2,0)} }\mov(9,0){\dashcirc(2)\mov(-1,0){\wavelin(-1,0)}\mov(1,0){\wavelin(1,0)}}
\ind(45,-15){b} \ind(90,-15){c}
}

%\vspace*{3mm}

These diagrams are applicable to the noncommutative supersymmetric QED as well as to the noncommutative supersymmetric Chern-Simons or Maxwell-Chern-Simons theories \cite{ours} whose actions can be obtained from the actions (\ref{sqed}) for the NC QED and (\ref{csna})  for the NC CS theories (and their sum for the NC Maxwell-Chern-Simons theory) in the section 3.4 by replacement of the algebraic products and commutators by the Moyal ones. The same operations must be carried out in the ghost action (\ref{gh}). Using the expressions for the superficial degree of divergence (\ref{sdd1}) and (\ref{sdd2}), one can show that there is no other linearly divergent graphs in the one-loop approximation (and, by the symmetry reasons, there is no one-loop logarithmic UV divergences as it should be in usual odd-dimensional field theories).

One can find that the leading, linearly divergent parts of these contributions, for all these theories, look like
\bea
S_a(p)&=&\frac{\xi}{2}\int d^2\theta A^{\alpha}(-p)A_{\alpha}(p)\int\frac{d^3k}{(2\pi)^3}\frac{\sin^2(k\wedge p)}{k^2};\nonumber\\
S_b(p)&=&\frac{1}{2}(1-\xi)\int d^2\theta A^{\alpha}(-p)A_{\alpha}(p)\int\frac{d^3k}{(2\pi)^3}\frac{\sin^2(k\wedge p)}{k^2};\nonumber\\
S_c(p)&=&-\frac{1}{2}\int d^2\theta A^{\alpha}(-p)A_{\alpha}(p)\int\frac{d^3k}{(2\pi)^3}\frac{\sin^2(k\wedge p)}{k^2}.
\eea
Sum of these contributions is equal to zero, hence the two-point function of the gauge superfield is one-loop finite and free of the UV/IR singularities (as for the possibility for logarithmic UV/IR singularities, all one-loop potentially logarithmically divergent contributions are proportional to $\frac{\tilde{p}^m}{\sqrt{\tilde{p}^2}}$, which is not a logarithmic divergence but a mild removable singularity which does not blow up, thus, these theories are one-loop finite, and NC supersymmetric QED and Maxwell-Chern-Simons theories are finite in three and higher loops explicitly as follows from their superficial degree of divergence, as we noted above).

\section[Fermionic noncommutativity]{On the noncommutativity in the fermionic sector of the superspace}

One of the possible extension of the noncommutativity concept is the deformation of the supersymmetry algebra carried out in the following way \cite{3dspace}.

Let us suppose that the spinor coordinates of the superspace, instead of the Grassmann algebra, form the Clifford algebra, thus obeying the following anticommutation relations:
\bea
\label{cliff}
\{ \theta^{\alpha},\theta^{\beta}\}=\Sigma^{\alpha\beta}.
\eea
It is clear that, to maintain the key relation of the supersymmetry algebra, that is, (\ref{anti}), we should deform the supersymmetry generators, which will imply in a need to deform the spinor supercovariant derivatives via introducing second-derivative terms into them, therefore the Leibnitz rule will not be satisfied. Therefore, we must try to extend the supersymmetry (indeed, in the four-dimensional case the similar deformation of the superspace had implied in a partial breaking of the supersymmetry \cite{Sei}). 

In the first way, we introduce an additional set of the supersymmetry generators thus considering the extended, $N=2$ supersymmetry, with the generators are 
\bea
Q_{\alpha}^i=i\partial^i_{\alpha}+\theta^{i\beta}\partial_{\beta\alpha}, 
\eea
with $i=1,2$ is a number of the set of spinor coordinates (and, hence -- of the supersymmetry generators). Then, we suppose that only in {\bf one} of the sets of the spinor coordinates, say $i=2$, the anticommutation relations are deformed in a manner (\ref{cliff}) while for $i=1$ they persist to be the same. Thus, only the unbroken generators $Q^1_{\alpha}$ satisfy the usual anticommutation relation $\{Q^1_{\alpha},Q^1_{\beta}\}=2i\partial_{\alpha\beta}$. The supercovariant derivatives anticommuting with them are
\bea
D_{\alpha}^i=\partial^i_{\alpha}+i\theta^{i\beta}\partial_{\beta\alpha}, 
\eea
with the only deformed anticommutation relation 
\bea
\{D^2_{\alpha},D^2_{\beta}\}=2i\partial_{\alpha\beta}-\Sigma^{\gamma\delta}\partial_{\alpha\gamma}\partial_{\beta\delta},
\eea
while all other anticommutation relations between the supercovariant derivatives persist to be the same.

The Moyal product compatible with this definition of the supersymmetry is defined as
\bea
\label{moy}
\Phi_1(z)*\Phi_2(z)=\exp(-\frac{1}{2}\Sigma^{\alpha\beta}D^2_{\alpha 1}D^2_{\beta 2})\Phi_1(z_1)\Phi_2(z_2)
|_{z_1=z_2=z}.
\eea
It is easy to see that in this case already the quadratic action will suffer noncommutative deformation which is a highly unusual situation (one should remind that for the usual bosonic Moyal product (\ref{moyal}), the quadratic action does not suffer any deformation, with the interaction vertices are deformed).

In the second way, we propose the following generators:
\bea
Q_{\alpha}^1&=&i\partial^1_{\alpha}+\theta^{2\beta}\partial_{\beta\alpha}, \nonumber\\
Q_{\alpha}^2&=&i\partial^2_{\alpha}+\theta^{1\beta}\partial_{\beta\alpha}.
\eea
Then, we again impose the nontrivial anticommutation relation $\{ \theta^{2\alpha},\theta^{2\beta}\}=\Sigma^{\alpha\beta}$. The Moyal product is again chosen in the form (\ref{moy}). It is easy to see that in this case the quadratic action is not deformed, and the only impact of the nontrivial anticommutation relations will present in the vertices of interaction, just as in \cite{Sei}. This approach has received further development in the case of ${\cal N}=2$ supersymmetry where the resulting SUSY algebra is rather similar to the four-dimensional ${\cal N}=1$ SUSY algebra \cite{GamaNC}.

\section{Conclusion}

In this chapter we gave a brief introduction to the properties of the three-dimensional supersymmetric field theories within the superfield approach. We described the structure of the most important three-dimensional supermultiplets, that is, scalar and spinor ones. We developed a detailed description of properties of the supercovariant derivatives and of the superfield approach for calculation of the quantum corrections, including the prescriptions for calculating the effective action in one-loop and higher-loop orders. The examples we presented show that supersymmetry indeed allows to improve essentially the renormalization properties of the field theories. Really, almost all supersymmetric field theories formulated in three-dimensional space-time are one-loop finite except of the exotic theories with an effective dynamics providing a nontrivial asymptotic behaviour of the effective propagators, such as, for example, the three-dimensional nonlinear sigma model \cite{sig} and $CP^{N-1}$ model \cite{ours}. Moreover, the three-dimensional supersymmetric QED is explicitly finite in all loop orders (it was argued in \cite{RR} and explicitly proved in \cite{oured}).

This improvement of the renormalization properties plays an important role in the context of the noncommutative field theories. Indeed, it is well known that the noncommutative theories suffer from arising of the nonintegrable infrared divergences due to the UV/IR mixing mechanism which converts part of the ultraviolet divergences to the infrared singularities whose presence in general case can break the perturbative expansion \cite{Minw}. At the same time, we have shown that the supersymmetry improves convergence of the Feynman diagrams leaving in many case only logarithmic UV divergences which in the noncommutative case are converted only to harmless logarithmic IR singularities whose presence does not break perturbative expansion. This is the key result shown in the papers \cite{ours}. At the same time, we have observed that the formalism of the supercovariant D-algebra is applicable in the noncommutative case as well as in the commutative one, for example, the calculations of the noncommutative analogue of the supergraphs studied in Section 3.4, differ only by multiplying of all contributions by the common phase factor, which is equal to 1 in the case of the coupling in the fundamental representation and $\frac{1}{4}\sin^2(k\wedge p)$, where $k\wedge p=k_m\Theta^{mn}p_n$, and $\Theta^{mn}$ is a noncommutativity matrix, in the adjoint representation case (see \cite{ours} for more details). 

We should note that neither the supersymmetric multiplets (scalar and spinor one) nor the actions presented in this section are not unique possible ones. The important example of the action for the spinor supersymmetric multiplet is the Dirac-like action whose properties were studied in \cite{ourdual}. Another important example of the multiplet and the corresponding action is the three-dimensional superfield supergravity action discussed in \cite{SGRS}. However, detailed perturbative study of this theory was not yet carried out.

Let us briefly discuss the main actual lines of studies in the three-dimensional supersymmetric theories. Presently, the main line of interest in the three-dimensional supersymmetry is related to the extended supersymmetric Chern-Simons theories, especially, $N=6$ and $N=8$ ones, which are known to be superconformal invariant and thus finite \cite{Aharony}. Another important line of study of these theories is based on the essentially $N=2$ supersymmetric description of such theories, whose properties are very similar to the usual $N=1$ supersymmetry in four-dimensional space-time \cite{BPl}.  Different aspects of the $N$-extended supersymmetric Chern-Simons theories are presented in \cite{Gaiotto}.

We close this chapter by mentioning of the noncommutative superspace where the anticommutation relations between fermionic superspace coordinates are the Clifford ones instead of the Grassmann ones. However, such a formulation seems to be realized only paying a price of a partial supersymmetry breaking, i.e. to proceed with this formulation in the three-dimensional space one should start with the $N$-extended supersymmetric theories, with further some of the supersymmetries are broken. A first attempt of developing such a formulation is presented in \cite{3dspace}, however, this issue certainly requires more study.

\newpage

\chapter{Four-dimensional superfield supersymmetry}

\section{General properties of the four-dimensional superspace}

Now, after we have described in details the superfield approach for the supersymmetric theories formulated in the three-dimensional space-time, let us go to the usual, four-dimensional case.

The essential difference of the four-dimensional space-time from the three-dimensional one consists in the fact that the Lorentz group $SO(1,3)$ characterizing the rotational symmetry of the space-time, possesses two mutually conjugated and linearly independent spinor representations, denoted as undotted and dotted one respectively. Each of these representations is realized by a group of the unimodular complex matrices $2\times 2$, that is, $SL(2,C)$, and two linear spaces on which ones these representations are acting are the spaces of undotted and dotted spinors $\psi^{\alpha}$, $\bar{\psi}^{\dot{\alpha}}$ respectively. It is easy to see that the invariant tensors of these two representations are the Levi-Civita symbols $\epsilon^{\alpha\beta}$, $\epsilon^{\dot{\alpha}\dot{\beta}}$. These tensors play the role of metric tensors for spinors and will be used to form the invariant scalar products of any spinors:
\bea
\psi^{\alpha}=\epsilon^{\alpha\beta}\psi_{\beta};\quad\, \psi_{\alpha}=\psi^{\beta}\epsilon_{\beta\alpha};\nonumber\\
\bar{\psi}_{\dot{\alpha}}=\epsilon_{\dot{\alpha}\dot{\beta}}\psi^{\dot{\beta}};\quad\, \bar{\psi}^{\dot{\alpha}}=\psi_{\dot{\beta}}\epsilon^{\dot{\beta}\dot{\alpha}}.
\eea
From a formal viewpoint, we can say that the Levi-Civita symbols ``act from a definite side" (these definitions have been used in the book \cite{BK0} and in most part of papers and will be used henceforth; note, however, that other definitions also can be introduced, they are used in some books, for example, in \cite{SGRS}). Unlike the previous section, in the four-dimensional case we will use the definitions of squares of spinors without $\frac{1}{2}$ factor, $\psi^2=\psi^{\alpha}\psi_{\alpha}$, which matches the conventions used in \cite{WB,BK0} and many other books and papers. It is easy to see that the indices $\alpha$, $\dot{\alpha}$ can take values 1,2 only. We have $\epsilon_{12}=\epsilon^{12}=\epsilon_{\dot{1}\dot{2}}=\epsilon^{\dot{1}\dot{2}}=1$ (we note that this definition differs from that one used in the previous chapter where one had $C_{12}=-C^{12}$).

As a next step, we introduce two types of the spinor (and therefore Grassmannian) variables $\theta^{\alpha}$, $\bar{\theta}^{\dot{\alpha}}$ which, together with the usual bosonic coordinates $x^a$ will be used as a coordinates on the new extended space which will be called the superspace. The spinors $\theta^{\alpha}$, $\bar{\theta}_{\dot{\alpha}}$ are mutually conjugated which is reasonable since they are transformed under mutually conjugated spinor representations of the Lorentz group. The conjugation is defined as
$(\theta^{\alpha})^{\dagger}=\bar{\theta}_{\dot{\alpha}}$, $(\theta^2)^{\dagger}=\bar{\theta}^2$. In the case of arbitrary spinors $\psi^{\alpha},\chi^{\beta}$ we can apply the following conjugation rule: $(\psi^{\alpha}\chi_{\alpha})^{\dagger}=\bar{\chi}_{\dot{\alpha}}\bar{\psi}^{\dot{\alpha}}$.

Therefore we can introduce the generalized coordinates $z^A=(x^a,\theta^{\alpha},\bar{\theta}^{\dot{\alpha}})$ parametrizing the superspace. In principle, it is possible to consider, instead of one set of Grassmann coordinates $(\theta^{\alpha},\bar{\theta}^{\dot{\alpha}})$, several sets of such coordinates. In the case where we have $N$ such sets, we can speak about the $N$-extended supersymmetry, where the Grassmann coordinates $(\theta^{i\alpha},\bar{\theta}^{i\dot{\alpha}})$ acquire additional index $i$ taking values from 1 to $N$. In the case of $N=2$, the superfield formalism is well formulated through the methodologies of the harmonic superspace \cite{GIKOS1,GIKOS2} and of the projective one \cite{projSSP}. The harmonic superspace formulation is also known in $N=3$ case \cite{Delduc}. However, in many cases the $N$-extended supersymmetric theories can be represented in terms of $N=1$ superfields, therefore we will mostly restrict ourselves to the $N=1$ supersymmetry case within this book. 

The supersymmetry transformations for coordinates are
\bea
\delta\theta^{\a}=\epsilon^{\a};\ \delta\bar{\theta}_{\ad}=\epsilon_{\ad};\ 
\delta x^a=-\epsilon\sigma^a\bar{\theta}+\bar{\epsilon}\sigma^a \theta.
\eea 
Here $\epsilon^{\a},\bar{\epsilon}^{\ad}$ are fermionic parameters. We use the definitions
$\epsilon\sigma^a\bar{\theta}\equiv\epsilon^{\alpha}\sigma^a_{\alpha\dot{\alpha}}\bar{\theta}^{\dot{\alpha}}$ and
$\bar{\epsilon}\bar{\sigma}^a \theta\equiv \bar{\epsilon}^{\dot{\alpha}}\bar{\sigma}^a_{\dot{\alpha}\alpha} \theta^{\alpha}$. Also, we can introduce the bispinor notation for the usual space-time derivatives: $\partial_{\alpha\dot{\alpha}}=\sigma^a_{\alpha\dot{\alpha}}\partial_a$.

The superfield is defined as a generic function of the superspace coordinates. We suggest it to have the form of a power series in spinor superspace coordinates $(\theta^{\alpha},\bar{\theta}^{\dot{\alpha}})$, which allows to present it as a following expansion (cf. f.e. \cite{WB,OM}):
\bea
\label{expfield}
F(x,\q,\bar{\q})&=&A(x)+\q^{\a}\psi_{\a}(x)+\bar{\q}_{\ad}\zeta^{\ad}(x)+
\q^2F(x)+\bar{\q}^2G(x)+i(\bar{\q}\sigma^a \q)A_a(x)+
\nonumber\\&+&\bar{\q}^2\q^{\a}
\chi_{\a}(x)+\q^2\bar{\q}_{\ad}\xi^{\ad}(x)+\q^2\bar{\q}^2H(x).
\eea
We note that this power series is finite due to anticommutation of
Grassmann spinor coordinates $\theta,\bar{\theta}$ which immediately annihilates third and higher degrees in $\theta$, $\bar{\theta}$. Further we will see that there are some restrictions on
structure of superfields caused by the form of representation of
supersymmetry algebra. Here $f(x),\psi_{\a}(x),\ldots$ are bosonic and
fermionic fields forming a component content of the superfield $F$ given by the expression above.
If a theory describing dynamics of these fields is supersymmetric, its
action should be invariant under supersymmetry transformations which are defined as the symmetry transformations with fermionic parameters.

{\bf Example.} The Wess-Zumino model \cite{WB} whose action, in the simplest case, that is, free massless theory, looks like
\bea
\label{WZ}
S=\int d^4x(\bar{\phi}\Box \phi-\frac{i}{2}\bar{\psi}^{\dot{\alpha}}\partial_{\dot{\alpha}\alpha}\psi^{\alpha}+\bar{F}F)
\eea
is invariant under the following transformations
\bea
\delta \phi(x)&=&\epsilon^{\a}\psi_{\a}(x);\nonumber\\
\delta \psi_{\a}(x)&=&\epsilon_{\a}\bar{F}(x)-\bar{\epsilon}^{\ad}i
\pa_{\a\ad}\phi(x);\nonumber\\
\delta F(x)&=&\bar{\epsilon}_{\ad}i\pa^{\a\ad}\psi_{\a},
\eea
with the analogous transformation for the conjugated fields $\bar{\phi}(x),\bar{\psi}_{\dot{\alpha}}(x),\bar{F}(x)$. Since $\epsilon^{\alpha}$, $\bar{\epsilon}^{\dot{\alpha}}$ are the (global) fermionic parameters, these transformations are the supersymmetry transformations.

Now, let us suggest that the variation of an arbitrary superfield $F(x,\q,\bar{\q})$ under the supersymmetry transformations has the form similar to the usual translations, i.e.
\bea
\label{svar}
\delta F(x,\q,\bar{\q})=
(\epsilon^{\a}Q_{\a}+\bar{\epsilon}_{\ad}\bar{Q}^{\ad})
F(x,\q,\bar{\q}).
\eea
Here we suppose that $Q_{\a},\bar{Q}_{\ad}$ are generators of supersymmetry possessing
anticommutation relations
\bea
\label{gencom}
\{Q_{\a},\bar{Q}_{\ad} \}&=&2i\sigma_{\a\ad}^m\partial_m ;\
\{Q_{\a},Q_{\b} \}=\{\bar{Q}_{\ad},\bar{Q}_{\bd}\}=0;\
[ Q_{\a},\partial_{m} ]=0.
\eea
The $\epsilon^{\alpha},\bar{\epsilon}^{\dot{\alpha}}$ are the infinitesimal parameters of the supersymmetry transformation. It is natural to treat the variation (\ref{svar}) as a translation in the superspace which we have already introduced (to be more precise, this is a translation of the fermionic sector of the superspace). This form of anticommutators is the simplest one allowing for nontrivial unification of the supersymmetry algebra with the Poincare algebra.

Translations on superspace are given by standard Poincare translations
and transformations (\ref{svar}) which can be called supertranslations. It is easy to see that (\ref{svar})
is a manifestly Lorentz covariant transformation. As we already have noted, the simplest, $N=1$ superspace is
parametrized by 4 bosonic coordinates $x^a$ and 4 fermionic ones 
$\q^{\a},\bar{\q}^{\ad}$ so it is 8-dimensional and is denoted as 
$R^{4|4}$. We will refer to it as to the four-dimensional superspace. It is natural to introduce the superfields, which will be ingredients of the new supersymmetric 
field theory as fields in the superspace.
The task of this chapter consists in development of the quantum theory for superfields in the four-dimensional superspace basing on principles of the standard quantum field theory methodology.

We introduce derivatives on the superspace in a manner similar to the three-dimensional case, that is, we use the same definition (\ref{derdef}), with we choose to use henceforth f.e. the left derivative, and the derivative with respect to $\bar{\theta}_{\dot{\alpha}}$ is defined analogously to the derivative with respect to $\theta^{\alpha}$.

Then, to introduce the integral we employ the definition
$\int d\theta \theta=1$, or, generally, 
$$\int d\q_{\a}\q^{\b}=\delta_{\a}^{\b}.$$ 
It is a convention. Note that $\q$ and $d\q$ have different
dimensions: the mass dimension of $\q$ is equal to $-\frac{1}{2}$, and of
$d\q$ -- to $\frac{1}{2}$, and variation $\delta\q$ never should be mixed
with differential $d\q$ since they have different dimensions. Then, an integral from a constant is zero,
$$\int d\q 1=0 
$$
this identity is caused by suggestion of the translation invariance due to which the
relation $\int d\q (\q+\lambda)= \int d\q\q$ for constant $\lambda$ 
must be satisfied, hence $\lambda\int d\q=0$. 
We introduce the following scalar measures for Grassmann integration:
\bea
d^2\q=-\frac{1}{4}d\q^{\a}d\q_{\a},\ d^2\bar{\q}=-\frac{1}{4}
\bar{d\q}_{\ad}\bar{d\q}^{\ad},\ d^4\q=d^2\q d^2\bar{\q}. 
\eea
These measures satisfy the relations
\bea
\int d^2\q \q^2=\int d^2\bar{\q} \bar{\q}^2=\int d^4 \q \q^4 =1
\eea
(here and further we denote $\q^4\equiv \q^2\bar{\q}^2$).

Since $\frac{\pa \q^{\a}}{\pa \q^{\b}}\equiv \pa_{\a}\q^{\b}=\delta^{\a}_{\b}$ as well as
$\int d\q_{\a}\q^{\b}=\delta_{\a}^{\b}$ (similarly, $\pa^{\ad}\bar{\q}_{\bd}=\int d\bar{\q}^{\ad}\bar{\q}_{\bd}=\delta^{\ad}_{\bd}$) we conclude that integration
and differentiation in Grassmann space are equivalent.
In particular, we see that
\bea
\int d^4\q F(x,\q,\bar{\q})&=&\frac{1}{16}\frac{\pa^2}{\pa\q^2}
\frac{\pa^2}{\pa\bar{\q}^2}F(x,\q,\bar{\q})=
\frac{1}{16}F(x,\q,\bar{\q})|_{\q^2\bar{\q}^2};\nonumber\\
\int d^2\q G(x,\q)&=&-\frac{1}{4}\frac{\pa^2}{\pa\q^2}G(x,\q)=
-\frac{1}{4}G(x,\q)|_{\q^2}.
\eea
Here $|_{\q^2}, |_{\q^2\bar{\q}^2}$ denotes the corresponding component
of the superfield.
Of course, differentiations with respect to Grassmann coordinates
anticommute.

The supersymmetry generators possess several realizations in terms of
$\frac{\pa}{\pa x^m}$ and 
$\frac{\pa}{\pa \q_{\a}}, \frac{\pa}{\pa\bar{\q}_{\ad}}$, 
f.e.
\bea
\label{gen}
\bar{Q}_{\ad}=i(\frac{\pa}{\pa\bar{\q}^{\ad}}-
i\q^{\a}(\sigma^m)_{\a\ad}\pa_m), \ 
Q_{\a}=i(\frac{\pa}{\pa\q^{\a}}+i\bar{\q}^{\bd}(\bar{\sigma}^m)_{\bd\a}\pa_m).
\eea
However, any possible realizations of the supersymmetry generators 
must satisfy relations (\ref{gencom}).

The spinor supercovariant derivatives $D_A$ also must be constructed from
$\frac{\pa}{\pa x^m}$ and 
$\frac{\pa}{\pa \q_{\a}}, \frac{\pa}{\pa\bar{\q}_{\ad}}$. 
They should anticommute with generators $Q_{\a},\bar{Q}_{\ad}$
which provides that $D_A\Phi$ is transformed covariantly, i.e. 
according to (\ref{svar}):
$$\delta (D_A\Phi)=(\epsilon Q+\bar{\epsilon}\bar{Q})D_A\Phi.
$$
F.e., if generators of the supersymmetry are realized in terms of
(\ref{gen})
supercovariant derivatives are realized as
\bea
\label{gend}
\bar{D}_{\ad}&=&-i\bar{Q}_{\ad}+2i\q^{\a}\pa_{\a\ad}
=\frac{\pa}{\pa\bar{\q}^{\ad}}+i\q^{\a}(\sigma^m)_{\a\ad}\pa_m, \nonumber\\ 
D_{\a}&=&-iQ_{\a}-2i\bar{\q}^{\ad}\pa_{\a\ad}
=\frac{\pa}{\pa\q^{\a}}-i\bar{\q}^{\bd}(\bar{\sigma}^m)_{\bd\a}\pa_m.
\eea
We can also use the notation $\pa_{\a\ad}=(\sigma^m)_{\a\ad}\pa_m$.
The spinor supercovariant derivatives satisfy the following
anticommutation relations
\bea
\label{salg}
\{D_{\a},\bar{D}_{\ad}\}=2i\pa_{\a\ad};
\ \{D_{\a},D_{\b}\}=
\{\bar{D}_{\ad},\bar{D}_{\bd}\}=0.
\eea
So we defined procedures of integration and differentiation 
in the superspace. Further we will use the integral measure for the complete superspace $d^8z=d^4xd^2\theta d^2\bar{\theta}$, the integral measure for the chiral subspace $d^6z=d^4x d^2\theta$ and the integral measure for the antichiral one $d^6\bar{z}=d^4x d^2\bar{\theta}$. We also can use the identities $D^2\theta^2=\bar{D}^2\bar{\theta}^2=-4$.

Now, let us introduce the delta
function. We suggest that it must satisfy the condition analogous to that one for a standard delta function
\bea
\int d^4\q' \delta^4(\q-\q')f(\q')=f(\q).
\eea 
This identity can be satisfied if we choose
\bea
\delta^4(\q-\q')=(\q-\q')^2(\bar{\q}-\bar{\q}')^2.
\eea
It is easy to see that this delta function satisfies the condition
\bea
\int d^4\q \delta^4(\q-\q')=1.
\eea
We note the identity
\bea
\delta^4(\q_1-\q_2)D^2_1\bar{D}^2_2\delta^4(\q_1-\q_2)=16\delta^4(\q_1-\q_2).
\eea
We introduce the notation $\delta_{12}\equiv\delta^4(\q_1-\q_2)$.
It is easy to see that $\delta_{12}\delta_{12}=
\delta_{12}D^{\a}\delta_{12}=\delta_{12}D^2\delta_{12}=
\delta_{12}\bar{D}_{\ad}\delta_{12}=0$.

A supermatrix is defined as a matrix $M=M^P_Q$ of the form
\bea
M=\left(\begin{array}{cc}
A&B\\
C&D
\end{array}\right). 
\eea
determining a quadratic form $z_P M^P_Q z^{'Q}$ with $z,z'$ are coordinates
on superspace. Here $A,B,C,D$ are even-even, even-odd, odd-even and
odd-odd blocks respectively.  
Superdeterminant of this matrix is introduced as
\bea
{\rm sdet} M=\int d^8 z_1 d^8 z_2 \exp (-z_1 M z_2).
\eea
It is equal to
\bea
{\rm sdet} M=\det A\, {\det}^{-1}(D-CA^{-1}B).
\eea
And a supertrace is equal to $\Str M=\sum\limits_A (-1)^{\epsilon_A}M^A_A=\tr A -
\tr D$. As usual, ${\rm sdet} M=\exp (\Str \log M)$. 

We can introduce change of variables in superspace. So, if the coordinates are changed as
\bea
x^{\prime a}=x^{\prime a}(x,\q,\bar{\q});\, 
\q^{\prime\a}=\q^{\prime\a}(x,\q,\bar{\q}),\,
\bar{\q}^{\prime\ad}=\bar{\q}^{\prime\ad}(x,\q,\bar{\q}),
\eea
the measure of integral over the superspace is transformed as
\bea
d^4 x' d^4 \q'= d^4 x d^4 \q \ {\rm sdet}(\frac{\partial z'}{\partial z}),
\eea
where supermatrix $(\frac{\partial z'}{\partial z})$ is 
\bea
\frac{\pa z'}{\pa z}=\left(\begin{array}{ccc}
\frac{\pa x'}{\pa x}&\frac{\pa x'}{\pa \q}&\frac{\pa x'}{\pa
  \bar{\q}}\\
\frac{\pa \q'}{\pa x}&\frac{\pa \q'}{\pa \q}&\frac{\pa \q'}{\pa
  \bar{\q}}\\
\frac{\pa \bar{\q}'}{\pa x}&\frac{\pa \bar{\q}'}{\pa \q}&
\frac{\pa \bar{\q}'}{\pa \bar{\q}}\\
\end{array}
\right).
\eea 

Now, let us characterize the typical superfields in $N=1$ superspace. There are two most important their examples. The first of them is the real scalar superfield whose form is given by (\ref{expfield}), but with the condition $V^{\dagger}=V$, i.e. 
\bea
\label{realfield}
V(x,\q,\bar{\q})&=&C(x)+\q^{\alpha}\chi_{\alpha}(x)+\bar{\q}_{\ad}\bar{\chi}^{\dot{\alpha}}(x)-
\q^2M(x)-\bar{\q}^2\bar{M}(x)+i(\bar{\q}\sigma^a \q)A_a(x)+
\nonumber\\&+&\bar{\q}^2\q^{\a}
\lambda_{\a}(x)+\q^2\bar{\q}_{\ad}\bar{\lambda}^{\ad}(x)+\q^2\bar{\q}^2{\cal D}(x).
\eea
The second one is the chiral superfield. It is defined in the following way:
the superfield $\Phi(z)$ is called chiral if and only if it satisfies the condition
$\bar{D}_{\ad}\Phi=0$. The choice of supercovariant derivatives in the form $\bar{D}_{\ad}=\frac{\partial}{\partial\bar{\theta}^{\ad}}$, $D_{\a}=\frac{\pa}{\pa\q^{\a}}-2i\bar{\q}^{\bd}(\bar{\sigma}^m)_{\bd\a}\pa_m$ allows one to reduce this condition to $\partial_{\ad}\Phi=0$, i.e. the chiral superfield $\Phi$ 
turns out to be $\bar{\theta}$-independent. It allows to represent it in the following form
\bea
\label{chirfield}
\Phi(x,\theta)=\phi(x)+\theta^{\alpha}\psi_{\alpha}(x)-\theta^2 F(x).
\eea
At the same time, one should introduce the antichiral superfield $\bar{\Phi}$ which is defined to satisfy the "conjugated" condition $D_{\alpha}\bar{\Phi}=0$. But, as the spinor supercovariant derivative $D_{\alpha}$ in form (\ref{gend}) is {\bf NOT} a straightforward analogue of the derivative $\bar{D}_{\dot{\alpha}}$, one {\bf CANNOT} use the analogue of the expression (\ref{chirfield}) for the antichiral field, whose form turns out to be more complicated, involving additional terms beside of the straightforward analogues of those ones in Eq. (\ref{chirfield}):
\bea
\bar{\Phi}(x,\theta,\bar{\theta})=\bar{\phi}(x)+\bar{\theta}_{\ad}\bar{\psi}^{\ad}(x)-\bar{\theta}^2\bar{F}(x)+\ldots,
\eea
where dots are for the $\theta$-dependent terms. However, one can use for the components of $\Phi$ the definition in terms of projections:
\bea
\label{projchir}
\phi(x)&=&\Phi(z)|;\nonumber\\
\psi_{\alpha}(x)&=&D_{\alpha}\Phi(z)|;\nonumber\\
F(x)&=&\frac{1}{4}D^2\Phi(z))|,
\eea
with the definitions for the components of $\bar{\Phi}$ can be obtained via a straightforward conjugation of these definitions. 
Here, similarly to the previous section, we suggest that $\Phi(z)|\equiv\Phi(z)|_{\theta=\bar{\theta}=0}$, with the condition $\theta=\bar{\theta}=0$ is imposed after the differentiation.

One can also define the components of the real scalar superfield in terms of projections:
\bea
\label{projreal}
C(x)&=&V(z)|,\nonumber\\
\chi_{\alpha}(x)&=&D_{\alpha}V(z)|;\quad\, \bar{\chi}^{\dot{\alpha}}(x)=\bar{D}^{\dot{\alpha}}V(z)|;\nonumber\\
M(x)&=&\frac{1}{4}D^2V(z))|;\quad\, \bar{M}(x)=\frac{1}{4}\bar{D}^2V(z)|;\nonumber\\
A_{\alpha\dot{\alpha}}&=&\frac{i}{2}[D_{\alpha},\bar{D}_{\dot{\alpha}}]V(z)|;\nonumber\\
\lambda_{\alpha}(x)&=&-\frac{1}{4}\bar{D}^2D_{\alpha}V(z)|;\quad\,\bar{\lambda}^{\dot{\alpha}}(x)=-\frac{1}{4}D^2\bar{D}^{\dot{\alpha}}V(z)|;\nonumber\\
{\cal D}(x)&=&\frac{1}{16}D^{\alpha}\bar{D}^2D_{\alpha}V(z)|.
\eea

To develop the functional integral approach, we also must introduce variational derivative.
In common field theory it is defined as
\bea
\frac{\delta}{\delta A(x)}\int d^4 y f(y) A(y) =f(x),
\eea
if $f(x)$ and $A(x)$ are functionally independent.
Just an analogous definition can be introduced for general (non-chiral)
superfield:
\bea
\frac{\delta}{\delta V(z)}\int d^8 z'f(z') V(z')=f(z).
\eea
Now, let us introduce the variational derivative with respect to a chiral superfield. As the chiral superfield effectively depends only on the one set of the Grassmann coordinates, that is, $\theta^{\alpha}$, the integral from a chiral function is
non-trivial only if it is calculated over chiral subspace, i.e. over 
$d^6z=d^4x d^2\q$. Hence we must introduce variational derivative with
respect to chiral superfield $\Phi$ as
\bea
\frac{\delta}{\delta\Phi(z)}\int d^6z'F(z')\Phi(z')=F(z).
\eea
And the variational derivative from integral over whole superspace with
respect to chiral superfield can be introduced as
\bea
\frac{\delta}{\delta\Phi(z)}\int d^8 z' G(z')\Phi(z')=
\frac{\delta}{\delta\Phi(z)}\int d^6 z' (-\frac{1}{4}\bar{D}^2)G(z')\Phi(z')
=-\frac{1}{4}\bar{D}^2G(z).
\eea
Therefore $\frac{\delta\Phi(z)}{\delta\Phi(z')}=
\delta_+(z-z')$ where
$\delta_+(z-z')=-\frac{1}{4}\bar{D}^2\delta^8(z-z')$
is a chiral delta function. It allows us to obtain useful relation
\bea
&&\frac{\delta^2}{\delta\Phi(z_1)\delta\bar{\Phi}(z_2)}\int d^8z\Phi\bar{\Phi}=\frac{1}{16}
\bar{D}^2_1D^2_2\delta^8(z_1-z_2)=\nonumber\\&=&
(-\frac{1}{4})D^2\delta_+(z_1-z_2)
=(-\frac{1}{4})\bar{D}^2\delta_-(z_1-z_2).
\eea
Here $\delta_-(z_1-z_2)=-\frac{1}{4}D^2\delta^8(z_1-z_2)$ is
antichiral delta function. Note the
relation $D^2_1\delta^8(z_1-z_2)=D^2_2\delta^8(z_1-z_2)$.

If we consider some differential operator $\Delta$ acting on superfields we can
introduce its functional supertrace and superdeterminant:
\bea
\Str \Delta=\int d^8 z_1 d^8 z_2 \delta^8(z_1-z_2) \Delta \delta^8(z_1-z_2).
\eea
If we introduce kernel of the $\Delta$ which has the 
form $\Delta(z_1,z_2)$ we can write
\bea
\Str\Delta=\int d^8 z \Delta(z,z).
\eea
Superdeterminant is introduced as
\bea
{\rm sdet} \Delta =\exp {\Str} (\log \Delta). 
\eea
Further we will be generally interested in theories describing
dynamics of chiral and real scalar superfield. Note that irreducible
representation of supersymmetry algebra is realized namely on these 
superfields \cite{WB}. The most important examples of such theories are Wess-Zumino
model,
general chiral superfield theory \cite{my2}, super-Yang-Mills
theory and four-dimensional dilaton supergravity \cite{my1}. All of them are constructed on the base of chiral (and antichiral) and real scalar superfields.
In this chapter we consider application of superfield approach to these models.

\section{Field theory models in the four-dimensional superspace}

In this section we will introduce some typical field theory models in the four-dimensional superspace. The key ingredients of these models are exactly the chiral and real scalar superfields introduced above. Their component contents are given by Eqs. (\ref{chirfield}), (\ref{realfield}) respectively.

\subsection{Chiral superfield models}

The simplest superfield model is the Wess-Zumino model \cite{WZ,WZ1,WB} describing the dynamics of the chiral superfield $\Phi$. Its explicit form in the components, in the particular (free massless) case is given by the expression (\ref{WZ}). Now, let us write down its superfield action. It has the form
\bea
\label{actsfWZ}
S=\int d^8z \Phi\bar{\Phi}+[\int d^6z (\frac{m}{2}\Phi^2+\frac{\lambda}{3!}\Phi^3)+h.c.]
\eea
First, let us briefly discuss the question of dimensions of the superfields. It is well known that the mass dimension of the coordinate $x^a$ is $-1$, and hence of the spatial derivative $\partial_a$ is $1$. We have already mentioned that the dimensions of the Grassmannian coordinates $\theta^{\alpha}$ and the corresponding derivatives $\partial_{\alpha}$ (and covariant derivatives $D_{\alpha}$ and differential measure $d\theta^{\alpha}$ as well) must be opposite. Taking into account the structure of the supercovariant derivatives (\ref{gend}), we can conclude that the dimension of the derivatives $\partial_{\alpha},\,D_{\alpha}$ and of the integral measure $d\theta^{\alpha}$ must be equal to $1/2$, and of the Grassmannian coordinate $\theta^{\alpha}$ itself must be equal to $-1/2$ (the dimensions of the conjugated coordinates $\bar{\theta}_{\dot{\alpha}}$ and the corresponding derivatives $\partial^{\dot{\alpha}}$, $\bar{D}^{\dot{\alpha}}$ are respectively the same). All this allows us to conclude that the dimension of the superfield $\Phi$ is equal to 1, hence the coupling constant $\lambda$ is dimensionless. Applying the usual argumentation of the quantum field theory, we can conclude that the Wess-Zumino model is renormalizable.

Second, let us obtain the component structure of the whole Wess-Zumino action. As we have already mentioned, the component structure of the first term of (\ref{actsfWZ}), that is, $\int d^8z\Phi\bar{\Phi}$, is given by (\ref{WZ}). The component structure of the term corresponding to the integral over chiral subspace can be easily obtained:
\bea
\int d^6z (\frac{m}{2}\Phi^2+\frac{\lambda}{3!}\Phi^3)=(-\frac{1}{4})\int d^4x (\frac{m}{2}D^2\Phi^2+\frac{\lambda}{3!}D^2\Phi^3)|,
\eea
then we employ the expressions (\ref{projchir}) and find
\bea
\int d^6z (\frac{m}{2}\Phi^2+\frac{\lambda}{3!}\Phi^3)=\int d^4x[-\frac{1}{4}(m+\lambda\phi)\psi^{\alpha}\psi_{\alpha}-F(m\phi+\frac{\lambda}{2}\phi^2)].
\eea
This allows us to write down the complete Wess-Zumino action in components:
\bea
\label{compwz}
S&=&\int d^4x \Big[\bar{\phi}\Box \phi-\frac{i}{2}\bar{\psi}^{\dot{\alpha}}\partial_{\dot{\alpha}\alpha}\psi^{\alpha}+\bar{F}F-\nonumber\\&-&
(\frac{1}{4}(m+\lambda\phi)\psi^{\alpha}\psi_{\alpha}+F(m\phi+\frac{\lambda}{2}\phi^2)+h.c.)
\Big].
\eea
One can see that the field $F$ has no dynamics even after adding the interaction term, therefore it is called the auxiliary field. Elimination the field $F$ with use of its equation of motion
\bea
F-(m\bar{\phi}+\frac{\lambda}{2}\bar{\phi}^2)=0
\eea
implies in the theory with $\phi^4$ interaction, this is the reason why the Wess-Zumino model is treated as a supersymmetric extension of the $\phi^4$ theory.

It is instructive to calculate the numbers of degrees of freedom in the Wess-Zumino model. It follows from (\ref{compwz}) that in this model there are four bosonic degrees of freedom (which correspond to two complex scalar fields) and four fermionic ones (which corresponds to two components of the complex spinor $\psi_{\alpha}$). In a generic case, in any supersymmetric theory numbers of bosonic and fermionic degrees of freedom are equal.

We note that the Wess-Zumino model is not an unique theory describing the quantum dynamics of the chiral superfield. Other important examples are the higher-derivative chiral superfield theories (in particular, dilaton supergravity) and general chiral superfield theories. The examples of actions of these theories in superfield form are respectively
\bea
S=\int d^8z \bar{\Phi}(\Box+M^2)\Phi+(\int d^6z W(\Phi)+h.c.)
\eea
and
\bea
S=\int d^8z K(\Phi,\bar{\Phi})+(\int d^6z W(\Phi)+h.c.).
\eea
Here $W(\Phi)$ is a holomorphic function of the chiral superfield (or of the set of chiral superfields) but not on its derivatives, and $K(\Phi,\bar{\Phi})$ is a real function of chiral and antichiral superfields. The component forms of these theories can be obtained in the same way as above, for example, the component expression for the higher-derivative chiral superfield action \cite{GSH,my1} looks like
\bea
\label{comphd}
S&=&\int d^4x \Big[\bar{\phi}\Box(\Box+M^2) \phi-\frac{i}{2}\bar{\psi}^{\dot{\alpha}}\partial_{\dot{\alpha}\alpha}(\Box+M^2)\psi^{\alpha}+\bar{F}(\Box+M^2)F-\nonumber\\&-&
\frac{1}{4}(4W_{\phi}F+W_{\phi\phi}\psi^{\alpha}\psi_{\alpha}+h.c.)
\Big].
\eea
and for the general chiral superfield model \cite{my2} --
\bea
\label{compgc}
S&=&\int d^4x \Big[-K_{\phi\bar{\phi}}(\pa^a\phi\pa_a\bar{\phi}-\bar{F}F-\frac{i}{2}
\bar{\psi}^{\dot{\alpha}}\partial_{\dot{\alpha}\alpha}\psi^{\alpha})-\nonumber\\&-&
\frac{1}{4}(K_{\phi\phi\bar{\phi}}(F\psi^{\alpha}\psi_{\alpha}+i\pa_a\phi
\bar{\psi}^{\dot{\alpha}}\partial_{\dot{\alpha}\alpha}\psi^{\alpha})+h.c.)
+ %\nonumber\\&+&
\frac{1}{16}K_{\phi\phi\bar{\phi}\bar{\phi}}\psi^{\alpha}\psi_{\alpha}\bar{\psi}_{\dot{\alpha}}\bar{\psi}^{\dot{\alpha}}-
\nonumber\\
&-&\frac{1}{4}(4W_{\phi}F+W_{\phi\phi}\psi^{\alpha}\psi_{\alpha}+h.c.)
\Big].
\eea
Here $K_{\phi}=\frac{\partial K(\Phi,\bar{\Phi})}{\partial\Phi}|_{\Phi=\phi,\bar{\Phi}=\bar{\phi}}$ etc., the similar definitions are applied to derivatives of $W$, i.e. all these functions depend on the scalar fields $\phi,\bar{\phi}$ only. We note that in the theory (\ref{comphd}) the auxiliary field $F$ acquires a nontrivial dynamics.

However, first of these models, involving higher derivatives, in principle can imply in ghost states, and the second one is non-renormalizable in a general case. 
We will study these theories within the superfield approach in details further.

\subsection{Abelian gauge superfield model}

Now, let us go to the real scalar superfield case. The key feature of this superfield consists in the fact that it allows for introducing the gauge symmetry on the superspace.

Indeed, let us consider the action of the real scalar superfield $V$ (see f.e. \cite{WB}):
\bea
\label{resca}
S=\frac{1}{2}\int d^8z V (\frac{D^{\alpha}\bar{D}^2D_{\alpha}}{8})V.
\eea
This action is evidently gauge invariant with respect to the transformations:
\bea
V\to V+i(\Lambda-\bar{\Lambda}),
\eea
where $\Lambda$ is a chiral superfield parameter, and $\bar{\Lambda}$ is an antichiral one (to show the invariance, we use the fact that $\bar{D}^2D_{\alpha}\Lambda=0$ for a chiral $\Lambda$). Further, to develop a consistent quantum description, we should fix the gauge.

This action is constructed with use of the operator $-\frac{D^{\alpha}\bar{D}^2D_{\alpha}}{8}\equiv \Pi_{1/2}\Box$, where $\Pi_{1/2}=-\frac{D^{\alpha}\bar{D}^2D_{\alpha}}{8\Box}$ is a projecting operator possessing the property
$\Pi_{1/2}^n=\Pi_{1/2}$ for any integer $n\geq 1$. One can see that there are two projecting operators more,
$\Pi_{0+}=\frac{\bar{D}^2D^2}{16\Box}$ and $\Pi_{0-}=\frac{D^2\bar{D}^2}{16\Box}$ satisfying the similar properties, i.e. $\Pi_{0\pm}^n=\Pi_{0\pm}$. One can say that $\Pi_{0+}$ is a projector on the chiral space, $\Pi_{0-}$ is a projector on the antichiral space since acting of $\Pi_{0+}$ on any superfield produces the chiral superfield, and of $\Pi_{0-}$ -- the antichiral superfield. The $\Pi_{1/2}$ projects on the so-called linear space since, for any superfield $\Sigma$, the new superfield $\Pi_{1/2}\Sigma\equiv \tilde{\Sigma}$ possesses the properties $D^2\tilde{\Sigma}=\bar{D}^2\tilde{\Sigma}=0$, and such a superfield is called the linear superfield (however, it is used very rarely, some results for it can be found in \cite{SGRS}). 

It is straightforward to show that the projecting operators $\Pi_{1/2}$, $\Pi_{0+}$ and $\Pi_{0-}$ satisfy the properties
\bea
&&\Pi_{1/2}+\Pi_{0+}+\Pi_{0-}=1;\nonumber\\
&&\Pi_{0+}\Pi_{0-}=\Pi_{0\pm}\Pi_{1/2}=\Pi_{0-}\Pi_{0+}=\Pi_{1/2}\Pi_{0\pm}=0,
\eea 
i.e. they form the complete and orthogonal set of the projecting operators. It implies that any superfield can be represented as a linear combination of chiral, antichiral and linear ones.

Now let us obtain the component structure of the action (\ref{resca}). In principle, it can be done via a straightforward use of the expression (\ref{resca}) and the projections (\ref{projreal}). However, such a way is very cumbersome. Therefore we use another manner: since the chiral and complete measures are related by the rule $\int d^8z=\int d^6z (-\frac{\bar{D}^2}{4})$, we can rewrite the action (\ref{resca}) in an equivalent form (cf. \cite{GRS}):
\bea
\label{resca1}
S=\frac{1}{64}\int d^6z W^{\alpha}W_{\alpha},
\eea
where 
\bea
\label{wab}
W_{\alpha}=-\bar{D}^2D_{\alpha}V
\eea
is an Abelian superfield strength. First, it is clear that $W_{\alpha}$ is chiral. Second, one can obtain its component expansion (in the chiral representation):
\bea
W_{\alpha}=4[\lambda_{\alpha}+\theta^{\beta}f_{\beta\alpha}+\theta_{\alpha}{\cal D}-\frac{i}{2}\theta^2\partial_{\alpha\dot{\beta}}\bar{\lambda}^{\dot{\beta}}].
\eea
or, in the form of projections,
\bea
\label{projw}
W_{\alpha}|&=&4\lambda_{\alpha};\nonumber\\
D_{(\beta}W_{\alpha)}|&=& 8f_{\alpha\beta};\nonumber\\
D^2W_{\alpha}|&=&8i\partial_{\alpha\dot{\beta}}\bar{\lambda}^{\dot{\beta}};\nonumber\\
-\frac{1}{2}D^{\alpha}W_{\alpha}|&=&4{\cal D}.
\eea
The $f_{\alpha\beta}=\partial_{\beta\dot{\beta}}A^{\dot{\beta}}_{\phantom{\beta}\alpha}+\partial_{\alpha\dot{\beta}}A^{\dot{\beta}}_{\phantom{\beta}\beta}$ here is the (symmetric) bispinor form of the usual stress tensor $F_{ab}$, i.e. $f_{\alpha\beta}=\frac{1}{2}(\sigma^{ab})_{\alpha\beta}F_{ab}$. In this case, the action (\ref{resca1}) is reduced to
\bea
\label{sqedcomp}
S=\frac{1}{4}\int d^4x(-\frac{1}{2}f^{\alpha\beta}f_{\alpha\beta}-i\bar{\lambda}^{\dot{\alpha}}\partial_{\dot{\alpha}\beta}\lambda^{\beta}+\frac{1}{2}{\cal D}^2),
\eea
that is, the action of the supersymmetric electrodynamics.

The key property of the action (\ref{sqedcomp}) is that it involves only higher components of the superfield $V$ (\ref{realfield}). From the formal viewpoint, it means that the lower components of this superfield, that is, $C(x)$, $\chi_{\alpha}(x)$, $\bar{\chi}^{\dot{\alpha}}(x)$, $M(x)$, $\bar{M}(x)$ can be removed via some gauge transformation. Indeed, if we consider the gauge parameter $\Lambda(z)$ whose components structure is
\bea
\Lambda(z)=i(\frac{1}{2}C(x)+\theta^{\alpha}\chi_{\alpha}(x)+\theta^2 M(x)),
\eea
with $\bar{C}=C$, we identically cancel the above-mentioned lower components of the real scalar superfield $V$. The gauge, in which the expansion of the $V$ begins with $i(\bar{\q}\sigma^a \q)A_a(x)$, is called the Wess-Zumino gauge, and it implies that $V^n=0$ for any $n\geq 3$. However, this gauge, although allowing to remove the nonpolynomiality of the action which is very useful in the non-Abelian case, breaks the superfield structure being thus very inconvenient within the superfield approach. Some attempts to conciliate this gauge with the superfield methodology have been carried out in \cite{noncov}.

\subsection{Non-Abelian gauge theories}

Now, let us generalize the gauge theory for the non-Abelian case, see f.e. \cite{BK0}. The key idea consists in using the action (\ref{resca1}) where the superfield strength (\ref{wab}) is promoted to a non-Abelian case. Indeed, let us suggest 
that it has the form
\bea
\label{wnab1}
W_{\alpha}=-\bar{D}^2(e^{-gV}D_{\alpha}e^{gV})
\eea
Here we suggest that $V$ is now a {\bf non-Abelian}, Lie-algebra valued real scalar superfield, $V=V^AT^A$, where $T^A$ are the Hermitian generators of some Lie group (the most popular examples of the gauge groups are $U(N)$ and $SU(N)$). It is clear that in the Abelian case this strength is reduced to the expression (\ref{wab}).

Let us suggest that the theory is invariant under the following gauge transformations:
\bea
\label{nonpol}
e^{gV}\to e^{-ig\bar{\Lambda}}e^{gV}e^{ig\Lambda},
\eea
where $\Lambda=\Lambda^AT^A$ is a Lie-algebra-valued chiral parameter of the gauge transformation, and $\bar{\Lambda}$ is an antichiral one. It is clear that the superfield strength $W_{\alpha}$, being Lie-algebra-valued, in this case is not invariant but transformed in a covariant manner:
\bea
W_{\alpha}\to e^{-ig\Lambda}W_{\alpha}e^{ig\Lambda}.
\eea
The action of the non-Abelian gauge theory can be obtained by a straightforward promotion of the action (\ref{resca1}) to the non-Abelian case (cf. \cite{GRS}):
\bea
\label{resca3}
S=\frac{1}{64g^2}{\rm tr}\int d^6z W^{\alpha}W_{\alpha}=-\frac{1}{16g^2}{\rm tr}\int d^8z (e^{-gV}D^{\alpha}e^{gV})\bar{D}^2(e^{-gV}D_{\alpha}e^{gV}).
\eea
The theory described by the expression (\ref{resca3}) is called the super-Yang-Mills theory.
Its action is essentially non-polynomial, however, its quadratic part reproduces the Abelian expression (\ref{resca}). In principle, one can expand this action in power series in $V$ and impose the Wess-Zumino gauge which allows to eliminate all $V^3$ and higher terms as well as in the Abelian case (for the discussion of the noncovariant gauges see \cite{noncov}). However, the covariant gauges are much more convenient for studying this theory.

Treating the component content, one must note that the nonpolynomial gauge transformations (\ref{nonpol}) are well applicable to eliminate the lower components of the superfield $V$. However, the non-Abelian theory (\ref{resca3}) is not free but nontrivially self-coupled. We suggest that the components of the non-Abelian strength $W_{\alpha}$ are again given by the expression (\ref{projw}), with the only modification that $f_{\alpha\beta}$ is now a non-Abelian stress tensor. Therefore, the action of the super-Yang-Mills theory, after carrying out the procedures similar to those ones realized above is rewritten in components as
\bea
\label{symcomp}
S=\frac{1}{4g^2}{\rm tr}\int d^4x(-\frac{1}{2}f^{\alpha\beta}f_{\alpha\beta}-i\bar{\lambda}^{\dot{\alpha}}{\nabla}_{\dot{\alpha}\beta}\lambda^{\beta}+\frac{1}{2}{\cal D}^2),
\eea
Here ${\nabla}_{\dot{\alpha}\beta}=\partial_{\dot{\alpha}\beta}+iA_{\dot{\alpha}\beta}$ is a gauge covariant space-time derivative.

The details of the supercovariant description of the super-Yang-Mills theories will be given further, when the quantum approach for it is discussed. Now, after we have formulated the theories on the classical level, we can start the perturbative description.

\section[Generating functional]{Generating functional and Green functions for superfields}

Now our aim consists of describing a method for calculation of 
generating functional and Green functions for superfields and
following application of this method to calculation of superfield
quantum corrections, i.e. in development of the Feynman supergraph 
technique. We note that during last years activity in 
development of nonperturbative approaches in superfield quantum theory
stimulated by paper \cite{SWgauge} essentially increased. Nevertheless, the importance of the results obtained through the
perturbative approach continues to be principal.

The generalization of path integral method for superfield theory turns
to be quite straightforward but a bit formal. Really, generating
functional is defined in terms of a path integral which is well-defined
only for some special cases. However, the case of Gaussian path
integral is, first, well-defined both in standard field theory and in
superfield theory, second, sufficient for the development of the superfield
perturbation technique.

Let us shortly describe the prescriptions for the introduction of Green functions in a common field
theory (see f.e. \cite{BSD}). Let the classical action $S[\phi_1\ldots\phi_n]$ be a local space-time
functional of $n$ superfields $\phi_i$. The equations of motion (with $S_i\equiv\frac{\delta S}{\delta\phi_i}$ etc.) are:
$S_i[\phi]=0|_{\phi=\phi_0}$.
The $\phi_0$ is a solution for this equation. We suppose that the Hessian is
non-singular at this point, i.e. ${\rm det}\, S_{ij}[\phi]_{\phi=\phi_0}\neq
0$ (or as is the same equation $S_{ij}|_{\phi=\phi_0}a^j=0$ is
satisfied if and only if $a^j=0$). If the Hessian is singular, we add to
the action some term to make it non-zero (in gauge theories such a term
is called the gauge-fixing one), after adding of this term all 
consideration is just the
same as if the Hessian would be non-zero from the very beginning. We suggest
that the action $S[\phi]$ is an analytic functional, i.e. it can be expanded
into power series in a neighborhood of $\phi_0$:
\bea
S[\phi]=S[\phi_0]+\sum_{n=2}^{\infty}\frac{1}{n!}S_{i_1\ldots i_n}
(\phi-\phi_0)^{i_n}\ldots (\phi-\phi_0)^{i_1}.
\eea
The term with $n=2$ is called a quadratic (or linearized) action:
\bea
S_0=\frac{1}{2}\tilde{\phi}^i S_{ij}[\phi_0]\tilde{\phi}^j.
\eea
Here and further $\tilde{\phi}^i=\phi^i-\phi_0^i $.
Terms with $n\geq 3$ are called interaction terms $S_{int}$, and the whole 
action is rewritten as
\bea
\label{inact}
S[\phi]=S[\phi_0]+S_0[\tilde{\phi};\phi_0]+S_{int}[\tilde{\phi};\phi_0].
\eea
The Green function $G^{ij}$ is determined on the base of the linearized action as
\bea
S_{ij}[\phi_0]G^{jk}=-\delta^k_i;\ G^{ij}S_{jk}[\phi_0]=-\delta^i_k.
\eea
The generating functional of Green functions usually is introduced as
\bea
\label{GF}
Z[J]=N\int D\phi \exp (\frac{i}{\hbar}(S[\phi]+J\phi)).
\eea
Here $J$ is a source, and $N$ is a normalization factor.

The Green functions can be obtained on the base of the generating
functional as
\bea
\label{i}
<\phi(z_1)\ldots\phi(z_n)>=\Big(\frac{1}{i}\frac{\delta}{\delta J (z_1)}\Big)
\ldots \Big(\frac{1}{i}\frac{\delta}{\delta J (z_n)}\Big)
N\int D\phi \exp (\frac{i}{\hbar}(S[\phi]+J\phi)).
\eea
We can calculate the path integral (\ref{GF}). To do it, after relabelling $\tilde{\phi}\to\phi$
in (\ref{inact}), we expand the complete action as 
$S[\phi]=S_0[\phi]+S_{int}[\phi]$  with $S_0[\phi]=\frac{1}{2}\int dz \phi\Delta\phi$ is a quadratic action, and $\Delta$ be
some operator. Of course, path integration is a quite formal
operation being well-defined only for the Gaussian integral and expressions
derived from it. However, both in the standard and in the superfield cases we mostly need
only Gaussian integrals.
As usual,
\bea
\label{ii}
& &\int D\phi \exp (\frac{i}{\hbar}(S[\phi]+J\phi))=
\int D\phi \exp (\frac{i}{\hbar}(\frac{1}{2}\phi\Delta\phi+S_{int}[\phi]+J\phi))
=\nonumber\\&=&
\exp(\frac{i}{\hbar}S_{int}(\frac{\hbar}{i}\frac{\delta}{\delta J}))
\int D\phi \exp (\frac{i}{\hbar}(\frac{1}{2}\phi\Delta\phi+J\phi)).
\eea
And (since this integral is Gaussian-like)
\bea
\label{iii}
\int D\phi \exp (\frac{i}{\hbar}(\frac{1}{2}\phi\Delta\phi+J\phi))=
\exp(-\frac{i}{2}J(\frac{\hbar}{\Delta})J)det^{-1/2}(\frac{\Delta}{\hbar}).
\eea
In these expression, the integration over space-time is assumed where it is necessary.
One concludes that all dependence of the expression on the sources is concentrated in the term
$\exp(-\frac{i}{2}J(\frac{\hbar}{\Delta})J)$. Constructing of Feynman
diagrams from expressions (\ref{i}, \ref{ii}, \ref{iii}) is quite 
straightforward.

Let us apply this approach to a superfield theory. Our example is the
Wess-Zumino model \cite{WB}, consideration of other theories is exactly
analogous. We do not address here the specifics of gauge theories in which one
must introduce gauge fixing and ghosts since after their introduction
all procedure is just the same.

The action of the Wess-Zumino model with chiral sources is
\bea
\label{actwz}
S_J[\Phi,\bar{\Phi};J,\bar{J}]
=\int d^8z\Phi\bar{\Phi}+(\int d^6z (\frac{\l}{3!}\Phi^3+
\frac{m}{2}\Phi^2+\Phi J)+h.c.)
\eea 
(as usual, conjugated terms to chiral superfields are antichiral ones). 
It can be rewritten in terms of integrals over chiral and antichiral
subspace only:
\bea
\label{only}
S_J[\Phi,\bar{\Phi};J,\bar{J}]
=\int d^6z (\frac{1}{2}\Phi(-\frac{\bar{D}^2}{4})\bar{\Phi}
+\frac{\l}{3!}\Phi^3+
\frac{m}{2}\Phi^2+\Phi J)+h.c.
\eea
The generating functional is
\bea
Z[J,\bar{J}]=\int D\Phi D\bar{\Phi}\exp(iS_J[\Phi,\bar{\Phi};J,\bar{J}]).
\eea
The action $S_J$ (\ref{only}) can be represented in matrix form
\bea
S_J&=&\frac{1}{2}\int dz_1 dz_2
\left(\begin{array}{cc}\Phi(z_1)&\bar{\Phi}(z_1)\end{array}
\right)
\left(\begin{array}{cc}
m&-\frac{1}{4}\bar{D}^2\\
-\frac{1}{4}D^2 & m
\end{array}\right)
\left(\begin{array}{cc}
\delta_+(z_1-z_2)&0\\
0&\delta_-(z_1-z_2)
\end{array}\right)\times\nonumber\\&\times&
\left(\begin{array}{c}\Phi(z_2)\\ \bar{\Phi}(z_2)
\end{array}\right)+\int d^6z\Phi(z) J(z)+\int d^6\bar{z}\bar{\Phi}(\bar{z})\bar{J}(\bar{z})+\frac{\lambda}{3!}(\int d^6 z \Phi^3 +h.c.).
\eea
In this expression, integration in all terms is assumed with taking into account the 
corresponding chirality, f.e. $\int dz_1dz_2\Phi(z_1)m\delta_+(z_1-z_2)\Phi(z_2)\equiv \int d^6z_1d^6z_2\Phi(z_1)m\delta_+(z_1-z_2)\Phi(z_2)$, etc. (this formalism has been developed in \cite{BK0}).
We see that the operator $\Delta$
determining quadratic part of the action (see (\ref{ii},\ref{iii})) looks
like
\bea
\Delta&=&
\left(\begin{array}{cc}
m&-\frac{1}{4}\bar{D}^2\\
-\frac{1}{4}D^2 & m
\end{array}\right).
\eea
The propagator is an operator inverse to this one:
\bea
G=\Delta^{-1}=\frac{1}{\Box-m^2}\left(\begin{array}{cc}
m&\frac{1}{4}\bar{D}^2\\
\frac{1}{4}D^2 & m
\end{array}\right).
\eea
In other words, propagator $G$ satisfies the equation
\bea
\Delta G=-
\left(\begin{array}{cc}
\delta_+(z_1-z_2)&0\\
0&\delta_-(z_1-z_2)
\end{array}\right).
\eea
The matrix ${\bf 1}=\left(\begin{array}{cc}
\delta_+(z_1-z_2)&0\\
0&\delta_-(z_1-z_2)
\end{array}\right)$ plays the role of a functional unit matrix within this description.

Thus, the generating functional can be presented as
\bea
\label{GF2}
Z[J,\bar{J}]&=&
\exp(i\frac{\l}{3!}\Big(\int d^6 z \left(\frac{1}{i}\frac{\delta}{\delta J(z)}\right)^3
+h.c.)\Big){\rm det}^{-1/2}\Delta\times\nonumber\\&\times&
\exp\Big\{-\frac{i}{2}\int dz_1 dz_2
\left(\begin{array}{cc}J(z_1)&\bar{J}(z_1)\end{array}
\right)\frac{1}{\Box-m^2}
\left(\begin{array}{cc}
m&\frac{1}{4}\bar{D}^2\\
\frac{1}{4}D^2 & m
\end{array}\right)\times\nonumber\\&\times&
\left(\begin{array}{cc}
\delta_+(z_1-z_2)&0\\
0&\delta_-(z_1-z_2)
\end{array}\right)
\left(\begin{array}{c}J(z_2)\\ \bar{J}(z_2)
\end{array}\right)\Big\}.
\eea
The equivalent form of this expression is
\bea
\label{GF2a}
Z[J,\bar{J}]&=&
\exp\left(i\frac{\l}{3!}\int d^6 z \left(\frac{1}{i}\frac{\delta}{\delta J(z)}\right)^3
+h.c.\right)({\rm det}^{-1/2}\Delta)\,
\exp(\frac{i}{2}A[J,\bar{J}])
\eea
where
\bea
\label{argexp}
A[J,\bar{J}]=-\Big(\int d^6 z J\frac{m}{\Box-m^2} J +2 \int d^6 z
J\frac{\frac{1}{4}\bar{D}^2}{\Box-m^2}\bar{J}+\int d^6 \bar{z} 
\bar{J}\frac{m}{\Box-m^2}\bar{J}
\Big).
\eea
We can introduce two-point free Green functions (that is, those ones corresponding to $\lambda=0$):
\bea
\label{2point}
G_{++}(z_1,z_2)&=&
\frac{1}{i^2}\frac{\delta^2 Z[J]}{\delta J(z_1)\delta J(z_2)}|_{J=0}=
i(-\frac{1}{4})^2\bar{D}^2_1\bar{D}^2_2K_{++}(z_1,z_2)\nonumber\\
G_{+-}(z_1,z_2)&=&
\frac{1}{i^2}\frac{\delta^2 Z[J]}{\delta J(z_1)\delta \bar{J}(z_2)}|_{J=0}=
i(-\frac{1}{4})^2\bar{D}^2_1D^2_2K_{+-}(z_1,z_2)\nonumber\\
G_{--}(z_1,z_2)&=&
\frac{1}{i^2}\frac{\delta^2 Z[J]}{\delta \bar{J}(z_1)\delta \bar{J}(z_2)}|_{J=0}=
i(-\frac{1}{4})^2 D^2_1 D^2_2K_{--}(z_1,z_2).
\eea
Here
$K_{+-}(z_1,z_2)=K_{-+}(z_1,z_2)=-\frac{1}{\Box-m^2}\delta^8(z_1-z_2)$,
$K_{++}(z_1,z_2)=\frac{m D^2}{4\Box(\Box-m^2)}\delta^8(z_1-z_2)$,
$K_{--}(z_1,z_2)=\frac{m \bar{D}^2}{4\Box(\Box-m^2)}\delta^8(z_1-z_2)$.

There is an alternative way to obtain the Green functions \cite{West}. Indeed, the quadratic part of the action (\ref{only}) can be rewritten as an integral over whole superspace:
\bea
\label{only1}
S_J[\Phi,\bar{\Phi};J,\bar{J}]
&=&\int d^8 z
\left(\Phi\bar{\Phi}+\frac{m}{2}\Phi(-\frac{D^2}{4\Box})\Phi+\frac{m}{2}\bar{\Phi}(-\frac{\bar{D}^2}{4\Box})\bar{\Phi}\right.\nonumber\\&+&\left.\Phi (-\frac{D^2}{4\Box})J+\bar{\Phi}(-\frac{\bar{D}^2}{4\Box})\bar{J}\right),
\eea
which has the matrix form
\bea
S_J[\Phi,\bar{\Phi};J,\bar{J}]
&=&\frac{1}{2}\int d^8z
\left(\begin{array}{cc}\Phi(z)&\bar{\Phi}(z)\end{array}
\right)
\left(\begin{array}{cc}
-m\frac{D^2}{4\Box}&1\\
1&-m\frac{\bar{D}^2}{4\Box} 
\end{array}\right)
%\times\nonumber\\&\times&
\left(\begin{array}{c}\Phi(z)\\ \bar{\Phi}(z)
\end{array}\right)+\nonumber\\&+&\int d^8z\left(\Phi(z) (-\frac{D^2}{4\Box})J(z)+\bar{\Phi}(\bar{z})(-\frac{\bar{D}^2}{4})\bar{J}(\bar{z})\right)
\eea
Then, since
\bea
\left(\begin{array}{cc}
-m\frac{D^2}{4\Box}&1\\
1&-m\frac{\bar{D}^2}{4\Box} 
\end{array}\right)^{-1}=
\frac{1}{\Box-m^2}\left(\begin{array}{cc}
\frac{m\bar{D}^2}{4}&\Box\\
\Box&\frac{mD^2}{4} 
\end{array}\right),
\eea
after the functional integration, we arrive at the expression similar to (\ref{GF2a}), that is,
\bea
\label{GF2a1}
Z[J,\bar{J}]&=&
\exp\left(i\frac{\l}{3!}\int d^6 z \Big(\frac{1}{i}\frac{\delta}{\delta J(z)}\Big)^3
+h.c.\right)({\rm det}^{-1/2}\tilde{\Delta})\,
\exp(-\frac{i}{2}\tilde{A}[J,\bar{J}]),
\eea
where $\tilde{\Delta}=\left(\begin{array}{cc}
-m\frac{D^2}{4\Box}&1\\
1&-m\frac{\bar{D}^2}{4\Box} 
\end{array}\right)$, and the argument of the exponential looks like
\bea
\label{argexp1}
\tilde{A}[J,\bar{J}]=\int d^8z
\left(\begin{array}{cc}(\frac{D^2}{4\Box})J(z)&(\frac{\bar{D}^2}{4\Box})\bar{J}(z)\end{array}
\right)
\left(\begin{array}{cc}
\frac{m\bar{D}^2}{4(\Box-m^2)}&\frac{\Box}{\Box-m^2}\\
\frac{\Box}{\Box-m^2}&\frac{mD^2}{4(\Box-m^2)} 
\end{array}\right)\left(\begin{array}{c}(\frac{D^2}{4\Box})J(z)\\(\frac{\bar{D}^2}{4\Box})\bar{J}(z)\end{array}
\right),
\eea
which can be reduced to (\ref{argexp}) by straightforward transformations, i.e. the expressions (\ref{argexp}) and (\ref{argexp1}) are equivalent. Therefore we have shown that these ways to obtain the Green functions (and hence the Green functions themselves) are equivalent.

We note that in theory of a standard (non-chiral, in particular, real scalar)
superfield the variational derivatives with respect to sources 
do not involve factors $D^2$,
$\bar{D}^2$. These factors are caused by chirality. For example, for a theory of the 
real scalar superfield $V$, with the action $S=-\frac{1}{2}\int d^8 z V\Box V$ (which emerges after 
an appropriate gauge fixing) the propagator is simply $G(z_1,z_2)=\frac{1}{\Box}\delta(z_1-z_2)$.

Different vacuum expectations can be expressed in terms of the generating 
functional (\ref{GF2}) as
\bea
\label{vacexp}
& &<\phi(x_1)\ldots \phi(x_n)\bar{\phi}(y_1)\ldots\bar{\phi}(y_m)>=\nonumber\\
&=&
(\frac{1}{i}\frac{\delta}{\delta J(x_1)})\ldots
(\frac{1}{i}\frac{\delta}{\delta J(x_n)})
(\frac{1}{i}\frac{\delta}{\delta \bar{J}(y_1)})\ldots
(\frac{1}{i}\frac{\delta}{\delta \bar{J}(y_m)})
\times\nonumber\\&\times&
\exp\Big\{i\frac{\l}{3!}\int d^6 z \left(\frac{1}{i}\frac{\delta}{\delta J(z)}\right)^3
+h.c.)
{\rm det}^{-1/2}\Delta\times\nonumber\\&\times&
\exp(-\frac{i}{2}\int dz_1 dz_2
\left(\begin{array}{cc}J(z_1)\bar{J}(z_1)\end{array}
\right)\frac{1}{\Box-m^2}
\left(\begin{array}{cc}
m&\frac{1}{4}\bar{D}^2\\
\frac{1}{4}D^2 & m
\end{array}\right)\times\nonumber\\&\times&
\left(\begin{array}{cc}
\delta_+(z_1-z_2)&0\\
0&\delta_-(z_1-z_2)
\end{array}\right)
\left(\begin{array}{c}J(z_2)\\ \bar{J}(z_2)
\end{array}\right)\Big\}.
\eea 
Of course, this expression contains all orders in the coupling $\l$. To
obtain vacuum expectations up to a some order in couplings we should
expand $\exp(i\frac{\l}{3!}\int d^6 z (\frac{\delta}{i\delta J(z)})^3
+h.c.)$ into power series. As a result as usual we arrive at some 
Feynman diagrams. In these diagrams, $n+m$ is the number of external
points, and the order in $\lambda$ is the number of internal points. Each vertex
evidently corresponds to integration over $d^6 z$ or $d^6\bar{z}$.
Therefore we can introduce Feynman diagrams for superfield
theory, i.e. Feynman supergraphs. Their value consists in the fact
that they allow one to preserve a manifest supersymmetry covariance at any
step of calculations.   

Generating functionals of arbitrary superfield models can be constructed 
by analogy with Wess-Zumino model:
\bea
Z[\vec{J}]=\exp(i(S[\vec{\phi}]+\vec{\phi}\vec{J})).
\eea
Here $\vec{\phi}$ is a column matrix denoting set of all superfields,
$\vec{J}$ is a column matrix denoting set of corresponding sources.  
The Green functions can be determined in analogy with (\ref{vacexp}). The generalization for the case of the  presence of the superfields of different natures, including not only chiral and real ones but other possible superfields, does not essentially differ.

\section{Feynman supergraphs}

Now, after we have introduced the generating functional (\ref{GF2}), we can start with formulation of the superfield Feynman diagram technique, or supergraph technique. It can be introduced as follows. 

One can easily read off from (\ref{2point}) that any
 $<\phi\bar{\phi}>$-propagator corresponds to $(\Box-m^2)^{-1}$, at a
 chiral vertex each propagator is associated with the factor
 $(-\frac{1}{4}\bar{D}^2)$, and at an antichiral one -- with
 $(-\frac{1}{4}D^2)$. However, each chiral (or antichiral) vertex
 involves an integration over $d^6 z$ (or $d^6\bar{z}$). Furthermore,
 since we deal with the delta function $\delta^8(z_1-z_2)$, for sake of unity it is
 more convenient to represent all contributions in the form of
 integrals over $d^8 z$ via the rule $d^6 z (-\frac{1}{4})\bar{D}^2 F
=\int d^8 z F$. As a result, if all superfields associated to a given $\int d^6 z\Phi^n$-vertex are contracted into propagators, this vertex is associated with $n-1$ $(-\frac{1}{4}\bar{D}^2)$ factors, and, similarly, any
$\int d^6\bar{z}\bar{\Phi}^m$-vertex of course, in the case when all superfields are contracted into propagators. -- with $m-1$
 $(-\frac{1}{4}D^2)$-factors. 
And the vertex $\int d^8 z \Phi^m \bar{\Phi}^n$, in the same case when all superfields are contracted into propagators, is associated with $m$
  factors and $n$
 $(-\frac{1}{4}D^2)$ factors. 
Here and further we refer to superfields contracted into propagators
as to the quantum ones. We will denote the quantum chiral (antichiral) superfields as $\phi$ ($\bar{\phi}$).
We see that the 
number of $D^2$, $\bar{D}^2$ factors for such vertices is the number
 of antichiral (chiral) {\bf quantum} superfields associated with this vertex.
There is no such $D^2,\bar{D}^2$-factors arising in propagators of a non-chiral (f.e. real) superfield, since the presence of such factors is motivated by the variational derivative with respect to a chiral superfield. 

The propagator $<\phi\phi>$ 
($<\bar{\phi}\bar{\phi}>$) corresponds to the $\frac{mD^2}{4\Box(\Box-m^2)}$
($\frac{m\bar{D}^2}{4\Box(\Box-m^2)}$) factor. However, only quantum fields
 (i.e. those ones contracted into propagators) carry $D^2$, 
$\bar{D}^2$ factors. External lines do not carry such a factor, and if
 one, two... $n$ chiral (antichiral) superfields associated with the
 vertex are external the number of $\bar{D}^2$ ($D^2$) factors
 corresponding to this vertex is less by one, two... $n$ than in the case
 when all superfields are contracted to propagators.

If we consider the theory of $N=1$ super-Yang-Mills (SYM) field, its quadratic
action being the sum of the action (\ref{resca1}) and the simplest gauge-fixing term $S_{gf}=-\frac{1}{2}{\rm tr}\int d^8z V\frac{\{D^2,\bar{D}^2\}}{16}V$ corresponding to the Feynman gauge (we consider more general gauge fixing in Section 4.11) looks like
\bea
S=-\frac{1}{2}{\rm tr}\int d^8 z V\Box V.
\eea
Here $tr$ is matrix trace (remind that the superfield $V$ is Lie-algebra valued).
There is no $D$ factors associated with this propagator but they are
associated with vertices. In a pure $N=1$ SYM theory vertices are given
by
\bea
S_{int}=\frac{g}{16}{\rm tr}\int d^8 z (\bar{D}^2D^{\alpha}V)[V,D_{\alpha}V]+\ldots
\eea
Here dots denote higher orders in $V$. 
The vertices of any order involve are two $D$ factors
and two $\bar{D}$ factors. The D-factors in vertices involving both
real and chiral (antichiral) superfields are arranged in a common way,
i.e. any vertex $\phi\bar{\phi} V^n$ involves one factor 
$(-\frac{1}{4}\bar{D}^2)$ acting to the $\Phi$ superfield when it is
contracted to $<\phi\bar{\phi}>$ propagator and one factor $(-\frac{1}{4}D^2)$
acting to the $\bar{\phi}$ superfield when it is
contracted to the $<\phi\bar{\phi}>$ propagator.
As a result, we can formulate Feynman rules.

The propagators look like
\bea
<\phi(z_1)\bar{\phi}(z_2)>&=&-\frac{1}{\Box-m^2}\delta^8(z_1-z_2);\\
<\phi(z_1)\phi(z_2)>&=&\frac{mD^2}{4\Box(\Box-m^2)}\delta^8(z_1-z_2);\nonumber\\
<V(z_1)V(z_2)>&=&\frac{1}{\Box}\delta^8(z_1-z_2),\nonumber
\eea
and the vertices (here $\phi$, $\bar{\phi}$ are quantum superfields) correspond to
\bea
&&\int d^6 z\phi^n \to (n-1)\, {\rm factors}\, (-\frac{1}{4})\bar{D}^2;\nonumber\\
&&\int d^8 z \phi\bar{\phi}V^m \to {\rm one\, factor}\, (-\frac{1}{4})D^2\, {\rm and\, one\, factor} \, (-\frac{1}{4})\bar{D}^2.
\eea
All derivatives in derivative depending vertices act on the propagators.
Any external chiral (antichiral) fields do not carry 
$\bar{D}\, (D)$-factors except  of those ones originated from the definition of the vertices.
One will find further that the difference of signs of $<VV>$ with respect to $<\Phi\bar{\Phi}>$-propagator plays an important role within the context of the super-Yang-Mills theory. However, in principle this difference of signs takes place even in the simple non-supersymmetric scalar QED (it follows from the requirement of positivity of energy of the free theories). As usual, within the supergraphs each propagator carries the factor $i$.

Of course, it is more suitable to make Fourier representation for all 
propagators (note that Fourier transformation is carried out with
respect to bosonic coordinates only) by the rule
\bea
\tilde{f}(k)=\int d^4xf(x)e^{ikx}.
\eea
The propagators in momentum representation look like
\bea
\label{primp}
<\phi(1)\bar{\phi}(2)>&=&\frac{1}{k^2+m^2}\delta^4_{12};\\
<\phi(1)\phi(2)>&=&\frac{mD^2}{4k^2(k^2+m^2)}\delta^4_{12};\nonumber\\
<V(1)V(2)>&=&-\frac{1}{k^2}\delta^4_{12}.
\eea
Here 1, 2 are numbers of arguments (actually we must write $V(1)\equiv V(-k,\theta_1)$ and $V(2)\equiv V(k,\theta_2)$ etc.), 
and $\delta^4_{12}\equiv
\delta^4(\q_1-\q_2)=(\q_1-\q_2)^2(\bar{\q}_1-\bar{\q}_2)^2$
is a delta function related to the Grassmann coordinates. The $D$-factors are introduced as
above. Note, however, that spinor derivatives depend after Fourier
transform on a momentum
of a propagator with which they are associated. 
The external superfields also can be represented in the form
of the Fourier integral. Each propagator is parametrized by a momentum, and any
vertex corresponds to an integration over $d^4\q$, a corresponding coupling and a delta
function over incoming momenta multiplied by $(2\pi)^4$. As usual, contribution of any supergraph
includes integration over all momenta and combinatoric factor which is
totally analogous to that one in standard quantum field theory.

Essentially new feature of superfield theories consists in the presence of the $D$-factors.
To evaluate $D$-algebra we can transport them via integration by
parts, then, we can use the identity 
\bea
\label{loop}
\delta^4_{12}D^2\bar{D}^2\delta^4_{12}=16\delta^4_{12}
\eea
To prove this identity we can use the expansion of supercovariant derivatives
(\ref{gend}) and note that due to the evident property
$\delta^4_{12}\frac{\pa}{\pa\q^{\a}}\delta^4_{12}=\frac{\pa}{\pa\q^{\a}}
\delta^4_{12}|_{\q_1=\q_2}=2(\q_{1\a}-\q_{2\a})
(\bar{\q}_1-\bar{\q}_2)^2|_{\q_1=\q_2}=0$, and some more similar relations, only terms of the form
$\delta^4_{12}(\frac{\pa}{\pa\q})^2(\frac{\pa}{\pa\bar{\q}})^2\delta^4_{12}
=16\delta^4_{12}$ survive.

We can prove the following {\bf theorem}. 

The final result for the contribution of any supergraph should have
the form
of {\bf one} integral over $d^4\q$ \cite{West}.

{\bf Proof:} Let us consider the supergraph with $L$ loops, $V$ vertices and $P$
propagators. Any vertex contains an integration over $d^4\q$, i.e. there
are $V$ such integrations.
Then, due to (\ref{primp}) any propagator carries
a delta function over Grassmann coordinates, i.e. there are $P$ delta
functions. Then, in any loop we can reduce the number of delta functions
by one using identity (\ref{loop}), i.e. there are $P-L$ independent
delta functions. As a result we can carry out $P-L$ integrations from
$V$, and after $D$-algebra transformations we stay with $V-(P-L)$
integrations. And, due to the famous topological identity, $V-(P-L)=1$, therefore the result contains one
integration over $d^4\q$. The theorem is proved. 

This theorem is often called the non-renormalization theorem. It means
that all quantum corrections are local in $\q$-space. This theorem is
often naively treated as a proof of absence of chiral corrections
(i.e. of those ones proportional to an integral over $d^2\q$). However, such an interpretation
is wrong since any contribution in the form of integral over chiral
subspace can be rewritten as an integral over whole superspace using
identity
\bea
\int d^6z f(\Phi)=\int d^8 z (-\frac{D^2}{4\Box})f(\Phi). 
\eea
This observation was firstly made in \cite{West2}, its consequences 
will be studied further.

Now let us study evaluation of contributions from supergraphs.
The algorithm of it, after rewriting all vertices as integrals over the whole superspace, is the following one.

1. We start with one of loops. 
If the number of $D$-factors in this loop is equal to 4 we turn to step 2.
If it is more than  4,
superfluous $D$-factors can be transported to external lines or another
loops via integration by parts, and some of them are converted into
internal momenta via identities $D^2\bar{D}^2D^2=16\Box D^2, 
\{D_{\alpha},\bar{D}_{\dot{\beta}}\}=2i\partial_{\alpha\dot{\beta}}$. 
As a result we stay with exactly 4 $D$-factors.
If the number of $D$-factors is less than 4 than contribution from the
entire supergraph is equal to zero.

2. We contract this loop into a point using identity (\ref{loop}) and
intergrate over one of $d^4\q$ via the delta function which is free of derivatives.  

3. This procedure is repeated for next loops.

4. We integrate over internal momenta.

The best way to study evaluation of supergraphs consists in
considering some examples.

%\newpage

{\bf Example 1. One-loop supergraph in the Wess-Zumino model.}

\vspace*{2mm}

\hspace{4.5cm}
\Lengthunit=1cm
%\Linewidth{1.2pt}
\GRAPH(hsize=3){\Circle(2)\mov(-1,0){\lin(-1,0)}\mov(1,0){\lin(1,0)}\ind(-10.5,1){-}\ind(10.5,1){-}
\ind(13,3){D^2}\ind(-13,3){\bar{D}^2}\ind(0,-13){Fig.2}
}

\vspace*{2mm}
  
The contribution of this supergraph is equal to
\bea
I_1&=&\frac{\lambda^2}{2}\int d^4\q_1
d^4\q_2\int\frac{d^4p}{(2\pi)^4}\Phi(-p,\q_1)\bar{\Phi}(p,\q_2)
\delta^4_{12}\frac{\bar{D}^2_1D^2_2}{16}\delta^4_{12}\times\nonumber\\&\times&
\int\frac{d^4k}{(2\pi)^4}\frac{1}{(k^2+m^2)((k+p)^2+m^2)}
\eea
The number of $D$-factors is just 4. $D$-algebra transformations are
trivial: we use identity (\ref{loop}) and write 
$\delta^4_{12}\frac{\bar{D}^2_1D^2_2}{16}\delta^4_{12}=\delta^4_{12}$.
The free delta function $\delta^4_{12}$ allows us to integrate over 
$d^4\q_2$, afterwards, we denote $\q_1=\q$. As a result we get
\bea
I_1=\frac{1}{2}\lambda^2\int d^4\q
\int\frac{d^4p}{(2\pi)^4}\Phi(-p,\q)\bar{\Phi}(p,\q)
\int\frac{d^4k}{(2\pi)^4}\frac{1}{(k^2+m^2)((k+p)^2+m^2)}
\eea
Integral over $k$ can be calculated via dimensional regularization,
the result for it is
\bea
\int\frac{d^4k}{(2\pi)^4}\frac{1}{(k^2+m^2)((k+p)^2+m^2)}=
\frac{1}{16\pi^2}
(\frac{1}{\epsilon}-\int_0^1 dt\log\frac{p^2t(1-t)+m^2}{\mu^2})
\eea 
As a result, contribution of this supergraph takes the form
\bea
\label{contri1}
I_1=\frac{1}{2}\lambda^2\int d^4\q
\int\frac{d^4p}{(2\pi)^4}\Phi(-p,\q)\bar{\Phi}(p,\q)
\frac{1}{16\pi^2}
(\frac{1}{\epsilon}-\int_0^1 dt\log\frac{p^2t(1-t)+m^2}{\mu^2})
\eea
However, using of the dimensional regularization in superfield theory in higher loops possesses
some peculiarities, in fact, in some cases it is ambiguous \cite{Jack}.

%\newpage

{\bf Example 2. Two-loop supergraph in the Wess-Zumino model.}

\vspace*{2mm}

\hspace{4.5cm}
\Lengthunit=1cm
%\Linewidth{1.2pt}
\GRAPH(hsize=3){\Circle(2)\mov(-1,0){\lin(2,0)}
\ind(-11,1){-}\ind(9,1){-}
\ind(11,3){D^2}\ind(-15,3){\bar{D}^2}
\ind(-13.5,-2){-}\ind(4,-1){|}
\ind(1.7,-5){D^2}\ind(-19,-4){\bar{D}^2}
\ind(-2,-14){Fig.3}
}

\vspace*{2mm}

The contribution of this supergraph is equal to 
\bea
I_2&=&\frac{\lambda^2}{6}\int\frac{d^4k d^4 l}{(2\pi)^8}\int d^4\q_1 d^4\q_2
(-\frac{D^2_1}{4})\delta^4_{12}\frac{D^2_1\bar{D}^2_2}{16}\delta^4_{12}
(-\frac{\bar{D}^2_2}{4})\delta^4_{12}\times\nonumber\\&\times&
\frac{1}{(k^2+m^2)(l^2+m^2)((k+l)^2+m^2)}
\eea
First we do $D$-algebra transformations: we can write
$$(-\frac{D^2_1}{4})\delta^4_{12}\frac{D^2_1\bar{D}^2_2}{16}\delta^4_{12}
(-\frac{\bar{D}^2_2}{4})\delta^4_{12}=
\delta^4_{12}
\frac{D^2_1\bar{D}^2_2}{16}\delta^4_{12}\frac{D^2_1\bar{D}^2_2}{16}
\delta^4_{12}
$$
Then we use identity (\ref{loop}) two times:
$$
\delta^4_{12}
\frac{D^2_1\bar{D}^2_2}{16}\delta^4_{12}\frac{D^2_1\bar{D}^2_2}{16}
\delta^4_{12}=\delta^4_{12}
$$ 
As a result we can integrate over $\q_2$ using the  delta function.
We get
\bea
I_2&=&\frac{\lambda^2}{6}\int\frac{d^4k d^4 l}{(2\pi)^8}\int d^4\q_1 
%(-\frac{D^2_1}{4})\delta^4_{12}\frac{D^2_1\bar{D}^2_2}{16}\delta^4_{12}
%(-\frac{\bar{D}^2_2}{4})\delta^4_{12}\times\nonumber\\&\times&
\frac{1}{(k^2+m^2)(l^2+m^2)((k+l)^2+m^2)}
\eea
This integral vanishes in the standard case since it is proportional to
an integral over $d^4\q$ from constant. However, if we suppose that $m$
is not a constant but $\q$-dependent superfield, this contribution will
not be equal to zero. Namely this case is studied when the effective action is considered
and $m$ is suggested to depend on background superfields (just this situation has been considered in \cite{my4}).

\newpage

{\bf Example 3. One-loop supergraph in dilaton supergravity.}

The action of the theory looks like \cite{my1}:
\bea
S&=&\int d^8z\Big[-\frac{Q^2}{16\pi^2}\bar{\sigma}\Box\sigma +\bar{D}^{\ad}\bar{\sigma}D^{\alpha}\sigma (\xi_1\partial_{\alpha\ad}(\sigma-\bar{\sigma})+\xi_2\bar{D}_{\ad}\bar{\sigma}D_{\alpha}\sigma)+\frac{m^2}{2}e^{\sigma+
\bar{\sigma}}
\Big]+\nonumber\\&+&[\lambda\int d^6z e^{3\sigma}+h.c.]
\eea

\vspace*{2mm}

One of the contributions to the wave function renormalization is given by the following supergraph:

\hspace{4.5cm}
\Lengthunit=1cm
%\Linewidth{1.2pt}
\GRAPH(hsize=3){\Circle(2)\mov(-1,0){\lin(-1,0)}\mov(1,0){\lin(1,0)}
\ind(-9,5){|}\ind(7,5){|}
\ind(12,9){\bar{D}^2D^{\beta}}\ind(-16,8){\bar{D}^{\dot{\alpha}}D^2}
\ind(-12,-6){|}\ind(4,-6){|}
\ind(9,-9){D^2\bar{D}^{\dot{\beta}}}\ind(-17,-11){D^{\alpha}\bar{D}^2}
\ind(-23,4){\partial_{\alpha\dot{\alpha}}}\ind(-23,0){|}
\ind(8,5){\partial_{\beta\dot{\beta}}}\ind(8,0){|}
\ind(-5,13){G(k)}\ind(-6,-14){G(k+p)}
\ind(-7,-22){Fig.4}
}

\vspace*{2mm}

The contribution of this supergraph is equal to
\bea
\label{i3}
I_3&=&\xi_1^2\int d^4\q_1 d^4\q_2\int\frac{d^4p}{(2\pi)^4}
\frac{d^4k}{(2\pi)^4}(\partial_{\alpha\dot{\alpha}}\sigma(-p,\q_1))
(\partial_{\beta\dot{\beta}}\bar{\sigma}(p,\q_2))\times\nonumber\\&\times&
\frac{D^{\alpha}\bar{D}^2D^2\bar{D}^{\dot{\beta}}}{16}\delta^4_{12}
\frac{\bar{D}^{\dot{\alpha}}D^2\bar{D}^2 D^{\beta}}{16}\delta^4_{12}
%\times\nonumber\\&\times&
G(k)G(k+p).
\eea
Here $G(k),G(k+p)$ are functions of momenta whose explicit form is not
essential here (they are exactly found in \cite{my1}). The derivatives
$\partial_{\a\dot{\a}},\partial_{\b\dot{\b}}$ are not transported from
external fields $\sigma,\bar{\sigma}$. Our aim here is to obtain terms
proportional to $\partial^m\sigma\partial^n\bar{\sigma}$ (we do not express derivatives acting on the externl fields in terms of the corresponding momenta since we will not more manipulate with these derivatives, so, we consider $\partial_{\alpha\dot{\alpha}}\sigma$ and
$\partial_{\beta\dot{\beta}}\bar{\sigma}$ as some independent external fields.
We suggest that spinor derivatives associated with one propagator
depend on momentum $k$, and with another -- to $k+p$.

Using commutation relations (\ref{salg}) we find that
\bea
\label{stru}
\frac{D^{\alpha}\bar{D}^2D^2\bar{D}^{\dot{\beta}}}{16}\delta^4_{12}
\frac{\bar{D}^{\dot{\alpha}}D^2\bar{D}^2 D^{\beta}}{16}\delta^4_{12}=-
\frac{4k^{\a\dot{\g}}\bar{D}_{\dot{\g}}D^2\bar{D}^{\bd}}{16}\delta^4_{12}
\frac{4(k+p)^{\g\ad}D_{\g}\bar{D}^2D^{\b}}{16}\delta^4_{12}.
\eea
We transport all spinor supercovariant derivatives to one propagator
(here the terms with spinor supercovariant derivatives moved to the
external lines are omitted as the irrelevant ones since they do not
contribute to the divergent part \cite{my1}).
As a result we arrive at
\bea
16k^{\a\dot{\g}}(k+p)^{\g\ad}\delta^4_{12}
\frac{\bar{D}^{\bd}D^2\bar{D}_{\dot{\g}}D_{\g}\bar{D}^2 D^{\b}}{256}
\delta^4_{12}.
\eea
We can use (\ref{salg}) several times. At the end we get
\bea
16k^{\a\dot{\g}}(k+p)^{\g\ad}(k+p)_{\g\dot{\g}}(k+p)^{\bd\delta}
\delta^4_{12}\frac{D_{\delta}\bar{D}^2D^{\beta}}{32}\delta^4_{12}.
\eea
Equations (\ref{salg}) and (\ref{loop}) allow one to write
$$
\delta^4_{12}D_{\delta}\bar{D}^2D^{\beta}\delta^4_{12}=-\frac{1}{2}
16\delta^{\b}_{\delta}\delta^4_{12}.
$$
We substitute this expression in (\ref{stru}). Using identity
$k^{\a\bd}k_{\g\bd}=\delta^{\a}_{\g}k^2$ we obtain the
contribution from (\ref{stru}) in the form
$$
k^{\a\ad}(k+p)^{\b\bd}(k+p)^2\delta^4_{12},
$$
which after integration over $\q_2$ leads to the 
following expression for $I_3$:
\bea
I_3&=&-4\xi_1^2\int d^4\q_1 \int\frac{d^4p}{(2\pi)^4}
\frac{d^4k}{(2\pi)^4}(\partial_{\alpha\dot{\alpha}}\sigma)(-p,\q_1)
(\partial_{\beta\dot{\beta}}\bar{\sigma})(p,\q_1)\times\nonumber\\&\times&
k^{\a\ad}(k+p)^{\b\bd}(k+p)^2G(k)G(k+p).
\eea
Detailed analysis carried out in \cite{my1} shows that this
correction is divergent.

Calculation of corrections from supergraphs in other superfield
theories is carried out on the base of analogous approach.

We demonstrated that supergraph technique is a very efficient method for
consideration of quantum corrections in superfield theories whereas
the component study is much more complicated since one supergraph
corresponds to several component diagrams (it is amusing that the
exact expression for the classical action of dilaton supergravity
occupies a whole page \cite{my1}). The next
step of its development consists in introducing renormalization in these theories.

\section{Superficial degree of divergence. Renormalization.}

We found that divergent quantum corrections arise in
superfield theories as well as in standard field theories. Therefore
we face two problems:

(i) to classify possible divergences;

(ii) to develop a procedure of renormalization in superfield theories.

It turns out that the technique for solving these problems is quite
analogous to that one used in standard field theory. First problem
can be solved on the base of superficial degree of divergence. The
natural way for solving second one is in introducing superfield
counterterms which are quite analogous to standard ones.

First of all let us consider the superficial degree of divergence
\cite{Collins}. 

{\bf Example.} The $N=1$ super-Yang-Mills (SYM) theory with chiral matter
(with Wess-Zumino self-interaction) \cite{BK0}. For all other models,
the consideration is quite analogous.
The action of the theory is
\bea
\label{actsym}
S&=&\int d^8 z\bar{\Phi}_i(e^{gV})^i_j\Phi^j+
(\int d^6 z(\frac{1}{2}m_{ij}\Phi_i\Phi_j+
\frac{\l_{ijk}}{3!}\Phi_i\Phi_j\Phi_k)+h.c.)-\\&-&
{\rm tr}\frac{1}{16g^2}\int d^8 z (e^{-gV}D^{\alpha}e^{gV})
\bar{D}^2(e^{-gV}D_{\a}e^{gV})+\nonumber\\&+&
\int d^8 z {\rm tr}\Big(\bar{c}'c-\bar{c}c'+
\frac{1}{2}g(\bar{c}'-c')[V,c+\bar{c}]+\ldots\Big).\nonumber
\eea
The dots are for the higher couplings involving ghosts. 
The triple vertices in this theory are (cf. \cite{GRS})
\bea
\label{vert}
&& \frac{\l_{ijk}}{3!}
\int d^6z\Phi_i\Phi_j\Phi_k+h.c.,\, 
g\int d^8 z \bar{\Phi}_i
V^A(T^A)_j^i \Phi^j,\, \nonumber\\
&& \frac{g}{16}{\rm tr}\int d^8 z (\bar{D}^2 D^{\alpha} V) [V,D_{\alpha} V],\, \frac{g}{2}{\rm tr}\int d^8z  (\bar{c}'-c')[V,c+\bar{c}].
\eea
Indices $i,j$ are matrix indices since $\Phi_i$ is an
isospinor, and $V\equiv V^A T^A$ is Lie-algebra valued as well as the ghosts are.
However, all vertices corresponding to pure SYM 
self-interaction contain exactly two chiral and two antichiral
derivatives. We have proved already that all corrections should be
proportional to one integral over $d^4\q$. 

As usual, the superficial degree of divergence (SDD) is the order of 
the integral over
internal momenta for the corresponding contribution, or, as is the same,
is a degree of homogeneity of diagram in momenta, considered after
evaluation of $D$-algebra transformations \cite{BK0}. 
The only difference of the SDD in the superfield case is the additional impact
from $D$-factors.

It is easy to see that contributions to the SDD are generated by 
momentum depending factors in propagators and vertices
(as usual, any internal momentum $k$ gives contribution 1), loop
integrations, or, in other words, by manifest momentum dependence
which is associated with propagators and loop integration, and by
$D$-factors which are associated with propagators and vertices
(note that due to identities $D^2\bar{D}^2D^2=16\Box D^2$,
$\{D_{\a},\bar{D}_{\ad}\}=2i\pa_{\a\ad}$, one chiral derivative
combined with an antichiral one can be converted to one momentum; therefore
any $D$-factor contribute to the SDD with $1/2$). If not all
spinor derivatives are converted to internal momenta, the SDD from
supergraph evidently decreases. 

Let us consider arbitrary supergraph with $L$ loops, $V$ vertices, $P$
propagators ($C$ of them are $<\phi\phi>$, $<\bar{\phi}\bar{\phi}>$
-propagators) and $E$ external lines ($E_c$ of them are chiral). 
We denote the SDD as $\omega$.

Any integration over internal momentum (i.e. over $d^4k$) contributes
to SDD with 4. Since the number of integrations over internal momenta is
the number of loops, the  total 
contribution from all such integrations is $4L$.  
Any propagator includes $\frac{1}{k^2+m^2}$ or $\frac{1}{k^2}$ 
(\ref{primp}), hence contribution of all propagators is equal to $-2P$. Since
$<\Phi\Phi>,<\bar{\Phi}\bar{\Phi}>$-propagator contains additional 
$\frac{1}{k^2}$ these propagators give additional contribution $-2C$.
Therefore manifest dependence of momenta gives contribution to
$\omega$ equal to $4L-2P-2C$.

Now let us consider contribution of $D$-factors to SDD.
Each {\bf vertex} (both pure gauge one and 
that one containing chiral superfields) 
without external chiral (antichiral) lines contains four 
$D$-factors (\ref{vert}) since any superfield $\phi$ (contracted to
propagator) corresponds to $\bar{D}^2$, and $\bar{\phi}$ -- to $D^2$.
Therefore each vertex gives contribution 2.
However, external chiral (antichiral) lines do not correspond to
$D$-factors. As a result, any external line decreases $\omega$ by 1, 
Each $<\phi\phi>,<\bar{\phi}\bar{\phi}>$-propagator contains a factor
$\bar{D}^2$ ($D^2$) with contribution 1. Then, due to the identity
(\ref{loop}) contracting any loop into a point decreases the number of
$D$-factors which can be converted to internal momenta by 4, and 
$\omega$ -- by 2.
As a result the total contribution of $D$-factors to $\omega$ is equal to
$2V-E_c-2L+C$ (remind that each $D$-factor contributes to $\omega$ with
$1/2$). 

Therefore SDD is equal to
\bea
\omega=4L-2P-2C+2V-E_c-2L+C=2L-2P+2V-C-E_c.
\eea
Using the well known topological identity $L+V-P=1$ we have
\bea
\label{ind}
\omega=2-C-E_c.
\eea
Really, the SDD can be lower than (\ref{ind}) if some of $D$-factors are
transported to external lines and do not generate internal momenta. If $N_D$
D-factors are moved to external lines the $\omega$ is equal to
\bea
\omega=2-C-E_c-\frac{1}{2}N_D.
\eea
This is the final expression for the SDD. As usual, at $\omega\geq 0$
 supergraph diverges, and at $\omega <0$ -- converges.
We note that:\\ 
1. $\omega\leq 2$ hence the SDD is restricted from above.\\
2. As the number of external lines grows, $\omega$ decreases. Therefore
the number of divergent structures is essentially restricted -- it is
finite (really, there can be no more than two external chiral legs
and no more than two $<\F\F>,<\bar{\F}\bar{\F}>$ propagators).
And since the number of divergent structures is finite the theory is
renormalizable. Hence we have just shown that the theory including chiral
superfields with Wess-Zumino-type interaction and gauge superfields
with action (\ref{actsym}) is renormalizable.
This is quite natural since the mass dimensions of all couplings in this
theory are equal to zero.

However, non-renormalizable superfield theories also exist.

{\bf Example.} General chiral superfield model \cite{my2}.

The action of the model is
\bea
\label{actgen}
S&=& \int d^8 z K(\Phi,\bar{\Phi})+(\int d^6 z W(\Phi) +h.c.)=\nonumber\\
&=&\int d^8 z \Phi\bar{\Phi}+
[\int d^6 z (\frac{1}{2}m\F^2+\frac{\l}{3!}\F^3)+h.c.]+\\&+&
\int d^8 z
\sum_{m,n=1}^{\infty}
 \frac{1}{m!n!}K_{mn}\F^n\bar{\F}^m+(\int d^6
 z\sum_{l=4}^{\infty}\frac{W_n}{n!}
\Phi^n+h.c.).\nonumber
\eea
Here $K_{ij}, W_l$ are constants with nontrivial (actually, negative) mass dimension, and we impose the restriction $K_{11}=0$ since the corresponding term is already taken into account as a kinetic term.

Propagators in the theory are just (\ref{primp}), their contribution
to the SDD is equal to $4L-2P-2C$ as above. However, the contribution from
$D$-factors differs. Any vertex $K_{nm}\Phi^n\bar{\F}^m$ corresponds
to $n$ $\bar{D}^2$-factors and $m$ $D^2$-factors.
The total contribution to $\omega$ from all such vertices is
$\sum\limits_{V_t}(n_v+m_v)$, i.e. the sum of $n$ and $m$ over all vertices 
corresponding to the integral over the total superspace.
Any vertex $W_l\F^l$ contains an integral over $d^6 z$ and
effectively corresponds to $(l-1)$ $\bD^2$-factors. Total
contribution from such vertices is $\sum\limits_{V_c}(l_c-1)$ (i.e. sum over
all purely chiral or antichiral vertices). Again, external lines
decrease the number of $D^2 \ (\bD^2)$-factors by $2E_c$ ($E_c$ is a
number of external lines), each $<\F\F>,<\bar{\F}\bar{\F}>$-propagator
carries one $\bD^2$ ($D^2$)-factor. Contraction of each loop to a
point decreases the number of $D$-factors by 4.
Hence the total number of $D$-factors is
\bea
2\sum_{V_t}(n_v+m_v)+2\sum_{V_c}(l_c-1)-2E_c-4L+2C.
\eea 
Contribution to the SDD from $D$-factors is their number divided by two.
Therefore the total SDD in this theory is equal to
\bea
\omega&=&4L-2P-2C+\frac{1}{2}
(2\sum_{V_t}(n_v+m_v)+2\sum_{V_c}(l_c-1)-2E_c-4L+2C)=
\nonumber\\&=&
2-2V-C-2E_c+[\sum_{V_t}(n_v+m_v)+\sum_{V_c}(l_c-1)].
\eea
Here we used $2L-2P=2-2V$. However, any vertex gives contribution
$-2$ to term $-2V$ and $l_c-1$ or $n_v+m_v$ to other terms of
$\omega$. It is evidently that either $l_c-1$ or $n_v+m_v$ 
can be more than 2 since either $l_c\geq 3$ or $n_v+m_v\geq
3$.
Hence in general case $\sum\limits_{V_t}(n_v+m_v)+\sum\limits_{V_c}(l_c-1)-2V\geq
0$,
thus the number of divergent structures is not restricted, and the theory is
non-renormalizable. This is quite natural since the constants $K_{ij}$
(if $i,j\geq 2$) and $W_l$ (if $l\geq 4$) have negative mass
dimensions.

The next problem consists in the introduction of regularization.
The most natural way of introducing regularization in supersymmetric
theories is the dimensional regularization. It can be introduced as usual:
an integral
$$
\int\frac{d^4k}{(2\pi)^4}\frac{1}{(k^2+m^2)^N}
$$
is replaced by
$$
\mu^{-\epsilon}\int\frac{d^{4+\epsilon}k}{(2\pi)^{4+\epsilon}}\frac{1}{(k^2+m^2)^N},
$$
so, all divergences are given by poles in $\epsilon$ (no more than
$\frac{1}{\epsilon^L}$ for $L$-loop correction).

However, there are some peculiarities. First of all, at the component
level any supersymmetric action includes spinors and hence
$\gamma$-matrices which are well defined if and only if the dimension of the
space-time is integer. Therefore we must use some modification of the
dimensional regularization called dimensional reduction. According to
it, all objects which are well-defined only at separate dimensions (such
as spinors and Dirac $\gamma$ matrices) are evaluated at these dimensions
(in our case -- at the dimension equal to 4), and integrals over momenta -- at
arbitrary dimensions.
However, dimensional reduction leads to some difficulties in
calculation of higher loop corrections since many supergraphs involve
contractions of essentially four-dimensional objects, such as
Levi-Civita tensor $\epsilon^{abcd}$, with $d$-dimensional objects,
and such contractions need additional definition. As a result
frequently the ambiguities arise. However, such phenomena are observed only
beyond two loops, see f.e. \cite{Jack}.

We also can use analytic regularization which corresponds to change
$$
\int\frac{d^4k}{(2\pi)^4}\frac{1}{(k^2+m^2)^n}\to
\mu^{2\epsilon}\int\frac{d^4k}{(2\pi)^4}\frac{1}{(k^2+m^2)^{n+\epsilon}}.
$$
However, this regularization also leads to some difficulties
(see discussion of questions connected to regularization in
supersymmetric theories in \cite{Jack}).

The technique for renormalization in superfield theories is quite
analogous to that one in common QFT. It is carried out via
introduction of counterterms.

{\bf Example.} Let us consider the one-loop contribution to the kinetic term in the
Wess-Zumino model. The corresponding supergraph is given by Fig. 2 (see
above), its contribution (\ref{contri1}) yields
\bea
I_1=\frac{1}{2}\lambda^2\int d^4\q
\int\frac{d^4p}{(2\pi)^4}\Phi(-p,\q)\bar{\Phi}(p,\q)
\frac{1}{16\pi^2}
(\frac{1}{\epsilon}-\int_0^1 dt\log\frac{p^2t(1-t)+m^2}{\mu^2}).
\eea 
We see that this divergence has the form of pole part proportional to 
$\frac{1}{\epsilon}$. To cancel it we must add to the initial kinetic
term
\bea
S=\int d^8 z \Phi(x,\q)\bar{\Phi}(x,\q) 
\eea
(which is just $\int \frac{d^4 p}{(2\pi)^4} d^4 \q \Phi (-p,\q)
\bar{\Phi}(p,\q)$) a counterterm
\bea
\Delta S_{countr}=-\frac{\lambda^2}{32\pi^2\epsilon}\int d^8 z
\Phi(z)\bar{\Phi}(z)
\eea
which corresponds to the replacement of $\int d^8 z \Phi\bar{\Phi}$ in
the classical action by $Z\int d^8 z \Phi(z)\bar{\Phi}(z)$ where
\bea
Z=1-\frac{\lambda^2}{32\pi^2\epsilon}
\eea
is a wave function renormalization.

The essential peculiarity of superfield theories is the fact that the
number of counterterms in these theories is less than in their
non-supersymmetric analogues. For example, Wess-Zumino model is a
supersymmetric generalization of $\phi^4$-theory, but it involves
only one renormalization constant corresponding to the renormalization of kinetic term and no renormalization of
couplings. The conclusion about absence of divergent correction to
coupling the $\lambda$ (or as is the same -- to chiral potential) is also
called non-renormalization theorem since it is treated as a consequence 
of the non-renormalization theorem discussed earlier, following which, all loop corrections have the form of the integral over $d^4\theta$. However, actually, the existence of the {\bf finite} corrections to the superpotential (also called the chiral or holomorphic potential) is not forbidden, for example, they are present in the
massless Wess-Zumino model \cite{West3,West4,my3,my4}.

There are also some interesting properties of renormalization in
superfield theories.

First, all tadpole-type contributions in the Wess-Zumino model vanish:
the supergraph

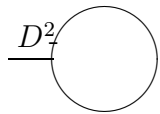
\unitlength=.4mm
\hspace*{5cm}
\begin{picture}(70,60)
\put(30,30){\circle{40}}
\put(13,30){\line(-1,0){15}}
\put(11.5,33){-}
\put(1,34){$D^2$}
\end{picture}

\noindent has a contribution proportional to
\noindent $D^2\delta_{11}=\delta_{12}D^2\delta_{12}=0$. The similar
situation can occur in other superfield models involving the Wess-Zumino
model as an ingredient.
However, in theories including vertices proportional to integral over
whole superspace (f.e. dilaton supergravity) tadpole contributions are
not equal to zero \cite{my1}.

Second, all contributions from vacuum supergraphs are proportional to
$\int d^4\q c$ (with $c$ is a constant) and also vanish. However, this
statement is not true for background dependent propagators. The methodology of
background dependent propagators naturally arises within the formalism of the effective action whose key features are the same in the common field theory (see Chapter 2) and in the supersymmetric field theory, hence, the concepts developed in the Chapter 2 can be straightforwardly applied for the superfield theories. In the next section we carry out this application.

\section[Effective action]{Effective action in superfield theories. Superfield proper-time technique}

Let us now discuss the problem of the effective action in superfield theories. 
In the Chapter 2, we have shown that the effective action $\Gamma[\Phi]$ depending on the background (super)field $\Phi$ (actually, in general case the $\Phi$ denotes a set of all background (super)fields, whose indices are suppressed) can be presented as $\Gamma[\Phi]=S[\Phi]+\bar{\Gamma}[\Phi]$, where $S[\Phi]$ is a classical action of the theory, and $\Gamma[\Phi]$ is a complete quantum contribution to the effective action which can be determined from the expression
\bea
\label{exph}
& &e^{\frac{i}{\hbar}\bar{\Gamma}[\Phi]}=\int D\phi e^{
\frac{i}{2}S^{\prime\prime}[\Phi]\phi^2}
\Big(
1+\frac{i\sqrt{\hbar}}{3!}S^{(3)}[\Phi]\phi^3+\frac{i\hbar}{4!}S^{(4)}
[\Phi]\phi^4+\nonumber\\&+&
\frac{1}{2}\left(\frac{i\sqrt{\hbar}}{3!}\right)^2(S^{(3)}[\Phi]\phi^3)^2+\ldots
\Big).
\eea 
Expanding the effective action in power series in $\hbar$ as $\bar{\Gamma}[\Phi]=\sum\limits_{L=1}^{\infty}\hbar^L\Gamma_L[\Phi]$, one can
express the one-loop contribution to $\Gamma[\Phi]$ in terms of the trace of the logarithm of the background dependent propagator $S^{\prime\prime}[\Phi]$ as
\bea
\Gamma^{(1)}=\frac{i}{2}{\rm Tr}\ln \Delta[\Phi],
\eea
where $\Delta[\Phi]\equiv S^{\prime\prime}[\Phi]$ is a background dependent operator characterizing the quadratic action of the quantum fields. This is the famous expression of the one-loop effective action in terms of the trace of the logarithm. Study of the one-loop correction is a starting point for any discussions of the effective action.

Now, it is instructive to discuss the following question: how the
definition of the one-loop correction in an effective action in terms of the
trace of the logarithm is related to the expression of the same correction in
terms of (super)graphs? 

To clarify this relation we give an example. The one-loop effective action
in the Wess-Zumino model is given by the following functional trace \cite{Buch1,BK0}:
\bea
\label{olwz}
\Gamma^{(1)}=\frac{i}{2}{\rm Tr}\log (\Box-\frac{1}{4}\Psi\bar{D}^2
-\frac{1}{4}\bar{\Psi}D^2).
\eea
Here $\Psi=m+\lambda\Phi$ is background chiral superfield. It is clear that the operator $\Delta$ in this case is $\Delta=\Box-\frac{1}{4}\Psi\bar{D}^2
-\frac{1}{4}\bar{\Psi}D^2$. The $\Gamma^{(1)}$ can be rewritten as
\bea
\Gamma^{(1)}=\frac{i}{2}{\rm Tr}\log[\Box(1-
\frac{1}{4\Box}(\Psi\bar{D}^2+\bar{\Psi}D^2)
)].
\eea
Expansion of the logarithm into power series leads to
\bea
\Gamma^{(1)}=-\frac{i}{2}{\rm Tr}\sum_{n=1}^{\infty}\frac{1}{n}
[\frac{1}{4\Box}(\Psi\bar{D}^2+\bar{\Psi}D^2)]^n.
\eea
This expression exactly reproduces the total contribution for 
the sum of the following supergraphs

\hspace{0.5cm}
\unitlength=.6mm
%\thicklines
%\GRAPH(hsize=3){%\ind(50,-30){Fig.1.}
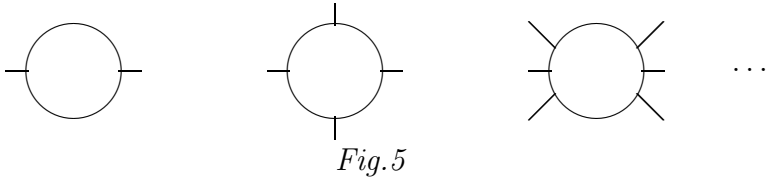
\begin{picture}(20,20)
\put(0,10){\circle{20}}\put(-10,10){\line(-1,0){5}}
%\put(-10,8.5){\line(-1,0){5}}
\put(10,10){\line(1,0){5}}%\put(10,8.5){\line(1,0){5}}
%\ind(24,0){W''}\ind(-26,0){\bar{W''}}
%\put(4.8,0){\GRAPH(hsize=3){
\end{picture}
\hspace{2cm}
\begin{picture}(20,20)
\put(0,10){\circle{20}}\put(-10,10){\line(-1,0){5}}
%\put(-10,8.5){\line(-1,0){5}}
\put(10,10){\line(1,0){5}}%\put(10,8.5){\line(1,0){5}}
\put(0,20){\line(0,1){5}}%\put(-1,20){\line(0,1){5}}
\put(0,-.1){\line(0,-1){5}}
%\put(-1,0){\line(0,-1){5}}
%}}
%\ind(-10,20){W''}\ind(-10,-20){W''}
%\ind(-49,0){\bar{W}''}\ind(5,0){\bar{W''}}
%\put(30,0){\ldots}
\put(0,-12){{\it Fig.5}}
\end{picture}
%}}}}
\hspace{2cm}
\begin{picture}(30,30)
\put(0,10){\circle{20}}
\put(-10,10){\line(-1,0){5}}%\put(-10,8.5){\line(-1,0){5}}
\put(10,10){\line(1,0){5}}%\put(10,8.5){\line(1,0){5}}
\put(-9,5){\line(-1,-1){6}}%\put(-8,4){\line(-1,-1){6}}
\put(9,5){\line(1,-1){6}}%\put(8,4){\line(1,-1){6}}
\put(-9,15){\line(-1,1){6}}%\put(-8,16){\line(-1,1){6}}
\put(9,15){\line(1,1){6}}%\put(8,16){\line(1,1){6}}
\put(30,10){\ldots}
\end{picture}

\vspace*{10mm}

\noindent External lines here are for alternating external $\Psi$ and
$\bar{\Psi}$ fields, with the $-\frac{D^2}{4}$ and $-\frac{\bar{D}^2}{4}$ factors are associated with the vertices as usual,
and internal ones are for the free propagator of the chiral superfield. At the same time, if we consider
theory of a real scalar superfield $u$ in the external chiral superfield
$\Psi$ with action
\bea
S=\frac{1}{2}\int d^8 z u (\Box-\frac{1}{4}\Psi\bar{D}^2-\frac{1}{4}\bar{\Psi}D^2)u,
\eea
it leads just to these supergraphs (if $\int d^8 z u(-\frac{1}{4}\Psi
\bar{D}^2) u$ and the conjugated term are treated as vertices), and
one-loop effective action for this theory is again given by (\ref{olwz}).

We can see that the expression of one-loop effective action in
the form of the trace of the logarithm of some operator 
allows to use some special technique which
is equivalent to supergraph approach, but more convenient in many
cases.
This technique is called proper-time technique.

As we have already proved, if the quadratic action of a 
quantum (super)field $\phi$
on classical background $\Phi$ has the form $\frac{1}{2}\int dx \phi \Delta[\Phi]\phi$
($\int dx$ here denotes integral over all (super)space), one-loop effective
action in this theory is $\Gamma^{(1)}=\frac{i}{2}Tr\int_0^{\infty}
\frac{ds}{s}e^{is\Delta}$. Therefore we face the problem of calculating
the operator $e^{is\Delta}$. In most important cases
$\Delta=\Box+\ldots$
where dots denote background dependent terms.
It is known \cite{BSD} that the best way
to find this operator in the case of common field theory is as follows.
We introduce the function $U(x,x'|s)=e^{is\Delta}\delta^4(x-x')$ called the Schwinger
kernel. Of course, the $U$
depends on background superfields. By the definition, it satisfies the equation:
\bea
i\frac{\partial U}{\partial s}=-U\Delta.
\eea
The $\Delta$ is supposed to have form of power series in
derivatives. And $U$ satisfies initial condition 
$$U(x,x')|_{s=0}=\delta^4(x-x').$$  
In a common case (especially, in the gravity theories) $U$ is represented in the form of infinite power
series in parameter $s$ (called the proper time) as \cite{BSD} 
\bea
U=-\frac{i}{(4\pi s)^2}\exp(\frac{i}{4s}(x-x')^2)
\sum_{n=0}^{\infty}a_n (is)^n.
\eea 
The (ultraviolet) divergences correspond to {\bf lower}
orders of this expansion (note that ultraviolet case corresponds to
$s\to 0$, infrared one -- to $s\to\infty$).
Coefficients $a_n$ depend on background superfields and their derivatives.
We note that if background superfields are put to zero, we arrive at
\bea
\label{u0}
U^{(0)}(x,x';s)=e^{is\Box}\delta^4(x-x')=-\frac{i}{(4\pi s)^2}
\exp(\frac{i}{4s}(x-x')^2),
\eea
which satisfies condition
\bea
i\int_0^{\infty} ds U^{(0)}(x,x';s)=\frac{1}{\Box}\delta^4(x-x').
\eea
The approach in case of superfield theories is quite
analogous. However, in the superfield case using of this approach is characterized by an essential advantage. 
Indeed, in this case it is more
convenient to expand Schwinger kernel $U(x,x';s)$ not in an infinite
power series in $s$ but in a power series in spinor
supercovariant derivatives which is {\bf finite} due to
anticommutation properties of spinor derivatives (we already know that the similar situation takes place in three-dimensional superfield theories).

Really, in most cases operator $\Delta$ in superfield theories looks
like
\bea
\Delta=\Box+\sum A_{nm}(D^{\a})^n(\bar{D}^{\ad})^m\equiv \Box+\tilde{\Delta}
\eea
with $\tilde{\Delta}$ is some background dependent operator
(in most cases it contains only even orders in spinor derivatives,
here we consider this case), 
$A_{nm}$ are background dependent coefficients.
We introduce the object (cf. \cite{BK0,Buch1}):
\bea
\label{u}
U(z,z';s)=\exp(is\Delta)\delta^8(z-z')\equiv
\exp(is\tilde{\Delta})\exp(is\Box)\delta^8(z-z').
\eea
The last identity is valid for studying of contributions which do not
depend
on space-time derivatives of superfields, i.e. for the contributions to the effective potential. 
Then, we suppose a natural initial condition 
$$
U(z,z';s)|_{s=0}=\delta^8(z-z').
$$ 
It is clear that
$\exp(is\Box)\delta^8(z-z')=\delta^4(\theta-\theta')U^{(0)}(x,x';s)$
where $U^{(0)}(x,x';s)$ is given by (\ref{u0}). Hence
\bea
U(z,z';s)=\exp(is\tilde{\Delta})U^{(0)}(x,x';s)\delta^4(\theta-\theta').
\eea 
Therefore we face the problem of calculating of the operator
$\Omega=\exp(is\tilde{\Delta})$.
The $\Omega$ satisfies the equation
\bea
\label{etu}
i\frac{\partial\Omega}{\partial s}=-\Omega\tilde{\Delta}.
\eea
It is easy to see that $\Omega|_{s=0}=1$.
We expand $\Omega$ into a finite power series in spinor supercovariant
derivatives (its finiteness is based on the anticommutation relations of the derivatives, cf. \cite{Buch1}; the terms linear in $D_{\alpha}$, $\bar{D}_{\dot{\alpha}}$ are not necessary for most of theories):
\bea
\label{tu}
\Omega&=&1+\frac{1}{16}A(s)D^2\bar{D}^2+\frac{1}{16}\tilde{A}(s)
\bar{D}^2D^2+\frac{1}{8}B^{\a}(s)D_{\a}\bar{D}^2+\frac{1}{8}\tilde{B}_{\ad}(s)
\bar{D}^{\ad}D^2+\nonumber\\&+&
\frac{1}{4}C(s)D^2+\frac{1}{4}\tilde{C}(s)\bar{D}^2,
\eea
and substitute (\ref{tu}) into the equation (\ref{etu}). As a result we
obtain  some power series in spinor derivatives in the right-hand side.
Comparing coefficients at analogous combinations of the derivatives in the right-hand side and
in the left-hand side of the equation (\ref{etu}) we get \cite{Buch1}
\bea
\label{sys0}
\frac{1}{16}\dot{A}&=&\Omega\tilde{\Delta}|_{D^2\bar{D^2}};\nonumber\\
\frac{1}{8}\dot{B}^{\alpha}&=&\Omega\tilde{\Delta}|_{D_{\alpha}\bar{D^2}};
\nonumber\\
\frac{1}{4}\dot{C}&=&\Omega\tilde{\Delta}|_{D^2}
\eea
and analogous equations for
$\tilde{A},\tilde{B}_{\ad},\tilde{C}$. Here the dot denotes
$\frac{1}{i}\frac{\partial}{\partial s}$, and $|_{D^2}$ etc. denotes coefficient
at $D^2$ etc. in the expansion of $\Omega\tilde{\Delta}$.
As a result we have a system of first-order differential equations on
coefficients determining structure of operator $\Omega$.
Since $\Omega|_{s=0}=1$, we have natural initial conditions
\bea
\label{con}
A|_{s=0}=\tilde{A}|_{s=0}=B^{\a}|_{s=0}=\tilde{B}_{\ad}|_{s=0}=
C|_{s=0}=\tilde{C}|_{s=0}=0.
\eea
The system (\ref{sys0}) with initial conditions (\ref{con}) can be
solved in a manner similar to a common system of differential equations. However, one must notice
that this solution can be exactly found only in special cases, for example,
for the dependence of the heat kernel only on background superfields but not on their derivatives, or
for the dependence on chiral background superfields only.

Then, the $\Omega(s)$ (sometimes it is also called a heat kernel) can be used for
calculation of Green function as
\bea
G(z_1,z_2)=i\int_0^{\infty}ds \Omega U^{(0)}(x,x';s)\delta^4(\q-\q') 
\eea
(note that $\Omega$ is a differential operator in superspace) and
for calculation of one-loop effective action as
\bea
\Gamma^{(1)}=\frac{i}{2}\int_0^{\infty}\frac{ds}{s}\int d^8 z d^8 z'
\delta^8(z-z')\Omega U^{(0)}(x,x';s)\delta^4(\q-\q').
\eea
As usual, $\int d^8 z =\int d^4 x d^4\q$, we also use the definition
(\ref{u0}).
Then, it is known that 
$\delta^4(\q-\bar{\q})D^2\bar{D}^2\delta^4(\q-\bar{\q})=
16\delta^4(\q-\bar{\q})$, and all products of less number of spinor
derivatives give a zero trace. Hence only coefficients of 
(\ref{tu}) giving non-zero contribution to one-loop effective action
are
$A$ and $\tilde{A}$. And one-loop effective action looks
like
\bea
\Gamma^{(1)}=\frac{i}{2}\int_0^{\infty}\frac{ds}{s}\int d^4 x d^4\theta
(A(s)+\tilde{A}(s))U^{(0)}(x,x';s)|_{x=x'}.
\eea
As a result we developed technique for calculating background
dependent propagators and one-loop effective action. Application of
this technique will be further considered for examples of several
theories. There exists an essential modification of this method for
supersymmetric gauge theories \cite{McA}. We discuss it further.

\section{Problem of superfield effective potential}

Effective potential in a standard quantum field theory is defined as
the effective Lagrangian considered at constant values of scalar fields,
and other fields are put to zero. The effective potential is used for
studying of spontaneous symmetry breaking and vacuum stability \cite{CW}. 

First, let us shortly describe effective potential in common quantum
field theory. The effective action has the form
\bea
\Gamma[\phi]=\int d^4 x
(-V_{eff}(\phi)+\frac{1}{2}Z(\phi)\pa_m\phi\pa^m\phi+
\ldots),
\eea
where $Z(\phi)$ is a some function of $\phi$, and $V_{eff}(\phi)$ is
effective potential. For slowly varying fields, therefore,
$$
\Gamma[\phi]=-\int d^4 x V_{eff}(\phi),
$$
therefore effective potential is a low-energy leading term. It can be
represented in the form of loop expansion
\bea
V_{eff}(\phi)=V(\phi)+\sum_{n=1}^{\infty}\hbar^n V^{(n)}(\phi).
\eea
For example, consider the theory with action
\bea
S=\int d^4x (-\frac{1}{2}\phi\Box\phi-V(\phi)).
\eea
After background-quantum splitting $\phi\to\Phi+\chi$ where $\Phi$ is
background superfield and $\chi$ is quantum one, we find the quadratic
action of quantum superfields
\bea
S_2=-\frac{1}{2}\int d^4 x \chi(\Box+V^{''}(\Phi))\chi,
\eea
which leads to one-loop effective action $\Gamma^{(1)}[\Phi]$ of the
form
\bea
\Gamma^{(1)}[\Phi]=\frac{i}{2}{\rm Tr}\log(\Box+V^{''}(\Phi)).
\eea
Following the previous studies, we can express this trace of logarithm in the
form of diagrams:

\hspace{0.5cm}
\unitlength=.6mm
%\thicklines
%\GRAPH(hsize=3){%\ind(50,-30){���.1.}
\begin{picture}(20,20)
\put(0,10){\circle{20}}\put(-10,10){\line(-1,0){5}}
%\put(-10,8.5){\line(-1,0){5}}
%\put(10,10){\line(1,0){5}}%\put(10,8.5){\line(1,0){5}}
%\ind(24,0){W''}\ind(-26,0){\bar{W''}}
%\put(4.8,0){\GRAPH(hsize=3){
\end{picture}
\hspace{2cm}
\begin{picture}(20,20)
\put(0,10){\circle{20}}%\put(-10,10){\line(-1,0){5}}
%\put(-10,8.5){\line(-1,0){5}}
\put(10,10){\line(1,0){5}}%\put(10,8.5){\line(1,0){5}}
\put(-10,10){\line(-1,0){5}}%\put(-1,20){\line(0,1){5}}
%\put(0,0.5){\line(0,-1){5}}
%\put(-1,0){\line(0,-1){5}}
%}}
%\ind(-10,20){W''}\ind(-10,-20){W''}
%\ind(-49,0){\bar{W}''}\ind(5,0){\bar{W''}}
%\put(30,0){\ldots}
\end{picture}
%}}}}
\hspace{2cm}
\begin{picture}(30,30)
\put(0,10){\circle{20}}
%\put(-10,10){\line(-1,0){5}}%\put(-10,8.5){\line(-1,0){5}}
%\put(10,10){\line(1,0){5}}%\put(10,8.5){\line(1,0){5}}
%\put(-9,5){\line(-1,-1){6}}%\put(-8,4){\line(-1,-1){6}}
\put(9,15){\line(1,1){6}}%\put(8,4){\line(1,-1){6}}
\put(-9,15){\line(-1,1){6}}%\put(-8,16){\line(-1,1){6}}
\put(0,-1){\line(0,-1){5}}%\put(8,16){\line(1,1){6}}
\put(30,10){\ldots}
\put(0,-15){{\it Fig. 6}}
\end{picture}

\vspace*{1cm}

\noindent where external lines are $V^{''}[\Phi]$. Internal lines
correspond to $-\frac{1}{\Box}\delta^4(x_1-x_2)$.

The complete effective potential is presented by a sum of contributions from these supergraphs:
\bea
U^{(1)}=\sum_{n=1}^{\infty}\frac{1}{2n}\int\frac{d^4k}{(2\pi)^4}
(\frac{V^{''}(\Phi)}{k^2})^n=-\int \frac{d^4k}{(2\pi)^4}
\log(1-\frac{V^{''}(\Phi)}{k^2}),
\eea
which after integration over $d^4k$ and subtraction of divergences is
equal to (cf. \cite{CW})
\bea
U^{(1)}=-\frac{1}{32\pi^2}(V^{''}(\Phi))^2(\log\frac{V^{''}(\Phi)}{\mu^2}+C),
\eea
where $C$ is some constant which can be fixed by imposing of some renormalization conditions (see \cite{CW} for details in the case of $\lambda\phi^4$ theory). The same result can be also obtained via
proper-time method.

Now we turn to a superfield case. Let $\Gamma[\Phi,\bar{\Phi}]$ be the
(renormalized) effective action for a theory of chiral and antichiral
superfields. We can represent it as \cite{Buch1}
\bea
\label{padrao}
%\begin{equation}
 \Gamma[\bar{\Phi},\Phi] &=& \int d^8z {\cal L}_{eff}
(\Phi,D_A\Phi,D_A D_B\Phi;\bar{\Phi},D_A\bar{\Phi},D_A D_B\bar{\Phi})
+\nonumber\\&+& (\int d^6z {\cal L}^{(c)}_{eff}(\Phi) + h.c.) +\ldots
%\end{equation}
\eea 
Here $D_A\Phi,D_AD_B\Phi,\ldots$ are spinor supercovariant
derivatives of superfields $\Phi,\bar{\Phi}$. The term ${\cal L}_{eff}$ is
called general effective Lagrangian, and ${\cal L}^{(c)}_{eff}$ is
called chiral effective Lagrangian. Both these effective Lagrangians
can be expanded into power series in supercovariant derivatives of
background superfields. The dots in this expression denote terms depending on
space-time derivatives of $\Phi$, $\bar{\Phi}$. Further, the structure of the effective action (\ref{padrao}) will be considered as a standard one for the superfield theories. 
We note that since chiral effective Lagrangian by definition
depends only on $\Phi$ but not on $\bar{D}^2\bar{\Phi}$ all
terms of the form
$$
\int d^6 z \Phi^n (\bar{D}^2\bar{\Phi})^m
$$
using relation $\int d^6z (-\frac{\bar{D}^2}{4})=\int d^8 z$ can be
rewritten as
$$
\int d^8 z \Phi^n\bar{\Phi} (\bar{D}^2\bar{\Phi})^{m-1}, 
$$
i.e. in the form corresponding to general effective
Lagrangian. Therefore here and further we consider all 
expressions which are formally chiral but involve $(\bar{D}^2\bar{\Phi})^m$ as contributions to
general effective Lagrangian.

We note that all chiral contributions can be also represented as
integral over whole superspace (this observation has been made first time in \cite{West2}):
\bea
\int d^6 z G(\Phi)=\int d^8 z (-\frac{D^2}{4\Box})G(\Phi).
\eea
Further, to recover the usual effective potential within the component approach we must put scalar
component fields to be constant, and spinor ones -- to zero, f.e. 
in the Wess-Zumino model we write
$$
A=const,\ F=const, \ \psi_{\a}=0.
$$
However, this condition is not supersymmetric, therefore we use
condition of superfield {\bf constant in space-time}:
\bea
\pa_a\Phi=0.
\eea
Since $\pa_a$ commutes with all generators of supersymmetry, this
condition is supersymmetric.

The effective potential is introduced as
\bea
\label{ep1}
V_{eff}=\Big\{-\int d^4\q {\cal L}_{eff}-(\int d^2\q {\cal
  L}^{(c)}_{eff}+h.c.)
\Big\}|_{\pa_a\Phi=\pa_a\bar{\Phi}=0}.
\eea
The minus sign is put by convention. We can introduce general effective
potential
${\cal L}_{eff}|_{\pa_a\Phi=\pa_a\bar{\Phi}=0}$ and chiral (or holomorphic) effective
potential
${\cal L}^{(c)}_{eff}|_{\pa_a\Phi=0}$.
It is easy to see that the general effective potential can be expressed as
\bea
\label{ep2}
{\cal L}_{eff}={\bf K}(\Phi,\bar{\Phi})+
{\bf F}(D_{\a}\Phi,
\bar{D}_{\ad}\bar{\Phi},D^2\Phi,\bar{D}^2\bar{\Phi};\Phi,\bar{\Phi})
\eea
with 
$
{\bf F}|_{D_{\a}\Phi,
\bar{D}_{\ad}\bar{\Phi},D^2\Phi,\bar{D}^2\bar{\Phi}=0}=0
$.
The ${\bf K}$ is called k\"{a}hlerian effective potential, and ${\bf
  F}$ is called auxiliary fields' effective potential, it is at least
of third order in auxiliary fields of $\Phi$ and $\bar{\Phi}$.
These objects can be represented in the form of loop expansion:
\bea
\label{ep3}
{\bf
  K}(\Phi,\bar{\Phi})&=&K_0(\Phi,\bar{\Phi})+\sum_{L=1}^{\infty}\hbar^L
K_L(\Phi,\bar{\Phi}),\\
{\bf F}&=&\sum_{L=1}^{\infty}\hbar^L F_L,
\eea
(the term corresponding to tree level, $L=0$, 
in the expression for ${\bf F}$ is absent for theories which do not
include derivative depending terms in the classical action, such as
the Wess-Zumino model), and
\bea
\label{ep5}
{\cal L}^{(c)}_{eff}(\Phi)={\cal L}^{(c)}(\Phi)+\sum_{L=1}^{\infty}
\hbar^L{\cal L}^{(c)}_L (\Phi).
\eea
Here $K_L,F_L,{\cal L}^{(c)}_L$ are quantum corrections.
For the Wess-Zumino model, ${\cal L}^{(c)}_1=0$, however, in some quantum
theories (f.e. in $N=1$ super-Yang-Mills theory with chiral matter)
a one-loop contribution to chiral effective potential exists
\cite{West4}.

The structure of the effective potential described by the equations
(\ref{ep1}--\ref{ep5}) is generic describing all theories of the chiral superfields (remind that within the phenomenological context, the matter is associated with chiral superfields since namely they involve the scalar fields with an usual dynamics) including the noncommutative
ones. However, we note that effective potential in theories including
gauge superfields must depend on them in a special way. Really, the
effective action in such theories should be expressed in terms of
some gauge convariant constructions, f.e. within background field method,
the gauge superfield is either incorporated to chiral superfields or
presents in supercovariant derivatives and gauge invariant superfield
strengths \cite{GRS,BK0}.

Let us give a few remarks about the method of calculation of the effective
potential. The best way for it is, of course, is based on the using of background
dependent propagators which are expressed in terms of common
propagators and background superfields. Background dependent
propagators can be  exactly found in certain cases. To calculate
k\"{a}hlerian effective potential and auxiliary fields' effective
potential one can straightforwardly omit all space-time derivatives,
moreover, to study k\"{a}hlerian effective potential one can omit {\bf all} supercovariant derivatives and treat background superfields as
constants until final integration. The calculation of chiral effective
potential, however, is characterized by some peculiarities. The best example for it is the Wess-Zumino model -- the simplest superfield theory. We will consider it in the next section.

\section[Wess-Zumino model]{The Wess-Zumino model and a problem of the chiral effective potential}

Now we turn to consideration of superfield effective potential in
Wess-Zumino model. Here we follow the papers \cite{my3,my4,Buch1} and
the book \cite{BK0}.

The superfield action of the Wess-Zumino model is given by
(\ref{actwz}). Following, as usual, the loop expansion approach, we carry out
background-quantum splitting by the rule
\bea
\Phi&\to&\Phi+\sqrt{\hbar}\phi;\nonumber\\
\bar{\Phi}&\to&\bar{\Phi}+\sqrt{\hbar}\bar{\phi}.
\eea
The standard expression defining effective action under such
changes takes the form (here
$\bar{\Gamma}=\sum\limits_{L=1}^{\infty}\hbar^L\Gamma_L$), with $\psi=m+\lambda\Phi$, $\bar{\psi}=m+\lambda\bar{\Phi}$:
\bea
e^{\frac{i}{\hbar}\bar{\Gamma}[\Phi,\bar{\Phi}]}&=&
\int D\phi
D\bar{\phi}
\exp\Big(\frac{i}{2}
\left(\begin{array}{cc}\phi\bar{\phi}\end{array}
\right)
\left(\begin{array}{cc}
\psi&-\frac{1}{4}\bar{D}^2\\
-\frac{1}{4}D^2 & \bar{\psi}
\end{array}\right)
%\times\nonumber\\&\times&
\left(\begin{array}{c}\phi\\ \bar{\phi}
\end{array}\right)+\nonumber\\&+&
i\sqrt{\hbar}(\frac{\l}{3!}\phi^3+h.c.)
\Big).
\eea
The quadratic action of quantum superfields looks like
\bea
S^{(2)}=\frac{1}{2}
\left(\begin{array}{cc}\phi\bar{\phi}\end{array}
\right)
\left(\begin{array}{cc}
\psi&-\frac{1}{4}\bar{D}^2\\
-\frac{1}{4}D^2 & \bar{\psi}
\end{array}\right)
%\times\nonumber\\&\times&
\left(\begin{array}{c}\phi\\ \bar{\phi}
\end{array}\right).
\eea
And the matrix superpropagator by definition is an operator inverse to
\bea
\left(\begin{array}{cc}
\psi&-\frac{1}{4}\bar{D}^2\\
-\frac{1}{4}D^2 & \bar{\psi}
\end{array}\right)
.
\eea
We can see that this matrix superpropagator can be represented in the form
\bea
G(z_1,z_2)=\left(
\begin{array}{cc}
G_{++}(z_1,z_2)&G_{+-}(z_1,z_2)\\
G_{-+}(z_1,z_2)&G_{--}(z_1,z_2)
\end{array}
\right).
\eea
where $+$ denotes chirality with respect to corresponding argument,
and $-$ correspondingly -- antichirality.

One can verify that in the Wess-Zumino model this matrix looks like
\bea
\label{gremat}
G(z_1,z_2)=
\frac{1}{16}\left(\begin{array}{cc}
\bar{D}^2_1\bar{D}^2_2G^{\psi}_v(z_1,z_2)&\bar{D}^2_1 D^2_2G^{\psi}_v(z_1,z_2)
\\
D^2_1\bar{D}^2_2G^{\psi}_v(z_1,z_2)& D^2_1 D^2_2 G^{\psi}_v(z_1,z_2)
\end{array}
\right).
\eea
where $G^{\psi}_v(z_1,z_2)=(\Box+\frac{1}{4}\psi\bar{D}^2+
\frac{1}{4}\bar{\psi}D^2)^{-1}\delta^8(z_1-z_2)$.
Really, let us consider the relation
\bea
\left(\begin{array}{cc}
\psi&-\frac{1}{4}\bar{D}^2\\
-\frac{1}{4}D^2 & \bar{\psi}
\end{array}\right)
\frac{1}{16}\left(\begin{array}{cc}
\bar{D}^2_1\bar{D}^2_2G^{\psi}_v(z_1,z_2)&\bar{D}^2_1 D^2_2G^{\psi}_v(z_1,z_2)
\\
D^2_1\bar{D}^2_2G^{\psi}_v(z_1,z_2)& D^2_1 D^2_2 G^{\psi}_v(z_1,z_2)
\end{array}
\right)=-\left(\begin{array}{cc}
\delta_+&0\\
0&\delta_-
\end{array}
\right)
\eea
and act on both parts of this relation with the operator
$$
\left(
\begin{array}{cc}
0&-\frac{1}{4}\bar{D}^2\\
-\frac{1}{4}D^2& 0
\end{array}
\right).
$$
We get the following system of equations on components of matrix
superpropagator $G$:
\bea
\Box G_{++}&-&\frac{1}{4}\bar{D}^2_1(\bar{\psi}G_{-+})=0;\nonumber\\
\Box G_{-+}&-&\frac{1}{4}D^2_1(\psi
G_{++})=\frac{1}{16}D^2_1\bar{D}^2_2\delta^8(z_1-z_2);
\nonumber\\
\Box G_{--}&-&\frac{1}{4}D^2_1(\psi G_{+-})=0;\nonumber\\
\Box G_{+-}&-&\frac{1}{4}\bar{D}^2_1(\bar{\psi}G_{--})=
\frac{1}{16}\bar{D}^2_1 D^2_2\delta^8(z_1-z_2).
\eea
A straightforward comparing shows that components
$G_{++},G_{+-},G_{-+},G_{--}$
given by (\ref{gremat}) satisfy this equation. 
Thus, we found matrix superpropagator (\ref{gremat}) which
will be used for calculation of loop corrections. 

Let us consider the one-loop effective action. Formally it has the form
$$
\Gamma^{(1)}=-\frac{i}{2}{\rm Tr}\log G
$$
where matrix superpropagator $G$ is given by (\ref{gremat}). However,
straightforward calculation of this trace is very complicated since the
elements of this matrix are defined in different subspaces.
The one-loop effective action $\Gamma^{(1)}$
can be obtained from relation
\bea
\label{eag}
e^{i\Gamma^{(1)}}=
\int D\phi
D\bar{\phi}
\exp\Big(\frac{i}{2}
\left(\begin{array}{cc}\phi\bar{\phi}\end{array}
\right)
\left(\begin{array}{cc}
\psi&-\frac{1}{4}\bar{D}^2\\
-\frac{1}{4}D^2 & \bar{\psi}
\end{array}\right)
%\times\nonumber\\&\times&
\left(\begin{array}{c}\phi\\ \bar{\phi}
\end{array}\right)
\Big).
\eea

Calculating of this path integral is essentially simplified by use of the trick \cite{Buch1} which is also applied
in many theories describing dynamics of chiral superfields.
We consider theory of real scalar superfield with action
\bea
S=-\frac{1}{16}\int d^8 z v D^{\a}\bar{D}^2 D_{\a}v.
\eea
The action is invariant under gauge transformations $\delta v=\Lambda-
\bar{\Lambda}$ (here $\Lambda$ is chiral, and $\bar{\Lambda}$ is
antichiral).
According to Faddeev-Popov approach, the effective action $W$ for this
theory can be introduced as
\bea
\label{eaw}
e^{iW}=\int Dv e^{-\frac{i}{16}\int d^8 z v D^{\a}\bar{D}^2 D_{\a}v}
\delta(\chi).
\eea
Here $\delta(\chi)$ is a functional delta function, and $\chi$ is a
gauge-fixing function. We choose $\chi$ in the form of column matrix
$$
\chi=\left(
\begin{array}{c}
\frac{1}{4}D^2v-\bar{\phi}\\
\frac{1}{4}\bar{D}^2 v-\phi
\end{array}
\right).
$$
Note that since supercovariant derivatives are not real we must impose
two conditions, and (\ref{eaw}) takes the form
\bea
\label{eaw1}
e^{iW}=\int Dv e^{-\frac{i}{16}\int d^8 z v D^{\a}\bar{D}^2 D_{\a}v}
\delta(\frac{1}{4}D^2v-\bar{\phi})\delta(\frac{1}{4}\bar{D}^2v-\phi)
{\rm det}\Delta,
\eea
where
$$
\Delta=\left(
\begin{array}{cc}
-\frac{1}{4}\bar{D}^2&0\\
0&-\frac{1}{4}D^2
\end{array}
\right)
$$
is a Faddeev-Popov matrix. We note that $W$ is constant by the construction.
We multiply left-hand sides and right-hand sides of (\ref{eag}) and 
(\ref{eaw1}) respectively, as a result we arrive at
\bea
e^{i\Gamma^{(1)}+W}&=&
\int D\phi
D\bar{\phi} Dv
\exp\Big(\frac{i}{2}
\left(\begin{array}{cc}\phi\bar{\phi}\end{array}
\right)
\left(\begin{array}{cc}
\psi&-\frac{1}{4}\bar{D}^2\\
-\frac{1}{4}D^2 & \bar{\psi}
\end{array}\right)
%\times\nonumber\\&\times&
\left(\begin{array}{c}\phi\\ \bar{\phi}
\end{array}\right)
-\frac{i}{16}vD^{\a}\bar{D}^2D_{\a}v
\Big)\times\nonumber\\&\times&
\delta(\frac{1}{4}D^2v-\bar{\phi})\delta(\frac{1}{4}\bar{D}^2v-\phi)
{\rm det}\Delta.
\eea
Integration over $\phi,\bar{\phi}$ with use of delta functions leads
to
\bea
e^{i\Gamma^{(1)}+W}&=&
\int D\phi
D\bar{\phi}
\exp\Big(\frac{i}{2}\int d^8z v(\Box-\frac{1}{4}\psi\bar{D}^2-\frac{1}{4}\bar{\psi}D^2)v
\Big)
{\rm det}\Delta.
\eea
However, $W$ and ${\rm det}\, \Delta$ are constants which can be
omitted. We also took into account
that
$\frac{1}{16}\{D^2,\bar{D}^2\}-\frac{1}{8}D^{\a}\bar{D}^2D_{\a}=\Box$, hence the one-loop effective action is equal to
\bea
\Gamma^{(1)}=\frac{i}{2}{\rm Tr}\log(
\Box-\frac{1}{4}\psi\bar{D}^2-\frac{1}{4}\bar{\psi}D^2)
\eea
Here as usual $\psi=m+\lambda\Phi,\bar{\psi}=m+\lambda\bar{\Phi}$.
This one-loop effective action can be expressed in form of Schwinger
expansion:
\bea
\label{gl0}
\Gamma^{(1)}=\frac{i}{2}{\rm Tr}\int_0^{\infty}\frac{ds}{s}\exp(is(
\Box-\frac{1}{4}\psi\bar{D}^2-\frac{1}{4}\bar{\psi}D^2)),
\eea
or, after manifest writing the trace,
\bea
\label{g11}
\Gamma^{(1)}=\frac{i}{2}\int d^8 z_1 d^8 z_2\int_0^{\infty}\frac{ds}{s}
\delta^8(z_1-z_2)
\exp(is(-\frac{1}{4}\psi\bar{D}^2-\frac{1}{4}\bar{\psi}D^2))
e^{is\Box}
\delta^8(z_1-z_2).
\eea
We consider the heat kernel 
$\Omega(\psi|s)=\exp(is(-\frac{1}{4}\psi\bar{D}^2-\frac{1}{4}\bar{\psi}D^2))
\equiv e^{is\Delta}$.
It evidently satisfies the equation
$$
\frac{\partial \Omega}{\partial s}=i\Omega\Delta.
$$
It turns out to be that if we calculate k\"{a}hlerian effective potential, when all supercovariant derivatives from background superfields 
$\psi,\bar{\psi}$ are omitted, this equation can be easily solved.
We express $\Omega$ in the form (\ref{tu}).
Then $\Omega\Delta$ is equal to
\bea
\Omega\Delta&=&-\frac{1}{4}\psi\bar{D}^2-\frac{1}{4}\bar{\psi}D^2-
\nonumber\\
&-&\frac{1}{4}\psi\tilde{A}\Box\bar{D}^2-\frac{1}{4}\bar{\psi}A\Box D^2+
\nonumber\\
&+&\frac{1}{4}\tilde{B}_{\ad}\psi\pa^{\a\ad}D_{\a}\bar{D}^2
-\frac{1}{4}B_{\a}\bar{\psi}\pa^{\a\ad}\bar{D}_{\ad}D^2-\nonumber\\
&-&\frac{1}{16}\bar{\psi}\bar{C}\bar{D}^2 D^2-\frac{1}{16}\psi C D^2\bar{D}^2.
\eea
Comparing coefficients at analogous derivatives in $\frac{1}{i}
\frac{\pa\Omega}{\pa s}$ and $\Omega\Delta$ we get the following
system of equations
\bea
\label{sys}
\dot{A}&=&-\psi C;\nonumber\\
\dot{B}^{\a}&=&2i\tilde{B}_{\ad}\psi\pa^{\a\ad};\nonumber\\
\dot{C}&=&-\bar{\psi}-\bar{\psi}A\Box.
\eea
System for $\tilde{A},\tilde{B},\tilde{C}$ has the analogous form with
changing $\psi\to\bar{\psi},A\to\tilde{A}$ etc.
Here dot denotes $\frac{1}{i}\frac{\pa}{\pa s}\equiv\frac{\pa}{\pa\tilde{s}}$, and $\tilde{s}=is$.
Since $\Omega|_{s=0}=1$, and all terms in expansion of $\Omega$
(\ref{tu}) are evidently linearly independent, natural initial
conditions are
\bea
\label{init}
A=\tilde{A}=B^{\a}=\tilde{B}_{\ad}=C=\tilde{C}|_{s=0}=0.
\eea
We find that the system of equations for $B^{\a}$ and $\tilde{B}_{\ad}$ is
closed
(it is isolated from the whole system (\ref{sys})) and
homogeneous. The initial conditions above make its only solution to be
zero,
$B^{\a},\tilde{B}_{\ad}=0$.
The remaining from (\ref{sys}) system for $A$ and $C$ (and analogous one 
for $\tilde{A}$ and $\tilde{C}$) can be easily solved like a standard 
system of common first-order differential equations.
Its solution looks like
\bea
C&=&-\sqrt{\frac{\bar{\psi}\Box}{\psi}}(A^1_0 \exp (i\omega s)-A^2_0
\exp (-i\omega s))\nonumber\\
A&=&A^1_0 \exp (i\omega s)+A^2_0 \exp (-i\omega s)-\frac{1}{\Box}
\eea
Here $\omega=\sqrt{\psi\bar{\psi}\Box}$.
Imposing initial conditions (\ref{init}) allows to fix coefficients
$A^1_0,A^2_0$. As a result we get
\bea
C&=&-\sqrt{\frac{\bar{\psi}}{\psi\Box}}\sinh(is\sqrt{\psi\bar{\psi}\Box})
\nonumber\\
A&=&\frac{1}{\Box}[\cosh(is\sqrt{\psi\bar{\psi}\Box})-1]
\eea
Since $A$ is symmetric with respect to change $\psi\to\bar{\psi}$ we
find that $A=\tilde{A}$. We note that only $A$ and $\tilde{A}$
contribute to trace in (\ref{g11}).
Therefore one-loop k\"{a}hlerian contribution to effective action is equal to
\bea
\label{ktrace}
\Gamma_K^{(1)}=i\int
d^4 x d^4\q\int_0^{\infty}\frac{d\tilde{s}}{\tilde{s}}
\frac{1}{\Box}[\cosh(\tilde{s}\sqrt{\psi\bar{\psi}\Box})-1]U^{(0)}(x,x';s)|_{x=x'}.
\eea
Here $U^{(0)}(x,x';s)$ is given by (\ref{u0}). This function satisfies the
equation (see section 4.6):
$$
\Box^n U^{(0)}(x,x';s)|_{x=x'}=-i
(\frac{\pa}{\pa \tilde{s}})^n\frac{1}{16\pi^2\tilde{s}^2}.
$$
We expand (\ref{ktrace}) into power series:
$$
\frac{1}{\Box}[\cosh(\tilde{s}\sqrt{\psi\bar{\psi}\Box})-1]=
\sum\limits_{n=0}^{\infty}\tilde{s}^{2n+2}\frac{(\psi\bar{\psi})^{n+1}}{(2n+2)!}
\Box^n.
$$
And 
\bea
\Gamma_K^{(1)}&=&
i\int
d^4 x d^4\q\int_0^{\infty}\frac{d\tilde{s}}{\tilde{s}}
\sum_{n=0}^{\infty}\tilde{s}^{2n+2}\frac{(\psi\bar{\psi})^{n+1}}{(2n+2)!}
\Box^nU^{(0)}(x,x';s)|_{x=x'}=\nonumber\\&=&
-i\int
d^4 x d^4\q\int_0^{\infty}\frac{d\tilde{s}}{\tilde{s}}
\sum_{n=0}^{\infty}\tilde{s}^{2n+2}\frac{(\psi\bar{\psi})^{n+1}}{(2n+2)!}
(\frac{\pa}{\pa \tilde{s}})^n\frac{-i}{16\pi^2\tilde{s}^2}=\nonumber\\
&=&
-\frac{1}{32\pi^2}\int d^8 z\int_{L^2}^{\infty}\frac{d\tilde{s}}{\tilde{s}^2}
\sum_{n=0}^{\infty}{(-1)}^n\frac{(\tilde{s}\psi\bar{\psi})^{n+1}(n+1)!}
{(2n+2)!}.
\eea
Here we cut the integral at the lower limit by introducing $L^2$ for
regularization.
We make the change $\tilde{s}\psi\bar{\psi}=t$. As a result, the one-loop
k\"{a}hlerian effective potential takes the form
\bea
\label{kahlpt}
K^{(1)}=-\frac{1}{32\pi^2}
\psi\bar{\psi}\int_{\psi\bar{\psi}L^2}^{\infty} \frac{dt}{t^2}
\sum_{n=0}^{\infty}\frac{(n+1)!t^{n+1}(-1)^n}{(2n+2)!}
\eea
Then, $\sum\limits_{n=0}^{\infty}\frac{(n+1)!t^{n+1}(-1)^n}{(2n+2)!}=
t\int_0^1 due^{-\frac{t}{4}(1-u^2)}$.
Hence 
\bea
K^{(1)}=-\frac{1}{32\pi^2}
\psi\bar{\psi}\int_{\psi\bar{\psi}L^2}^{\infty}\frac{dt}{t}
\int_0^1 due^{-\frac{t}{4}(1-u^2)}.
\eea
At $L^2\to 0$ this integral tends to
\bea
K^{(1)}=-\frac{1}{32\pi^2}\psi\bar{\psi}\log(\mu^2
L^2)
-\frac{1}{32\pi^2}\psi\bar{\psi}(\log\frac{\psi\bar{\psi}}{\mu^2}-\xi).
\eea 
where $\xi$ is some constant which can be absorbed into redefinition
of $\mu$. We can add the counterterm
$\frac{1}{32\pi^2}\psi\bar{\psi}\log(\mu^2 L^2)$ to
cancel the divergence. Such a
counterterm corresponds to a usual renormalization of kinetic
term by the rule
\bea
\Phi\to Z^{1/2}\Phi;\ Z=1+\frac{\lambda^2}{32\pi^2}\log(\mu^2 L^2).
\eea
And the renormalized k\"{a}hlerian effective potential is
\bea
\label{kren}
K^{(1)}=-\frac{1}{32\pi^2}\psi\bar{\psi}(\log\frac{\psi\bar{\psi}}{\mu^2}-\xi).
\eea

Another way for calculating of the k\"{a}hlerian effective potential
consists in
summarizing of contributions from supergraphs given by Fig. 5.
Sum of these contributions looks like \cite{PW}
\bea
K^{(1)}=\int \frac{d^4k}{(2\pi)^4}
\int d^4\q_1\ldots d^4\q_{2n}
\sum_{n=1}^{\infty}\frac{1}{2n}
\Big(\frac{\psi\bar{\psi}}{k^4}\Big)^n\frac{D^2}{4}\delta_{12}\frac{\bar{D}^2}{4}
\delta_{23}\ldots\frac{D^2}{4}\delta_{n-1,n}\frac{\bar{D}^2}{4}\delta_{n1}.
\eea
which after $D$-algebra transformations and summation looks like
\bea
K^{(1)}=\mu^{\epsilon}\int\frac{d^{4-\epsilon} k}{(2\pi)^{4-\epsilon}}\frac{1}{2k^2}
\log(1+\frac{\psi\bar{\psi}}{k^2})
\eea
(here we carried out a dimensional regularization by introducing the
parameter $\epsilon$). Integration leads to
\bea
K^{(1)}=\frac{1}{32\pi^2}[\frac{\psi\bar{\psi}}{\epsilon}-\psi\bar{\psi}\log
\frac{\psi\bar{\psi}}{e\mu^2}]
\eea
where $e=\exp(1)$. A subtraction of the divergence and a redefinition of $\mu$
leads to the result (\ref{kren}). 

Now we turn to calculation of the chiral effective potential.
It is not equal to zero for massless theories. Really, as it was noted
by West \cite{West2} the mechanism of arising chiral corrections is
the following one. If the theory describes dynamics of chiral and
antichiral superfields, then quantum correction of the form
\bea
\label{n1}
\int d^8 z f(\Phi)(-\frac{D^2}{4\Box})g(\Phi)
\eea
can be rewritten as
\bea
\int d^6 z f(\Phi)g(\Phi).
\eea
Here we used properties $\int d^8 z =\int d^6 z (-\frac{D^2}{4})$ and
$\bar{D}^2D^2\Phi=16\Box\Phi$ (last identity is true for any chiral
superfield $\Phi$), and $f(\Phi),\, g(\Phi)$
are arbitrary functions of chiral superfield $\Phi$. However, presence
of factor $\Box^{-1}$ is
characteristic for massless theories, in massive theories where we
have $(\Box-m^2)^{-1}$ instead of $\Box^{-1}$, and this mechanism of
arising contributions to chiral effective potential does not work. We note that this situation is rather generic, that is, the perturbative contributions to the chiral effective potential can arise only if all propagators are massless (except of very rare situations where the integral over massive propagators completely factorizes out giving only a constant with no dependence on external momenta). Actually, namely this effect is crucial to prove the Goldstone theorem in noncommutative superfield theories \cite{ourGold} which is related with the statement of absence of $\Phi^2$ corrections in theories with chiral self-couplings.

In the case of massless theory we can find matrix superpropagator (\ref{gremat})
exactly: first,
\bea
G^{\psi}_v(z_1,z_2)&=&(\Box+\frac{1}{4}\psi\bar{D}^2)^{-1}\delta^8(z_1-z_2)=
\frac{1}{\Box_1}\delta^8(z_1-z_2)-\nonumber\\&-&
\frac{1}{\Box_1}\psi(z_1)
\frac{\bar{D}^2_1}{4\Box_1}\delta^8(z_1-z_2).
\eea
The higher terms in this expansion are equal to zero because they are
proportional to $\bar{D}^2\psi=0$ or $\bar{D}^4=0$. It follows straightforwardly from this expression that ${\rm Tr}\ln G^{\psi}_v=0$, therefore, the one-loop effective potential in the Wess-Zumino model identically vanishes.
The components of matrix superpropagator (\ref{gremat}) look like
\bea
G_{++}&=&0;
G_{+-}=G^*_{-+}=\frac{\bar{D}^2_1D^2_2}{16\Box}\delta^8(z_1-z_2)
\nonumber\\
G_{--}&=&-\frac{D^2_1}{4\Box_1}[\psi(z_1)\frac{\bar{D}^2_1
  D^2_2}{16\Box}\delta^8(z_1-z_2)].
\eea
Here $*$ denotes complex conjugation. We note that background chiral
superfield $\psi$ is not constant, otherwise when we arrive at
expression proportional to $D^2\psi$ we get singularity $\frac{0}{0}$
\cite{my3}.
The only two-loop contribution to chiral effective potential is given
by the following supergraph

\hspace{4.5cm}
\Lengthunit=1.5cm
%\Linewidth{1.2pt}
\GRAPH(hsize=3){\ind(0,-16){Fig.7}%\thicklines
\mov(.5,0){\Circle(2)\mov(-1,0){\lin(2,0)}
\ind(-2,10){|}
\ind(-2,-3){\bar{D}^2}\ind(-2,0){|}\ind(-2,-10){|}\ind(-2,-13){\bar{D}^2}
\ind(-2,7){\bar{D}^2}
\ind(-9,2){D^2} \ind(-10,-2){D^2} \ind(8,2){D^2} \ind(8,-2){D^2}
\ind(-18,2){-} \ind(-18,-2){-} \ind(0,2){-} \ind(0,-2){-}
%\Linewidth{0.3pt}
\mov(-1,1){\lin(-.7,.7)}%\mov(-1.1,1){\lin(-.7,.7)}
\mov(-1,-1){\lin(.7,-.7)}%\mov(-1.1,-1){\lin(.7,-.7)}
\mov(-1,0){\lin(-.7,.7)}%\mov(-1.1,0){\lin(-.7,.7)}}
}}

\vspace*{2mm}

\noindent External lines are chiral. We use representation in which
$\Phi(z)=\Phi(x,\theta)$, and
$\bar{D}_{\ad}=-\frac{\pa}{\pa\bar{\q}^{\ad}}$.
Remind that in this case $\psi=\lambda\Phi$.

The contribution of the supergraph given in the Fig.7 looks like
\begin{eqnarray}
\label{I1}
I&=&\frac{\lambda^5}{12}
\int \frac{d^4p_1 d^4p_2}{{(2\pi)}^8}\frac{d^4k d^4l}{{(2\pi)}^8}
\int d^4\theta_1 d^4\theta_2 d^4\theta_3 d^4\theta_4 d^4\theta_5
\Phi(-p_1,\theta_4)
\Phi(-p_2,\theta_5)\times\nonumber\\&\times&
\Phi(p_1+p_2,\theta_3)
\frac{1}{k^2 l^2 {(k+p_1)}^2{(l+p_2)}^2{(l+k)}^2{(l+k+p_1+p_2)}^2}
\times\nonumber\\&\times&
\delta_{13}\frac{\bar{D}^2_3}{4}\delta_{32}
\frac{D^2_1 \bar{D}^2_4}{16}\delta_{14}\frac{D^2_4}{4}\delta_{42}
\frac{D^2_1 \bar{D}^2_5}{16}\delta_{15}\frac{D^2_5}{4}\delta_{52}.
\end{eqnarray}
After $D$-algebra transformation this expression can be written as
\begin{eqnarray}
\label{app}
I&=&\frac{\lambda^5}{12}
\int \frac{d^4p_1 d^4p_2}{{(2\pi)}^8}\frac{d^4k d^4l}{{(2\pi)}^8}
\int d^2\theta
\Phi(-p_1,\theta)
\Phi(-p_2,\theta)%\times\nonumber\\&\times&
\Phi(p_1+p_2,\theta)
\times\nonumber\\&\times&
\frac{k^2 p_2^2+ l^2 p_1^2 +2 (k l)(p_1 p_2)}
{k^2 l^2 {(k+p_1)}^2{(l+p_2)}^2{(l+k)}^2{(l+k+p_1+p_2)}^2}.
\end{eqnarray}
Here we made transformation $\int d^4\q=\int
d^2\q(-\frac{1}{4}\bar{D}^2)$
and took into account that\\ $\bar{D}^2D^2\Phi(p,\q)=-16p^2\Phi(p,\q)$.

As we know, the effective potential is the
effective Lagrangian for superfields slowly varying in space-time.
Let us study behaviour of the expression (\ref{app}) in this case.
The contribution (\ref{app}) can be expressed as
\begin{eqnarray}
\label{cont}
I&=&\frac{\lambda^5}{12}\int d^2\theta \int \frac{d^4p_1 d^4p_2}{{(2\pi)}^8}
\Phi(-p_1,\theta)
\Phi(-p_2,\theta)%\times\nonumber\\&\times&
\Phi(p_1+p_2,\theta)
S(p_1, p_2).
\end{eqnarray}
Here $p_1, p_2$ are external momenta, and
\begin{eqnarray}
S(p_1,p_2)=\int\frac{d^4 k d^4 l}{{(2\pi)}^8}\frac{k^2 p_2^2+ l^2
p_1^2 +2 (kl)(p_1 p_2)} {k^2 l^2
{(k+p_1)}^2{(l+p_2)}^2{(l+k)}^2{(l+k+p_1+p_2)}^2}.
\end{eqnarray}
After Fourier transform eq. (\ref{cont}) has the form
\begin{eqnarray}
\label{cont1}
I&=&\frac{\lambda^5}{12}\int d^2\theta \int d^4 x_1 d^4x_2 d^4 x_3
\int\frac{d^4p_1 d^4p_2}{{(2\pi)}^8}
\Phi(x_1,\theta)
\Phi(x_2,\theta)\times\nonumber\\&\times&
\Phi(x_3,\theta)
%\times\nonumber\\&\times&
\exp[i(-p_1 x_1- p_2 x_2+(p_1+p_2)x_3)]
S(p_1, p_2).
\end{eqnarray}
Since superfields in the case under consideration are slowly varying in
space-time we can put
$
\Phi(x_1,\theta)
\Phi(x_2,\theta)
\Phi(x_3,\theta)
\simeq \Phi^3(x_1,\theta).
$.
As a result one gets
\begin{eqnarray}
I&=&\frac{\lambda^5}{12}\int d^2\theta \int d^4 x_1 d^4 x_2 d^4 x_3
\int\frac{d^4p_1 d^4p_2}{{(2\pi)}^8}
\Phi^3(x_1,\q)
\times\nonumber\\&\times&
\exp[i(-p_1 x_1- p_2 x_2+(p_1+p_2)x_3)]
S(p_1, p_2).
\end{eqnarray}
Integration over $d^4 x_2 d^4 x_3$ leads to delta-functions
$\delta(p_2)\delta(p_1+p_2)$. Hence the eq. (\ref{cont1}) takes the form
\begin{equation}
I=\frac{\lambda^5}{12}\int d^2\theta \int d^4 x_1 
\Phi^3(x_1,\q)
S(p_1,p_2)|_{p_1,p_2=0}.
\end{equation}
Therefore final result for two-loop correction to chiral 
(frequently called holomorphic) effective
potential looks like
\begin{equation}
\label{l2c}
W^{(2)}=
\frac{\lambda^5}{2{(16\pi^2)}^2}\zeta(3)\Phi^3(z),
\end{equation}
where we took into account that (cf. \cite{Us,West3})
\begin{eqnarray}
\label{usint}
S(p_1,p_2)|_{p_1,p_2=0}&=&\int \frac{d^4k d^4l}{{(2\pi)}^8}
%\times\nonumber\\&\times&
\frac{k^2 p_2^2+ l^2 p_1^2 +2 (k l)(p_1 p_2)}
{k^2 l^2 {(k+p_1)}^2{(l+p_2)}^2{(l+k)}^2{(l+k+p_1+p_2)}^2}|_{p_1=p_2=0}=\nonumber\\
&=&
\frac{6}{{(4\pi)}^4}\zeta(3).
\end{eqnarray}
We see that the correction (\ref{l2c}) is finite and does not require any renormalization. 
Actually, it follows from \cite{Us} that a whole class of two-loop finite supergraphs yields contributions proportional to $\zeta(3)$. In principle, all finite two-loop contributions to the chiral effective potential in the super-Yang-Mills theory which we consider below have just this structure hence they are proportional to $\zeta(3)$.

The situation in $N=1$ super-Yang-Mills theory, however, possesses
some peculiarities. First of all, in this theory the one-loop contribution to the chiral effective potential is nontrivial. It is described by the supergraph

\vspace*{1mm}

\hspace*{2cm}
\Lengthunit=1.5cm
\GRAPH(hsize=3){\ind(20,-19){Fig.8}
\mov(2,0){\halfcirc(2)[u]%\wavelin(-1,0)\wavelin(1,0)
\mov(-1,0){\arcto(.6,-.8)[-0.7]}\mov(1,0){\arcto(-.6,-.8)[0.7]}\mov(-.4,-.8){\wavearcto(.8,0)[-.7]}
%\mov(-.1,-1){\arcto(1,1)[-0.7]\wavearcto(-1,1)[0.7]}
\ind(-2,10){|}
%\ind(-2,-3){\bar{D}^2}\ind(-2,0){|}
\ind(-8,-7){-}\ind(-10,-9){D^2}\ind(4,-7){-}\ind(7,-9){D^2}
\ind(-2,7){\bar{D}^2}
%\ind(-9,2){D^2} \ind(-10,-2){\bar{D}^2} \ind(8,2){D^2} \ind(8,-2){\bar{D}^2}
%\ind(-18,2){-} \ind(-19,-2){-} \ind(0,2){-} \ind(-1,-2){-}
%\Linewidth{0.3pt}
\mov(-.6,1){\lin(0,.5)}
\mov(-.9,-.8){\lin(-.7,-.7)}
\mov(-.2,-.8){\lin(.7,-.7)}
}}

\vspace*{2mm}

It was shown in \cite{West4} that the contribution of this diagram, after simple D-algebra transformations, looks like
\bea
\label{ch1}
{\cal L}^{(1)}_c=-\frac{1}{16\pi^2}C_0\lambda_{dec}g^2(T^I)_a^d(T^I)_b^e\int d^6z \Phi^a(z)\Phi^b(z)\Phi^c(z),
\eea
where 
\bea
\label{c0}
C_0=\int_0^1d\alpha\frac{\ln\alpha(1-\alpha)}{1-\alpha(1-\alpha)}
\eea
is a finite constant.
There is no other one-loop contributions to the chiral effective potential. 

As for the two-loop order, both finite and divergent two-loop chiral contributions in this theory are possible. First, the finite ones depicted at Figs. 9a-9g.

%\vspace*{4mm}

\hspace*{2cm}
\Lengthunit=1.5cm
\GRAPH(hsize=3){\ind(0,-20){Fig.9a}
\mov(.5,0){\halfcirc(2)[u]\lin(-1,0)\wavelin(1,0)
\mov(-.1,-1){\arcto(1,1)[-0.7]\wavearcto(-1,1)[0.7]}
\ind(-2,10){|}
\ind(-9,-3){\bar{D}^2}\ind(-4,0){|}\ind(-2,-10){|}\ind(-2,-13){D^2}
\ind(-2,7){\bar{D}^2}\ind(-4,-3){D^2}
\ind(-12,2){D^2} \ind(8,2){D^2} \ind(8,-2){\bar{D}^2}
\ind(-18,2){-} \ind(-16,0){|} \ind(-0.5,2){-} \ind(-1,-2){-}
%\Linewidth{0.3pt}
\mov(-1.3,1){\lin(-.7,.7)}
\mov(-1.35,-1){\lin(.7,-.7)}
\mov(-1.35,0){\lin(-.7,.7)}
}}
\hspace*{2cm}
\GRAPH(hsize=3){\ind(0,-20){Fig.9b}
\mov(.5,0){%\halfcirc(2)[u]
\lin(1.1,0)\wavelin(-1,0)
\mov(0,-1){\arcto(1,1)[-0.7]\wavearcto(-1,1)[0.7]}
\mov(-.1,1){\arcto(1,-1)[0.7]\wavearcto(-1,-1)[-0.7]}
\ind(-2,10){|}
\ind(-2,-3){D^2}\ind(-2,0){|}\ind(-2,-10){|}\ind(-2,-13){D^2}
\ind(-2,7){D^2}
\ind(-9,2){\bar{D}^2D^{\alpha}} \ind(-10,-2){D_{\alpha}} \ind(8,2){\bar{D}^2} \ind(8,-2){\bar{D}^2}
\ind(-18,2){-} \ind(-18,-2){-} \ind(-0.5,2){-} \ind(-1,-2){-}
%\Linewidth{0.3pt}
\mov(-1.3,1){\lin(-.7,.7)}
\mov(-1.35,-1){\lin(.7,-.7)}
\mov(-1.35,0){\lin(-.7,.7)}
}}

%\newpage

\hspace*{2cm}
\Lengthunit=1.5cm
\GRAPH(hsize=3){\ind(0,-18){Fig.9c}
\mov(.5,0){\halfcirc(2)[u]\wavelin(-1,0)\wavelin(1,0)\mov(-1,0){\lin(-1,0)}
\mov(-.1,-1){\arcto(1,1)[-0.7]\wavearcto(-1,1)[0.7]}
\ind(-2,10){|}
%\ind(-2,-3){\bar{D}^2}\ind(-2,0){|}
\ind(-2,-10){|}\ind(-2,-13){D^2}
\ind(-2,7){\bar{D}^2}
\ind(-9,2){D^2} \ind(8,2){D^2} \ind(8,-2){\bar{D}^2}
\ind(-17,2){-} \ind(2,2){-}\ind(2,-2){-}%\ind(-18,-2){-}  
%\Linewidth{0.3pt}
\mov(-1.1,1){\lin(-.7,.7)}
\mov(-1,-1){\lin(-.7,-.7)}
%\mov(-1,0){\lin(-.7,.7)}
}}
\hspace*{2cm}
\Lengthunit=1.5cm
\GRAPH(hsize=3){\ind(0,-18){Fig.9d}
\mov(.5,0){%%\halfcirc(2)[u]
\lin(-1,0)\lin(1,0)\mov(-1,0){\lin(-1,0)}
\mov(-.1,-1){\arcto(1,1)[-0.7]\wavearcto(-1,1)[0.7]}
\mov(-.2,1){\arcto(1,-1)[0.7]\wavearcto(-1,-1)[-0.7]}
\ind(-2,10){|}
%\ind(-2,-3){\bar{D}^2}\ind(-2,0){|}
\ind(-2,-10){|}\ind(-2,-13){D^2}
\ind(-2,7){D^2}
\ind(-9,2){D^2}  \ind(8,2){\bar{D}^2} \ind(8,-2){\bar{D}^2}%\ind(-10,-2){D^2}
\ind(-14,0){|} \ind(1,2){-}\ind(1,-2){-}%\ind(-18,-2){-}  
%\Linewidth{0.3pt}
\mov(-.9,1){\lin(-.7,.7)}
\mov(-1,-1){\lin(-.7,-.7)}
%\mov(-1,0){\lin(-.7,.7)}
}}

\hspace*{2cm}
\Lengthunit=1.5cm
\GRAPH(hsize=3){\ind(0,-19){Fig.9e}
\mov(.5,0){\halfcirc(2)[u]\wavelin(-1,0)\wavelin(1,0)
\mov(-1.1,0){\arcto(.6,-.8)[-0.7]}\mov(.9,0){\arcto(-.6,-.8)[0.7]}\mov(-.5,-.8){\wavearcto(.8,0)[-.7]}
%\mov(-.1,-1){\arcto(1,1)[-0.7]\wavearcto(-1,1)[0.7]}
\ind(-2,10){|}
\ind(-9,-7){-}\ind(4,-7){-}
\ind(-9,-10){D^2}\ind(7,-10){D^2}
\ind(-2,7){\bar{D}^2}
\ind(-9,2){D^2} \ind(-10,-3){\bar{D}^2} \ind(8,2){D^2} \ind(8,-2){\bar{D}^2}
\ind(-19,2){-} \ind(-19,-2){-} \ind(-1.5,2){-} \ind(-1.5,-2){-}
%\Linewidth{0.3pt}
\mov(-1,1){\lin(0,.7)}
\mov(-1.7,-.8){\lin(-.7,-.7)}
\mov(-.8,-.7){\lin(.7,-.7)}
}}
\hspace*{2cm}
\Lengthunit=1.5cm
\GRAPH(hsize=3){\ind(0,-19){Fig.9f}
\mov(.5,0){\lin(-1,0)\lin(1,0)\mov(0,1){\arcto(1,-1)[0.7]\wavearcto(-1,-1)[-0.7]}
\mov(-1.1,0){\arcto(.6,-.8)[-0.7]}\mov(.9,0){\wavearcto(-.6,-.8)[0.7]}\mov(-.5,-.8){\arcto(.8,0)[-.7]}
%\mov(-.1,-1){\arcto(1,1)[-0.7]\wavearcto(-1,1)[0.7]}
\ind(1,10){|}
%\ind(-2,-3){\bar{D}^2}\ind(-2,0){|}
\ind(-6.5,-8){|}\ind(-7,-13){\bar{D}^2}\ind(0,-8){|}\ind(2,-13){D^2}
\ind(2,7){D^2}
\ind(-9,2){\bar{D}^2} \ind(-18,-3){D^2} \ind(8,2){\bar{D}^2} \ind(7,-4){D^2}
\ind(-16,0){|} \ind(-19,-2){-} \ind(-1,2){-} \ind(-3,-1){|}
%\Linewidth{0.3pt}
\mov(-1,1){\lin(0,.7)}
\mov(-1.7,-.8){\lin(-.7,-.7)}
\mov(-.9,-.8){\lin(.7,-.7)}
}}

%\vspace*{4mm}

\hspace*{2cm}
\Lengthunit=1.5cm
\GRAPH(hsize=3){\ind(0,-19){Fig.9g}
\mov(.5,0){\lin(-1,0)\lin(1,0)\mov(0,1){\arcto(-1,-1)[-0.7]\wavearcto(1,-1)[0.7]}
\mov(-1.1,0){\wavearcto(.6,-.8)[-0.7]}\mov(.9,0){\arcto(-.6,-.8)[0.7]}\mov(-.5,-.8){\arcto(.8,0)[-.7]}
%\mov(-.1,-1){\arcto(1,1)[-0.7]\wavearcto(-1,1)[0.7]}
\ind(-3,10){|}
%\ind(-2,-3){\bar{D}^2}\ind(-2,0){|}
\ind(-6.5,-8){|}\ind(-7,-13){D^2}\ind(1.5,-8){|}\ind(3,-13){\bar{D}^2}
\ind(-2,6){D^2}
\ind(-15,3){\bar{D}^2} \ind(-10,-3){D^2} \ind(2,2){\bar{D}^2} \ind(8,-3){D^2}
\ind(-18,2){-} \ind(-16,0){|} \ind(0,-2){-} \ind(-3,-1){|}
%\Linewidth{0.3pt}
\mov(-1.2,1){\lin(0,.7)}
\mov(-1.7,-.8){\lin(-.7,-.7)}
\mov(-.9,-.75){\lin(.7,-.7)}
}}

\vspace*{2mm}

Second, the divergent ones are depicted at Figs. 9h--9m.

\vspace*{2mm}

\hspace*{4cm}
\Lengthunit=1cm
\GRAPH(hsize=3){\ind(20,-18){Fig.9h}
\mov(.5,0){\halfcirc(2)[u]\mov(-1,0){\lin(-1,0)}
\mov(-.1,-1){\wavearcto(1,1)[-0.7]\arcto(-1,1)[0.7]\lin(0,-.7)}
\mov(1.8,0){\halfcirc(2)[u]\halfwavecirc(2)[d]}\mov(2.8,0){\lin(1,0)}
\ind(-14.5,2){-}\ind(-17,2){\bar{D}^2}\ind(4.5,2){-}\ind(1.5,2){D^2}
\ind(7,2){-}\ind(11.5,2){\bar{D}^2}\ind(25.5,2){-}\ind(28.5,2){D^2}
\ind(-8,-9){|}\ind(-8,-14){D^2}
}}

\vspace*{2mm}
\Lengthunit=1.5cm
\hspace*{5mm}\GRAPH(hsize=3){\ind(0,-19){Fig.9i}
\mov(.5,0){\halfcirc(2)[u]\lin(-1,0)\lin(1,0)
\mov(-1.1,0){\wavearcto(.6,-.8)[-0.7]}\mov(.9,0){\wavearcto(-.6,-.8)[0.7]}\mov(-.5,-.8){\arcto(.4,-.2)[-.7]}\mov(.3,-.8){\arcto(-.4,-.2)[.7]}
%\mov(-.1,-1){\arcto(1,1)[-0.7]\wavearcto(-1,1)[0.7]}
\ind(-2,-10){|}
%\ind(-2,-3){\bar{D}^2}\ind(-2,0){|}
\ind(-6.5,-8){|}\ind(-7,-13){D^2}\ind(.5,-10){|}\ind(2,-13){D^2}
\ind(-2,-7){\bar{D}^2}
\ind(-16,2){D^2} \ind(-11,-3){\bar{D}^2} \ind(8,2){\bar{D}^2} \ind(1,-3){D^2}
\ind(-16,0){|} \ind(-3,-1){|}
\ind(-19,2){-} \ind(-1,2){-} 
%\Linewidth{0.3pt}
\mov(-1.3,-1){\lin(0,-.7)}
\mov(-1.8,-.8){\lin(-.7,-.7)}
\mov(-1,-.8){\lin(.7,-.7)}
\hspace*{1.5cm}
%\Lengthunit=1.5cm
\GRAPH(hsize=2){\ind(0,-19){Fig.9j}
\mov(.5,0){\halfdashcirc(2)[u]\dashlin(-1,0)\dashlin(1,0)
\mov(-1.1,0){\wavearcto(.6,-.8)[-0.7]}\mov(.9,0){\wavearcto(-.6,-.8)[0.7]}\mov(-.5,-.8){\arcto(.4,-.2)[-.7]}\mov(.3,-.8){\arcto(-.4,-.2)[.7]}
%\mov(-.1,-1){\arcto(1,1)[-0.7]\wavearcto(-1,1)[0.7]}
\ind(-2,-10){|}
%\ind(-2,-3){\bar{D}^2}\ind(-2,0){|}
\ind(-6.5,-8){|}\ind(-7,-13){D^2}\ind(.5,-10){|}\ind(2,-13){D^2}
\ind(-2,-7){\bar{D}^2}
\ind(-16,2){D^2} \ind(-11,-3){\bar{D}^2} \ind(8,2){\bar{D}^2} \ind(1,-3){D^2}
%\ind(-9,2){D^2} \ind(-10,-2){\bar{D}^2} \ind(8,2){\bar{D}^2} \ind(8,-2){D^2}
%\ind(-20,2){-} \ind(-20,-2){-} \ind(-2,2){-} \ind(-3,-2){-}
\ind(-18,2){-} \ind(0,2){-} 
\ind(-16,0){|} \ind(-3,-1){|}
%\ind(-19,-2){-} \ind(-1,-2){-}
%\Linewidth{0.3pt}
\mov(-1.3,-1){\lin(0,-.7)}
\mov(-1.8,-.8){\lin(-.7,-.7)}
\mov(-1,-.8){\lin(.7,-.7)}
}}
\hspace*{1.5cm}
%\Lengthunit=1.5cm
\GRAPH(hsize=2){\ind(0,-19){Fig.9k}
\mov(.5,0){\halfwavecirc(2)[u]\wavelin(-1,0)\wavelin(1,0)
\mov(-1.1,0){\wavearcto(.6,-.8)[-0.7]}\mov(.9,0){\wavearcto(-.6,-.8)[0.7]}\mov(-.5,-.8){\arcto(.4,-.2)[-.7]}\mov(.3,-.8){\arcto(-.4,-.2)[.7]}
%\mov(-.1,-1){\arcto(1,1)[-0.7]\wavearcto(-1,1)[0.7]}
\ind(-2,-10){|}
%\ind(-2,-3){\bar{D}^2}\ind(-2,0){|}
\ind(-6.5,-8){|}\ind(-7,-13){D^2}\ind(.5,-10){|}\ind(2,-13){D^2}
\ind(-2,-7){\bar{D}^2}
\ind(-9,2){D_{\beta}} \ind(-10,-2){\bar{D}^2D^{\beta}} \ind(8,2){\bar{D}^2D^{\alpha}} \ind(8,-2){D_{\alpha}}
%\ind(-20,2){-} \ind(-20,-2){-} \ind(-2,2){-} \ind(-3,-2){-}
\ind(-18,2){-} \ind(-19,-2){-} \ind(0,2){-} \ind(-1,-2){-}
%\Linewidth{0.3pt}
\mov(-1.3,-1){\lin(0,-.7)}
\mov(-1.8,-.8){\lin(-.7,-.7)}
\mov(-1,-.8){\lin(.7,-.7)}
}}
}}

%\newpage 

\vspace*{2mm}

\hspace*{2cm}
\Lengthunit=1.5cm
\GRAPH(hsize=3){\ind(0,-19){Fig.9l}
\mov(.5,0){\halfcirc(2)[u]\wavelin(-1,0)\wavelin(1,0)
\mov(-1.1,0){\arcto(.6,-.8)[-0.7]}\mov(.9,0){\arcto(-.6,-.8)[0.7]}\mov(-.5,-.8){\wavearcto(.4,-.2)[-.7]}\mov(.3,-.8){\arcto(-.4,-.2)[.7]}
%\mov(-.1,-1){\arcto(1,1)[-0.7]\wavearcto(-1,1)[0.7]}
\ind(-2,-10){|}
%\ind(-2,-3){\bar{D}^2}\ind(-2,0){|}
\ind(-6.5,-8){|}\ind(-7,-13){D^2}\ind(3.5,-8){|}\ind(4,-13){D^2}
\ind(-2,-7){\bar{D}^2}
\ind(-9,2){D^2} \ind(-10,-2){\bar{D}^2} \ind(8,2){\bar{D}^2} \ind(8,-2){D^2}
\ind(-19,2){-} \ind(-19.5,-2){-} \ind(-.5,2){-} \ind(-.5,-2){-}
%\ind(-18,2){-} \ind(-19,-2){-} \ind(0,2){-} \ind(-1,-2){-}
%\Linewidth{0.3pt}
\mov(-1.3,-1){\lin(0,-.7)}
\mov(-1.8,-.8){\lin(-.7,-.7)}
\mov(-1,-.8){\lin(.7,-.7)}
}}
%\hspace*{4cm}
\Lengthunit=1.5cm
\GRAPH(hsize=3){\ind(0,-19){Fig.9m}
\mov(.5,0){\halfcirc(2)[u]\lin(-1,0)\lin(1,0)
\mov(-1.1,0){\arcto(.6,-.8)[-0.7]}\mov(.9,0){\arcto(-.6,-.8)[0.7]}\mov(-.5,-.8){\arcto(.4,-.2)[-.7]}\mov(.3,-.8){\wavearcto(-.4,-.2)[.7]}
%\mov(-.1,-1){\arcto(1,1)[-0.7]\wavearcto(-1,1)[0.7]}
\ind(-2,-10){|}
%\ind(-2,-3){\bar{D}^2}\ind(-2,0){|}
\ind(-6.5,-8){|}\ind(-7,-13){\bar{D}^2}\ind(3.5,-8){|}\ind(4,-13){D^2}
\ind(-2,-7){D^2}
\ind(-9,2){D^2} \ind(-10,-2){D^2} \ind(8,2){\bar{D}^2} \ind(8,-2){\bar{D}^2}
%\ind(-20,2){-} \ind(-20,-2){-} \ind(-2,2){-} \ind(-3,-2){-}
\ind(-18,2){-} \ind(-19,-2){-} \ind(0,2){-} \ind(-1,-2){-}
%\Linewidth{0.3pt}
\mov(-1.3,-1){\lin(0,-.7)}
\mov(-1.8,-.8){\lin(-.7,-.7)}
\mov(-1,-.8){\lin(.7,-.7)}
}}

\vspace*{2mm}

Here we use the Feynman gauge for the propagators of the gauge field, to avoid the infrared singularities.
Contributions of all finite diagrams to the effective potential can be shown to be
proportional to $\frac{\zeta(3)}{(4\pi)^4}\Phi^3$ (the corresponding integral over momenta looks similarly to (\ref{usint}) and yields just this result) being analogous to the case of
Wess-Zumino model and matching the class of contributions described in \cite{Us}. 

As an example of the divergent contribution we will consider the supergraph given by Fig. 9h. It is, after D-algebra
transformation, proportional to
\bea
\label{kdiv}
& &\int d^2\q\int\frac{d^4p_1 d^4p_2}{(2\pi)^8}
(p_1+p_2)^2\int\frac{d^4k  d^4l}{(2\pi)^8}\frac{1}{k^2(k+p_1)^2(k+p_2)^2}
\times\nonumber\\&\times&
\frac{1}{l^2(l+p_1+p_2)^2}\Phi(-p_1,\q)
\Phi(-p_2,\q)\Phi(p_1+p_2,\q).
\eea  
To obtain the low-energy leading contribution we must consider the limit at 
$p_1,p_2\to 0$. It is known \cite{West4} that
$$
\lim_{p_1,p_2\to 0}
(p_1+p_2)^2\int\frac{d^4k}{(2\pi)^4}\frac{1}{k^2(k+p_1)^2(k+p_2)^2}=
\frac{1}{16\pi^2}\int_0^1 d\a \frac{\log [\a(1-\a)]}{1-\a(1-\a)}=
\frac{C_0}{16\pi^2},
$$
where $C_0$ is the finite constant given by (\ref{c0}). The integral over $l$ is divergent, and
after dimensional regularization it is equal to
$$
\int\frac{d^{4-\epsilon}l}{(2\pi)^{4-\epsilon}}\frac{1}{l^2(l+p_1+p_2)^2}=
\frac{1}{16\pi^2}(\frac{2}{\epsilon}+\log\frac{(p_1+p_2)^2}{\mu^2}).
$$
After cancellation of divergence with help of 
the appropriate one-loop counterterm and
transforming
to coordinate representation we see that expression (\ref{kdiv}) for
slowly varying in space-time superfields takes the form
\bea
\int d^6 z \frac{1}{(16\pi^2)^2}\Phi^2(z)\log(-\frac{\Box}{\mu^2})\Phi(z).
\eea

Carrying out the calculations for all supergraphs above we conclude that the leading chiral correction in $N=1$ super-Yang-Mills theory
with chiral matter is nonlocal.
Its complete form, after subtracting of the corresponding counterterms, is
\bea
\label{res1a}
\tilde{\Gamma}^{(2)}_c&=&\frac{g^2}{(4\pi)^4}\{\zeta(3)[6^5g^2(\lambda_{imn}(T^KT^IT^K)^n_k(T^I)^m_j+\lambda_{lnp}(T^IT^K)^p_k(T^I)^l_i(T^K)^n_j+\nonumber\\&+&\lambda_{imn}(T^IT^K)_k^n\{T^K,T^I\}^m_j)+4\times 6^3(T^IT^J)^l_i(T^J)^m_k
(\lambda_{lmp}(T^I)^p_j+\nonumber\\&+&\lambda_{ljp}(T^I)^p_m)]+3\times 6^4[\frac{\lambda_{iml}g^2}{3!}(\frac{1}{64}-4)(T^K)^a_b(T^I)^c_d(T^K)^m_j(T^I)^l_k\times\nonumber\\&\times&
[(T^L),(T^N)]^b_a[(T^L),(T^N)]^d_c+\frac{\lambda_{rnl}\lambda_{rns}\lambda_{skm}}{(3!)^3}(T^I)^l_i(T^I)^m_j+\nonumber\\&+&\lambda_{snp}g^2(T^K)^s_i(T^K)^p_k(T^I)^n_m(T^I)^m_j+\nonumber\\&+&\lambda_{irp}g^2(T^I)_n^m(T^K)_m^n(T^I)^p_k(T^K)^r_j](1-\gamma)C_0+\nonumber\\&+&2\times 6^4\frac{\lambda_{iml}}{2!3!}g^2(T^K)^s_p(T^K)^p_k(T^I)^l_s(T^I)^m_j(1-\frac{\gamma}{2}C_0)\}\times\nonumber\\&\times&
\int d^6z\Phi^i(z)\Phi^j(z)\Phi^k(z)
-\nonumber\\&-&\frac{2g^4}{(4\pi)^4}6^4\frac{\lambda_{iml}}{2!3!}(T^K)^s_p(T^K)^p_k(T^I)^l_s(T^I)^m_j\int d^6z\Phi^i(z)\Phi^j(z)[\ln\left(-\frac{\Box}{\mu^2}\right)]\Phi^k(z)-\nonumber\\&-&3\times 6^4\{\int d^2\theta\frac{d^4p_1d^4p_2}{(2\pi)^8}[\frac{\lambda_{iml}g^4}{3!}(\frac{1}{64}-4)(T^K)^a_b(T^I)^c_d(T^K)^m_j(T^I)^l_k\times\nonumber\\&\times&[(T^L),(T^N)]^b_a[(T^L),(T^N)]^d_c+\lambda_{irp}g^4(T^I)^m_n(T^K)^n_m(T^I)^p_k(T^K)^r_j+\nonumber\\&+&\frac{\lambda_{rnl}\lambda_{rns}\lambda_{skm}g^2}{(3!)^3}(T^I)^l_i(T^I)^m_j+\lambda_{smp}g^4(T^I)^m_n(T^I)^n_j(T^K)^p_k(T^K)^s_i]\}\times\nonumber\\&\times&
\int_0^1d\beta d\alpha[\ln\frac{p^2_1\beta(1-\beta)+p^2_2\alpha(1-\alpha)-2p_1p_2\alpha\beta}{\mu^2(1-\alpha-\beta)}]\times\\&\times&\frac{(p_1+p_2)^2}{p^2_1\beta(1-\beta)+p^2_2\alpha(1-\alpha)-2p_1p_2\alpha\beta}\Phi^i(-p_1,\theta)\Phi^j(-p_2,\theta)\Phi^k(p_1+p_2,\theta).\nonumber
\eea
This result is found for the super-Yang-Mills theory with an arbitrary non-Abelian gauge group. In the particular case of the $SU(N)$ group, the result reduces to
\bea
\label{efsta}
\tilde{\Gamma}^{(2)}_c&=&a_1\lambda_{ijk}g^4\int d^6z\Phi^i(z)\Phi^j(z)\Phi^k(z)+
\nonumber\\&+&a_2\lambda_{ijk}g^4\int d^6z\Phi^i(z)\Phi^j(z)[\ln(-\frac{\Box}{\mu^2})]\Phi^k(z)+\nonumber\\&+&
\lambda_{rnj}\lambda_{rns}\lambda_{ski}g^2\Big(a_3\int d^6z\Phi^i(z)\Phi^j(z)\Phi^k(z)+
\nonumber\\&+&a_4\int d^6z\Phi^i(z)\Phi^j(z)[\ln(-\frac{\Box}{\mu^2})]\Phi^k(z)\Big).
\eea
Here the constants $a_1,a_2,a_3,a_4$ are equal to
\bea
a_1&=&\frac{1}{(4\pi)^4}\{\zeta(3)[6^5(-\frac{N-1}{2N^2}+\frac{1}{4}(1-\frac{1}{N})^2+\frac{1}{4}(N+1-\frac{2}{N}))+\nonumber\\&=&2\times 6^3(1-\frac{1}{N})^2]+3\times 6^4[\frac{1}{3!}\frac{N-1}{2}(\frac{1}{256}-1)+\nonumber\\&+&\frac{(N^2-1)(N-1)}{4N^2}+\frac{N-1}{4N}](1-\frac{\gamma}{2})C_0+\nonumber\\&+&6^3\frac{(N^2-1)(N-1)}{4N^2}(1-\frac{\gamma}{2})C_0
\}-\nonumber\\&-&
\frac{C}{(4\pi)^4}3\times 6^4\{\frac{1}{3!}(\frac{1}{256}-1)\frac{N-1}{2}+\frac{(N-1)(N^2-1)}{4N^2}+\frac{N-1}{4N}
\};\nonumber\\
a_2&=&-6^3\frac{1}{(4\pi)^4}\frac{(N-1)(N^2-1)}{4N^2}-\\&-&\frac{C_0}{(4\pi)^4}3\times 6^4\{\frac{1}{3!}(\frac{1}{256}-1)\frac{N-1}{2}+\frac{(N-1)(N^2-1)}{4N^2}+\frac{N-1}{4N}\};\nonumber\\
a_3&=&(1-\frac{1}{N})\frac{9}{(4\pi)^4}(C+(1-\frac{\gamma}{2})C_0);\nonumber\\
a_4&=&(1-\frac{1}{N})\frac{9}{(4\pi)^4}C_0,\nonumber
\eea 
and
\bea
C&=&\int_0^1 d\beta d\alpha[\ln\frac{p^2_1\beta(1-\beta)+p^2_2\alpha(1-\alpha)-2p_1p_2\alpha\beta}{(p_1+p_2)^2(1-\alpha-\beta)}]\times\nonumber\\&\times&\frac{(p_1+p_2)^2}{p^2_1\beta(1-\beta)+p^2_2\alpha(1-\alpha)-2p_1p_2\alpha\beta}|_{p_1,p_2\to 0}.
\eea
We note that nonlocal corrections in the super-Yang-Mills theory arise also in
the pure gauge sector \cite{KuzYar}.

This nonlocality is a natural consequence of the presence of the UV divergence. However, treating the problem of the possible renormalization of the chiral effective potential (to the best of our knowledge, it was not discussed yet in scientific literature), one must, first, note that the divergence in the two-loop chiral effective potential is caused by the one-loop divergent contributions to the two- and three-point functions involving the gauge superfield. In the maximal, $N=4$ supersymmetric case, these divergences must mutually cancel (for the two-point function, both of gauge and chiral superfields, such a cancellation has been explicitly shown in \cite{SGRS}, see also \cite{Kovacz}). It allows to conclude that in the ``complete'', $N=4$ super-Yang-Mills theory, no such a nonlocality arises. Another way for interpretation of this situation can be related with the concept of the Wilsonian effective action \cite{Wil} whose use naturally introduces an infrared cutoff parameter $\Lambda$ with a subsequent replacement of the nonlocal factor $\ln(-\frac{\Box}{\mu^2})$ in (\ref{efsta}) by a constant $\ln(\frac{\Lambda^2}{\mu^2})$, thus, this expression acquires an usual local form of the chiral effective potential.

Chiral contributions to effective action arise also in other theories
describing dynamics of chiral superfields. F.e. in general chiral
superfield theory the leading chiral contribution is also chiral effective
potential (see next section), in the higher-derivative field theory leading chiral
contribution is of second order in space-time derivatives of chiral
superfield (see section 4.10), and these corrections are
finite. 

We conclude that the presence of quantum contributions to chiral effective Lagrangian is
quite characteristic for theories including chiral superfields.

\section{General chiral superfield model}

In the previous section, we have studied the Wess-Zumino model, that is, the simplest example of the theory describing a dynamics of the chiral multiplet. The natural question is the possibility for constructing of the most generic model for the chiral superfield. Such a model naturally arises within the context of the superstring theory, from which viewpoint, low-energy effective field theory models represent themselves as effective theories where the integration over massive
string modes is carried out, and the ten-dimensional background manifold, where the dynamics of the superstring takes place, has the form $M^4\times K$ where $M^4$ is the usual four-dimensional Minkowski space,
and $K$ is some six-dimensional compact manifold. The theory  reduced to $M^4$ is described by the action 
\cite{GSH}
\bea
\label{genact}
S[\Phi,\bar{\Phi}]=\int d^8 z K(\Phi^i,\bar{\Phi}^i)+
(\int d^6 z W(\Phi^i)+h.c.).
\eea  
We can use matrix notations via introduction of column vector
$\vec{\Phi}=\{\Phi^i\}$, after which the consideration in the 
case of several chiral superfield is analogous to the case of one
chiral superfield (some extensions of this model involving the gauge
superfields are given in \cite{Cvet}). 
We can consider this theory for arbitrary functions
$K$ and $W$. Note that there is no higher derivatives in the classical
action. Therefore the theory with the action (\ref{genact}) is the
most general theory without higher derivatives describing a dynamics of a
chiral superfield. There are a lot of phenomenological applications of
this model in string theory (see \cite{Cvet} and references therein).
In general case this theory is nonrenormalizable; however, it must be treated as
an effective theory aimed for studying of the low-energy domain. Therefore
all integrals over momenta effectively involve  natural cutoff by the condition $p\ll
M_{String}$ where $p$ is momentum, and $M_{String}=10^{17} GeV
\sim 10^{-2} M_{Pl}$ is a characteristic string mass.

The effective action in the theory, as well as that one in the
Wess-Zumino model, can be presented as a series
in supercovariant derivatives
$D_A=(\partial_a,D_{\alpha},\bar{D}_{\dot{\alpha}})$ in the form (\ref{ep1}--\ref{ep5}) being described by the same objects, that is, k\"{a}hlerian effective potential $K_{eff}$, auxiliary fields' effective potential $F_{eff}$ and the chiral (or holomorphic) effective potential $W_{eff}$.

The one-loop contribution to the effective action is totally
determined by the quadratic part of
expansion of
$\frac{1}{\hbar}S[\bar{\Phi}+\sqrt{\hbar}\bar{\phi},\Phi+\sqrt{\hbar}\phi]$
in quantum fields $\phi,\bar{\phi}$ which looks like
\begin{eqnarray}
\label{qua}
S_2=\frac{1}{2}\int d^8 z \left(\begin{array}{cc}\phi&\bar{\phi}
\end{array}\right)
\left(\begin{array}{cc}
K_{\Phi\Phi}&K_{\Phi\bar{\Phi}}\\
K_{\Phi\bar{\Phi}}&K_{\bar{\Phi}\bar{\Phi}}
\end{array}\right)
\left(\begin{array}{c}\phi\\
\bar{\phi}
\end{array}
\right)+[\int d^6 z \frac{1}{2}W^{''}\phi^2+h.c.]
\end{eqnarray}
and defines the propagators and the the higher terms of 
expansion define the vertices.
Here
$K_{\Phi\bar{\Phi}}=\frac{\partial^2 K(\bar{\Phi},\Phi)}
{\partial\Phi\partial\bar{\Phi}}$,
$K_{\Phi\Phi}=\frac{\partial^2 K(\bar{\Phi},\Phi)}
{\partial\Phi^2}$ etc,  $W^{''}=\frac{d^2 W}{d\Phi^2}$.

The corresponding matrix superpropagator has the form
\bea
G(z_1,z_2)=\left(
\begin{array}{cc}
G_{++}(z_1,z_2)&G_{+-}(z_1,z_2)\\
G_{-+}(z_1,z_2)&G_{--}(z_1,z_2)
\end{array}
\right)
\eea
where $+$ denotes chirality with respect to corresponding argument,
and $-$ correspondingly -- antichirality.
This propagator, in the case when all
derivatives of superfields $\Phi,\bar{\Phi}$ are omitted (that is just the case of the k\"{a}hlerian effective potential), satisfies the equation
\bea
\left(\begin{array}{cc}
W^{''}&-\frac{1}{4}K_{\Phi\bar{\Phi}}\bar{D}^2\\
-\frac{1}{4}K_{\Phi\bar{\Phi}} D^2& \bar{W}^{''}
\end{array}\right)
\left(\begin{array}{cc}
G_{++}(z_1,z_2)&G_{+-}(z_1,z_2)\\
G_{-+}(z_1,z_2)& G_{--}(z_1,z_2)
\end{array}
\right)=-
\left(\begin{array}{cc}
\delta_+&0\\
0&\delta_-
\end{array}
\right).
\eea
The solution of this equation looks like
\bea
\label{grematg}
G=
\frac{1}{K^2_{\Phi\bar{\Phi}}\Box-W^{''}\bar{W}^{''}}
\left(\begin{array}{cc}
\bar{W}^{''}&\frac{1}{4}K_{\Phi\bar{\Phi}}\bar{D}^2\\
\frac{1}{4}K_{\Phi\bar{\Phi}} D^2& W^{''}
\end{array}\right)
\left(\begin{array}{cc}
\delta_+&0\\
0&\delta_-
\end{array}
\right).
\eea
Now we turn to studying of quantum contributions to k\"{a}hlerian effective
potential depending only on
superfields $\Phi$, $\bar{\Phi}$ but not on their derivatives.

The one-loop diagrams contributing to k\"{a}hlerian effective potential are

\hspace{0.5cm}
\unitlength=.6mm
\begin{picture}(20,20)
\put(0,10){\circle{20}}\put(-10,10){\line(-1,0){5}}\put(-10,8.5){\line(-1,0){5}}
\put(10,10){\line(1,0){5}}\put(10,8.5){\line(1,0){5}}
%\ind(24,0){W''}\ind(-26,0){\bar{W''}}
%\put(4.8,0){\GRAPH(hsize=3){
\end{picture}
\hspace{2cm}
\begin{picture}(20,20)
\put(0,10){\circle{20}}\put(-10,10){\line(-1,0){5}}
\put(-10,8.5){\line(-1,0){5}}
\put(10,10){\line(1,0){5}}\put(10,8.5){\line(1,0){5}}
\put(0,20){\line(0,1){5}}\put(-1,20){\line(0,1){5}}
\put(0,0){\line(0,-1){5}}
\put(-1,0){\line(0,-1){5}}
\put(0,-13){Fig.10}
%}}
%\ind(-10,20){W''}\ind(-10,-20){W''}
%\ind(-49,0){\bar{W}''}\ind(5,0){\bar{W''}}
%\put(30,0){\ldots}
\end{picture}
%}}}}
\hspace{2cm}
\begin{picture}(30,30)
\put(0,10){\circle{20}}
\put(-10,10){\line(-1,0){5}}\put(-10,8.5){\line(-1,0){5}}
\put(10,10){\line(1,0){5}}\put(10,8.5){\line(1,0){5}}
\put(-9,5){\line(-1,-1){6}}\put(-8,4){\line(-1,-1){6}}
\put(9,5){\line(1,-1){6}}\put(8,4){\line(1,-1){6}}
\put(-9,15){\line(-1,1){6}}\put(-8,16){\line(-1,1){6}}
\put(9,15){\line(1,1){6}}\put(8,16){\line(1,1){6}}
\put(30,10){\ldots}
\end{picture}

\vspace{1cm}

Double external lines correspond to alternating $W''$ and $\bar{W}''$.
Internal lines are the background dependent $<\phi\bar{\phi}>$-propagators of
the form
\bea
\label{propc}
G_0\equiv<\phi\bar{\phi}>=-\frac{\bar{D}^2 D^2}{16 K_{\Phi\bar{\Phi}}\Box}\delta^8(z_1-z_2).
\eea
We note that this propagator is valid for calculating not only of the K\"{a}hlerian potential but also of the chiral one. It can be obtained in two ways.
The first way consists in a summation:

\hspace{2cm}
\Lengthunit=1.5cm
\Linewidth{1.2pt}
\GRAPH(hsize=3){\Linewidth{1.2pt}\lin(1,0)\ind(12,0){=}\Linewidth{.3pt}\mov(1.5,0){\lin(1,0)}\ind(27,0){+}
\mov(3,0){\lin(2,0)}\mov(4,0){\dashdotlin(0,1)}\ind(52,0){+}\ind(56,0){\ldots}
\ind(0,0){|}\ind(7,0){|}\ind(1,3){D^2}\ind(8,3){\bar{D}^2}
\ind(13,0){|}\ind(21,0){|}\ind(15,3){D^2}\ind(22,3){\bar{D}^2}
\ind(28,0){|}\ind(36,0){|}\ind(30,3){D^2}\ind(35,3){\bar{D}^2}
\ind(38,0){|}\ind(45,0){|}\ind(40,3){D^2}\ind(46,3){\bar{D}^2}
}
 
\vspace*{1mm}

\noindent 

Here dashed-and-dotted vertical line is the external field $K_{\Phi\bar{\Phi}}-1$ taken in the chiral case $\bar{\Phi}=0$, and the propagator is usual $-\bar{D}^2_1 D^2_2
\frac{\delta^8(z_1-z_2)}{16\Box}$: both if the external field is constant or if it is chiral, we have that the sum of the contributions (with integrals in internal points are assumed where it is necessary) above is
\bea
\sum\limits_{n=0}^{\infty}-\frac{\bar{D}^2 D^2}{16\Box}
[-(K_{\Phi\bar{\Phi}}-1)
\frac{\bar{D}^2 D^2}{16\Box}]^n\delta^8(z_1-z_2)=-
\frac{\bar{D}^2_1 D^2_2}{16 K_{\Phi\bar{\Phi}}(z_1)\Box}\delta^8(z_1-z_2),
\eea
which proves (\ref{propc}). 

In another manner, the result (\ref{propc}) can be proved through the definition  of the new chiral field $\phi'=K_{\Phi\bar{\Phi}}\phi$ (indeed, in this case $K_{\Phi\bar{\Phi}}$ is either constant or chiral, so, $\phi'$ can be treated to be chiral as well, so that the free action acquires the form $S_f=\int d^8z\phi'\bar{\phi}$ which yields $<\phi'(z_1)\bar{\phi}(z_2)>=-\frac{\bar{D}^2_1 D^2_2}{16\Box}\delta^8(z_1-z_2)$, which, in its part, yields (\ref{propc}).

Supergraph of the structure depicted at Fig. 10, with $2n$
legs represents itself as a ring containing $n$ links of the following form

\hspace{2cm}
\unitlength=2mm
%\GRAPH(hsize=3){
\begin{picture}(40,15)
\put(10,4){\line(1,0){20}}\put(10,4){\line(0,1){10}}\put(10.5,4){\line(0,1){10}}
\put(20,4){\line(0,1){10}}\put(20.5,4){\line(0,1){10}}
%\ind(20,-10){���.2}
\put(12,1){$\bar{D}^2$}\put(12,4){$|$}\put(22,1){$D^2$}\put(22,4){$|$}
\put(11,14){$W''$}\put(21,14){$\bar{W}''$}
\end{picture}
%\vspace*{-2mm}

Carrying out the calculations in the same way as in the previous section (see for the details \cite{my5,my6}), we find that the total contribution of all these diagrams
after $D$-algebra transformations, summation, integration over momenta
and subtraction of divergences is equal to
\begin{equation}
\label{k1}
K^{(1)}=-
\frac{1}{32\pi^2}\, {\rm tr}\, \frac{W^{\prime\prime}\bar{W}^{\prime\prime}}{K^2_{\Phi\bar{\Phi}}}
\ln\Big(\frac{W^{\prime\prime}\bar{W}^{\prime\prime}}{\mu^2K^2_{\Phi\bar{\Phi}}}\Big).
\end{equation}
This form is more convenient for analysis of many-field model than
that one given in \cite{my2,my5}, and ${\rm tr}$ denotes trace of
product of the given matrices.
It is easy to show that the present result reduces to the known
expression for the Wess-Zumino model (\ref{kren}), for the choice $W''=\lambda\Phi$.

Let us consider the chiral (holomorphic) effective potential
$W_{eff}(\Phi)$. The mechanism of its arising is just the same than in
Wess-Zumino model.
We note again that the chiral contributions to effective action can be
generated by supergraphs containing massless propagators only.
To find chiral corrections to effective action we put
$\bar{\Phi}=0$ in eq. (\ref{qua}).
Therefore here and further all derivatives of $K$, $W$ and $\bar{W}$ in (\ref{qua}) will
be taken at $\bar{\Phi}=0$. We  call the theory massless if $W^{''}|_{\Phi=0}=0$.
Further we consider only massless theory and follow \cite{Buch5}.

To construct the supergraphs which yield chiral contributions
one splits the action (\ref{qua}) into sum of free
part and vertices of interaction.
As a free part we take the action $S_0=\int d^8 z K_{\Phi\bar{\Phi}}\phi\bar{\phi}$.
And the term $S[\bar{\phi},\phi,\Phi]-S_0$
will be treated as vertices where $S[\bar{\phi},\phi,\Phi]$
is given by eq. (\ref{qua}). Our purpose is to find the 
first leading contribution to
$W_{eff}(\Phi)$. The straightforward inspection shows that there is no one-loop contributions to the chiral effective potential.
As we will show, chiral loop contributions begin with two
loops. Therefore we keep in eq. (\ref{qua})  only the terms of second, 
third and fourth
orders in quantum fields. Also, the vertex $K_{\Phi\Phi}\phi^2$ evidently yields zero contribution to the chiral effective potential. As for the vertices $K_{\bar{\Phi}\bar{\Phi}}\bar{\phi}^2$, they cannot contribute to the chiral effective potential as it follows from a straightforward calculation of numbers of quantum fields $\phi,\bar{\phi}$ (which should be equal) and D-factors \cite{Buch5}.

As a result we find that the only two-loop supergraph contributing to
chiral effective potential looks like

\vspace*{4mm}

\hspace{4.5cm}
\Lengthunit=1.5cm
\Linewidth{1.2pt}
\GRAPH(hsize=3){\ind(0,-16){Fig.11}%\thicklines
\mov(.5,0){\Circle(2)\mov(-1,0){\lin(2,0)}
\ind(-2,10){|}
\ind(-2,-3){\bar{D}^2}\ind(-2,0){|}\ind(-2,-10){|}\ind(-2,-13){\bar{D}^2}
\ind(-2,7){\bar{D}^2}
\ind(-9,2){D^2} \ind(-10,-2){D^2} \ind(8,2){D^2} \ind(8,-2){D^2}
\ind(-18,2){-} \ind(-18,-2){-} \ind(0,2){-} \ind(0,-2){-}
\Linewidth{0.3pt}
\mov(-1,1){\lin(-.7,.7)}\mov(-1.1,1){\lin(-.7,.7)}
\mov(-1,-1){\lin(.7,-.7)}\mov(-1.1,-1){\lin(.7,-.7)}
\mov(-1,0){\lin(-.7,.7)}\mov(-1.1,0){\lin(-.7,.7)}}
}

\vspace*{3mm}

In the Fig. 11, double external lines are $W''$. Here internal lines are propagators $<\phi\bar{\phi}>$ depending on background
chiral superfields which have the form (\ref{propc}), where, however, the superfield $K_{\Phi\bar{\Phi}}$ is not restricted to be a constant more.

After $D$-algebra transformations and loop integrations completely analogous to those ones of the previous section
we find that two-loop contribution to holomorphic effective
potential in this model looks like
\begin{equation}
\label{l2cg}
W^{(2)}=
\frac{1}{2{(16\pi^2)}^2}\zeta(3) \lambda^2
{\Big\{\frac{W''(z)}{K^2_{\Phi\bar{\Phi}}(z)}\Big\}}^3.
\end{equation}
One reminds that $\lambda=\bar{W}^{'''}(\bar{\Phi})|_{\bar{\Phi}=0}$ and
$K_{\Phi\bar{\Phi}}(z)=\frac{\partial ^2 K(\bar{\Phi},\Phi)}
{\partial\Phi\partial\bar{\Phi}}
|_{\bar{\Phi}=0}$ here.
We see that the correction (\ref{l2cg}) is finite and does not require
renormalization in any case despite the theory is non-renormalizable
in general case. In the case of the Wess-Zumino model it reduces to (\ref{l2c}).

We note that the calculation of two-loop k\"{a}hlerian effective
potential can be carried out with help of matrix superpropagator 
(\ref{grematg}). The results are given in \cite{my2,my5}.

Now let us consider some phenomenological applications of the theory
characterized by the action (\ref{genact}). Let us suppose that the column
vector $\vec{\Phi}$ describes two superfields: the light (massless) one
$\phi$ and the heavy one $\Phi$, $\vec{\Phi}=\left(\begin{array}{c}\phi\\ \Phi
\end{array}\right)$. For this case, we find the one-loop effective
action and eliminate heavy superfields with use of effective equations
of motion. As a result we arrive at the effective action of light
superfields. There is a decoupling theorem \cite{Collins,Syma}
according to which this effective action after redefining of parameters, such as fields, masses, couplings,
can be expressed in the form of a sum of effective action
of the theory obtained from initial one by putting heavy fields to zero
and terms proportional to different powers of $\frac{1}{M}$ where $M$
is mass of heavy superfield (which in the case under consideration is
put, by phenomenological reasons,
to be equal to $M_{String}$ \cite{Cvet,my5}). 

We study such a theory in one-loop approximation.
The low-energy leading one-loop contribution to effective action is
given by (\ref{k1}). In principle, one can proceed with diagonalization of the matrices $K_{\Phi\bar{\Phi}}=
\frac{\pa^2 K}{\pa\Phi^i\pa\bar{\Phi}^j}$ and $W''=
\frac{\pa^2 W}{\pa\Phi^i\pa\Phi^j}$. However, within this review let us restrict ourselves by a qualitative description of the situation only.
 
We consider, as a didactic example, the minimal theory a bit different from that one used \cite{my6} with
\bea
K&=&\phi\bar{\phi}+\Phi\bar{\Phi};\ 
W=\frac{M}{2}\Phi^2+\frac{\lambda}{2}\phi\Phi^2+\frac{g}{3!}\phi^3.
\eea
Let us for simplicity consider the situation when the heavy superfield $\Phi$ is a purely quantum one, and the light superfield $\phi$ is a purely background one. In this case the quadratic action of $\Phi$ looks like
\bea
S_2=\int d^8z \Phi\bar{\Phi}+(\frac{1}{2}\int d^6z(M+\lambda\phi)\Phi^2+h.c.)
\eea
It is evident that this action identically replays the quadratic action of the quantum fields in an usual Wess-Zumino model after the background-quantum splitting   (besides of (\ref{qua}), see also the previous section and \cite{Buch1}).
So, the renormalized one-loop contribution to the low-energy effective action is given by the result (\ref{k1}) at $K(\Phi,\Phi)=\Phi\bar{\Phi}$, which takes the form
\bea
K^{(1)}_{\Phi}=-\frac{1}{32\pi^2}(M+\lambda\phi)(M+\lambda\bar{\phi})\ln\frac{(M+\lambda\phi)(M+\lambda\bar{\phi})}{\mu^2}.
\eea
We expand it in power series in $\frac{1}{M}$, with $\mu'=\alpha\mu$, and $\alpha$ is a some number:
\bea
\label{hea}
K^{(1)}_{\Phi}=-\frac{1}{32\pi^2}\lambda^2\phi\bar{\phi}\ln\frac{M^2}{{\mu'}^2}+O(\frac{1}{M}).
\eea
We see that this expression involves the term $\ln\frac{M^2}{\mu^2}$. 
It is clear that this term is not suppressed at $M\to\infty$ but instead of this, increases as $M$ grows. Moreover, if one suggests that the light field $\phi$ is not purely external but possesses a nontrivial quantum dynamics, its one-loop K\"{a}hlerian effective potential is
\bea
\label{lig}
K^{(1)}_{\phi}=-\frac{1}{32\pi^2}g^2\phi\bar{\phi}\ln\frac{g^2\phi\bar{\phi}}{\mu^2},
\eea
and if we fix the normalization parameter as $\mu=\alpha M$, we see that in this case, instead of the contribution (\ref{hea}), the (\ref{lig}) will grow as $M\to\infty$. The same situation can be verified as well for the models involving light and heavy fields in different situations (presence of nonminimal couplings, nontrivial quantum and background parts both for light and heavy fields etc.) \cite{my6}.

To argue the existence of these contributions increasing with growth of $M$ in a general case, one can notice as well the following facts.
First, if the theory involve a cubic self-coupling of the light (massless) chiral superfield $\phi$ with a coupling $g$, the one-loop effective action will involve the term $c\int d^8z g^2\phi\bar{\phi}\ln\frac{g^2\phi\bar{\phi}}{\mu^2}$, where $c$ is a some finite number, and $\mu$ is a renormalzation scale (if this self-coupling is not cubic but a generic function $V(\phi)$, the theory turns out to be non-renormalizable, but the corresponding contribution to the one-loop effective action can be obtained through a straightforward replacement $\phi\to V^{\prime\prime}(\phi)$). At the same time, the sector in which the light $\phi$ is coupled to the heavy $\Phi$, in the case of the theory with divergences, certainly will yield, among other terms, the contribution $b\int d^8z \phi\bar{\phi}\ln\frac{M^2}{\mu^2}$, where $b$ is also finite (this contribution is always present if the corresponding supergraphs involve massive propagators), as we already showed. Therefore, either this contribution increases with growth of $M$, or, if one fixes the parameter $\mu$ imposing the condition $M=\mu$ (or $M=b\mu$, with  constant $b$), the one-loop contribution of the light fields only takes the form $c\int d^8z g^2\phi\bar{\phi}\ln\frac{g^2\phi\bar{\phi}}{M^2}$ which increases with growth of $M$. Actually we showed that, such increasing contributions emerge everywhere we have propagators of the heavy fields, if one has a dependence on $\mu$ (i.e. if the theory involves divergences), which is common if the theory does ot involve higher derivatives or extended supersymmetry.. 

\section{The higher-derivative chiral superfield models}

The next generalization for the models of the chiral superfield involves including the higher derivatives. 
Originally, the higher derivatives extension has been introduced for the purposes of the regularization in
\cite{Ili}.  Further, the higher-derivative extension was carried out in gravity, first in
\cite{stelle}  where this extension was shown to improve the
renormalization properties of field theories.  Moreover, it turns out that the
higher-derivative additive modifications of the gravity action 
arise due to the presence of the conformal anomaly of matter fields in
curved space \cite{AM}. In \cite{BK}, a supersymmetric analogue of this anomaly and the corresponding additive modification of the supergravity action were obtained, and  in  \cite{my1} the quantum dynamics of the conformal sector (dilaton)
was studied. 

Actually, the higher-derivative field theories are studied in
different contexts, including different gravity modifications which
are intensively applied to obtain the cosmic acceleration \cite{Dol},
and the Horava model of  gravity \cite{Horava}. In the context of
supersymmetry, the interest in the higher-derivative superfield
theories was recently recovered due to the paper \cite{Ant}. Treating the well known problem of arising of the ghosts whose presence is typical for the higher-derivative field theories, we note that an attempt to solve this problem was carried out in \cite{Smilga} where the hypothesis that the contribution of the ghosts can be decoupled in certain cases (so-called ``benign ghosts'') was argued.  At the same time, one can treat the higher-derivative theories as effective theories for studying of the low-energy scale effects.

We start with the simplest example of a higher-derivative superfield theory \cite{ourhigh}:
\bea
\label{firstex}
S[\Phi,\bar{\Phi}]=\int d^8z \Phi\Box\bar{\Phi}+(\int d^6z W(\Phi)+h.c.).
\eea
Here $\Phi$ is a chiral superfield, and $W(\Phi)$ is an  arbitrary
function. A particular form of this action was studied in \cite{my6}, for the case $W(\Phi)=\Lambda e^{3\Phi}$, it corresponds to the four-dimensional anomaly-modified dilaton supergravity in the infrared limit (where, in particular, all derivative-dependent interaction terms of the classical action simply vanish).
This action, being reduced to the components, contains fourth order in space-time derivatives.
We will refer to this theory as to the minimal higher-derivative theory.

The effective action $\Gamma[\Phi,\bar{\Phi}]$, as usual, can be represented as a generating functional of the one-particle-irreducible vertex Green functions:
\bea
\label{def0hd}
e^{i\Gamma[\Phi,\bar{\Phi}]}=\int D\phi D\bar{\phi} \exp(iS[\Phi+\phi,\bar{\Phi}+\bar{\phi}])|_{1PI}.
\eea
Here the $\Phi,\bar{\Phi}$ are background (classical) fields and $\phi,\bar{\phi}$ are  quantum fields.
As usual, we can represent the structure of the effective action in this theory in the form given by (\ref{ep1}--\ref{ep5}).

To obtain the one-loop effective action, one should expand the right-hand side of the equation (\ref{def0hd}) up to the second order in the quantum superfields $\phi$, $\bar{\phi}$ (cf. \cite{BO}). As a result, the one-loop effective action is defined from the expression:
\bea
\label{def1}
e^{i\Gamma^{(1)}[\Phi,\bar{\Phi}]}=\int D\phi D\bar{\phi} \exp(i[\int d^8z \phi\Box\bar{\phi}+(\frac{1}{2}\int d^6z W^{\prime\prime}(\Phi)\phi^2+h.c.)]),
\eea
and this effective action can be represented in the form of the functional supertrace:
\bea
\label{st}
\Gamma^{(1)}[\Phi,\bar{\Phi}]=\frac{i}{2}{\rm Tr}\ln\left(
\begin{array}{cc}
W^{\prime\prime}& -\Box\frac{\bar{D}^2}{4}\\
-\Box\frac{D^2}{4} &\bar{W}^{\prime\prime}
\end{array}
\right).
\eea
Again, the elements of this matrix are defined in different subspaces of the
superspace and mix the chiralities. To find a more simple equivalent form, we again, as in the previous sections, use the trick based on the Faddeev-Popov methodology.

Let us consider the free higher-derivative theory of the real scalar superfield whose action is
\bea
S_v=-\frac{1}{16}\int d^8z vD^{\alpha}\bar{D}^2D_{\alpha}\Box v.
\eea
This action is evidently invariant under the usual gauge transformations $\delta v=\Lambda+\bar{\Lambda}$, where $\Lambda$ is a chiral superfield, and $\bar{\Lambda}$ is an antichiral one. Following general prescriptions of the Faddeev-Popov method, one can define the effective action $W_v$ of this theory  as
\bea
\label{efv}
e^{iW_v}=\int Dv \exp(-\frac{i}{16}\int d^8z vD^{\alpha}\bar{D}^2D_{\alpha}\Box v)\delta(\frac{1}{4}D^2v-\bar{\phi})\delta(\frac{1}{4}\bar{D}^2v-\phi)\Delta_{FP},
\eea
where the $\frac{1}{4}D^2v-\bar{\phi}$, $\frac{1}{4}\bar{D}^2v-\phi$
play the role of the gauge fixing functions and $\phi,\bar{\phi}$ are the same as in (\ref{def1}), and $\Delta_{FP}$ is a Faddeev-Popov determinant. One should notice that
the $W_v$ is a constant.

Then, let us multiply the expressions (\ref{def1}) and (\ref{efv}). The functional integration over $\phi,\bar{\phi}$ is straightforward, and after omitting irrelevant constants, the one-loop effective action takes the form
\bea
\label{tracelog}
\Gamma^{(1)}=\frac{i}{2}{\rm Tr}\ln (\Box^2-\frac{1}{4}W^{\prime\prime}(\Phi)\bar{D}^2-\frac{1}{4}\bar{W}^{\prime\prime}(\bar{\Phi})D^2).
\eea
So, the expression for the one-loop effective action is radically simplified.

Therefore we face the problem of calculating of trace of the logarithm
of the higher-derivative operator. The most convenient way to do it is based on the use of the proper-time representation (see Sec. 4.6, 4.8):
\bea
\Gamma^{(1)}&=&\frac{i}{2}{\rm Tr}\int\frac{ds}{s}\exp[is (\Box^2+\frac{1}{4}\Psi\bar{D}^2+\frac{1}{4}\bar{\Psi}D^2)].
\eea
Here we denoted $W^{\prime\prime}(\Phi)=-\Psi$, $\bar{W}^{\prime\prime}(\bar{\Phi})=-\bar{\Psi}$ for the convenience. One should remind that $\Psi$ is a chiral superfield, and $\bar{\Psi}$ is an antichiral one.

Disregarding the terms involving the space-time derivatives of $\Phi,\bar{\Phi}$, which will not contribute to lower orders of the derivative expansion of the effective action, corresponding to fourth and higher orders in space-time derivatives of the scalar components of these superfields, we can rewrite this expression as 
\bea
\Gamma^{(1)}&=&\frac{i}{2}\int d^8z_1\int\frac{ds}{s}\exp[is (\frac{1}{4}\Psi\bar{D}^2+\frac{1}{4}\bar{\Psi}D^2)]e^{is\Box^2}\delta^8(z_1-z_2)|_{z_1=z_2}.
\eea
Now, let us proceed in a way similar to that one used in Sec. 4.8. As a first step, we introduce operators
\bea
\Delta=\frac{1}{4}\Psi\bar{D}^2+\frac{1}{4}\bar{\Psi}D^2;\quad\, \Omega(\Psi,\bar{\Psi},s)=e^{is\Delta}, 
\eea 
where $\Omega$ can be again expanded in the form (\ref{tu}) and satisfies the superfield heat conductivity equation
\bea
\frac{1}{i}\frac{d\Omega}{ds}=\Omega\Delta.
\eea
The initial condition is evidently $\Omega|_{s=0}=1$, hence $A(s=0)=\tilde{A}(s=0)=B_{\alpha}(s=0)=\tilde{B}_{\dot{\alpha}}(s=0)=C(s=0)=\tilde{C}(s=0)=0$.
The system involving these coefficients turns out to be exactly the same as in the Wess-Zumino case (see Sec. 4.8), hence the coefficients $A$ and $\tilde{A}$ (they are again the only ones contributing to the one-loop effective potential) reproduce the results obtained in the Wess-Zumino model \cite{Buch1}. In our case, unlike Sec. 4.8, we give these coefficients up to the fourth order in the spinor supercovariant derivatives of superfields:
\bea
\label{aplus}
A(s)+\tilde{A}(s)&=&\frac{2}{\Box}[\cosh(\tilde{s}U)-1]+\nonumber\\&+&
\tilde{s}\frac{D^2\Psi\bar{D}^2\bar{\Psi}}{64\Box}(\tilde{s}\cosh(\tilde{s}U) -\frac{1}{U}\sinh(\tilde{s}U))+\nonumber\\&+&
\frac{\tilde{s}}{64U^2}[\bar{\Psi}\bar{D}^2\bar{\Psi}(D^{\alpha}\Psi)(D_{\alpha}\Psi)+\Psi D^2\Psi(\bar{D}_{\dot{\alpha}}\bar{\Psi})(\bar{D}^{\dot{\alpha}}\bar{\Psi})]\times\nonumber\\&\times&
(\frac{1}{3}\tilde{s}^2U\sinh(\tilde{s}U)-\tilde{s}\cosh(\tilde{s}U)+\frac{1}{U}\sinh(\tilde{s}U))+\nonumber\\&+&
\frac{\tilde{s}}{256}(D^{\alpha}\Psi)(D_{\alpha}\Psi)(\bar{D}_{\dot{\alpha}}\bar{\Psi})(\bar{D}^{\dot{\alpha}}\bar{\Psi})[\frac{1}{2}\tilde{s}^3\cosh(\tilde{s}U)-\frac{5}{3}\frac{\tilde{s}^2}{U}\sinh(\tilde{s}U)+\nonumber\\&+&
\frac{7}{2U^2}(\tilde{s}\cosh(\tilde{s}U)-\frac{1}{U}\sinh(\tilde{s}U))].
\eea 
Here $\tilde{s}=is$, $U=\sqrt{\Psi\bar{\Psi}\Box}$. The higher orders in supercovariant derivatives of $\Psi$, $\bar{\Psi}$ in principle also can be found, however, the complete result would be extremely cumbersome. 

The one-loop effective action can be expressed as 
\bea
\label{gammaa}
\Gamma^{(1)}&=&-\frac{i}{2}\int d^4\theta d^4x_1\int\frac{ds}{s}[A(s)+\tilde{A}(s)]e^{is\Box^2}\delta^4(x_1-x_2)|_{x_1=x_2}.
\eea
The differences with the Wess-Zumino model will arise since the d'Alembertian operators from the expansion of $A(s)+\tilde{A}(s)$, will act not on the usual function $e^{is\Box}\delta^8(z_1-z_2)$, as it occurs in that case, but on the function $e^{is\Box^2}\delta^8(z_1-z_2)$. 

It remains to substitute (\ref{aplus}) into (\ref{gammaa}) and to expand  it in the power series in $\Box$.
The contribution to the one-loop k\"{a}hlerian effective action is given by the first line of (\ref{aplus}), i.e.
\bea
\Gamma^{(1)}_K=-i\int d^4\theta d^4x_1\int\frac{ds}{s}\frac{1}{\Box}[\cosh(\tilde{s}U)-1]e^{is\Box^2}\delta^4(x_1-x_2)|_{x_1=x_2},
\eea
which, after expanding in series in $\Box$ yields
\bea
\Gamma^{(1)}_K=\int d^4\theta d^4x_1\int\frac{dt}{t}\sum\limits_{n=0}^{\infty}\frac{1}{(2n+2)!}(t^2\Psi\bar{\Psi})^{n+1}\Box^n e^{-t\Box^2}\delta^4(x_1-x_2)|_{x_1=x_2},
\eea
where we  carried out the Wick rotation $s=it$ (with $t=-\tilde{s}$) and $x_0=ix_{0E}$ for convenience. 

It is convenient to split the indices $n$ into odd, $n=2l+1$ and even, $n=2l$, ones. As a result, we can write $K^{(1)}$ as
\bea
\label{dsum}
\Gamma^{(1)}_K&=&\int d^4\theta d^4x_1\int\frac{dt}{t}\sum\limits_{l=0}^{\infty}\left[\frac{1}{(4l+2)!}(t^2\Psi\bar{\Psi})^{2l+1}\Box^{2l}+
\frac{1}{(4l+4)!}(t^2\Psi\bar{\Psi})^{2l+2}\Box^{2l+1}\right]\times\nonumber\\&\times& e^{-t\Box^2}\delta^4(x_1-x_2)|_{x_1=x_2}.
\eea
Now, let us consider the structure $\Box^n e^{-t\Box^2}\delta^4(x_1-x_2)|_{x_1=x_2}$. It is clear that the function $V(t;x_1,x_2)=e^{-t\Box^2}\delta^4(x_1-x_2)$ which we will call the free heat kernel satisfies the equation
\bea
\Box^2V(t;x_1,x_2)=-\frac{d}{dt}V(t;x_1,x_2),
\eea
hence
\bea
\Box^{2l}V(t;x_1,x_2)=(-\frac{d}{dt})^lV(t;x_1,x_2); \quad\,\Box^{2l+1}V(t;x_1,x_2)=(-\frac{d}{dt})^l\Box V(t;x_1,x_2).
\eea
In this subsection, the above  expressions will be considered only in the limit $x_1=x_2$. One can find that
\bea
&&V(t;x_1,x_2)|_{x_1=x_2}=\int\frac{d^4k}{(2\pi)^4}e^{-tk^4}=\frac{1}{32\pi^2t};\nonumber\\
&&\Box V(t;x_1,x_2)|_{x_1=x_2}=\int\frac{d^4k}{(2\pi)^4}(-k^2)e^{-tk^4}=-\frac{1}{32\pi^{3/2}t^{3/2}},
\eea
therefore
\bea
\label{boxl}
&&\Box^{2l}V(t;x_1,x_2)|_{x_1=x_2}=(-\frac{d}{dt})^l\frac{1}{32\pi^2t}=\frac{(-1)^ll!}{32\pi^2t^{l+1}};\nonumber\\ &&\Box^{2l+1}V(t;x_1,x_2)|_{x_1=x_2}=(-\frac{d}{dt})^l(-\frac{1}{32\pi^{3/2}t^{3/2}})=-\frac{(-1)^{l+1}(2l+1)!!}{32\pi^{3/2}2^lt^{l+3/2}}.
\eea
Replacing all this into (\ref{dsum}), we arrive at
\bea
\label{dsum2}
\Gamma^{(1)}_K&=&\frac{1}{32\pi^2}\int d^8z\int dt\sum\limits_{l=0}^{\infty}(-1)^l\Big[t^{3l}\frac{l!(\Psi\bar{\Psi})^{2l+1}}{(4l+2)!}- \nonumber\\&-&
t^{3l+3/2}\frac{(\Psi\bar{\Psi})^{2l+2}}{(4l+4)!}\frac{\sqrt{\pi}(2l+1)!!}{2^l}\Big].
\eea
To simplify this expression, let us make the change $t(\Psi\bar{\Psi})^{2/3}=u$ (note that $u$ is dimensionless). We find
\bea
\label{dsum3}
\Gamma^{(1)}_K&=&\frac{1}{32\pi^2}\int d^8z(\Psi\bar{\Psi})^{1/3}\int du\sum\limits_{l=0}^{\infty}\Big[\frac{(-1)^lu^{3l}l!}{(4l+2)!}- \nonumber\\&-&
\frac{(-1)^lu^{3l+3/2}}{(4l+4)!}\frac{\sqrt{\pi}(2l+1)!!}{2^l}\Big],
\eea
so the one-loop K\"{a}hlerian effective potential is
\bea
K^{(1)}&=&\frac{c_0}{32\pi^2}\int d^8z(\Psi\bar{\Psi})^{1/3},
\eea
where
\bea
c_0=\int du\sum\limits_{l=0}^{\infty}\left[\frac{(-1)^lu^{3l}l!}{(4l+2)!}-
\frac{(-1)^lu^{3l+3/2}}{(4l+4)!}\frac{\sqrt{\pi}(2l+1)!!}{2^l}\right]
\eea
is a finite constant. It is easy to see that the result for dilaton supergravity \cite{my1}, being a particular case of this result, is explicitly reproduced.

Now, let us calculate the one-loop auxiliary fields' effective action. To do it, let us consider all derivative dependent terms in (\ref{aplus}). After their expansion in power series in $\Box$, we find
\bea
\label{f1}
\Gamma^{(1)}_F&=&-i\int d^4\theta d^4x_1\int\frac{dt}{t}\sum\limits_{n=0}^{\infty}\Big[\frac{D^2\Psi\bar{D}^2\bar{\Psi}}{64}t^{2n+4}(\Psi\bar{\Psi})^{n+1}[\frac{1}{(2n+2)!}-\frac{1}{(2n+3)!}]+\nonumber\\&+&\frac{1}{64}[\bar{\Psi}\bar{D}^2\bar{\Psi}D^{\alpha}\Psi D_{\alpha}\Psi+h.c.]
t^{2n+4}(\Psi\bar{\Psi})^n[\frac{1}{3(2n+1)!}-\frac{1}{(2n+2)!}+\frac{1}{(2n+3)!}]+\nonumber\\&+&
\frac{1}{256}D^{\alpha}\Psi D_{\alpha}\Psi\bar{D}_{\dot{\alpha}}\bar{\Psi}\bar{D}^{\dot{\alpha}}\bar{\Psi}t^{2n+6}(\Psi\bar{\Psi})^n
\times\nonumber\\&\times&[\frac{1}{2(2n)!}-\frac{5}{3(2n+1)!}+\frac{7}{2(2n+2)!}-\frac{7}{2(2n+3)!}]
\Big]\times\nonumber\\&\times&\Box^n
e^{-t\Box^2}\delta^4(x_1-x_2)|_{x_1=x_2}.
\eea
Then, we apply the same scheme as above. By its essence, this expression looks like
\bea
\Gamma^{(1)}_F=i\int d^4\theta d^4x_1\int\frac{dt}{t}\sum\limits_{n=0}^{\infty} A_n(\Psi,\bar{\Psi},t)\Box^ne^{-t\Box^2}\delta^4(x_1-x_2)|_{x_1=x_2}.
\eea
Here $A_n$ are some functions of fields whose explicit form can be
read off from (\ref{f1}). After carrying out the transformations we used above, we find the auxiliary fields' effective potential to be
\bea
F^{(1)}&=&C_1\frac{\bar{D}^2\bar{\Psi}D^2\Psi}{\Psi\bar{\Psi}}+C_2[\bar{\Psi}\bar{D}^2\bar{\Psi}D^{\alpha}\Psi D_{\alpha}\Psi+h.c.]\frac{1}{(\Psi\bar{\Psi})^2}+\nonumber\\&+&
C_3D^{\alpha}\Psi D_{\alpha}\Psi\bar{D}_{\dot{\alpha}}\bar{\Psi}\bar{D}^{\dot{\alpha}}\bar{\Psi}\frac{1}{(\Psi\bar{\Psi})^2},
\eea
where $C_1,C_2,C_3$ are some numbers.
We note that, in principle, this form can be predicted without explicit calculations. Indeed, this form should involve exactly two $D_{\alpha}$ derivatives and two $\bar{D}_{\dot{\alpha}}$ derivatives. Also, by dimensional reasons, the numbers of fields $\Psi$ (and similarly $\bar{\Psi}$ should be equal in a numerator and a denominator of any contribution to this expression, which also must be symmetric with respect to change $\Psi\to\bar{\Psi}$. Hence we can have only the terms listed in the expression above.

To close the consideration of the one-loop effective action for this
model, let us discuss the one-loop chiral contributions to the effective
action. It is clear that they differ from zero only if  
$\bar{W}^{\prime\prime}(\bar{\Phi})|_{\bar{\Phi}=0}=\Lambda\neq 0$
(essentially it means that $\Lambda$ is related with the mass of the theory; one should notice that the case $\Lambda=0$ gives zero one-loop chiral corrections). One should note that since the theory is therefore massive, there is no contributions to the chiral effective potential. It was argued in \cite{my7} that there are two possible types of chiral contributions in this theory, that is, $f_1(\Psi)\Box\Psi$ and $f_2(\Psi)\pa^m\Psi\pa_m\Psi$, however, the first of them can be reduced to the second one through integration by parts. Hence, we have the only possible form for the chiral quantum correction, that is,
\bea
\label{chir0}
{\cal L}^{(1)}_c=f(\Psi)\pa^m\Psi\pa_m\Psi.
\eea 
By dimensional reasons, one can write this expression as
\bea
\label{chir0a}
{\cal L}^{(1)}_c=af(\Psi/\Lambda)\Psi^{-5/3}\pa^m\Psi\pa_m\Psi,
\eea 
where $a$ is a some number. Actually, it is convenient to rewrite this expression in terms of $\Phi$ and suggest that $W^{\prime\prime}(\Phi)|_{\Phi=0}=\Lambda\neq 0$ as well, in this case we can suggest f.e. 
$W^{\prime\prime}(\Phi)=\Lambda e^{3\Phi}$, as it occurs in the dilaton supergravity, we have
\bea
\label{chir0aa}
{\cal L}^{(1)}_c=af(e^{3\Phi})e^{\Phi}\pa^m\Phi\pa_m\Phi.
\eea 
Namely this result is found in a dilaton supergravity \cite{my7}.

We note however that these expressions, first, do not contribute to the chiral effective potential, second, can be treated (and perhaps it is more natural) as contributions to the auxiliary fields' effective potential although they are chiral, since, being expressed in the form of the integral over whole superspace $d^8z$, they contain no nonlocalities being of the form $\int d^8 z (f_1(\Psi)D^{\alpha}\Psi D_{\alpha}\Psi+f_2(\Psi)D^2\Psi)$, with $f_1(\Psi)$ and $f_2(\Psi)$ are some functions of $\Psi$ only, with no derivatives or nonlocal factors.

Now, after we have calculated these contributions, let us study a bit
different form of the higher-derivative theory whose kinetic term is
\bea
S_K=\int d^8z \Phi(\Box-M^2)\bar{\Phi}.
\eea
This kinetic term is equivalent to the one of the Wess-Zumino model with a higher-derivative regulator \cite{Ili}.
The importance of theories with such a kinetic term follows from the observation made in \cite{Ant} where the higher-derivative superfield theory namely with this kinetic term has been shown to be classically equivalent to the chiral superfield theory which does not involve higher derivatives, but, instead of this, describes dynamics of a set of chiral superfields. 
The key idea of \cite{Ant} consists in introducing  new chiral fields $\chi$ and $\tilde{\chi}$, which are subsequently expressed  in terms of $\Phi$ and $\bar{D}^2\bar{\Phi}$, with $\Phi$ being the  basic chiral superfield described by the initial higher-derivative theory. This  is  done by the  use of a linear transformation, which however, becomes singular in the case $M=0$. So, the studies carried out in the paper \cite{Ant} are applicable only for the theory with $M\neq 0$.

However, the analysis of the effective action for such  theory within the proper-time method is more complicated than for the theory studied above. The calculation of the Schwinger coefficients $A(s)$ and $\tilde{A}(s)$ is the same as above. The analogue of the free heat kernel function $V(t;x_1,x_2)$, after introducing of the same trick as above, can be shown to be equal to  
\bea
\label{kern}
V(s;x_1,x_2)=e^{-s(\Box^2-M^2\Box)}\delta^4(x_1-x_2).
\eea
However, even the evaluation of the  case $x_1=x_2$, which is only interesting for us in the one-loop approximation, is a nontrivial problem which can be reasonably solved only for very large mass $M$. Let us proceed with this calculation.

After Fourier transform and Wick rotation, the function $V(t;x_1,x_2)|_{x_1=x_2}$ looks like
\bea
I(s)\equiv V(s;x_1,x_2)|_{x_1=x_2}=\int\frac{d^4k}{(2\pi)^4}e^{-s(k^4+k^2M^2)}.
\eea
Changing variables,  $k^2=u$, we find
\bea
I(s)=\frac{1}{16\pi^2}e^{\frac{tM^4}{4}}\int_0^{\infty} duue^{-s(u+\frac{M^2}{2})^2}.
\eea
Replacing  then $u+\frac{M^2}{2}=u'$ and integrating over $u$ where it is possible, we find
\bea
I(s)=\frac{1}{32\pi^2s}-\frac{M^2}{32\pi^2}e^{\frac{sM^4}{4}}\int_{M^2/2}^{\infty}due^{-su^2}.
\eea
We find that this expression for the heat kernel function can be
expressed through the probability integral function
\bea
\Phi(x)=\frac{2}{\sqrt{\pi}}\int_0^x dt e^{-t^2}.
\eea
The presence of such a function seems to make impossible finding  the explicit one-loop k\"{a}hlerian potential in the general case.
It is clear that $\Phi(x\to\infty)\to 1$.
Indeed,
\bea
I(s)=\frac{1}{16\pi^2}(\frac{1}{2s}-\frac{M^2}{2}e^{sM^2/4}(\int_0^{\infty}due^{-su^2}-\int_0^{M^2/2}due^{-su^2}))
\eea
Substituting $su^2=w^2$, we get
\bea
I(s)&=&\frac{1}{16\pi^2}(\frac{1}{2s}-\frac{M^2}{2}e^{sM^2/4}(\frac{1}{2}\sqrt{\frac{\pi}{s}}-\frac{1}{\sqrt{s}}\int_0^{M^2\sqrt{s}/2}dwe^{-w^2}))=
\nonumber\\&=&
\frac{1}{16\pi^2}[\frac{1}{2s}-\frac{M^2}{2}e^{sM^2/4}\frac{1}{2}\sqrt{\frac{\pi}{s}}(1-\Phi(M^2\sqrt{s}/2)].
\eea
To evaluate this expression, we employ the asymptotics of the probability integral $\Phi(y)$ at large arguments \cite{Abr}:
\bea
\Phi(y)|_{y\to\infty}=1-\frac{1}{\pi}e^{-y^2}\sum\limits_{k=0}^{\infty}\frac{(-1)^k\Gamma(k+\frac{1}{2})}{y^{2k+1}}.
\eea
We find that the term with $k=0$ identically cancels the ``usual'' term $\frac{1}{2s}$. Taking into account only the $M\to\infty$ dominant term (remind that the limit of very high masses was studied earlier in \cite{my6}), one finds
\bea
I(s)=\frac{1}{16\pi^2s^2M^4},
\eea 
which  differs from the case $M=0$ considered earlier where the analogue of this function was proportional to
$\frac{1}{s}$. One could note that such behaviour of the heat kernel seems to be similar to that one occurring in the Wess-Zumino model \cite{Buch1,my3}. Nevertheless, the presence of a large mass in the denominator gives a hope that the corrections to the effective action will be suppressed in a $M\to\infty$ limit.

To give a description of the principal difference of new theory, we restrict ourselves only to calculation of the k\"{a}hlerian effective potential in a new theory. Let it action be
\bea
\label{twoex}
S[\Phi,\bar{\Phi}]=\int d^8z \Phi(\Box-M^2)\bar{\Phi}+(\int d^6z W(\Phi)+h.c.).
\eea
Here $M$ is a large parameter related to the physical mass.
Using the insertion of the effective action of the free real scalar superfield whose classical action looks like
\bea
S_v=-\frac{1}{16}\int d^8z vD^{\alpha}\bar{D}^2D_{\alpha}(\Box-M^2) v,
\eea
one can show that the one-loop effective action corresponding to the theory (\ref{twoex}) can be expressed through the following Schwinger representation
\bea
\Gamma^{(1)}&=&\frac{i}{2}{\rm Tr}\int\frac{ds}{s}\exp[is (\Box(\Box-M^2)+\frac{1}{4}\Psi\bar{D}^2+\frac{1}{4}\bar{\Psi}D^2)].
\eea
Since we restrict ourselves here to the k\"{a}hlerian part of the effective potential, we can express the one-loop effective action as
\bea
\Gamma^{(1)}&=&\frac{i}{2}{\rm Tr}\int d^8z\int\frac{ds}{s}
\exp[is (\frac{1}{4}\Psi\bar{D}^2+\frac{1}{4}\bar{\Psi}D^2)]e^{is\Box(\Box-M^2)}\delta^8(z-z')|_{z=z'}.
\eea
The relevant terms from the operator $\exp(is (\frac{1}{4}\Psi\bar{D}^2+\frac{1}{4}\bar{\Psi}D^2))$ again have the form (\ref{aplus}),
and, after Wick rotation $s=it$, the one-loop k\"{a}hlerian effective action looks like
\bea
\Gamma^{(1)}_K=-i\int d^4\theta d^4x_1\int\frac{dt}{t}\frac{1}{\Box}[\cosh(t\sqrt{\Psi\bar{\Psi}\Box})-1]e^{-t\Box(\Box-M^2)}\delta^4(x_1-x_2)|_{x_1=x_2}.
\eea
Expanding this in series in $\Box$, after Wick rotation we find
\bea
\Gamma^{(1)}_K=\int d^4\theta d^4x_1\int\frac{dt}{t}\sum\limits_{n=0}^{\infty}\frac{1}{(2n+2)!}(t^2\Psi\bar{\Psi})^{n+1}\Box^n V(t;x_1,x_2)|_{x_1=x_2}.
\eea
Here the function $V(t;x_1,x_2)$ can be read off from the (\ref{kern}).
As we already noted, this expression can be found in a closed form
only 
in the limit $M\to\infty$.
It follows from (\ref{kern}) that
\bea
\Box^n V(t;x_1,x_2)=\frac{1}{t^n}(\frac{d}{d(M^2)})^nV(t;x_1,x_2),
\eea
so that, after taking $x_1=x_2$ , 
\bea
\Box^n V(s;x_1,x_2)|_{x_1=x_2}=\frac{(-1)^n(n+1)!}{16\pi^2(M^2t)^{n+2}}.
\eea
Putting all together, we find
\bea
\Gamma^{(1)}_K=\frac{1}{32\pi^2}\int d^8z\int\frac{dt}{M^2t^2}\sum\limits_{n=0}^{\infty}\frac{(-1)^n(n+1)!}{(2n+2)!}\left(\frac{t\Psi\bar{\Psi}}{M^2}\right)^{n+1}.
\eea
This expression is similar to Eq. (\ref{kahlpt}) obtained for the Wess-Zumino model. As a result, we have
\bea
K^{(1)}=\frac{1}{32\pi^2} \frac{\Psi\bar{\Psi}}{M^4}\sum\limits_{n=0}^{\infty}\int_{\frac{\Psi\bar{\Psi}L^2}{M^4}}^{\infty}\frac{du}{u}\frac{(-1)^nu^n(n+1)!}{(2n+2)!}.
\eea
To avoid divergence of the integral, we introduced the cutoff $L^2$ at the lower limit. As $L^2\to 0$, one obtains
\bea
\label{k1kp}
K^{(1)}&=&-\frac{1}{32\pi^2}\frac{\Psi\bar{\Psi}}{M^4}\ln(\mu^2L^2)-\frac{1}{32\pi^2}\frac{\Psi\bar{\Psi}}{M^4}(\ln\frac{\Psi\bar{\Psi}}{M^4\mu^2}-\xi).
\eea
Here $ \xi $ is some finite constant which can be absorbed into a redefinition of $ \mu^2 $.
This contribution is divergent but turns out to be suppressed in the large $M$ limit.
This divergence can be eliminated by adding a counterterm
\bea
K^{(1)}_{countr}=\frac{1}{32\pi^2}\frac{\Psi\bar{\Psi}}{M^4}\ln(\mu^2L^2).
\eea
Thus, the renormalized k\"{a}hlerian effective potential is
\bea
K^{(1)}&=&-\frac{1}{32\pi^2}\frac{\Psi\bar{\Psi}}{M^4}(\ln\frac{\Psi\bar{\Psi}}{M^4\mu^2}-\xi).
\eea

It is interesting also to proceed with the calculations in terms of the Feynman supergraphs, using the method similar to \cite{PW}. The one-loop effective action is described by the same supergraphs as at Fig. 10, with the external legs correspond to alternative $\Psi$ and $\bar{\Psi}$. Their sum, calculated along the same lines as in the Section 4.8, after the Wick rotation is given by
\bea
\Gamma^{(1)}&=&-\frac{1}{2}\sum\limits_{n=1}^{\infty}\frac{1}{n}\int d^4\theta\int\frac{d^4k_E}{(2\pi)^4}(\Psi\bar{\Psi}\frac{D^2\bar{D}^2}{16k^4_E(k^2_E+M^2)^2})^n
\delta_{12}|_{\theta_1=\theta_2}.
\eea
The D-algebra transformations are simple, so we can easily sum the series and obtain
\bea
\label{ghd}
\Gamma^{(1)}&=&-\frac{1}{2}\int d^4\theta\int\frac{d^4k_E}{(2\pi)^4}\frac{1}{k^2_E}\ln\Big(1+\frac{\Psi\bar{\Psi}}{k^2_E(k^2_E+M^2)^2}\Big).
\eea
This integral can be exactly calculated only at $M=0$, giving
\bea
\Gamma^{(1)}=c\int d^4\theta (\Psi\bar{\Psi})^{1/3},
\eea
with $c$ is a finite number.
We note that this result can be obtained even without calculations. Indeed, by the symmetry reasons, the one-loop effective action can be only a function of $\Psi\bar{\Psi}$, and by the dimensional restrictions, the only answer for it is just the expression above. However, if $M\neq 0$, the $\Gamma^{(1)}$ (\ref{ghd}) cannot be expressed in terms of the elementary functions and can be evaluated only in the limits $M\to 0$ or $M\to\infty$ as it has been done above. This methodology can be applied as well to the case of the presence of the higher-derivative gauge fields \cite{highed}.

We considered the one-loop effective potential for two different versions of
the higher-derivative chiral superfield models. 
It turns out  that, in the case when the mass term is purely
chiral (a similar situation with the mass term takes place in the
Wess-Zumino model), the theory is finite. At the same time, if the
mass term arises in the general Lagrangian (that is the situation
considered in \cite{Ant}), the theory displays divergences though
being super-renormalizable. We note, however, that the equivalence of
the higher-derivative theory of the chiral superfield and the theory
without higher derivatives but with an extended number of chiral
superfields described in \cite{Ant} occurs only in the case when the
mass term belongs to the general Lagrangian (that is, the second case
considered in the section). Therefore, the presence of these divergences can
be considered as a sign in favour of the equivalence established in \cite{Ant}. Indeed, the expression (\ref{k1kp}), after relabelling $\frac{\Psi}{M^2}\to\psi$, identically reproduces the one-loop k\"{a}hlerian effective potential in the Wess-Zumino model, with $\psi=m+\lambda\Phi$. We can  therefore  conclude that the higher-derivative theory (\ref{twoex}), in the limit $M\to\infty$, yields the same quantum contribution to the effective potential as the Wess-Zumino model. 

\section{Supergauge theories}

This section is a brief review of results on supergauge theories. 
Unfortunately, the restricted volume of this review does not allow to
discuss all essential results of last years in this sphere hence we
only give here main ones.

\subsection{General description of supergauge theories}

The starting point of our consideration is an action of ${\cal N}=1$
super-Yang-Mills theory (cf. \cite{GRS}):
\bea
\label{symact}
S_{SYM}=\frac{1}{64g^2}\int d^6 z\, {\rm tr}\, W^{\a}W_{\a},
\eea
where
\bea
\label{w}
W_{\a}=-\bar{D}^2(e^{-gV}D_{\alpha}e^{gV}); V(z)=V^I(z) T^I.
\eea
The $V(z)=V^I(z)T^I$ is a real scalar Lie-algebra-valued superfield.
We can expand the action (\ref{symact}) into power series in coupling
$g$. As a result we get
\bea
\label{gauge1}
S=\frac{1}{16}\int d^8 z\, {\rm tr}\, (V D^{\a}\bar{D}^2 D_{\a}V+\ldots).
\eea
Here dots denote higher orders in $g$.
The action (\ref{symact}) is invariant under gauge transformations
\bea
\label{gtr}
e^{gV}\to e^{-ig\bar{\L}}e^{gV}e^{ig\L}.
\eea
where $\bar{D}_{\ad}\L=0$. The equivalent form of this transformation
\cite{BK0}
is
\bea
\label{gauge2}
\delta V=iL_{gV/2}(\L+\bar{\L}+{\rm coth}L_{gV/2}(\L-\bar{\L})).
\eea
Here $L_{gV}A=[gV,A]$ is a Lie derivative.
It is easy to see that strengths $W_{\a},\bar{W}_{\ad}$ transform covariantly 
under such transformations (in the Abelian case they are invariant).
The leading order in (\ref{gauge2}) is
\bea
\delta V=i(\Lambda-\bar{\Lambda}).
\eea

Since the theory is gauge invariant we must introduce gauge-fixing
functions for quantization. The most natural form of them is (cf. section 4.8
where these gauge-fixing functions were used for calculation of
one-loop effective action in Wess-Zumino model)
\bea
\chi(V)&=&-\frac{1}{4}\bar{D}^2V+f(z)\\
\bar{\chi}(V)&=&-\frac{1}{4}D^2 V+\bar{f}(z).\nonumber
\eea
Here $f(z)$ is an arbitrary chiral superfield.
The variation of these gauge fixing functions under transformations
(\ref{gauge2}) is
\bea
\label{vargau}
\delta\left(
\begin{array}{c}
\chi(V)\\
\bar{\chi}(V)
\end{array}
\right)
=
\left(
\begin{array}{cc}
0&-\frac{1}{4}\bar{D}^2\\
-\frac{1}{4}D^2& 0
\end{array}
\right)
\left(
\begin{array}{c}
\delta V\\
\delta V
\end{array}
\right).
\eea
According to Faddeev-Popov approach we can introduce the ghost action
\bea
S_{GH}={\bf c}'\delta\chi|_{g\Lambda=c,g\bar{\Lambda}=\bar{c}}
\eea
i.e. parameters of the gauge transformation $g\L,g\bar{\L}$ in this case are replaced by ghosts.
Here $\delta\chi\equiv\delta\left(
\begin{array}{c}
\chi(V)\\
\bar{\chi}(V)
\end{array}
\right)
$ from (\ref{vargau}), ${\bf c}'$ is a line $(c' \bar{c}')$ and since 
$\L$ is chiral $c,c'$ are also chiral ones. Here $c,c'$ are chiral
ghosts and $\bar{c},\bar{c}'$ are antichiral ones. As usual, ghosts
are fermions.

Therefore
\bea
S_{GH}=\int d^6 z\, {\rm tr}\, c'\frac{\delta\chi}{\delta V}\delta V+
\int d^6 \bar{z}\, {\rm tr}\, \bar{c}'\frac{\delta\bar{\chi}}{\delta
  V}\delta V,
\eea
where 
\bea
\delta V=L_{gV/2}(c+\bar{c}+{\rm coth}L_{gV/2}(c-\bar{c})).
\eea
Hence, the action of ghosts looks like
\bea
S_{GH}=\int d^8 z {\rm tr}\, (\bar{c}'-c')L_{gV/2}(c+\bar{c}+{\rm
  coth}L_{gV/2} (c-\bar{c})).
\eea
Then, the generating functional for this theory at zero sources  
according to Faddeev-Popov approach looks like
\bea
\label{gf0}
Z[J]|_{J=0}=\int DV D\{c\}e^{i(S_{SYM}+S_{GH})}
\delta_+(\frac{1}{4}\bar{D}^2V-f)\delta_-(\frac{1}{4}D^2V-\bar{f}).
\eea
Here we use the notation $D\{c\}\equiv Dc Dc' D\bar{c} D\bar{c}'$ for an integral over all types of ghosts.
We can average over functions $f$ and $\bar{f}$ with weight
\bea
\exp(\frac{i}{\xi}\int d^8 z (f\bar{f}+b\bar{b})),
\eea
where $\xi$ is a some number (actually, it is a gauge parameter). The $b,\bar{b}$ are
Nielsen-Kallosh ghosts (in this case their contribution to effective
action is a constant, but in background-covariant formulation it is
non-trivial). 
As a result, the (\ref{gf0}) takes the form
\bea
Z[J]_{J=0}=\int DV D\{c\}e^{i(S_{SYM}+S_{GH}+S_{GF})},
\eea 
where
\bea
S_{GF}=\frac{1}{16\xi}\int d^8 z {\rm tr} (\bar{D}^2 V) (D^2 V)
\eea
is a gauge-fixing action \cite{BK0}.

We introduce the total action
\bea
\label{total}
S_{total}=S_{SYM}+S_{GF}+S_{GH}
\eea
and the generating functional
\bea
Z[J,\{\eta\}]&=&\int DV D\{c\}\exp(i(S_{total}+\int d^8 z {\rm tr} JV
+\int d^6 z {\rm tr}(\eta'c'+ \eta c)+\nonumber\\&+&\int d^6\bar {z}
(\bar{\eta}'\bar{c}'+\bar{\eta}\bar{c}) )).
\eea
Here $\{\eta\}$ is the set of all sources: $\eta,\eta',\bar{\eta},\bar{\eta}'$.
To develop a diagram technique we must split action $S_{total}$ into a  sum of the
free (quadratic) part and vertices. It is easy to see \cite{BK0,Kovacz}
that
\bea
\label{expsum}
e^{-gV}D_{\a}e^{gV}=g D_{\a}V-\frac{1}{2}g^2 [V,D_{\a}V]+\frac{1}{6}
g^3[V,[V,D_{\a}V]]+\ldots.
\eea
Therefore (\ref{symact}) looks like
\bea
\label{ssym}
S_{SYM}&=&\int d^8 z\, {\rm tr}\, \Big(\frac{1}{16}V D^{\a} \bar{D}^2 D_{\a}
V+\frac{1}{16}g (\bar{D}^2 D^{\a} V)[V, D_{\a}V]-\nonumber\\&-&
\frac{1}{64}g^2
[V,D^{\a}V]\bar{D}^2[V,D_{\a}V]-\frac{1}{48}g^2 (\bar{D}^2D^{\a}V)
[V,[V,D_{\a}V]]+\ldots
\Big).
\eea
And the ghost action is
\bea
\label{sgh}
S_{GH}&=&\int d^8 z\, {\rm tr}\, \Big(\bar{c}'c-\bar{c}c'+
\frac{1}{2}g(\bar{c}'-c')[V,c+\bar{c}]+\frac{g^2}{12}(c'-\bar{c}')[V,[V,c-\bar{c}]]
\Big)+\nonumber\\
&+&\ldots.
\eea
This expression is sufficient in one- and two-loop calculations.

The quadratic action, with the added gauge-fixing term, is
\bea
S_0=-\frac{1}{2}\int d^8 z \, {\rm tr}\, V\Big(-\frac{1}{8}D^{\alpha}\bar{D}^2D_{\alpha}
+\frac{1}{16\xi}\{D^2,\bar{D}^2\}
\Big)V+\int d^8 z {\rm tr}(\bar{c}'c-\bar{c}c').
\eea
Vertices can be read off from (\ref{ssym},\ref{sgh}).
Propagators look like
\bea
<V^I(z_1)V^J(z_2)>&=&\delta^{IJ}\frac{1}{\Box}\Big(-\frac{1}{8\Box}D^{\a}\bar{D}^2
D_{\a}+\xi\frac{\{D^2,\bar{D}^2\} }{16\Box}
\Big)\delta^8(z_1-z_2);\\
<\bar{c}^{\prime I}(z_1)c^{J}(z_2)>&=&<c^{\prime I}(z_1)\bar{c}^J(z_2)>=-\delta^{IJ}\frac{1}{\Box}
\delta^8(z_1-z_2).\nonumber
\eea
We note that ghosts are fermions hence any ghost loop corresponds to
minus sign. Then, $D$-factors are associated with vertices containing
ghosts just by the same rule as with vertices containing any chiral 
superfields. The algebraic indices are suggested where it is necessary.
We note that if we choose $\xi=1$ (Feynman gauge) the propagator of
gauge superfield takes the simplest form
\bea
\label{pfe}
<V^I(z_1)V^J(z_2)>=\delta^{IJ}\frac{1}{\Box}\delta^8(z_1-z_2).
\eea
Note that its sign is opposite to the sign of propagator of chiral 
superfield (this difference of signs plays an important role for some cancellations of divergences, see f.e. \cite{SGRS}).

If we want to introduce interaction of a chiral superfield with the gauge
one, the quadratic part in $\Phi,\bar{\Phi}$ looks like
\bea
\label{chint}
S=\int d^8 z \bar{\Phi}_i (e^{gV})^i_j\Phi^j,
\eea
if chiral superfield $\Phi_i$ is transformed under some representation
of the gauge group (i.e. it is an isospinor) or
\bea
S=\int d^8 z {\rm tr} (\bar{\Phi}e^{gV}\Phi e^{-gV}),
\eea
if chiral superfield $\Phi=\Phi^a T^a$ is Lie-algebra-valued. Note
that under gauge transformations (\ref{gtr}) the chiral superfield is
transformed as
\bea
\Phi\to e^{-ig\L}\Phi
\eea
for isospinor chiral superfield and as
\bea
\Phi\to e^{-ig\L}\Phi e^{ig\L}
\eea
for Lie-algebra-valued chiral superfield. Note that $\L,\bar{\L}$ are 
Lie-algebra-valued parameters in both cases.
The vertices can be easily obtained by expanding into power series
expressions corresponding to interaction: in first case
\bea 
\int d^8 z [\bar{\Phi}_i (e^{gV})^i_j\Phi^j-\Phi_i\bar{\Phi}^i]=
\int d^8 z \sum_{n=1}^{\infty}\frac{1}{n!}\bar{\Phi}_i {(gV^n)}^i_j\Phi^j,
\eea
and in second one --
\bea
{\rm tr}\, \int d^8 z ( (\bar{\Phi}e^{gV}\Phi e^{-gV})-\bar{\Phi}\Phi)=
{\rm tr}\, \int d^8 z (g\bar{\Phi}[V,\Phi]+\frac{g^2}{2}
\bar{\Phi}[V,[V,\Phi]]+\ldots).
\eea

The diagram technique derived now is very suitable for calculations in
sector of background $\Phi,\bar{\Phi}$ only and for calculation of 
divergences.

Let us consider an example.
The $N=2$ super-Yang-Mills theory with a chiral matter is described by the 
action (see f.e. \cite{Gio,Kovacz}):
\bea
\label{n2sym}
S&=&\frac{1}{64g^2}\int d^6 z\, {\rm tr}\, W^{\a} W_{\a} +{\rm tr} \int d^8 z
\bar{\Phi}e^{gV}\Phi e^{-gV}+\nonumber\\&+&
\sum_{i=1}^n\Big[ig(\int d^6 z Q_i \Phi
\tilde{Q}_i+h.c.)+\int d^8 z \bar{\tilde{Q}}_i e^{-gV}\tilde{Q}_i
+\int d^8 z \bar{Q}_i e^{gV} Q_i
\Big].
\eea
Here $\Phi$ is Lie-algebra-valued chiral superfield, and $Q^i,\tilde{Q}_i$
are chiral superfields transformed under mutually conjugated
representations of Lie algebra. They are often called matter hypermultiplets. In the particular case where we have only one superfield $Q$ and one superfield $\tilde{Q}$, we get the $N=4$ super-Yang-Mills theory (the extended supersymmetry manifests itself through the so-called hidden SUSY transformations relating the chiral and real superfields between themselves, see f.e. \cite{SGRS}).

We note that there is another equivalent formulation of the $N=4$ super-Yang-Mills theory in which instead of $\Phi$, $Q^i$ and $\tilde{Q}_i$, one has three Lie-algebra-valued chiral superfields $\Phi_i$, with $i=1\ldots 3$, with the following action (cf. \cite{SGRS}):
\bea
\label{n4sym}
S&=&\frac{1}{64g^2}\int d^6 z\, {\rm tr}\, W^{\a} W_{\a} +{\rm tr} \int d^8 z
\sum_{i=1}^3\bar{\Phi}_ie^{gV}\Phi_i e^{-gV}+\nonumber\\&+&
\frac{g}{3!}\epsilon_{ijk}{\rm tr}\Big[(\int d^6 z \Phi_i\Phi_j\Phi_k+h.c.)
\Big].
\eea

Let us consider the structure of one-loop divergences in the theory (\ref{n2sym}). For
simplicity we choose Feynman gauge $\xi=1$ in which the 
propagator has the most
simple structure (\ref{pfe}), therefore all tadpole diagrams given in
\cite{Kovacz} evidently vanish.

First we consider contributions to wave function renormalization of
$\Phi$ field 

\vspace*{3mm}

\Lengthunit=1cm
\hspace*{3cm}\GRAPH(hsize=3)
{\Linewidth{1.2pt}\Circle(2)\Linewidth{.5pt}\mov(-1,0){\lin(-.5,0)}\ind(30,-15){Fig. 12}
\mov(.9,0){\lin(.5,0)}}
\hspace*{3cm}\GRAPH(hsize=3)
{\halfcirc(2)[u]\halfwavecirc(2)[d]\mov(-1,0){\lin(-.5,0)}
\mov(.9,0){\lin(.5,0)}
}

\vspace*{3mm}

Here the thin line is propagator of the Lie-algebra valued chiral superfield $\Phi$, the thick one -- 
of hypermultiplets 
$Q,\tilde{Q}$, the wavy one -- of real superfield $v$, the dashed one 
-- of ghosts.

One-loop divergent contributions from these supergraphs are
respectively (see f.e. \cite{Gio})
\bea
2\sum_M\int\frac{d^4k}{(2\pi)^4}\frac{1}{k^2(k+p)^2}\Phi^a\bar{\Phi}^b {\rm
  tr}_M
(T^a T^b)
\eea
and
\bea
-2\int\frac{d^4k}{(2\pi)^4}\frac{1}{k^2(k+p)^2}\Phi^a\bar{\Phi}^b
{\rm tr}_{ad} (T^a T^c T^d){\rm tr}_{ad} (T^b T^c T^d).
\eea
Here $tr_M$ denotes trace in representation under which
hypermultiplets are transformed. Coefficient 2 is caused by presence
of
two chiral hypermultiplets $Q$ and $\tilde{Q}$ in the first term, and by two different contraction in the second term.
We see that if $\sum_M{\rm tr}_M(T^a T^b)=
{\rm tr}_{ad} (T^a T^c T^d){\rm tr}_{ad} (T^b T^c T^d)$
there is no divergent contributions to wave function renormalization. In other words, the divergences are cancelled when the hypermultiplets are transformed under the specific representations of the gauge groups.

Contributions to the hypermultiplet wave function renormalization look like 

\vspace*{3mm}

\Lengthunit=1cm
\hspace*{3cm}\GRAPH(hsize=3)
{\Linewidth{.5pt}\halfcirc(2)[u]\Linewidth{1.2pt}\halfcirc(2)[d]\ind(30,-15){Fig. 13}
\mov(-1.1,0){\lin(-.5,0)}\mov(.9,0){\lin(.5,0)}\Linewidth{.5pt}
}
\hspace*{3cm}\GRAPH(hsize=3)
{\halfwavecirc(2)[u]\Linewidth{1.2pt}\halfcirc(2)[d]
\mov(-1.1,0){\lin(-.5,0)}\mov(.9,0){\lin(.5,0)}\Linewidth{.5pt}
}

\vspace*{3mm}

The one-loop divergent contributions from these supergraphs are
respectively \cite{Gio}
\bea
\int\frac{d^4k}{(2\pi)^4}\frac{1}{k^2(k+p)^2}\bar{Q}_i Q_l
(T^a)^{ij} (T^a)^{jl}
\eea
and
\bea
-\int\frac{d^4k}{(2\pi)^4}\frac{1}{k^2(k+p)^2}\bar{Q}_i Q_l
(T^a)^{ij} (T^a)^{jl}.
\eea
These corrections evidently cancel each other, hence the hypermultiplet wave function renormalization is trivial.
In both these cases cancellation is caused by the difference in signs of
propagators of gauge superfield and chiral superfields (both Lie-algebra valued and the hypermultiplet ones). The same is correct for $\tilde{Q}$ and $\bar{\tilde{Q}}$.

Then, let us turn to the gauge sector of the theory.
One-loop contributions to wave function renormalization for gauge
superfields are represented by the following supergraphs: 

\vspace*{2mm}

\Lengthunit=1cm
\hspace*{3cm}\Linewidth{.5pt}
\GRAPH(hsize=3)
{\Circle(2)\mov(-1,0){\wavelin(-.5,0)}
\mov(.9,0){\wavelin(.5,0)}}
\hspace*{6cm}\GRAPH(hsize=3)
{\wavecirc(2)\mov(-1,0){\wavelin(-.5,0)}
\mov(.9,0){\wavelin(.5,0)}
}

\vspace*{3mm}

\Lengthunit=1cm
\hspace*{3cm}\GRAPH(hsize=3)
{\Linewidth{1.2pt}\Circle(2)\Linewidth{.5pt}\mov(-1,0){\wavelin(-.5,0)}\ind(30,-15){Fig. 14}
\mov(.9,0){\wavelin(.5,0)}}
\hspace*{6cm}\GRAPH(hsize=3)
{\dashcirc(2)\mov(-1,0){\wavelin(-.5,0)}\mov(1,0){\wavelin(.5,0)}}

\vspace*{2mm}

To make a brief description of the implications of the extended supersymmetry, we discuss now only the mutual cancellation of the quadratic divergences. Nevertheless, one can show \cite{SGRS}, that the mutual cancellation of the logarithmic divergences within the background field method occurs for the same relations for the gauge group generators.
So, the quadratically divergent contributions of these four supergraphs are respectively given by
(see f.e. \cite{SGRS,Gio})
\bea
&&\int\frac{d^4k}{(2\pi)^4}\frac{1}{k^2} V^a V^b
{\rm tr}_{ad} (T^a T^c T^d) {\rm tr}_{ad} (T^b T^c T^d);\nonumber\\
&&\int\frac{d^4k}{(2\pi)^4}\frac{1}{k^2} V^a V^b{\rm tr}_{ad} 
(T^a T^c T^d) {\rm tr}_{ad} (T^b T^c T^d);\nonumber\\
2\sum_M&&\int\frac{d^4k}{(2\pi)^4}\frac{1}{k^2} V^a V^b{\rm tr}_M 
(T^a T^b);\nonumber\\
-4&&\int\frac{d^4k}{(2\pi)^4}\frac{1}{k^2} V^a V^b{\rm tr}_{ad} (T^a T^c T^d) {\rm tr}_{ad} (T^b T^c T^d).
\eea
Hence the same condition for cancellation of divergences for the two-point function of chiral superfields, that is, 
\bea
\label{regen}
{\rm tr}_{ad} (T^a T^c T^d) {\rm tr}_{ad} (T^b T^c T^d)={\rm
  tr}_M (T^a T^b),
\eea  
implies cancellation of the quadratic divergences for the two-point function of the gauge superfield, and there is
no quadratically divergent correction to $<VV>$-propagator. As we have already mentioned, the same relation (\ref{regen}) implies the cancellation of the logarithmically divergent contributions. This condition is satisfied for an appropriate gauge group, in particular, $SU(N)$. We note that the tadpole graphs (for the massless chiral superfield) give contributions identically equal to zero. This mechanism explicating the vanishing of divergences is discussed, f.e. in
\cite{SGRS}, \cite{HST}, where it is shown to be caused by the $N=2$ 
superconformal symmetry.
The most important example of such theories is $N=4$ super-Yang-Mills
theory (\ref{n4sym}) which is equivalent to (\ref{n2sym}) with only one pair of hypermultiplets $Q,\tilde{Q}$ which are transformed under the adjoint representation of the Lie
algebra. This theory is known to be finite (see f.e. \cite{Kovacz,SGRS}).

\subsection{Background field method}

The approach to study of the gauge theories described up to this place is very useful for
consideration of divergences and corrections in the sector of chiral
superfields $\Phi,Q,\tilde{Q}$ only. To study contributions depending
on gauge superfields we must develop a method allowing to preserve
manifest gauge invariance at any step, whereas earlier we have obtained
contributions in terms of superfield $V$ which are in general case not
gauge invariant. Therefore we must introduce an approach in which
external lines are background strengths $W_{\a},\bar{W}_{\ad}$ and
their {\bf gauge covariant} derivatives.
This method was developed in \cite{GRS} (see also \cite{GZ} and
references therein), here we give its description.

The problem of calculation of the effective action in the super-Yang-Mills theory
described by the action (\ref{symact}) is much more complicated than in
other field theories. The main difficulties are the following ones. First, the
nonpolynomiality of the action (\ref{symact}) implies in the infinite
number of vertices which seems to result in infinite number of types
of the divergent quantum corrections (such a situation is treated 
in common cases as the non-renormalizability of the theory), second, it is easy
to see that the common
background-quantum splitting $V\to V_0+v$ where $V_0$ is a background
field and $v$ is a quantum field cannot provide manifest gauge covariance of
the quantum corrections. Really, because of the nonpolynomiality of $W^{\a}$
(\ref{w}), to get a covariant quantum correction (which by definition
must be expressed in terms of the strengths $W_{\a},\bar{W}_{\ad}$
which are the only objects transforming in a covariant way under the gauge
transformations unlike of the superfield $V$ itself)
we need to summarize an infinite number of supergraphs with different 
numbers of external $V_0$ legs (to the best of our knowledge, the explicit form 
of this summation never has been performed in the literature).
The background field method provides an effective solution for both
these problems.

The starting point of the method under discussion is a {\bf nonlinear}
background-quantum splitting for the superfield $V$ defining the
action (\ref{symact}) \cite{GRS}:
\bea
\label{split}
e^{gV}\to e^{g\O}e^{gv}e^{g\bar{\O}}. 
\eea
Here the $v$ is a quantum field, the $\O,\bar{\O}$ are the background
superfields (they are not necessary chiral/antichiral ones, the only restriction is $e^{g\O}e^{g\bar{\O}}=e^{gV}$, with $V$ is a background gauge field, so, in principle one can have f.e. $\O=\bar{\O}$, or f.e. $\O=0$).  
After such a background-quantum splitting (unfortunately, the complete proof of this statement is very tedious), the classical action (\ref{symact})
takes the form:
\bea
\label{actqua}
S=-\frac{1}{16g^2}{\rm tr}
\int d^8 z (e^{-gv}{\D}^{\a}e^{gv})\bar{\D}^2 (e^{-gv}{\D}_{\a}e^{gv}).
\eea
In this expression (which describes a theory of real scalar superfield
$v$ coupled to background superfields $\O,\bar{\O}$) 
the $W^{\alpha}$ is a background strength (we kept it in the explicit form), and the ${\D}^{\a},\bar{\D}^{\ad}$ are the
background covariant derivatives defined by the expressions \cite{SGRS,GRS}:
\bea
\label{der}
{\D}^{\a}&=&e^{-g\O}D^{\a}e^{g\O},\nonumber\\
\bar{\D}^{\ad}&=&e^{g\bar{\O}}\bar{D}^{\ad}e^{-g\bar{\O}}.
\eea
Our further aim consists in study of the action (\ref{actqua}). To do
it let us first describe the properties of the background covariant
derivatives given by (\ref{der}).
The $D^{\a},\bar{D}^{\ad}$ in (\ref{der}) act on all
on the right. Then, as well as the covariant derivatives in usual
differential geometry, the
background covariant derivatives
${\D}^{\a},\bar{\D}^{\ad}$ can be represented in the following
``standard'' form
\bea
{\D}^{\a}=D^{\a}-i\Gamma^{\a},\quad \,
\bar{\D}^{\ad}=\bar{D}^{\ad}-i\bar{\Gamma}^{\ad},
\eea
where
\bea
\Gamma^{\a}=ie^{-g\O}(D^{\a}e^{g\O}), \quad \, 
\bar{\Gamma}^{\ad}=ie^{g\bar{\O}}(\bar{D}^{\ad}e^{-g\bar{\O}})
\eea
are the superfield connections.

Let us study the (anti)commutation relations for the
${\D}^{\a},\bar{\D}^{\ad}$. 
We start with imposing the following constraint
\bea
\label{constr}
{\D}_{\a\ad}=-\frac{i}{2}\{{\D}_{\a},\bar{\D}_{\ad}\}
\eea
which represent itself as a background covariant analogue of the common
anticommutation relation $\partial_{\a\ad}=-
\frac{i}{2}\{D_{\a},\bar{D}_{\ad}\}$. 
Then, it is easy to verify straightforwardly the following definition
of the background strength $W_{\a}$ (cf. \cite{GS}):
\bea
\label{defw}
W_{\a}=[\bar{\D}^{\ad},\{\bar{\D}_{\ad},{\D}_{\a}\}]=
2i[\bar{\D}^{\ad},{\D}_{\a\ad}].
\eea
Really (it is straightforward to check this relation for $\Omega=V$, and $\bar{\Omega}=0$; also, one should remind that derivatives act on all on the right, not only on the adjacent field), after we substitute expressions (\ref{der}) for the
background-covariant derivatives to (\ref{defw}), and take into account
that $e^{g\O}e^{g\bar{\O}}=e^{gV}$ in the case of absence of the
quantum field $v$ (cf. (\ref{split})), we get just the definition
(\ref{w}). 

The relation (\ref{defw}) is crucial. Its treating consists of the fact 
that the background-covariant
space-time derivative ${\D}_{\a\ad}$ has non-zero commutators with
spinor background-covariant derivatives unlike of the common covariant
derivatives, and, moreover, that 
the background strengths could arise during the
D-algebra transformations.
In particular, the identity (\ref{defw}) implies in the following expressions
(cf. \cite{GZ}):
\bea
\label{dd2}
[{\D}_{\a},\bar{\D}^2]=-W_{\a}+4i\bar{\D}^{\ad}{\D}_{\a\ad}=%\nonumber\\&=&
W_{\a}+4i{\D}_{\a\ad}\bar{\D}^{\ad}.
\eea

We also need in definition of the (background) covariantly chiral
superfields to describe coupling of the gauge superfields to the matter.
By definition, the superfield $\Phi$ is referred as the (background)
covariantly chiral one if it satisfies the condition
$\bar{\D}_{\ad}\Phi=0$. It is easy to see that
the $\Phi$ is related
to the common chiral field $\Phi_0$ as
\bea
\label{chidef}
\Phi=e^{g\bar{\O}}\Phi_0.
\eea
Really, condition of chirality $\bar{D}_{\ad}\Phi_0=0$ implies in
\bea
e^{g\bar{\O}}\bar{D}_{\ad}e^{-g\bar{\O}}e^{g\bar{\O}}\Phi_0=0.
\eea
Using the definitions (\ref{der},\ref{chidef}) we arrive just to the
condition $\bar{\D}_{\ad}\Phi=0$.

Now we are in position to develop the perturbative approach for the
theory 
with action 
(\ref{actqua}). First, we note that 
this theory possesses the symmetry with respect to the
following gauge transformations:
\bea
e^{gv}\to e^{ig\bar{\L}}e^{gv}e^{-ig\L},
\eea
where $\L$ is a {\bf covariantly} chiral parameter, i.e. it satisfies
the condition $\bar{\D}_{\ad}\L=0$ (similarly, ${\D}_{\a}\bar{\L}=0$).
Therefore we need to introduce a gauge fixing.
The most natural background covariant gauge fixing term looks like
\bea
\label{gfcv}
S_{gf}=-\frac{1}{32}{\rm tr}\int d^8 z v\{{\D}^2,\bar{\D}^2\}v,
\eea
which is a covariant generalization of the common gauge fixing term in
the Feynman gauge.
Summarizing the (\ref{actqua}) and (\ref{gfcv}), we get the following action of the quantum $v$ field:
\bea
\label{ac2q}
S_t=S+S_{gf}=-\frac{1}{2}{\rm tr}\int d^8 z v \Box v+S_{int},
\eea
where the $S_{int}$ in the expression above is an interaction part.
In principle, one can fix other gauges (by putting the factor
$\xi^{-1}$ in $S_{gf}$), however, even the problem of finding the
propagator for $v$ field appears to be very complicated for an arbitrary gauge (up to this time
nobody found this propagator in a closed form). Hence the Feynman
gauge is the most convenient one.
The propagator of the $v$ field in the Feynman gauge is
\bea
\label{p1}
<v^I(z_1)v^J(z_2)>=\Box^{-1}\delta^{IJ}\delta^8(z_1-z_2)
\eea
(for the sake of the uniqueness of the consideration we relate all
terms involving $W_{\a},\bar{W}_{\ad}$ to the interaction part).

The interaction part of the total action of $v$
involves both covariant generalizations of the ``common'' vertices 
and the new vertices involving the background strengths 
$W_{\a},\bar{W}_{\ad}$ manifestly (these expression are the analogues to those ones used in \cite{GRS,GZ}, where, however, other conventions are used):
\bea
\label{s1}
S_{int}&=&{\rm tr}\int d^8
z\Big(\frac{1}{2}v(W^{\a}{\D}_{\a}+\bar{W}_{\ad}{\D}^{\ad})v+
\frac{1}{16}g 
v\{ {\D}^{\a}v,\bar{\D}^2{\D}_{\a}v \}-\nonumber\\ &-&
\frac{1}{48}g[[{\D}^{\a}v,v],v]W_{\a}-
\frac{1}{64}g^2[v,{\D}^{\a}v]\bar{\D}^2[v,{\D}_{\a}v]-
\nonumber\\&-&
\frac{1}{48}g^2\bar{\D}^2{\D}^{\a}v[v,[v,{\D}_{\a}v]]-\frac{1}{192}g^2
[[[{\D}^{\a}v,v],v],v]W_{\a}
\Big)+\ldots.
\eea
We note that the terms proportional to $W_{\alpha}$ in the triple- and higher-order vertices in $v$ arose from the anticommutation of the gauge covariant derivatives in (\ref{actqua}) between themselves, with use of the expression (\ref{defw}).

We also can prove the following important relation (cf. \cite{GRS}):
\bea
\label{boxpm}
& &{\D}^{\alpha}\bar{\D}^2{\D}_{\alpha}-\frac{1}{2}\{{\D}^2,\bar{\D}^2\}-W^{\alpha}{\D}_{\alpha}-\frac{1}{2}({\D}^{\alpha}W_{\alpha})=
-8(\Box+W^{\alpha}{\D}_{\alpha}+\bar{W}_{\alpha}\bar{\D}^{\ad});\nonumber\\
& &{\D}^2\bar{\D}^2{\D}^2=
16(\Box+\bar{W}_{\ad}\bar{\D}^{\ad}+
\frac{1}{2}(\bar{\D}_{\ad}\bar{W}^{\ad})){\D}^2
\equiv 16\Box_-{\D}^2;\nonumber\\
& &\bar{\D}^2{\D}^2\bar{\D}^2=
16(\Box+W^{\a}{\D}_{\a}+
\frac{1}{2}({\D}^{\a}{W}_{\a}))\bar{\D}^2
\equiv 16\Box_- \bar{\D}^2.
\eea

Because of the gauge symmetry, one needs to introduce ghosts. Since
the form of the
gauge transformations and gauge fixing action is very similar to the
``common'' superfield case with only difference consisting in covariant
chirality instead of common one, the action of ghosts in this case
also will be analogous to the common case with only difference
consisting in the fact of the 
{\bf covariant} chirality of the ghosts $c,c'$ or {\bf covariant} 
antichirality of the ghosts $\bar{c},\bar{c}'$:
\bea
S_{gh}={\rm tr}\int d^8 z (\bar{c}'-c')L_{gv/2}(c+\bar{c}+
{\rm cth}L_{gv/2}(c-\bar{c}))
\eea
which after expansion in the power series gives:
\bea
\label{s2}
S_{gh}={\rm tr}\int d^8 z(\bar{c}'c+c'\bar{c}+\frac{1}{2}(\bar{c}'-c')[v,c+\bar{c}]
+\frac{1}{12}(c'-\bar{c}')[v,[v,c-\bar{c}]]+\ldots ).
\eea
The action of the matter after the background-quantum splitting of the
gauge fields takes the form
\bea
\label{s3}
S_m=\int d^8 z \bar{\Phi}e^{gv}\Phi,
\eea
where $\Phi,\bar{\Phi}$ are the {\bf covariantly} chiral and
antichiral superfields.
The propagators of covariantly chiral superfields and ghosts are
\bea
\label{p2}
<\bar{\phi}\phi>=-\Box^{-1}_+\delta^8(z_1-z_2),\quad\,
<\bar{c}^{\prime I}c^J>=<\bar{c}^Ic^{\prime J}>=-\delta^{IJ}\Box^{-1}_+\delta^8(z_1-z_2),
\eea
with ${\cal D}^2$, $\bar{\cal D}^2$ factors are associated with the vertices in the same manner as $D^2$ and $\bar{D}^2$ for the propagators of the usual chiral superfields. 
We note that such propagators can be expanded into power series in 
$W_{\a},\bar{W}_{\ad}$, see (\ref{boxpm}).
The expressions (\ref{s1}, \ref{s2}, \ref{s3}, \ref{p1}, \ref{p2}) can be 
used for constructing the supergraphs. Some examples of the application of
the method for the supergraph calculations (in the pure SYM theory without matter) can be found in \cite{GZ}.

Now let us give a comparative characteristics for the two methods of
superfield calculations -- the background field method and the
``common'' method. 

The crucial difference is the following one. In the framework of the
``common'' method the quadratic and linear divergences could arise for
the supergraph with arbitrary any number of the external legs. 
Really, it is easy to show that the superficial degree of
divergence for the ``common'' supergraph is
\bea
\omega=2-\hf N_D - E_{\phi},
\eea 
where $N_D$ is a number of spinor supercovariant derivatives
associated to the external legs, $E_{\phi}$ is a number of external
chiral (antichiral) legs. We note that the quadratic and/or linear
divergences are possible for any number of external $v$ legs.  
At the same time, in the framework of the background field method by
the construction of the background-quantum splitting the only external
lines are the background strengths  and/or their covariant
derivatives. The superficial degree of divergence in this case 
can be shown to have
the form
\bea
\omega=2-\frac{3}{2}N_W-\frac{1}{2}N_D+\e-E_{\phi},
\eea
where $N_W$ is a number of the background strength legs, $\e=1$ for
the chiral (antichiral) contribution (whose only possible structure is
$\int d^6 z W^2$, $\int d^6\bar{z}\bar{W}^2$), otherwise $\e=0$; the $N_D$ is
the number of derivatives acting on external $W_{\a},\bar{W}_{\ad}$
legs (the derivatives presenting in each $W_{\a},\bar{W}_{\ad}$ by
definition must not be taken into account!). We see that in the
framework of this approach only logarithmic overall divergences are
possible, they arise for the terms proportional to $W^2$, with the
quadratic and linear subdivergences (which are important if we make a
noncommutative generalization) can arise only in subgraphs which are
not associated to the external $W_{\a},\bar{W}_{\ad}$ legs. 

From formal viewpoint such a difference has the following
origin. Really, the superfield strength by construction contains three
spinor derivatives, hence arising of any superfield strength leg in
the framework of ``common'' formalism
decreases by three the number of the $D$-factors which could be
converted to momenta (we note that use of the background field method
allows to sum automatically the infinite number of ``common'' graphs
and forbids existence of the supergraphs with the superficial
quadratic or linear divergence),
as a result, the convergence of the supergraph is improved. It is
essential in the noncommutative field theory since it means that the
only problems could be generated by subgraphs only as each
contribution to the effective action is in worst case only
logarithmically divergent. In part, it means that there is no
contradiction between the result of Bichl et al. \cite{Bichl}
according to which the different contributions to the one-loop
two-point function of $v$ field in the $U(1)$ NC SYM theory 
possess quadratic divergences (only their sum is free of dangerous
UV/IR mixing) and the result of Zanon et al. \cite{Zan}
according to which all one-loop contributions to the effective action
in the same theory are free of the dangerous UV/IR mixing (notice that
the calculations in the last paper were carried out in the framework
of the background field method); we should mention also use of the
background field method in the papers \cite{Zan} devoted to study of
the one-loop effective action in the noncommutative super-Yang-Mills
theories. We note that the background field
method allows to preserve the gauge covariance at all steps of calculations.

However, the background field method has one disadvantage -- presence
of nontrivial commutators of the background covariant derivatives makes all
calculations extremely difficult from the technical viewpoint even in the case of the absence of the external chiral matter fields. 

\subsection{Proper-time method for the supergauge theories}

Nevertheless, we should note that there exists a powerful tool allowing for the application of the background field formalism, at least in the one-loop order, in a manner based on the proper-time method, for the supersymmetric gauge theories. This method has been developed in \cite{McA}. To illustrate it, let us consider the following one-loop effective action of the gauge theories:
\bea
\label{foll}
\Gamma^{(1)}=\frac{i}{2}\ln\det({\cal D}^a{\cal D}_a+W^{\alpha}{\cal D}_{\alpha}+\bar{W}_{\dot{\alpha}}\bar{{\cal D}}^{\dot{\alpha}}+
|\Phi|^2).
\eea
Actually, this expression emerges when one introduces to the super-Yang-Mills theory, whose action is given by the sum of (\ref{ac2q}) and (\ref{s1}), the coupling to the external chiral matter.
It has been discussed in \cite{BKT}. Here we illustrate how the result for it can be obtained with use of the method developed in \cite{McA}.

We start with the definition based on the well-known zeta function regularization procedure (see f.e. \cite{Eliz} for a review on this methodology):
\bea
\ln\det\Delta=-\zeta^{\prime}(0),
\eea
where the zeta function corresponding to the operator $\hat{K}=e^{t\Delta}$, where $t$ is a proper time (see \cite{BSD} for the general review on the proper time methodology) is defined as
\bea
\zeta(s)=\frac{1}{\Gamma(s)}\int_0^{\infty}dt t^{s-1}K(t),
\eea
where $K(t)$ is a functional trace of $\hat{K}$.
Alternatively, one can use straightforwardly Schwinger-De Witt representation which allows to write
\bea
\ln\det\Delta=\int_0^{\infty}\frac{dt}{t}{\rm tr}e^{t\Delta}\equiv\int_0^{\infty}\frac{dt}{t}K(t).
\eea
Here, the kernel $K(t)$ corresponding to the operator $\hat{K}$, as usual, looks like
\bea
K(t)=\int d^8z\lim_{z\to z'}e^{t\Delta}\delta^8(z-z').
\eea
We note that here we follow the methodology developed in \cite{McA} and use the notations from that paper.
In our case, when the effective action is given by (\ref{foll}), the $K(t)$ reads as
\bea
\label{thekern}
K(t)=\int d^8z\lim_{z\to z'}e^{t({\cal D}^a{\cal D}_a+W^{\alpha}{\cal D}_{\alpha}+\bar{W}_{\dot{\alpha}}\bar{{\cal D}}^{\dot{\alpha}}+|\Phi|^2)}\delta^8(z-z').
\eea
Then, we use the Fourier representation of the complete delta function $\delta^8(z-z')$, both for its bosonic and fermionic parts:
\bea
\delta^8(z-z')=\int\frac{d^4k}{(2\pi)^4}e^{ik(x-x')}\int d^4\epsilon e^{i\epsilon^{\alpha}(\theta_{\alpha}-\theta^{\prime}_{\alpha})}e^{i\bar{\epsilon}_{\dot{\alpha}}(\bar{\theta}^{\dot{\alpha}}-
\bar{\theta}^{\prime\dot{\alpha}})}.
\eea
To simplify the calculations, we factorize out the chiral matter superfields (which are consisted as constants within this calculations), writing $K(t)=e^{t{|\Phi|^2}}\tilde{K}(t)$. 
Then, the key point of the methodology proposed in \cite{McA} follows: we take into account that the $K(t)$ (\ref{thekern}) is an operator multiplied by the delta function which further must act on some other function. Since the delta function is expanded into the Fourier series through the expansion above, one can verify that when the operator whose kernel is given by (\ref{thekern}) acts on an arbitrary function, the following objects will emerge:
\bea
X_a={\cal D}_a+ik_a,\, X_{\alpha}={\cal D}_{\alpha}+i\epsilon_{\alpha}, \, \bar{X}_{\dot{\alpha}}=\bar{{\cal D}}_{\dot{\alpha}}+i\epsilon_{\dot{\alpha}}.
\eea
Indeed, for example, the purely spatial derivatives part in $K(t)$ (remind that $K(t)$ is a kernel of the operator), acting on an arbitrary function $F(x')$ (in this case -- the function of the bosonic coordinates only), will evidently produce the result
\bea
e^{t{\cal D}^a{\cal D}_a}e^{ik(x-x')}F(x')|_{x=x'}=e^{tX^aX_a}F(x).
\eea
Thus, acting of each derivative on the object $e^{ik(x-x')}F(x')$ or the similar one involving the Grassmannian delta function will augment this derivative with a term equal to the corresponding moment multiplied by $i$.  
Repeating this arguments for the spinor derivatives, we will see that the kernel of the operator which would act on the arbitrary function of all superspace coordinates is
\bea
\label{ktil}
\tilde{K}(t)=\int d^8z\int\frac{d^4k}{(2\pi)^4}\int d^4\epsilon\lim_{z\to z'}e^{t(X^aX_a+W^{\alpha}X_{\alpha}+\bar{W}_{\dot{\alpha}}\bar{X}^{\dot{\alpha}})}.
\eea
Now, let us differentiate this kernel with respect to the proper time $t$.
It is straightforward to see that the derivative of $\tilde{K}(t)$ (\ref{ktil}) looks like
\bea
\label{hk}
\frac{d\tilde{K}(t)}{dt}=K^a_a+W^{\alpha}K_{\alpha}(t)+\bar{W}_{\dot{\alpha}}\bar{K}^{\dot{\alpha}},
\eea
where
\bea
\label{moments}
K_{A_1\ldots A_n}=\int\frac{d^4k}{(2\pi)^4}\int d^4\epsilon X_{A_1}\ldots X_{A_n}e^{t\tilde{\Delta}}
\eea
have the role of $n$-th momenta of the generalized Gaussian (cf. \cite{McA}),
and $\tilde{\Delta}=X^aX_a+W^{\alpha}X_{\alpha}+\bar{W}_{\dot{\alpha}}\bar{X}^{\dot{\alpha}}$. The expression (\ref{hk}) will be treated by us as a main equation of this study, similarly to the equation (\ref{etu}) for the Wess-Zumino model, aimed for calculating the corresponding heat kernel.

Then, following \cite{McA}, we use some identities representing themselves as integrals from total derivatives over the whole space of the momenta conjugated to superspace coordinates:
\bea
\label{idip}
&&\int\frac{d^4k}{(2\pi)^4}\int d^4\epsilon\frac{\partial}{\partial \epsilon_{\alpha}}(X_{A_1}\ldots X_{A_n}e^{t\tilde{\Delta}})=0;\nonumber\\
&&\int\frac{d^4k}{(2\pi)^4}\int d^4\epsilon\frac{\partial}{\partial \bar{\epsilon}_{\dot{\alpha}}}(X_{A_1}\ldots X_{A_n}e^{t\tilde{\Delta}})=0;\nonumber\\
&&\int\frac{d^4k}{(2\pi)^4}\int d^4\epsilon\frac{\partial}{\partial k_a}(X_{A_1}\ldots X_{A_n}e^{t\tilde{\Delta}})=0.
\eea
Actually, we will need to consider the terms with only one $X_{\alpha}$ (or $X_{\dot{\alpha}}$) and no more than two $X_a$.
So, these identities imply in need to consider the following expression:
\bea
\label{adjs}
\frac{\partial}{\partial \epsilon_{\alpha}}e^{t\tilde{\Delta}}=-it\sum\limits_{n=0}^{\infty}\frac{t^n}{(n+1)!}ad^{(n)}(\tilde{\Delta})(W^{\alpha})
e^{t\tilde{\Delta}},
\eea
where $ad(\tilde{\Delta})(W_{\beta})=[\tilde{\Delta},W_{\beta}]$,  $ad^{(2)}(\tilde{\Delta})(W_{\beta})=[\tilde{\Delta},[\tilde{\Delta},W_{\beta}]]$, etc. The expression for the derivative with respect to $\bar{\epsilon}_{\dot{\alpha}}$ is a straightforward analogue of this one, and that one involving the derivative with respect to the $k_a$ will be introduced further. 

If we want to restrict ourselves to the expressions involving, at most, first derivatives of any superfield strengths (note that just these expressions, being projected to the components, give different degrees of the stress tensor $F_{ab}$, whereas the higher derivatives of $W_{\alpha}$, $\bar{W}_{\dot{\alpha}}$ imply in the terms involving the derivatives of $F_{ab}$), we must consider the only nontrivial commutators:
\bea
[X_a,X_b]&=&-\frac{1}{2}(\bar{{\cal D}}\bar{\sigma}_{ab}\bar{W}-{\cal D}\sigma_{ab}W)\equiv -\frac{1}{2}(\bar{M}_{ab}-M_{ab});\quad\,
[X_a,X_{\alpha}]=i(\sigma_a)_{\alpha\dot{\alpha}}\bar{W}^{\dot{\alpha}};\nonumber\\
\{X_{\alpha},W_{\beta}\}&=&({\cal D}_{\alpha}W_{\beta})=N_{\alpha\beta};\quad\,
\{\bar{X}_{\dot{\alpha}},\bar{W}_{\dot{\beta}}\}=(\bar{{\cal D}}_{\dot{\alpha}}\bar{W}_{\dot{\beta}})=N_{\dot{\alpha}\dot{\beta}}.
\eea 
Now, it is the crucial moment that the background fields belong to Abelian phase (cf. \cite{BKT}). So, one can write $ad(\tilde{\Delta})(W_{\beta})=[\tilde{\Delta},W_{\beta}]=W^{\alpha}N_{\alpha\beta}$. Repeating the calculation of the commutator $n$ times, we find that 
\bea
\label{nadj}
ad^{(n)}(\tilde{\Delta})(W_{\beta})=W^{\alpha}(N^n)_{\alpha\beta}.
\eea

Then, we make use of the identities (\ref{idip}). The first one looks like
\bea
\label{idip1}
0=\int\frac{d^4k}{(2\pi)^4}\frac{\partial}{\partial\epsilon_{\beta}}(X_{\alpha}e^{t\tilde{\Delta}})=i\delta_{\alpha}^{\beta}
\tilde{K}(t)-\int\frac{d^4k}{(2\pi)^4}X_{\alpha}\frac{\partial}{\partial\epsilon_{\beta}}e^{t\tilde{\Delta}}.
\eea
We employ the expression (\ref{adjs}) to find the derivative with respect to $\epsilon_{\beta}$, and, afterwards (\ref{nadj}), to find $n$-th adjoint of $W_{\alpha}$. As a consequence, we find that 
\bea
\label{adjs1}
\frac{\partial}{\partial \epsilon_{\beta}}e^{t\tilde{\Delta}}=-it\sum\limits_{n=0}^{\infty}\frac{t^n}{(n+1)!}W^{\gamma}(N^n)_{\gamma}^{\phantom{\gamma}\beta}=-iW^{\gamma}(\frac{e^{tN}-1}{N})_{\gamma}^{\phantom{\gamma}\beta}.
\eea
We substitute these expressions to (\ref{idip1}). Then, it remains to carry out the anticommutation between $X_{\alpha}$ and $W^{\gamma}$, that is, $\{X_{\alpha},W^{\gamma}\}=N_{\alpha}^{\phantom{\alpha}\gamma}$. We arrive at
\bea
0=\delta_{\alpha}^{\beta}\tilde{K}(t)+N_{\alpha}^{\phantom{\alpha}\gamma}(\frac{e^{tN}-1}{N})_{\gamma}^{\phantom{\gamma}\beta}
\tilde{K}(t)-W^{\gamma}(\frac{e^{tN}-1}{N})_{\gamma}^{\phantom{\gamma}\beta}K_{\alpha}(t).
\eea
Moving the term with $W^{\gamma}K_{\alpha}(t)$ to the left-hand side of the expression, multiplying by the matrix inverse to $(\frac{e^{tN}-1}{N})$,  and calculating the trace (with imposing the restriction that $N_{\alpha}^{\alpha}=0$), we arrive at
\bea
W^{\alpha}K_{\alpha}(t)&=&{\rm tr}(\frac{N}{e^{tN}-1})\tilde{K}(t).
\eea
Proceeding in a similar way for the identity conjugated to (\ref{idip1}), we find
\bea
\bar{W}_{\dot{\alpha}}\bar{K}^{\dot{\alpha}}(t)&=&{\rm tr}(\frac{\bar{N}}{e^{t\bar{N}}-1})\tilde{K}(t).
\eea
We note that this term was absent in \cite{McA} where only the contribution dependent on $W_{\alpha}$ but not $\bar{W}_{\dot{\alpha}}$ was considered.

Finally, one can identically repeat the calculation of \cite{McA}, to obtain the $K_{ab}(t)$. One starts with the identity
\bea
\label{idip3}
0=\int\frac{d^4k}{(2\pi)^4}\frac{\partial}{\partial k_b}(X_a e^{t\tilde{\Delta}})=i\delta_{ab}\tilde{K}(t)+\int\frac{d^4k}{(2\pi)^4}
X_a\frac{\partial}{\partial k_b} e^{t\tilde{\Delta}},
\eea
and., similarly to the calculations above, one finds 
\bea
\frac{\partial}{\partial k_b} e^{t\tilde{\Delta}}=\sum\limits_{n=0}^{\infty}\frac{t^n}{(n+1)!}ad^{(n)}(X\cdot X)(X_b)=2itB_{bc}(t)X_c, 
\eea
where
\bea
B_{bc}=\left(\frac{e^{-t(\bar{M}-M)}-1}{-t(\bar{M}-M)}\right)_{bc}.
\eea
Therefore, restoring the $K_{ac}$ with use of its definitions from (\ref{moments}), one finds that the identity (\ref{idip3}) leads to
\bea
0=i\delta_{ab}\tilde{K}(t)+2itB_{bc}(t)K_{ac}(t)
\eea
(the term involving $K_a(t)$ will be irrelevant just as in \cite{McA}), so, one has
\bea
K_{ab}(t)=-\frac{1}{2t}(B^{-1})_{ba}(t)\tilde{K}(t)=\frac{1}{2}\left(\frac{\bar{M}-M}{e^{-t(\bar{M}-M)}-1}\right)_{ba}\tilde{K}(t).
\eea

For the sake of simplicity, we suggested within these calculations that $N_{\alpha}^{\alpha}=0$, as well as $N_{\dot{\alpha}}^{\dot{\alpha}}=0$. These identities can be imposed since these expressions do not contribute to the degrees of freedom of the stress tensor $F_{ab}$; actually, for the Abelian background superfield they are just equivalent to the Bianchi identities $D^{\alpha}W_{\alpha}=0$, $\bar{D}_{\dot{\alpha}}\bar{W}^{\dot{\alpha}}=0$.
Substituting these expressions to (\ref{hk}), we arrive at
\bea
\label{hk1}
\frac{d\tilde{K}(t)}{dt}=\left[\frac{1}{2}\left(\frac{\bar{M}-M}{e^{-t(\bar{M}-M)}-1}\right)_a^a+
{\rm tr}\left(\frac{N}{e^{tN}-1}\right)+{\rm tr}\left(\frac{\bar{N}}{e^{t\bar{N}}-1}\right)\right]\tilde{K}(t).
\eea
The solution of this equation is
\bea
\tilde{K}(t)=\frac{W^2\bar{W}^2}{16\pi^2}{\rm det}\left(\frac{e^{-tN}-1}{N}\right){\rm det}\left(\frac{e^{t\bar{N}}-1}{\bar{N}}\right){\rm det}\left(\frac{1-e^{-t(\bar{M}-M)}}{M-\bar{M}}\right)^{-1/2}.
\eea
The overall constant factor is fixed from the fact that, just as in \cite{McA}, the solution of this equation at $M=\bar{M}=N=0$ must be $K(t)=\frac{W^2\bar{W}^2}{16\pi^2}$, that is, the expression which yields the well-known result for the four-point function of $W_{\alpha}$ and $\bar{W}_{\dot{\alpha}}$ but not on their derivatives \cite{SGRS,bug1}.
On the base of this kernel, one can write down the following one-loop effective action:
\bea
\Gamma^{(1)}&=&\int d^8z\int\frac{dt}{t}e^{-t|\Phi|^2}\frac{W^2\bar{W}^2}{16\pi^2}{\rm det}\left(\frac{e^{tN}-1}{N}\right){\rm det}\left(\frac{e^{t\bar{N}}-1}{\bar{N}}\right)\times\nonumber\\&\times&
{\rm det}\left(\frac{1-e^{-t(\bar{M}-M)}}{M-\bar{M}}\right)^{-1/2}.
\eea
 We note that due to the identity \cite{McA}: 
\bea
{\rm det}(\frac{1-e^{-2tF}}{F})^{-1/2}=\frac{1}{4t^2}{\rm det}(\frac{tF}{\sinh tF})^{1/2},
\eea
this expression can be rewritten in an alternative form:
\bea
\Gamma^{(1)}&=&\int d^8z\int\frac{dt}{t^3}e^{-t|\Phi|^2}\frac{W^2\bar{W}^2}{16\pi^2}{\rm det}\left(\frac{e^{tN}-1}{N}\right){\rm det}\left(\frac{e^{t\bar{N}}-1}{\bar{N}}\right)\times\nonumber\\&\times&
{\rm det}\left(\frac{t(\bar{M}-M)}{\sinh t(\bar{M}-M)}\right)^{1/2}.
\eea
This is just the result obtained in \cite{BKT} for the $SU(2)$ gauge group broken to $U(1)$. We note that if one will suggest that the derivatives of the strengths are zero, i.e. $M=\bar{M}=N=0$, one will have ${\rm det}(\frac{e^{tN}-1}{N})|_{N=0}= t^2$ (remind that $N$ is $2\times 2$ matrix since the spinor indices take values 1 and 2), one recovers the well-known result \cite{SGRS,bug1}:
\bea
\Gamma^{(1)}=\frac{1}{16\pi^2}\int d^8z\frac{W^2\bar{W}^2}{(\Phi\bar{\Phi})^2}.
\eea
 
This study can be easily generalized for the $SU(n)$ gauge group broken to its maximal Abelian subgroup (Abelian torus) $U(1)^{n-1}$. Indeed, following \cite{bug1}, we can write the one-loop effective action in this case as 
\bea
\label{foll1}
\Gamma^{(1)}=\frac{i}{2}\sum\limits_{k<l}\ln\det(-{\cal D}^a{\cal D}_a-(W^{\alpha}_k-W^{\alpha}_l){\cal D}_{\alpha}-
(\bar{W}_{\dot{\alpha}k}-\bar{W}_{\dot{\alpha}l})\bar{{\cal D}}^{\dot{\alpha}}-
|\Phi_k-\Phi_l|^2).
\eea
Here $W^{\alpha}_k-W^{\alpha}_l$ etc. are the superfield roots of the $su(n)$ algebra (see the detailed discussion of the algebra roots for $su(n)$ and structure of this expression in \cite{bug1}).
Explicitly repeating the calculation above for (\ref{foll1}), we arrive at
\bea
\Gamma^{(1)}&=&\sum\limits_{k<l}\int d^8z\int\frac{dt}{t}e^{-t|\Phi_{kl}|^2}\frac{W_{kl}^2\bar{W}_{kl}^2}{16\pi^2}{\rm det}\left(\frac{e^{tN_{kl}}-1}{N_{kl}}\right)
{\rm det}\left(\frac{e^{t\bar{N_{kl}}}-1}{\bar{N_{kl}}}\right)\times\nonumber\\&\times&
{\rm det}\left(\frac{1-e^{-t(\bar{M}_{kl}-M_{kl})}}{M_{kl}-\bar{M}_{kl}}\right)^{-1/2},
\eea
where $W^{\alpha}_{kl}=W^{\alpha}_k-W^{\alpha}_l$, $\Phi_{kl}=\Phi_k-\Phi_l$ etc. We close this section with the conclusion that the one-loop effective action of the super-Yang-Mills theory, in the approximation of constant background fields $F_{ab}$ and $\phi$ (that is, the principal components of $W_{\alpha}$ and $\Phi$), has been successfully calculated for the gauge group $SU(n)$, with an arbitrary $n$. In principle, the calculation for other gauge groups will not essentially differ.

\chapter{Supersymmetry breaking}

Throughout these lecture notes, we have discussed the superfield methodology for studying the supersymmetric field theories. However, it is known that, actually the supersymmetry is broken at the observed scales of energy. There are two ways to describe breaking of any symmetry including the supersymmetry, those are -- explicit and spontaneous breaking. The detailed discussions of the supersymmetry breaking are presented in famous review papers, such as \cite{Luty,Intri}, and, of course, in classical supesymmetry textbooks like \cite{SGRS,West,WB}. Here we do not suggest to give a detailed discussion of the supersymmetry breaking, following a more modest aim -- to present a brief description of some ways to describe these phenomena in terms of the superfield methodology (here we concentrate on the theories defined in the four-dimensional space-time).

\section{Explicit supersymmetry breaking}

Within this methodology we suggest that the classical action of the theory involves a small additive term whose presence breaks the supersymmetry. Dealing within the superfield approach, at least formally, we can introduce such an additive term through an introduction of a special extra ``superfield'' with a broken component structure, with a corresponding non-trivial component of this superfield is a constant. Actually such a superfield, called a {\bf spurion}, represents itself as a some generalization of the coupling constant (actually, this is a soft supersymmetry breaking since the divergences continue to be logarithmic; we also note that additive supersymmetry-breaking terms are small). There is a natural restriction on the structure of such terms: it is easy to see that only the logarithmic divergences can emerge in the quantum corrections in Wess-Zumino and super-Yang-Mills theories (while for the Wess-Zumino model it follows directly form the study of the superficial degree of divergence, for the super-Yang-Mills theory the corresponding divergences can be shown to be forbidden by the gauge symmetry). Therefore, we suppose the spurion terms to be in the form which does not introduce quadratic divergences. In particular, it is necessary for this that the constant defining the spurion to have a non-negative mass dimension.

The key idea of the supersymmetry breaking is the following one \cite{Luty}: it is easy to see that the above-mentioned constant cannot be the lower ($\theta,\bar{\theta}$-independent) component of the superfield since its variation under the supersymmetry transformations vanishes, and the supersymmetry is not broken. Therefore, there are three possible monomial forms of the spurion (we note that the constant fields $\mu$, $\bar{\mu}$, $\nu$ must be scalar, to maintain Lorentz invariance), cf. \cite{SGRS}:
\bea
\label{spur}
\chi=\mu^2\theta^2,\quad\, \bar{\chi}=\bar{\mu}^2\bar{\theta}^2, \quad\, U=\nu^2\theta^2\bar{\theta}^2.
\eea
Two first types of spurions can emerge in chiral and antichiral sectors of the action respectively, and the third one -- only in a general sector. The corresponding additional terms in the action of the Wess-Zumino model (and models including it as an ingredient) will be respectively
\bea
\Delta S_1=\int d^6 z \chi \Phi^2+h.c.= \int d^4x (\mu^2\varphi^2+h.c.), \quad\, \Delta S_2=\int d^8 z U\Phi\bar{\Phi}= \int d^4x \nu^2\varphi\bar{\varphi},
\eea
where $\varphi$ is a scalar component of the chiral superfield $\Phi$.
A brief inspection of the component structure shows that these additive terms, although do not generate new types of counterterms, destroy the equality of masses of bosonic and fermonic fields which is known to be characteristic, thus, the supersymmetry in these cases is broken. We note that the mass dimension of $\mu^2$ is 2, and of $\nu^2$ is zero, i.e. in both cases it is non-negative, and the renormalizability of the theory is not jeopardized. 

It is easy to verify that if the mass dimension of the spurion superfield is non-negative, no new divergent terms can emerge. Indeed, in this case only the vertices with degrees of superfields and/or derivatives no higher than those ones present in the initial action of the theory are present, therefore there is no possibility for non-renormalizable interactions. 

However, in principle, there are non-renormalizable spurion couplings, like f.e. \\ $\int d^8 z U D^{\alpha}\Phi D_{\alpha} \Phi$ \cite{SGRS}. In this case the spurion has a negative dimension. From the viewpoint of the Feynman supergraphs, each spurion vertex in  this case involves two extra spinor derivatives which increases degree of divergence of the corresponding supergraph. Such a manner of supersymmetry breaking is evidently not soft.

We close this section with a mentioning that the soft supersymmetry breaking has wide phenomenological and cosmological applications (in the context of the dark matter problem), see f.e. \cite{Nath}.

\section{Spontaneous supersymmetry breaking}

Within the study of this concept we suggest that the classical action of the theory is supersymmetric, thus, it is naturally formulated in terms of the superfields. However, the vacuum is not supersymmetric, so, the supersymmetry is broken in a spontaneous manner. This way of supersymmetry breaking seems to be more delicate being a principal subject of many studies.

The key observation which gave rise to many discussions of the spontaneous supersymmetry breaking is the following one (it is explained f.e. \cite{SGRS,WB} and many other textbooks): in any supersymmetric theory, the Hamiltonian, that is, the energy operator, can be written in terms of the supersymmetry generators as
\bea
H\equiv P^0=\frac{1}{4}(\sigma^0)^{\alpha\dot{\beta}}\{Q_{\alpha},\bar{Q}_{\dot{\beta}}\}=\frac{1}{4}(\{Q_1,\bar{Q}_{\dot{1}}\}+
\{Q_2,\bar{Q}_{\dot{2}}\}),
\eea
so, the Hamiltonian of the supersymmetric theories is positive, with the supersymmetry generators play the role of the creation and annihilation operators. As a result, action of the annihilation operator on the vacuum states $|0>$ should give zero, $\bar{Q}_{\dot{\alpha}}|0>=0$. Similarly, one has $Q_{\alpha}|0>=0$. Therefore, the variation of the vacuum under the supersymmetry transformations is zero, $(\epsilon^{\alpha}Q_{\alpha}+\bar{\epsilon}_{\dot{\alpha}}\bar{Q}^{\alpha})|0>=0$. It is also clear that the minimum of the Hamiltonian of the field theory is a minimum of its potential since the kinetic energy is essentially non-negative. Therefore, we have a natural criteria: if the vacuum (that is, the lower value of the potential) of the supersymmetric theory is zero, the supersymmetry is not spontaneously broken. 

Now, let us see examples. First, one of the simplest examples of the models involving the spontaneous supersymmetry breaking is the super-QED extended by the additive, gauge invariant Fayet-Iliopoulos term \cite{FI}:
\bea
S_{FI}=-\xi\int d^8 z V.
\eea
This term is linear in the superfield. Moreover, in components it has the simple form $\xi \int d^4 x {\cal D}(x)$, where ${\cal D}$ is higher component of the gauge superfield (\ref{realfield}). It is clear that this term breaks the parity $V\to -V$ (and, consequently, ${\cal D}\to -{\cal D}$). Since this field enters the action of the super-QED (cf. (\ref{sqedcomp})) only through the term
\bea
S_{{\cal D}}=\int d^4x(\frac{1}{2}{\cal D}^2-\xi {\cal D}),
\eea
it is clear that the equations of motion for ${\cal D}$, besides of the usual solution ${\cal D}=0$, yield also the solution ${\cal D}=\xi$ which does not possess the symmetry ${\cal D}\to -{\cal D}$, so, this symmetry is broken. The supersymmetry is also evidently broken since the supersymmetry transformations near this vacuum, for some components of the superfield $V$ (in particular, for the vector field $A_a$) will be $\xi$-independent, but for some (in particular, for the photino $\lambda_{\alpha}$) -- $\xi$-dependent. Introducing of interaction of the gauge superfield with the chiral matter will not essentially modify the situation \cite{SGRS}.

Other important example of the models displaying the spontaneous supersymmetry breaking are the O'Raifeartaigh models \cite{Raif}. In models of this class, one has a set of chiral superfields with usual kinetic terms for them, but with more sophisticated (although renormalizable) potential. The simplest example of such models is (see \cite{Raif,Shih}):
\bea
S=\int d^8z (\bar{X}X+\bar{\phi}_1\phi_1+\bar{\phi}_2\phi_2)+[\int d^6z (m\phi_1\phi_2+hX\phi^2_1+fX)+h.c.],
\eea
where $X$ and $\phi_{1,2}$ are chiral superfields. 
If one would obtain the equations of motion for this theory and then put $D^2\phi_{1,2}\simeq 0$, $D^2X\simeq 0$, in order to consider only slowly varying superfields (that is, the k\"{a}hlerian potential), one arrive at the following system of equations:
\bea
&&f+h\phi^2_1=0;\nonumber\\
&& m\phi_2+hX\phi_1=0;\nonumber\\
&& m\phi_1=0.
\eea
Such a system is evidently inconsistent, therefore, the vacuum for this theory simply does not exist, and the supersymmetry is broken. It was shown in \cite{Shih} that a whole class of superfield theories possesses a similar behaviour.

We close this section with recommending of the brilliant review \cite{Intri} for a further reading on this subject.

\chapter{Conclusions}
We considered superfield method in supersymmetric field theory, both in three- and four-dimensional space-times. This
method allows to preserve manifest supersymmetry at any step of
calculations, and the calculations within it turns to be much more
compact than within the framework of the component approach.

We studied several examples of superfield theories and presented in details
quantum calculations for them. In three dimensions, the examples were a scalar superfield theory and supersymmetric gauge theories. In four dimensions, we considered the Wess-Zumino model,
the general chiral superfield model, higher-derivative chiral superfield model and $N=1$
super-Yang-Mills theory with chiral matter. In all these theories we
developed supergraph technique, studied general form of superfield
effective action and calculated low-energy leading contributions
to effective action. It is natural to expect that development of
superfield quantum calculations in other  superfield models formulated in
terms of $N=1$ superfields including different supergravity models is
in principle no more difficult. We also discussed the background field
method in the supergrauge theories.

Let us briefly discuss other applications and generalizations of the 
superfield approach in the quantum field theory.
In the last years the following most important ways of applications of
superfield supersymmetry were developed.

1. {\bf Studying of theories with extended supersymmetry}. It is known that
theories with extended supersymmetry possess better renormalization
properties, f.e. as $N=1$ super-Yang-Mills theory is renormalizable,
the $N=4$ super-Yang-Mills theory is finite. 
The most important examples of theories with extended supersymmetry
are $N=2$ and $N=4$ super-Yang-Mills theories. During last years
numerous results in studying of these theories were obtained (see f.e.
\cite{BKT,bug1,bug2} and references therein).

It is natural to expect that the most adequate method for
consideration of such theories must possess manifest $N=2$
supersymmetry. Such a method is a harmonic superspace approach
developed in \cite{GIKOS1,GIKOS2}. 
This method is based on consideration of superfields being the
functions of bosonic space-time coordinates $x^a$, two sets of
Grassmann coordinates $\theta^{i\a},\bar{\theta}^{i\ad}$ with 
$i=1,2$ and spherical harmonics $u^{\pm i}$. Introducing of analytic
superfield \cite{GIKOS1} allows to develop formulation in terms of
unconstrained $N=2$ superfield and to avoid arising of component
fields with higher spins. The formulations of $N=2$ and
$N=4$ super-Yang-Mills theories in harmonic superspace is given
in \cite{GIKOS1,GIKOS2}, background field method for these theories is
developed in \cite{bbiko,bbko,bko}, and examples of quantum
calculations are given in \cite{bug1,bug2,bs,km,kt}. The most important results
presented in these papers are 
calculation of holomorphic action of $N=2$ matter hypermultiplets
in external $N=2$ gauge superfield, calculation of one-loop
nonholomorphic effective potential in $N=4$ super-Yang-Mills theory
and proof of its absence in higher loops, calculation of one-loop
effective action for $N=4$ super-Yang-Mills theory for constant
strength tensor $F_{ab}$, calculation of superconformal anomaly of
$N=2$ matter interacting with $N=2$ supergravity. During last years
other important results achieved in these investigations were calculation of
contributions depending on derivatives of $N=2$ super-Yang-Mills
strength ${\cal W}$ \cite{bpt,k1},  calculation of contributions depending
on background matter hypermultiplet fields \cite{bip}, and
development of quantum approach for 
$N=3$ super-Yang-Mills theory \cite{bsiz} (the manifestly $N=3$ supersymmetric
approach based on the harmonic superspace technique for the 
$N=3$ supersymmetric theory was introduced in the paper \cite{Delduc}). Also, it is necessary to mention the intensive studies of the three-dimensional extended supersymmetric theories, especially $N=6$ and $N=8$ Chern-Simons theories (see f.e. \cite{Aharony}), where also the harmonic superspace approach has been successfully developed and applied (see f.e. \cite{Buchd3}).

2. {\bf Noncommutative supersymmetric theories}. Noncommutative
theories have been intensively studied during last years. Concept of
space-time noncommutativity was introduced to quantum field theory due
to some consequences of D-branes theory \cite{SW2} and to
consideration of quantum theories on very small distances where
quantum fluctuations of geometry are essential. Consideration of
supersymmetric noncommutative theories is quite natural. During last
years some interesting results in studying of noncommutative
supersymmetric theories were obtained but they were mostly based on
component approach. The first superfield results were
calculation of leading ($\sim F^4$) correction to one-loop effective
action for $N=4$ super-Yang-Mills theory \cite{Zan} and formulation of
supergraph technique for noncommutative Wess-Zumino model \cite{Popp}.
Further, the quantum superfield studies for the Wess-Zumino model
\cite{ourwz} and
four-dimensional superfield QED \cite{ourqed} and super-Yang-Mills
theories \cite{oursym} were
carried out. There are also a list of examples of calculations in three-dimensional supersymmetric field theories \cite{ours,sigcomp}, 
and some of them have been considered in this book. These theory were shown to be consistent in the sense of
absence of the nonintegrable UV/IR infrared divergences. Thus, we can
speak about construction of the consistent noncommutative
generalizations of the supersymmetric theories of electromagnetic,
strong and weak interactions. 
Therefore, the next most important problem could consist in
development of the noncommutative supersymmetric generalization for
the last fundamental interaction -- the gravitational one. However,
this problem is certainly extremely difficult (see discussion of the
problem f.e. in \cite{Chams,Cardella}). 

3. {\bf Noncommutative superspace.} One more approach in the
superfield quantum theory is based on use of the noncommutative
superspace \cite{Sei}. Within it, the fermionic superspace coordinates form the
Clifford algebra instead of the Grassmann algebra, which, in part,
leads to the modified construction of the Moyal product. In the
framework of this approach, the generalizations of the Wess-Zumino
(see f.e. \cite{PenGri} and reference therein), gauge  
(see f.e. \cite{PenRom} and reference therein) and general chiral
superfield models \cite{az} were studied. A three-dimensional version of the noncommutative superspace \cite{3dspace} has been presented in this work.

Then, there are a lot of applications of superfields approach to
problems of supersymmetric quantum field theory (f.e. to studying of
AdS/CFT correspondence which was carried out mostly on base of
component approach), and of course consideration of many problems
originated from superstrings and branes theory.

As a final conclusion, we can suppose that superfield approach in
quantum field theory is a very perspective one, and there are a lot of
ways for its development and more applications. 

\newpage

{\bf Acknowledgements.} Author is grateful to Profs. E. A. Asano, L. C. T. Brito, I. L. Buchbinder, M. Cvetic, A. F. Ferrari, H. O. Girotti, M. Gomes, S. M. Kuzenko, J. R. Nascimento, A. A. Ribeiro, V. O. Rivelles,    A. J. da Silva,  E. O. Silva,  for fruitful collaboration and interesting discussions. The work has been partially supported by CNPq.

\end{document}